\documentclass[preprint,12pt]{elsarticle}

\usepackage{amssymb}
\usepackage{mathbbol}

\usepackage{amsmath,amsthm,amsfonts,amscd,amsmath,amsfonts} 
\usepackage{upgreek}
\usepackage{mdframed}
\usepackage{multicol}
\usepackage{graphicx}
\usepackage{nomencl}
\usepackage[nottoc]{tocbibind}
\usepackage{setspace}
\usepackage{subfig}
\usepackage{verbatim}   	
\usepackage{makeidx}    	
\usepackage{color}
\usepackage{xcolor}
\usepackage{float}
\usepackage{enumerate}
\usepackage{url}	
\usepackage{algpseudocode}
\usepackage{tcolorbox}
\usepackage{xcolor}
\usepackage{multirow}
\usepackage{array}
\usepackage[linesnumbered,ruled,vlined]{algorithm2e}
\usepackage{ulem}
\usepackage[pdftex,bookmarks]{hyperref}
\usepackage{multirow}
\usepackage{bbm, dsfont}
\usepackage[percent]{overpic}
\usepackage[table]{xcolor}
\usepackage{booktabs} 
\usepackage{algpseudocode} 

\newcommand{\blue}[1]{\textcolor{black}{#1}}

\newcommand{\bs}{\boldsymbol}

\usepackage[margin=1.2in]{geometry} 

\usepackage{etoolbox} 
\makeatletter
\patchcmd{\@startsection}
  {\@afterindenttrue}
  {\@afterindentfalse}
  {}{}
\makeatother

\journal{Biomechanics and modeling in mechanobiology}

\begin{document}

\begin{frontmatter}



\title{
\textbf{Dynamic Image-Informed Selection of Biomechanical Tumor Growth Models}
}


\author[inst1]{Abdullah Al Noman}
\author[inst1]{Pratyush Kumar Singh}
\author[inst2]{David A Hormuth, II}
\author[inst1]{\\ Danial Faghihi\corref{cor1}}

\affiliation[inst1]{organization={Department of Mechanical and Aerospace Engineering, \\University at Buffalo},
            city={Buffalo},
            state={NY},
            country={USA}}
            
\affiliation[inst2]{organization={
Oden Institute for Computational Engineering and Sciences, Livestrong Cancer Institutes, \\
University of Texas at Austin},
            city={Austin},
            state={TX},
            country={USA}}

\cortext[cor1]{Corresponding Author, \texttt{danialfa@buffalo.edu} (D. Faghihi) \\
Submitted to \textit{Biomechanics and modeling in mechanobiology}}            

\begin{abstract}
Glioblastoma progression is strongly influenced by evolving mechanical interactions between the tumor and surrounding brain tissue. However, the extent to which finite-deformation mechanics and constitutive assumptions improve \blue{subject-specific} prediction as tumor burden evolves remains unclear.
%
We introduce a sequential Bayesian inference and dynamic model selection framework that assimilates longitudinal murine magnetic resonance imaging (MRI) data to calibrate spatially varying tumor diffusivity, proliferation rate, and tissue stiffness in biomechanical tumor growth models. Competing formulations were compared \blue{at each imaging time}, including reaction--diffusion without mechanics and reaction--diffusion coupled to linear elasticity or hyperelastic mechanics, using posterior model plausibility to \blue{adapt model choice for individualized one-scan-ahead prediction as new MRI scans are acquired}.
%
Across the studied animals, mechanically coupled models were consistently
more plausible than the uncoupled reaction-–diffusion model, and the evolution of model plausibility indicated an increasing role of mass effect and stress-mediated feedback of tumor growth during progression. 
While linear and hyperelastic coupled tumor growth models often produced similar tumor morphology, they yield \blue{distinct} stress, deformation, and inferred stiffness fields, \blue{with the hyperelastic formulation often receiving higher posterior plausibility at later imaging times}.
%
These results indicate that, \blue{within the present longitudinal murine dataset, mechanical coupling is favored for image-informed glioma growth prediction and that constitutive assumptions should be evaluated sequentially for each subject rather than fixed a priori}.

\end{abstract}


\end{frontmatter}



\section{Introduction}
\label{sec:introduction}
Glioblastoma (GBM), the most aggressive primary brain tumor, exhibits irregular
and strongly subject-specific growth patterns driven by heterogeneous proliferation
and invasion. A growing body of work in mechanobiology and biomechanics indicates
that these patterns are governed by mechanical interaction with the confined brain and an evolving microenvironmental
stiffness \cite{lucci2022coupling,ghahramani2026biomechanical,mpekris2015stress,goriely2015mechanics}.
As GBM expands within the confined cranial cavity, solid stresses develop due to internal
growth heterogeneity and external constraint from surrounding tissue. These stresses
have been shown to modulate proliferation, compress vasculature, and promote invasive
phenotypes through mechanotransduction \cite{helmlinger1997solid,stylianopoulos2013coevolution,jain2014role,mascheroni2016predicting}.
Capturing this coupled tumor-tissue interaction is therefore central to predictive,
biomechanical models of GBM progression and to mechanistic interpretation
of growth patterns observed in longitudinal imaging \cite{lucci2022coupling,ghahramani2026biomechanical}.

Many current continuum tumor growth models for predicting GBMs rely \blue{on} reaction--diffusion (RD) formulations \cite{swanson2000}
that represent proliferation and invasion but do not explicitly account for mass
effect. Mechanically coupled extensions have incorporated deformation by coupling
RD dynamics to small-strain linear elasticity mechanics and calibrating model
parameters from imaging data \cite{tuncc2021modeling,hormuth2017mechanically,hogea2007modeling, lima2016selection}.
More recent biomechanics models have introduced more comprehensive
descriptions, including multiphase models coupled with finite-deformation elasticity, to better
represent soft-tissue mechanics and quantify growth-induced stress and deformation
\cite{faghihi2020coupled,wong2014tumor,lucci2022coupling}.
Despite recent progress in image-driven GBM modeling, a key scientific gap remains is the choice of constitutive assumptions that are typically imposed \textit{a priori}. Most studies calibrate a single chosen formulation and assess performance primarily through morphology-based agreement with the imaging data, without dynamically updating the underlying mechanical description. 
\blue{In particular, existing Bayesian calibration studies have primarily focused on estimating parameters within a prescribed tumor growth model, whereas the present question is whether the biomechanical model itself remains plausible as subject-specific longitudinal data are assimilated. Thus, the gap addressed here is not Bayesian calibration alone, but sequential, subject-specific comparison of competing biomechanical models using posterior model plausibility.}
Given the subject-specific and evolving nature of GBM growth, it remains unclear \blue{how strongly longitudinal imaging data support mechanical coupling, and whether linear elastic or hyperelastic mechanical assumptions are more plausible for a given subject as tumor burden evolves}. These limitations motivate a dynamic model selection framework capable of choosing between competing biomechanical \blue{models} as longitudinal data accumulate.
\blue{Here, the dynamic model selection refers to sequential updating of posterior model plausibility as new longitudinal imaging data are assimilated, not to automatic discovery of new model forms. At each assimilation time, candidate competing models specified \textit{a priori} are recalibrated using the accumulated MRI data from a given animal, their posterior model probabilities are recomputed, and the most plausible model can be used for individualized one-scan-ahead prediction.}

In this work, we address these gaps by developing an image-driven framework for sequential Bayesian inference and dynamic selection of biomechanical glioma growth models. \blue{The present study builds on prior murine glioma imaging studies in which MRI-constrained reaction--diffusion models were personalized to individual animals and subsequently extended through linear-elastic mechanical coupling to improve prediction under mass-effect constraints \cite{hormuth2015predicting,hormuth2017}. It also extends our earlier Bayesian inference framework for estimating spatially varying tumor-growth parameters from MRI data \cite{liang2023bayesian}. While that work focused on calibrating heterogeneous parameter fields within a prescribed tumor-growth model, the present study develops a scalable formulation for computing posterior model plausibility and uses it to compare competing biomechanical model hypotheses sequentially as new longitudinal MRI data are assimilated.}
Longitudinal MRI data from preclinical murine glioma subjects are used to update subject-specific fields of tumor diffusivity, proliferation rate, and tissue stiffness, while also quantifying uncertainty in the inferred high-dimensional parameter space. Instead of prescribing a fixed biomechanical formulation, we compare three candidate models: reaction--diffusion without mechanics, reaction--diffusion coupled with linear elasticity, and reaction--diffusion coupled with hyperelastic mechanics. 
\blue{We focus on these parsimonious reaction--diffusion-based formulations because the available MRI-derived tumor volume fraction data directly constrain proliferation, invasion, and mechanically mediated deformation more strongly than additional biological compartments such as necrosis or apoptosis.} Posterior model plausibility is recomputed at each assimilation time, enabling subject-specific model selection for one-scan-ahead prediction as new MRI scans become available.
\blue{Using longitudinal murine glioma data, we show that mechanically coupled models are favored over the uncoupled RD model in the present dataset, while the relative support for linear elastic versus hyperelastic mechanics evolves across animals and imaging times. These results position posterior model plausibility as a mechanics-informed criterion for adapting biomechanical tumor growth predictions from longitudinal imaging data, beyond morphology agreement alone.}

The remainder of the paper is organized as follows. Section~\ref{sec:methods}
describes the longitudinal murine glioma MRI data, the biomechanical tumor
growth model, the
MRI-informed sequential Bayesian inference framework, and the dynamic model
selection theory based on posterior plausibility, including scalable solution and computation of model evidence. 
Section~\ref{sec:results} reports
the prior specification and hyper-parameter selection, presents sequential
subject-specific calibration and sequential predictions, and 
time-resolved model plausibility across the competing
biomechanical computational models. 
Finally, Section~\ref{sec:conclusions} discusses the
biomechanical implications of the results, limitations, and directions for
future extensions toward more comprehensive mechanics-informed tumor growth
modeling.

\section{Methods}\label{sec:methods}

\subsection{Imaging data in murine glioma}\label{sec:mri}
Longitudinal magnetic resonance imaging (MRI) data of glioma growth in female Wistar rats \blue{were obtained from previously published murine glioma imaging and modeling studies \cite{hormuth2017,hormuth2015predicting,hormuth2016dis,hormuth2014comparison}}. Tumors were established via stereotaxic injection of $10^5$ C6 glioma cells into the neocortex. \blue{MRI data were acquired at multiple time points after tumor implantation using a 9.4T preclinical MRI system with a 38-mm-diameter Litz quadrature coil (Doty Scientific).} 
For each scan, the imaged volume was $32,\mathrm{mm} \times 32,\mathrm{mm} \times 16,\mathrm{mm}$, discretized on a $128 \times 128 \times 16$ image grid, corresponding to voxels of size $0.25,\mathrm{mm} \times 0.25,\mathrm{mm} \times 1,\mathrm{mm}$. All longitudinal scans for a given animal were rigidly registered to the earliest scan using mutual information to maintain spatial correspondence across time.

Multiple MRI modalities were used to construct the computational model and define the data for sequential Bayesian inference. $T_2$-weighted images \blue{(fast spin echo sequence with $TR=2000$ ms and $TE=30$ ms)} were used to define the brain geometry and computational domain, while contrast-enhanced $T_1$-weighted images \blue{(spoiled gradient echo sequence with $TR=45$ ms, $TE=1.4$ ms, and $\alpha=20^\circ$)} were used to delineate tumor burden. Diffusion-weighted MRI \blue{($TR=2000$ ms, $TE=30$ ms, $b$-values of 0, 300, 500, 700, 800, and 1100 s/mm$^2$, $\Delta=25$ ms, and $\delta=2$ ms)} was used to derive apparent diffusion coefficient (ADC) maps \cite{padhani2009}. \blue{The primary imaging outputs used in the present analysis were segmented tumor regions, ADC maps, contrast-enhanced $T_1$-weighted images, and $T_2$-weighted images. Additional MRI-derived quantities, including $T_1$ maps and dynamic contrast-enhanced MRI maps, were acquired in the original studies but were not used here. Full details of the animal experiments, image acquisition, preprocessing, and original image-based modeling analyses are provided in \cite{hormuth2017,hormuth2015predicting,hormuth2016dis}.}
ADC maps were used as an indirect measure of local tumor cellularity. \blue{ADC quantifies the diffusion of free water molecules within tissue and is restricted by microstructural barriers such as cell membranes. Consequently, an inverse relationship between ADC and cellular density has been reported in tumor imaging studies \cite{sugahara1999usefulness,anderson2000effects,barnes2015correlation,atuegwu2011integration}, supporting the use of ADC as a first-order surrogate for local tumor cell burden within the segmented tumor region.} Following established protocols \cite{hormuth2017,hormuth2015predicting}, ADC measurements were converted into spatially resolved estimates of tumor cell density. The resulting cell-density estimates were normalized by a voxel-dependent carrying capacity, taken as the local upper bound on tumor cell concentration. This normalization yields a dimensionless tumor volume fraction field at multiple time points, which serves as the primary data quantity for calibrating spatiotemporal tumor growth and assessing the evolving mechanical interaction between the growing tumor and host tissue.

In the present work, longitudinal imaging data from four animals are considered. The imaging time points, measured in days after tumor implantation, are as follows: Rat~I: 12, 14, 15, 16, 19, and 21 days; Rat~II: 10, 12, 14, 15, 16, and 18 days; Rat~III: 10, 12, 14, 15, 16, and 19 days; \blue{and Rat~IV: 10, 12, 14, 16, 18, and 20 days.}
\blue{The 10 to 21 day imaging window was determined by the earliest time at which tumors were consistently visible on MRI and by mandatory humane endpoints associated with rapid C6 glioma growth.} All animal procedures were approved by the appropriate Institutional Animal Care and Use Committee, as detailed in the original experimental studies \cite{hormuth2017,hormuth2015predicting}.
\blue{The present analysis is conducted on a selected axial slice containing the primary tumor burden. Because the MRI slice thickness is approximately four times larger than the in-plane resolution, the selected axial slice represents a finite-thickness slab through the tumor region. The resulting two-dimensional analysis is therefore interpreted as a slice-based approximation, with explicit through-plane deformation and full volumetric tumor mechanics left to future three-dimensional inference.}

\subsection{Biomechanical tumor growth models}\label{sec:model}
The widely used continuum models to describe tissue-scale glioma growth is reaction--diffusion (RD) models consisting of the proliferation-infiltration representation of tumor evolution through the combined effects of local proliferation and spatial invasion \cite{swanson2000}. However, such models neglect the mechanical interactions between the growing tumor and the surrounding brain tissue. Mechanical stresses generated during tumor expansion can suppress proliferation, alter transport processes, and induce deformation of healthy tissue, commonly referred to as mass effect. 
\blue{Previous mechanically coupled models have often represented these effects using small-strain linear elasticity \cite{hormuth2017mechanically, tuncc2021modeling, hogea2007modeling}, incorporating tumor-induced tissue deformation and stress-mediated growth modulation. 
In contrast, the present work extends this class of mechanically coupled RD models by incorporating finite-strain tissue mechanics through a hyperelastic constitutive relation.}
Our ultimate goal is the calibration of biomechanical tumor growth models using longitudinal MRI data. Since all MRI scans acquired at later time points are rigidly registered to the first imaging time point and the finite element mesh is constructed from the corresponding initial brain geometry, it is natural and computationally convenient to formulate and solve the coupled governing equations in a Lagrangian (reference) configuration.

Let $\phi(\mathbf{X},t) \in [0,1]$ denote the tumor volume fraction defined over the reference configuration $\Omega_0$, where $\mathbf{X}$ is the material coordinate. The coupled system governing tumor growth and tissue deformation is given by
\begin{eqnarray}
J \frac{\partial \phi}{\partial t} &=&
\nabla_0 \cdot \left(
J D \, e^{-H p_{\tau}}
\mathbf{F}^{-1}\mathbf{F}^{-T} \nabla_0 \phi
\right)
+ J G (1-\phi)\phi
\quad \text{in } \Omega_0 \times (0,T], \label{eq:rd_PDE_REF} \\
\nabla_0 \cdot \mathbf{P} &=&
C J \mathbf{F}^{-T} \nabla_0 \phi
\quad \text{in } \Omega_0 \times (0,T], \label{eq:mech_PDE_REF}
\end{eqnarray}
together with the boundary and initial conditions
\begin{eqnarray}\label{eq:rd_hyper_bcs}
\left( J D \, e^{-H p_{\tau}} \mathbf{F}^{-1}\mathbf{F}^{-T} \nabla_0 \phi \right)\!\cdot \mathbf{N} &=& 0
\quad \text{on } \partial\Omega_0 \times (0,T], \\
\mathbf{u} &=& \mathbf{0}
\quad \text{on } \partial\Omega_0 \times (0,T], \nonumber\\
\phi(\mathbf{X},0) &=& \phi_0(\mathbf{X})
\quad \text{in } \Omega_0. \nonumber
\end{eqnarray}
Here, $\mathbf{u}$ is the displacement field, $\mathbf{F} = \mathbf{I} + \nabla_0 \mathbf{u}$ is the deformation gradient, and Jacobian $J = \det(\mathbf{F})$ represents the local volume change. The tensor $\mathbf{P}$ denotes the first Piola--Kirchhoff stress, $\nabla_0(\cdot)$ is the gradient with respect to the reference configuration, and $\mathbf{N}$ is the outward unit normal on $\partial\Omega_0$. Mechanical stress enters the tumor growth model through the scalar quantity
$
p_{\tau} := -\frac{1}{3}\,\mathrm{tr}(\boldsymbol{\tau}), 
$
where $\boldsymbol{\tau}$ is the Kirchhoff stress tensor. The quantity $p_{\tau}$ represents a measure of local mean compressive stress and provides a scalar, frame-invariant description of mechanical loading suitable for use in a Lagrangian formulation.

\blue{The first partial differential equation (PDE) in \ref{eq:rd_PDE_REF} is the Lagrangian form of a reaction--diffusion balance law for tumor volume fraction, where the factors involving $J$ and $\mathbf{F}^{-1}\mathbf{F}^{-T}$ arise from mapping diffusion from the deformed configuration to the reference configuration. The logistic source term $G(1-\phi)\phi$ represents local tumor proliferation adopted from \cite{swanson2000, hormuth2015predicting}.
The present formulation does not explicitly include necrosis, apoptosis, or additional cellular compartments. It is intentionally restricted to a parsimonious model whose spatially varying parameters can be constrained and informed by the available MRI-derived tumor volume fraction data.
The second PDE in \ref{eq:mech_PDE_REF} represents quasi-static mechanical equilibrium of the deforming tissue, where tumor expansion induces a mass-effect driving term proportional to the spatial gradient of tumor volume fraction.
For the associated boundary and initial conditions in \eqref{eq:rd_hyper_bcs}, the zero-flux condition for $\phi$ represents no net tumor-cell flux across the outer boundary of the computational brain domain. The displacement condition $\mathbf{u}=0$ on $\partial\Omega_0$ provides a simplified representation of the mechanical confinement imposed by the surrounding cranial environment and skull. The initial condition $\phi_0$ is obtained from the first MRI-derived tumor volume fraction field.
}
The model parameters are $D$ denoting the baseline tumor cell diffusivity, $H$ controls the sensitivity of diffusion to mechanical stress, $G$ is the proliferation rate, and $C$ is a coupling parameter governing the magnitude of tumor-induced mass effect. The reaction--diffusion and mechanical partial differential equations are coupled through two mechanisms: (i) \textit{stress-modulated diffusivity}, where cell motility decreases exponentially with increasing compressive stress, modeled through the Kirchhoff stress invariant $p_{\tau}$, which captures the inhibitory effect of mechanical confinement on cell migration; and (ii) \textit{mass effect}, represented \blue{phenomenologically} by a pressure-like body force proportional to the tumor volume fraction gradient, which drives deformation of the surrounding tissue.

\blue{To enable calibration from single-slice MRI data, we adopt a plane-strain approximation of a three-dimensional compressible Neo-Hookean material. This approximation permits controlled comparison of constitutive assumptions in the selected slice, but does not capture out-of-plane deformation.}
The strain-energy density function is defined as
\[
W(\mathbf{B}) =
\frac{S}{2}\left(\mathrm{tr}\!\left(J^{-2/3}\mathbf{B}\right) - 3\right)
+ \frac{K}{2}(J-1)^2,
\]
where $\mathbf{B} = \mathbf{F}\mathbf{F}^T$ is the left Cauchy--Green deformation tensor. The parameters $S$ and $K$ denote the shear and bulk moduli, respectively, and are related to Young’s modulus $E$ and Poisson’s ratio $\nu$ by
$
S = \frac{E}{2(1+\nu)},
K = \frac{E}{3(1-2\nu)}.
$
The first Piola--Kirchhoff stress tensor is obtained from the strain-energy density function as
$
\mathbf{P} = \frac{\partial W}{\partial \mathbf{F}}.
$
For the compressible Neo-Hookean model defined above, this yields the explicit expression
\begin{equation}\label{eq:PK1}
\mathbf{P}
=
S J^{-2/3}
\left(
\mathbf{F}
- \frac{1}{3}\,\mathrm{tr}(\mathbf{B})\,\mathbf{F}^{-T}
\right)
+ K J (J-1)\,\mathbf{F}^{-T}.
\end{equation}

\subsection{Image-informed sequential Bayesian inference}\label{sec:bayes_filt}
Longitudinal MRI provides subject-specific observations of tumor evolution over time, enabling individualized prediction of tumor growth and biophysical parameters dynamics using computational models. As imaging measurements become available sequentially, Bayesian filtering offers a principled framework for updating model parameters, propagating uncertainty through the coupled growth--mechanics model, and quantifying the reliability of subject-specific predictions.
Since the inferred tumor volume fraction field reflects both mechanically mediated proliferative and infiltration growth, accurate prediction of tumor shape and spatial extent is essential for resolving biomechanical heterogeneity. To this end, we formulate an image-informed sequential Bayesian inference framework with spatially varying parameters, allowing longitudinal MRI data to update both the temporal evolution and spatial structure of growth and mechanical parameter fields.

Let $\bs d^{(1:k)} = \{ d(\mathbf{X}, t_1), \ldots, d(\mathbf{X}, t_k) \}$ denote the sequence of imaging-derived tumor volume fraction fields acquired up to time $t_k$, with all images registered to the reference configuration $\Omega_0$ defined by the first imaging time point. Let $\boldsymbol{\theta}^{(k)} = \boldsymbol{\theta}(\mathbf{X}, t_k) \in \Theta$ denote the model parameter vector at time $t_k$, which includes spatially varying growth, diffusion, and mechanical coefficients of the biomechanical tumor growth model.
\blue{Here, the superscript \(k\) denotes the assimilation step and conditioning on data available up to \(t_k\), and no explicit temporal evolution law is prescribed for the parameter fields.}
The sequential Bayesian filtering consists of updating the posterior distribution of the model parameters as new imaging data become available. At time $t_{k}$, the posterior probability distribution function (PDF) is given by,
\begin{equation}
\label{eq:bayes_filter}
\pi_{\mathrm{post}}\!\left( \boldsymbol{\theta}^{(k)} \mid \bs d^{(1:k)} \right)
=
\frac{
\pi_{\mathrm{like}}\!\left( \bs d^{(1:k)} \mid \boldsymbol{\theta}^{(k)} \right)
\, \pi_{\mathrm{pr}}\!\left( \boldsymbol{\theta} \right)
}{
\pi_{\mathrm{evid}}\!\left( \bs d^{(1:k)} \right)
},
\end{equation}
where $\pi_{\mathrm{pr}}(\boldsymbol{\theta})$ denotes the prior PDF, $\pi_{\mathrm{like}}(\bs d^{(1:k)} \mid \boldsymbol{\theta}^{(k)})$ is the likelihood function, and $\pi_{\mathrm{evid}}(\bs d^{(1:k)})$ is the evidence PDF.
Given the posterior distribution at time $t_k$, a predictive distribution for the tumor state at the next time point $t_{k+1}$ is obtained by propagating parameter uncertainty through the biomechanical tumor growth model,
\begin{equation}
\label{eq:predictive}
\pi\!\left( \phi^{(k+1)} \mid \bs d^{(1:k)} \right)
=
\int_{\Theta}
\pi\!\left( \phi^{(k+1)} \mid \boldsymbol{\theta}^{(k)} \right)
\pi_{\mathrm{post}}\!\left( \boldsymbol{\theta}^{(k)} \mid \bs d^{(1:k)} \right)
\, \mathrm{d}\boldsymbol{\theta},
\end{equation}
where $\phi^{(k+1)} = \phi(\mathbf{X}, t_{k+1})$ denotes the tumor volume fraction field at the future prediction time.
This \textit{one-scan-ahead} prediction is continued as new longitudinal imaging data are acquired, enabling sequential refinement of model parameters and tumor growth predictions.

%

To enable spatially varying Bayesian inference, the model parameter fields are discretized over the finite element mesh defined on the reference configuration $\Omega_0$. A Gaussian random field prior is imposed on the discretized parameter vector
$\boldsymbol{\theta} \sim \mathcal{N}\!\left( \bar{\boldsymbol{\theta}}, \boldsymbol{\Gamma}_{\mathrm{pr}} \right)$,
where $\bar{\boldsymbol{\theta}}$ denotes the prior mean field and $\boldsymbol{\Gamma}_{\mathrm{pr}}$ is a covariance matrix corresponding to a Mat\'ern class.
Following \cite{roininen2014whittle, tan2024scalable, liang2023bayesian, singh2025chance}, such Gaussian random field can be equivalently defined as the solution of the following stochastic partial differential equation,
\begin{eqnarray}
\label{eq:matern}
-\gamma \Delta_0 \boldsymbol{\theta} + \delta \boldsymbol{\theta} = \boldsymbol{\dot{W}}, & \text{in } \Omega_0, \nonumber\\[4pt]
\boldsymbol{\theta} + \beta \nabla_0 \boldsymbol{\theta} \cdot \mathbf{N} = 0, & \text{on } \partial\Omega_0 ,
\end{eqnarray}
where $\dot{\bs W}$ denotes spatial Gaussian white noise,  $\Delta_0$ is the Laplacian operators with respect to the reference configuration, and $\mathbf{N}$ is the outward unit normal on $\partial\Omega_0$.
The parameters $\delta$ and $\gamma$ control the marginal variance $\sigma^2$ and correlation length $\rho$ of the prior field through
$\delta = \frac{\sqrt{2}}{\sigma \rho \sqrt{\pi}}$ and
$\gamma = \frac{1}{4\sqrt{2\pi}} \frac{\rho}{\sigma}$,
while the Robin boundary coefficient is set as
$\beta = \frac{\sqrt{\delta\gamma}}{1.42}$,
which mitigates boundary artifacts arising from truncation of the spatial domain \cite{daon2016mitigating}.
\blue{The choice of Gaussian random field prior the spatially varying parameter fields is because it provides a tractable way to encode expected spatial smoothness, marginal variability, and correlation length scales of the biophysical parameters. This prior also regularizes the high-dimensional inverse problem and reduces nonphysical spatial oscillations in regions weakly informed by the imaging data. Although the true parameter fields need not to be Gaussian, the Gaussian prior provides a computationally efficient and widely used representation for Bayesian inverse problems with spatially correlated parameters \cite{alghamdi2021bayesian, liang2023bayesian, VillaPetraGhattas21}.}
The likelihood function is defined through an additive noise model relating imaging-derived observations to model predictions. 
Assuming zero-mean, Gaussian noise with variance $\sigma_{\mathrm{noise}}^2$ to account for both model and measurement errors, the likelihood function associated with the cumulative imaging data $\bs d^{(1:k)}$ is given by
\begin{equation}
\label{eq:like}
\pi_{\mathrm{like}}\!\left( \bs d^{(1:k)} \mid \boldsymbol{\theta}^{(k)} \right)
\propto
\exp\!\left(
-\sum_{i=1}^{k}
\frac{1}{2\sigma_{\mathrm{noise}}^2}
\int_{\Omega_0}
\left[
\phi(\mathbf{X},t_i;\boldsymbol{\theta}^{(k)})
-
d(\mathbf{X},t_i)
\right]^2
\,\mathrm{d}\mathbf{X}
\right).
\end{equation}
\blue{The MRI-derived tumor volume fraction fields $d$ are therefore not treated as exact ground truth observation of tumor. The Gaussian noise variance $\sigma_{\mathrm{noise}}^2$ represents an aggregate observation uncertainty that includes MRI measurement noise, image registration and segmentation errors, and residual model discrepancy.}

\subsection{Dynamic tumor growth model selection}\label{sec:selection}
A variety of tumor growth models have been proposed to describe GBM evolution, differing in their biological realism, mechanical fidelity, and number of parameters. The biomechanical tumor growth model described in Section~\ref{sec:model} represents one such formulation. Alternative choices include more detailed hyperelastic constitutive descriptions of brain tissue, e.g., \cite{budday2015physical, mihai2017family}, simplified mechanical representations based on linear elasticity \cite{hormuth2017mechanically}, and purely reaction-diffusion models that neglect tissue deformation altogether \cite{swanson2000}. Each model emphasizes different aspects of tumor progression and may be more appropriate under different stages of growth and mechanically mediated mass effect.
\blue{
However, the adequacy of a prescribed biomechanical formulation may change as longitudinal data are assimilated and tumor burden evolves. Mechanical effects may be limited during early infiltrative growth but become more influential as stress builds and deformation of the surrounding tissue emerges. Relying on a fixed model without reassessing its support can therefore reduce mechanistic fidelity and predictive reliability. To address this limitation, we adopt a dynamic Bayesian model selection framework in which longitudinal MRI data from an individual subject are used to quantify the relative plausibility of competing tumor growth models.
}

Let $\{M_1,\ldots,M_m\}$ denote a set of competing tumor growth models. At imaging time $t_k$, model plausibility is quantified through the posterior probability of the discrete model variable conditioned on the cumulative imaging data $\bs d^{(1:k)}$. Given a prior probability mass function over models, $\pi_{\mathrm{pr}}(M)$, the posterior model probabilities are given by
\begin{equation}
\label{eq:model_post}
\pi\!\left(M_i \mid \bs d^{(1:k)}\right)
=
\frac{
\pi_{\mathrm{evid}}\!\left(\bs d^{(1:k)} \mid {M}_i\right)\,
\pi_{\mathrm{pr}}\!\left({M}_i\right)
}{
\sum_{j=1}^{m}
\pi_{\mathrm{evid}}\!\left(\bs d^{(1:k)} \mid {M}_j\right)\,
\pi_{\mathrm{pr}}\!\left({M}_j\right)
},
\qquad i=1,\ldots,m.
\end{equation}
Here, $\pi_{\mathrm{evid}}(\bs d^{(1:k)} \mid {M}_i)$ denotes the model evidence (marginal likelihood) under model ${M}_i$, obtained by marginalizing over the corresponding model parameters,
\begin{equation}
\label{eq:model_evidence}
\pi_{\mathrm{evid}}\!\left(\bs d^{(1:k)} \mid {M}_i\right)
=
\int_{\Theta_i}
\pi_{\mathrm{like}}\!\left(\bs d^{(1:k)} \mid \boldsymbol{\theta}_i, {M}_i\right)\,
\pi_{\mathrm{pr}}\!\left(\boldsymbol{\theta}_i \mid {M}_i\right)\,
\mathrm{d}\boldsymbol{\theta}_i,
\end{equation}
where $\Theta_i$ denotes the parameter space associated with model $\mathcal{M}_i$. 
In this setting,
as additional imaging data become available, the posterior model probabilities $\pi\!\left(M_i^{(k)} \mid \bs d^{(1:k)}\right)$ are updated sequentially,  allowing longitudinal MRI measurements \blue{to change the relative support for candidate models over time.}
\blue{In particular, for each animal, this evidence calculation is repeated at each assimilation time using all MRI data available up to that time. The resulting posterior model probabilities define a subject-specific, time-updated ranking of the candidate models, which can be used to select the model for the subsequent one-scan-ahead prediction.}

\subsection{Scalable Bayesian filtering and model plausibility}
\label{sec:solution}

Finite element discretization of the spatially varying growth and mechanical parameter fields yields a high-dimensional parameters $\boldsymbol{\theta} \in \mathbb{R}^{N_p}$, rendering Bayesian inference computationally intractable with standard methods. To address this challenge, we employ scalable Bayesian solution strategies whose computational cost in terms of number of model solves is independent of the parameter dimension.

\paragraph{1) Laplace approximation (LA) of the posterior}
The primary solution strategy is a Laplace approximation of the Bayesian filtering posterior, obtained by locally approximating the posterior distribution in \eqref{eq:bayes_filter} with a Gaussian centered at the maximum a posteriori (MAP) estimate. 
At each assimilation step, \blue{defined here as the Bayesian update performed when a new MRI scan becomes available,} the posterior is approximated as,
$
\pi_{\mathrm{post}}\!(\boldsymbol{\theta}^{(k)} \mid \bs d^{(1:k)})
\approx
\mathcal{N}\!(
\boldsymbol{\theta}_{\mathrm{MAP}}^{(k)},
\boldsymbol{\Gamma}_{\mathrm{post}}^{(k)}
).
$
The MAP estimate is computed by minimizing the negative log-posterior,
\begin{equation}
\label{eq:map_main}
\boldsymbol{\theta}_{\mathrm{MAP}}^{(k)}
=
\arg\min_{\boldsymbol{\theta}}
\left\{
\frac{1}{2\sigma_{\mathrm{noise}}^{2}}
\sum_{i=1}^{k}
\int_{\Omega_0}
\left[
\phi(\mathbf{X},t_i;\boldsymbol{\theta})
-
d(\mathbf{X},t_i)
\right]^2
\,\mathrm{d}\mathbf{X}
+
\frac{1}{2}
\big\|
\boldsymbol{\theta}-\bar{\boldsymbol{\theta}}
\big\|^2_{\boldsymbol{\Gamma}_{\mathrm{pr}}^{-1}}
\right\},
\end{equation}
which balances data misfit with the Gaussian random field prior. This optimization problem is solved using a globalized inexact Newton--CG method with adjoint-based gradient and Hessian--vector evaluations, combined with a backtracking line search to ensure global convergence \cite{VillaPetraGhattas21}.
The posterior covariance is given by
\begin{equation}
\label{eq:post_cov_main}
\boldsymbol{\Gamma}_{\mathrm{post}}^{(k)}
=
\left(
\boldsymbol{H}_{\mathrm{misfit}}\!\left(\boldsymbol{\theta}_{\mathrm{MAP}}^{(k)}\right)
+
\boldsymbol{\Gamma}_{\mathrm{pr}}^{-1}
\right)^{-1},
\end{equation}
where $\boldsymbol{H}_{\mathrm{misfit}}$ denotes the Hessian of the data misfit term. Exploiting the rapid decay of the generalized eigenvalues $\{\lambda_i\}$ of the data misfit Hessian with respect to the prior precision, we retain only the leading $r \ll N_p$ modes, yielding the low-rank approximation
\begin{equation}
\label{eq:post_cov_lowrank_main}
\boldsymbol{\Gamma}_{\mathrm{post}}^{(k)}
\approx
\boldsymbol{\Gamma}_{\mathrm{pr}}
-
\boldsymbol{W}_r
\boldsymbol{D}_r
\boldsymbol{W}_r^{T},
\end{equation}
where $\boldsymbol{D}_r = \mathrm{diag}\!\left(\lambda_i/(1+\lambda_i)\right)$ and the columns of $\boldsymbol{W}_r$ are the dominant generalized eigenvectors.
This low-rank structure enables scalable Bayesian filtering whose dominant cost depends only weakly on the parameter dimension and is governed primarily by the number of retained Hessian modes and the underlying forward and adjoint solves \cite{isaac2015scalable, kim2023hippylib, VillaPetraGhattas21}.

%

In this work, the Laplace approximation framework is further extended to enable scalable evaluation of the posterior model plausibility defined in \eqref{eq:model_post}. For each candidate model $\{{M}_i\}_{i=1}^m$, the model evidence conditioned on the cumulative imaging data $\bs d^{(1:k)}$ is approximated as
\begin{equation}
\label{eq:evidence_main}
\small
\log \pi_{\mathrm{evid}}\!\left(\bs d^{(1:k)} \mid {M}_i \right)
\approx
- \Phi_i\!\left(\boldsymbol{\theta}_{\mathrm{MAP},i}^{(k)}\right)
- \frac{1}{2}
\big\|
\boldsymbol{\theta}_{\mathrm{MAP},i}^{(k)} - \bar{\boldsymbol{\theta}}_{i}
\big\|^2_{\boldsymbol{\Gamma}_{\mathrm{pr},i}^{-1}}
- \frac{N_d}{2}\log\!\left(2\pi\sigma_{\mathrm{noise}}^{2}\right)
- \frac{1}{2}\sum_{j=1}^{r_i}\log(1+\lambda_{i,j}),
\end{equation}
where $\boldsymbol{\theta}_{\mathrm{MAP},i}^{(k)}$ denotes the MAP estimate under model ${M}_i$, $\Phi_i$ is the corresponding data misfit functional, $N_d$ is number of data points, $\boldsymbol{\Gamma}_{\mathrm{pr},i}$ is the prior covariance associated with ${M}_i$, and $\{\lambda_{i,j}\}_{j=1}^{r_i}$ are the dominant generalized eigenvalues of the data misfit Hessian. Full derivations of the Laplace-approximated posterior and model evidence are provided in the \textit{Supplementary Information}.
Substituting the approximated evidences \eqref{eq:evidence_main} into \eqref{eq:model_post} yields a scalable approximation of the posterior model plausibility. In the absence of prior preference among competing models, i.e.,
$\pi_{\mathrm{pr}}\!\left({M}_i\right) = \frac{1}{m},$
the posterior model probabilities reduce to the normalized model evidences,
\begin{equation}
\label{eq:model_post_uniform}
\pi\!\left(M_i \mid \bs d^{(1:k)}\right)
=
\frac{
\pi_{\mathrm{evid}}\!\left(\bs d^{(1:k)} \mid {M}_i\right)
}{
\sum_{j=1}^{m}
\pi_{\mathrm{evid}}\!\left(\bs d^{(1:k)} \mid {M}_j\right)
},
\qquad i=1,\ldots,m.
\end{equation}

\paragraph{2) Dimension-independent Markov chain Monte Carlo (MCMC)}
Although the Laplace approximation enables scalable Bayesian filtering, its reliance on local linearization about the MAP estimate may lead to inaccuracies when the posterior distribution exhibits strong non-Gaussian structure. To assess the accuracy of the LA, we compare its results against a sampling-based reference solution using dimension-independent MCMC.
We employ the generalized preconditioned Crank--Nicolson (gpCN) algorithm. The pCN method~\cite{cotter2013mcmc} is well-defined in infinite-dimensional parameter spaces and exhibits dimension-independent mixing when sampling Gaussian measures, but in PDE-constrained inverse problems it is often sensitive to step-size selection, leading to poor acceptance rates and high autocorrelation. The gpCN algorithm~\cite{pinski2015algorithms} overcomes these limitations by incorporating local posterior geometry through the MAP estimate and posterior covariance obtained from the Laplace approximation. This results in improved sampling efficiency while retaining dimension-independent scalability~\cite{alghamdi2021bayesian, kim2023hippylib}. 
The gpCN algorithm used to sample the posterior associated with one Bayesian update in \eqref{eq:bayes_filter} is summarized in Algorithm~\ref{alg:gpcn}.
\blue{In Algorithm~\ref{alg:gpcn}, the posterior ratio is evaluated with respect to the Gaussian reference measure $\mathcal{N}(\boldsymbol{\theta}_{\mathrm{MAP}},\boldsymbol{\Gamma}_{\mathrm{post}})$ used in the gpCN proposal, equivalently including the corresponding proposal density correction in the acceptance ratio.}

\begin{algorithm}[t]
\caption{Generalized preconditioned Crank-Nicolson (gpCN) sampler for one update of the Bayesian filter.}
\label{alg:gpcn}
\begin{algorithmic}[1]
\Require MAP estimate $\boldsymbol{\theta}_{\mathrm{MAP}}$, posterior covariance $\boldsymbol{\Gamma}_{\mathrm{post}}$, step size $s\in(0,1)$, chain length $L$, initial state $\boldsymbol{\theta}^{0}$
\State Set $\ell \gets 0$
\State \textbf{while} $\ell \le L-1$ \textbf{do}
\State \hspace{0.5em} Draw $\boldsymbol{\xi}^{\ell} \sim \mathcal{N}(\mathbf{0}, \boldsymbol{\Gamma}_{\mathrm{post}})$
\State \hspace{0.5em} $\boldsymbol{\theta}^{\star} \gets \boldsymbol{\theta}_{\mathrm{MAP}} + \sqrt{1-s^2}\,(\boldsymbol{\theta}^{\ell}-\boldsymbol{\theta}_{\mathrm{MAP}}) + s\,\boldsymbol{\xi}^{\ell}$
\State \hspace{0.5em} $\alpha \gets \min\!\left(1, \dfrac{\pi_{\mathrm{post}}(\boldsymbol{\theta}^{\star}\mid \bs d)}{\pi_{\mathrm{post}}(\boldsymbol{\theta}^{\ell}\mid \bs d)}\right)$
\State \hspace{0.5em} Draw $u \sim \mathcal{U}(0,1)$.
If $u \le \alpha$ set $\boldsymbol{\theta}^{\ell+1} \gets \boldsymbol{\theta}^{\star}$; else set $\boldsymbol{\theta}^{\ell+1} \gets \boldsymbol{\theta}^{\ell}$
\State \hspace{0.5em} $\ell \gets \ell + 1$
\State \textbf{end while}
\State \Return $\{\boldsymbol{\theta}^{\ell}\}_{\ell=1}^{L}$
\end{algorithmic}
\end{algorithm}

\section{Results}\label{sec:results}
We present results from sequential Bayesian inference of the
biomechanically coupled tumor growth model introduced in Section~\ref{sec:model}, using longitudinal MRI data from murine glioma subjects described in Section~\ref{sec:mri}. The focus is to quantify how mechanical coupling and constitutive
assumptions influence model plausibility, inferred parameter fields,
and predictive performance.
All simulations are performed in two spatial dimensions (2D) using the central axial
slice along the cranial--caudal direction. 
\blue{This slice was selected because it contained the dominant MRI-derived tumor burden over the longitudinal imaging window. A slice selection analysis and an adjacent slice inference check are reported in the \textit{Supplementary Information}, supporting that the tumor area and model plausibility trends are not artifacts of the selected slice.}
\blue{Although MRI-driven 3D Bayesian inference has been demonstrated previously \cite{liang2023bayesian},} restricting the present analysis to 2D enables a controlled and computationally feasible comparison between LA and gpCN Bayesian inference, \blue{and the resulting mechanics should therefore be interpreted as slice-based rather than fully three-dimensional}.
For each subject, $T_2$-weighted MRI are used to construct
the finite element mesh over the domain $\Omega_0$. 
The tumor volume
fraction at the first scan defines the initial condition
$\phi_0(\boldsymbol{X})$, with DW-MRI data interpolated to mesh nodes.
At each imaging time, data up to that scan are assimilated via
Bayesian filtering, followed by a \textit{one-scan-ahead} prediction to
the subsequent MRI time point. 
\blue{The main analysis focuses on one-scan-ahead prediction because the framework is designed for sequential data assimilation; a multi-step prediction experiment without intermediate data assimilation is provided in the \textit{Supplementary Information} to assess degradation over longer forecast horizons.}
Dynamic model selection is performed \blue{separately for each animal by evaluating the posterior plausibility of three candidate formulations specified \textit{a priori}: reaction--diffusion without mechanics, reaction--diffusion coupled with linear elasticity, and reaction--diffusion coupled with hyperelastic mechanics}.
\blue{Thus, model plausibility is updated as each animal's MRI time series is assimilated and can be used to adapt the one-scan-ahead prediction model.} 
Predictive agreement is quantified using the Dice
similarity coefficient and normalized tumor area (NTA).
The Dice coefficient ranges from 0 (no spatial overlap) to 1 (perfect overlap)
and characterizes the geometric agreement between predicted and MRI-derived tumor
shapes.
\blue{For Dice, NTA, and tumor boundary calculations, the tumor region was defined by thresholding the normalized tumor volume fraction at $\phi=0.25$ for both MRI-derived fields and model predictions. A threshold sensitivity check for Rat~III at Day~19 showed only modest changes across $\phi=0.20$, $0.25$, and $0.30$, with NTA values of $0.27$, $0.26$, and $0.25$ and Dice values of $0.91$, $0.93$, and $0.93$, respectively. These small variations indicate that the reported predictive trends are not sensitive to moderate changes in the tumor volume fraction cutoff.}
Finite element simulations are performed using FEniCS \cite{fenics2015}, and both LA and
gpCN Bayesian inference, including scalable evaluation of posterior model
plausibility, are implemented using hIPPYLib \cite{VillaPetraGhattas21, VillaPetraGhattas18, VillaPetraGhattas16}.

\begin{figure}[!ht]
\centering
\hspace*{-0.2\textwidth}
\begin{minipage}[c]{0.48\textwidth}
    \centering
    \includegraphics[width=.7\textwidth]{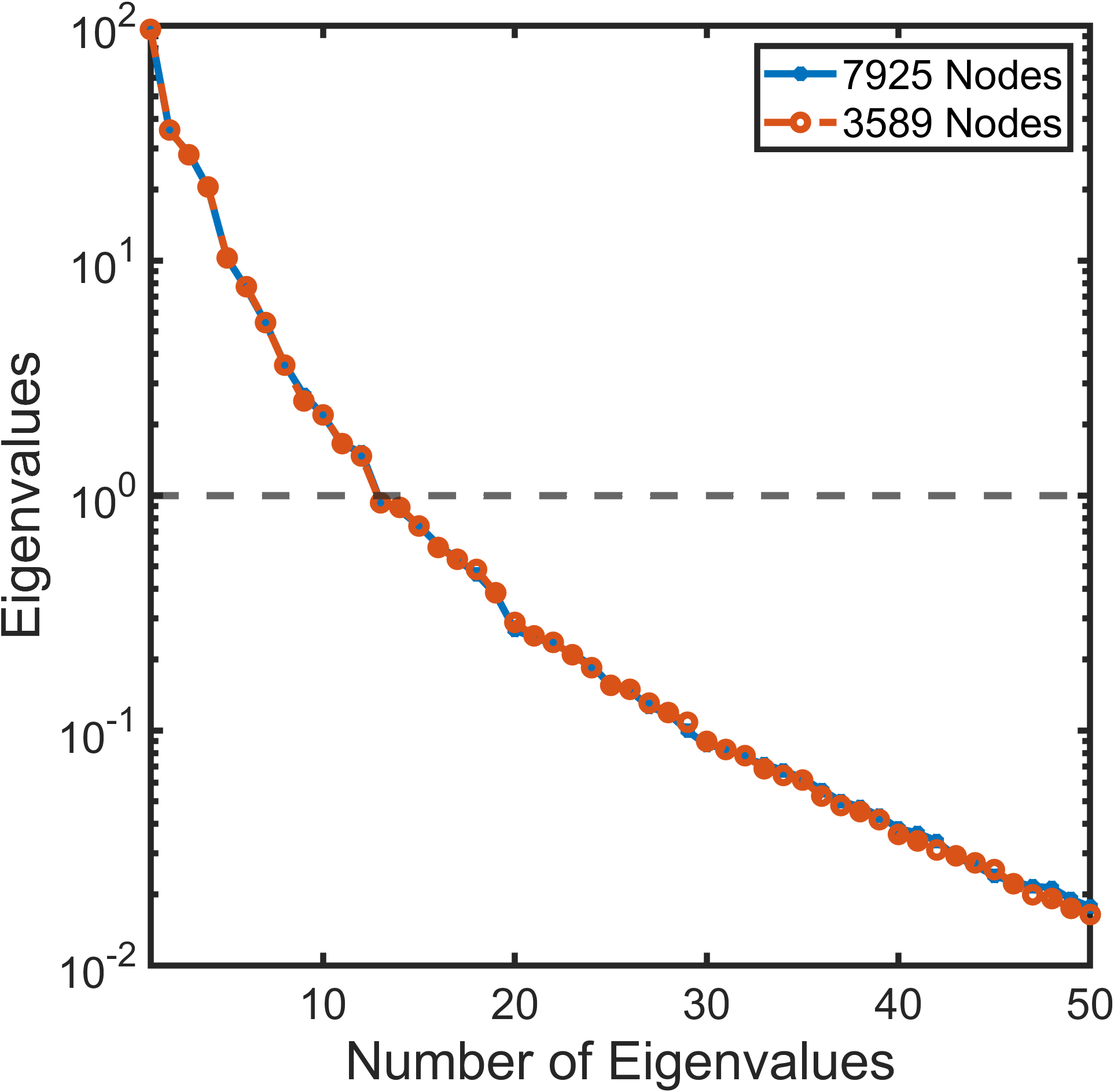}
\end{minipage}
\hspace{-0.05\textwidth}
\begin{minipage}[c]{0.48\textwidth}
    \centering
    \setlength{\tabcolsep}{3pt}
    \begin{tabular}{ccc}
        \includegraphics[width=0.4\textwidth]{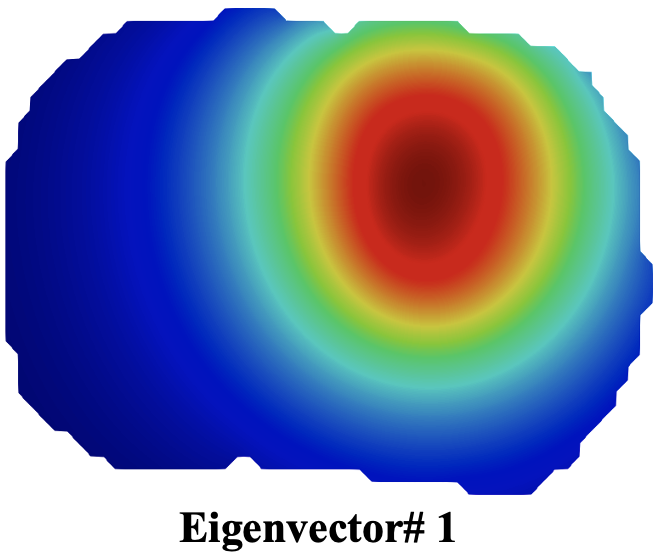} &
        \includegraphics[width=0.4\textwidth]{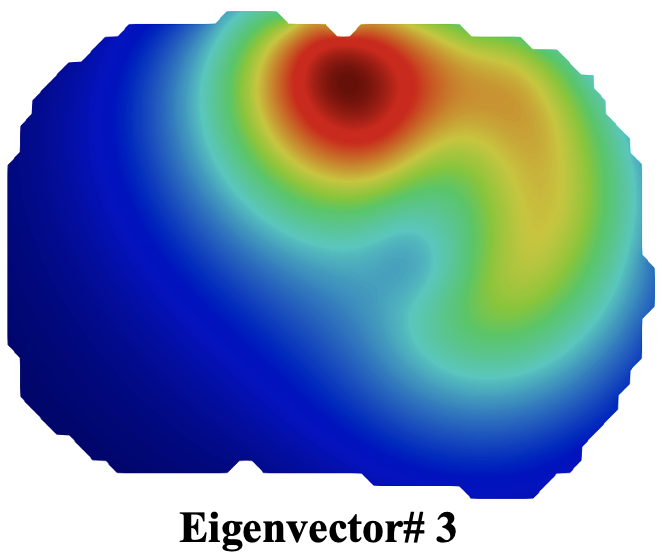} &
        \includegraphics[width=0.4\textwidth]{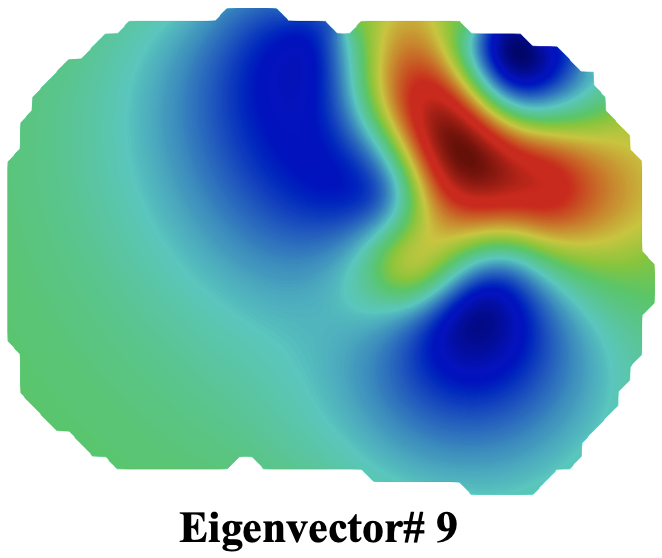} \\
        \includegraphics[width=0.4\textwidth]{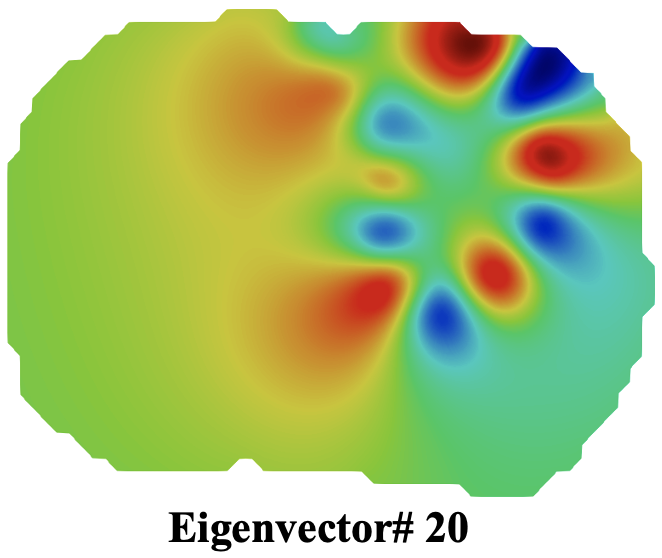} &
        \includegraphics[width=0.4\textwidth]{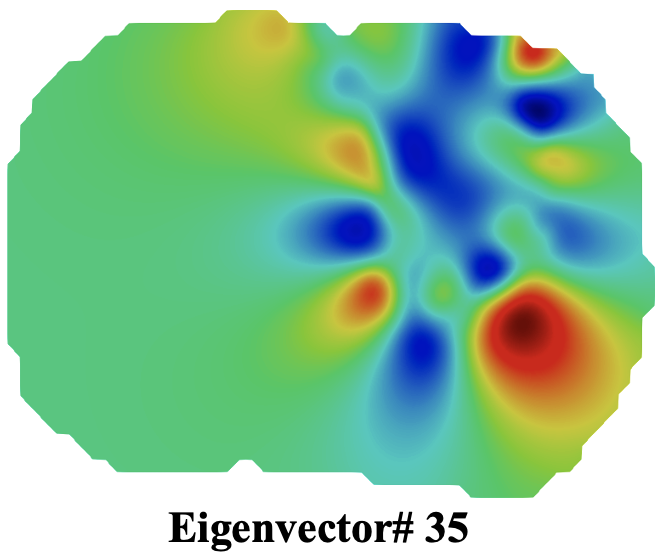} &
        \includegraphics[width=0.4\textwidth]{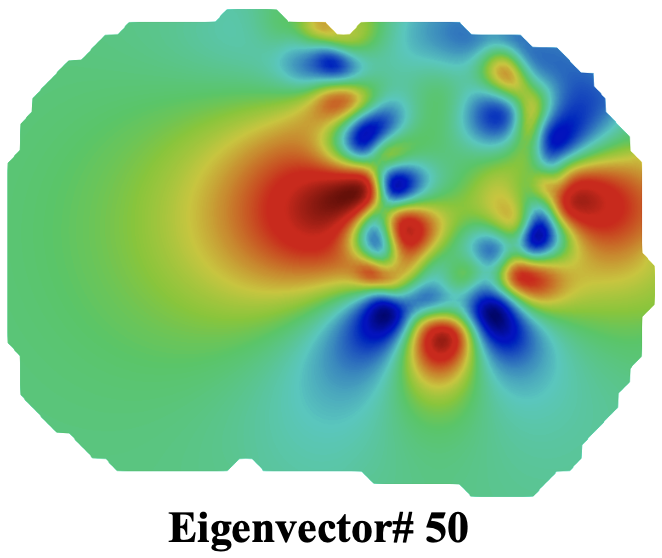}
    \end{tabular}
\end{minipage}
\vspace{-1pt}
\caption{
Scalability of the LA solution of Bayesian inference for Rat~III from MRI data of days 10-16. Eigenvalue decay and selected eigenvectors of the prior-preconditioned data misfit Hessian demonstrate rapid spectral decay, mesh independence, and spatial localization of data-informed modes.
\blue{The leading 50 Hessian eigenmodes were retained in the low-rank posterior approximation, while a rank-sensitivity check is reported in the \textit{Supplementary Information}.}
}
\label{fig:eigen_analysis}
\end{figure}

\subsection{Calibration parameters and hyper-parameter specification}
\begin{table}[t]
\centering
\small
\caption{Hyper-parameters and fixed biophysical parameters
used in the sequential Bayesian inference for \blue{four} Wistar rats.}
\label{tab:parameters_values}
\begin{tabular}{l c c c c}
\toprule
\multicolumn{5}{l}{\textbf{Prior mean and variance of inferred parameters}} \\
\midrule
 & $\log(E)$ & $\log(D)$ & $\log(G)$ &  \\
Units & log(kPa) & log(mm$^2$/day) & log(day$^{-1}$) &  \\
\midrule
Mean & $-0.1562$ & $-0.3492$ & $-1.8965$ &  \\
Variance & 0.01 & 0.09 & 0.0064 &  \\
\addlinespace[1.5mm]

\multicolumn{5}{l}{\textbf{Prior correlation lengths and noise variance}} \\
\midrule
 & $\rho_E$ (mm) & $\rho_D$ (mm) & $\rho_G$ (mm) & $\sigma^2_{\rm noise}$ \\
Value & 5.5 & 6.5 & 3.5 & $1.23 \times 10^{-3}$ \\
\addlinespace[1.5mm]

\multicolumn{5}{l}{\textbf{Fixed biophysical model parameters}} \\
\midrule
Parameter & $H$ (kPa$^{-1}$) & $C$ (kPa) & $\nu$ (--) &  \\
Value & 1.5 & 0.65 & 0.45 &  \\
\bottomrule
\end{tabular}
\end{table}
Among the biophysical parameters defining the 
\blue{biomechanical tumor growth model
\eqref{eq:rd_PDE_REF} and \eqref{eq:mech_PDE_REF},}
the calibration parameters are tumor diffusivity, proliferation rate, and elastic modulus $\bs \theta = (\log D(\bs X, t), \log G(\bs X, t), \log E(\bs X, t))$ that are considered as subject-specific, and spatially varying and represented in logarithmic to ensure their positive values within the sequential Bayesian inference. 
These fields
are sequentially updated via Bayesian filtering using longitudinal MRI data, as
they directly control tumor infiltration, volumetric growth, and growth-induced
mechanical resistance.
The stress-mediated diffusivity decay coefficient $H$,
the growth--mechanics coupling coefficient $C$, and Poisson’s ratio $\nu$ are
treated as spatially homogeneous and fixed. The biophysical motivation of this choice is that
$C$ governs the magnitude of tumor-induced mass effect, whose spatial variation
cannot be reliably inferred from imaging data alone, while $H$ globally modulates
stress-induced growth inhibition and together with $\nu$ are weakly identifiable given MRI data of tumor volume fraction.
\blue{To assess whether this choice affects model selection, a sensitivity analysis with $\pm 25\%$ perturbations of $C$ and $H$ is reported in the \textit{Supplementary Information}, indicating the posterior plausibility trends were preserved under these perturbations.}
Prior means and variances for the inferred parameters, along with values of the
fixed parameters, are taken from spatially homogeneous posterior estimates reported
for Wistar rat glioma growth
\cite{lima2016selection,hormuth2017mechanically,hormuth2015predicting} and are
summarized in Table~\ref{tab:parameters_values}. The table also reports the prior
correlation lengths $\rho_E$, $\rho_D$, and $\rho_G$, and the observation noise
variance $\sigma_{\rm noise}$, determined via the cross-validation procedure described in
\cite{liang2023bayesian}. In particular, MRI data from Rat~III over days 10--16 are used in a grid
search, with Pareto-optimal hyper-parameters selected by \blue{maximizing Dice agreement and minimizing NTA prediction error} at the final prediction time at day~19.
These hyper-parameters are then fixed and applied to the remaining subjects.

\begin{figure}[!ht]
    \centering
    \raggedright
    \renewcommand{\arraystretch}{0}
    \setlength{\fboxsep}{0pt} 
    \setlength{\tabcolsep}{1pt} 
    
    \begin{tabular}{@{}l@{}l@{}}
        
        \begin{tabular}{@{} *{6}{>{\centering\arraybackslash}p{0.148\textwidth}} @{}} 
            \colorbox{black}{\makebox[0.148\textwidth][c]{\vrule width 0pt height 0.4cm \textcolor{white}{\small Day 12}}} & 
            \colorbox{black}{\makebox[0.148\textwidth][c]{\vrule width 0pt height 0.4cm \textcolor{white}{\small Day 14}}} & 
            \colorbox{black}{\makebox[0.148\textwidth][c]{\vrule width 0pt height 0.4cm \textcolor{white}{\small Day 15}}} & 
            \colorbox{black}{\makebox[0.148\textwidth][c]{\vrule width 0pt height 0.4cm \textcolor{white}{\small Day 16}}} & 
            \colorbox{black}{\makebox[0.148\textwidth][c]{\vrule width 0pt height 0.4cm \textcolor{white}{\small Day 19}}} & 
            \colorbox{black}{\makebox[0.148\textwidth][c]{\vrule width 0pt height 0.4cm \textcolor{white}{\small Day 21}}} \\ [3pt]
            
            \includegraphics[width=0.15\textwidth]{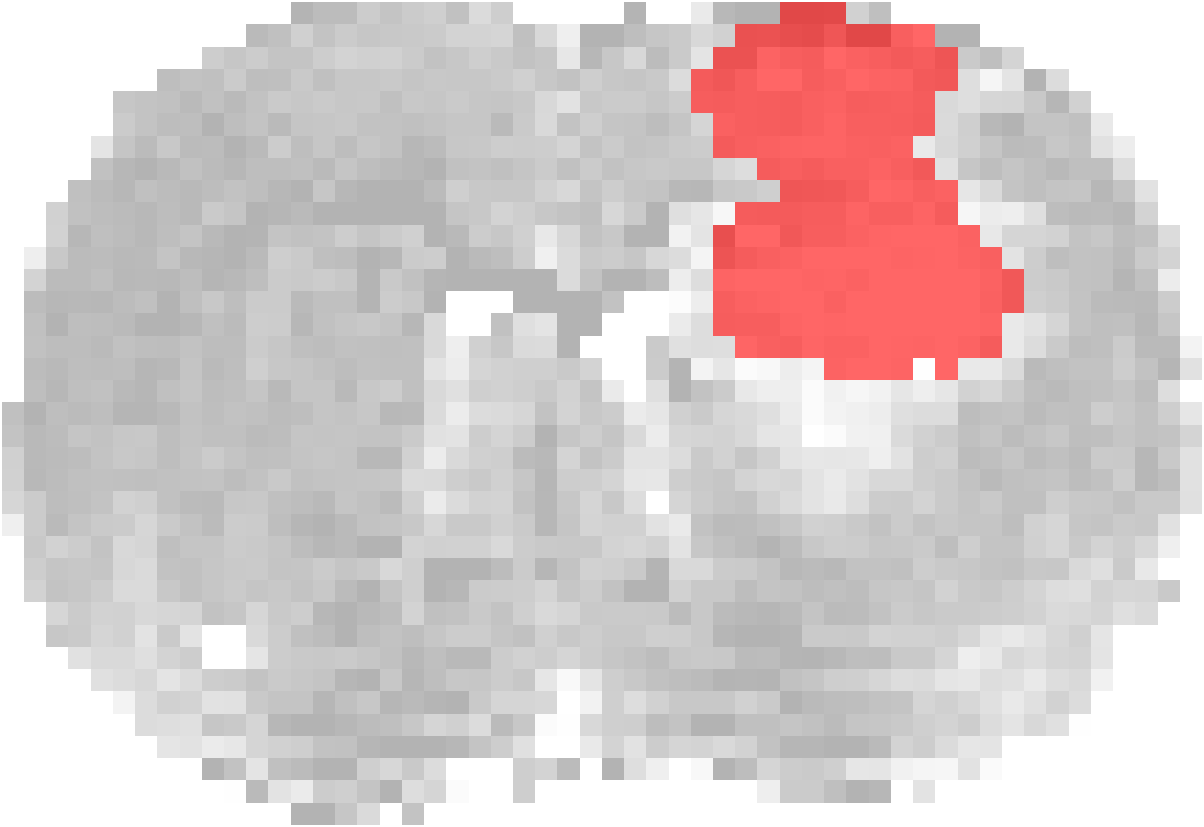} & 
            \includegraphics[width=0.15\textwidth]{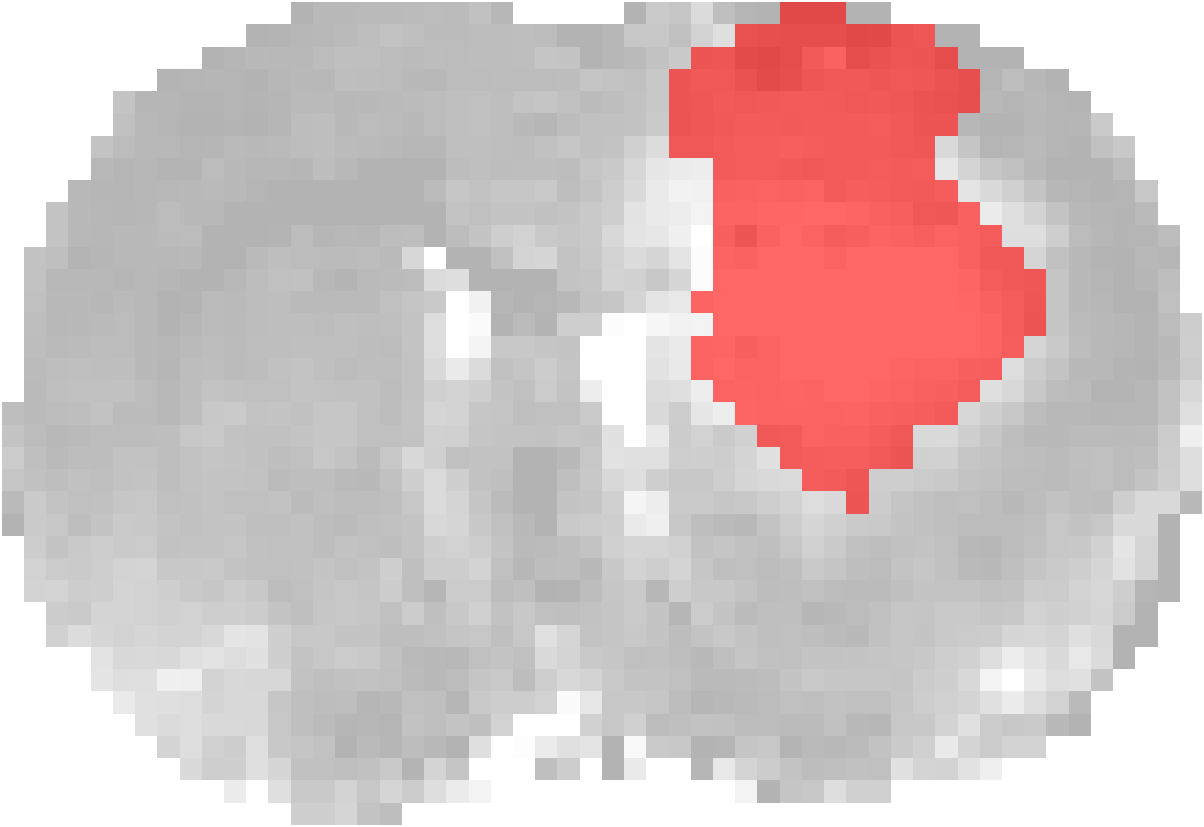} & 
            \includegraphics[width=0.15\textwidth]{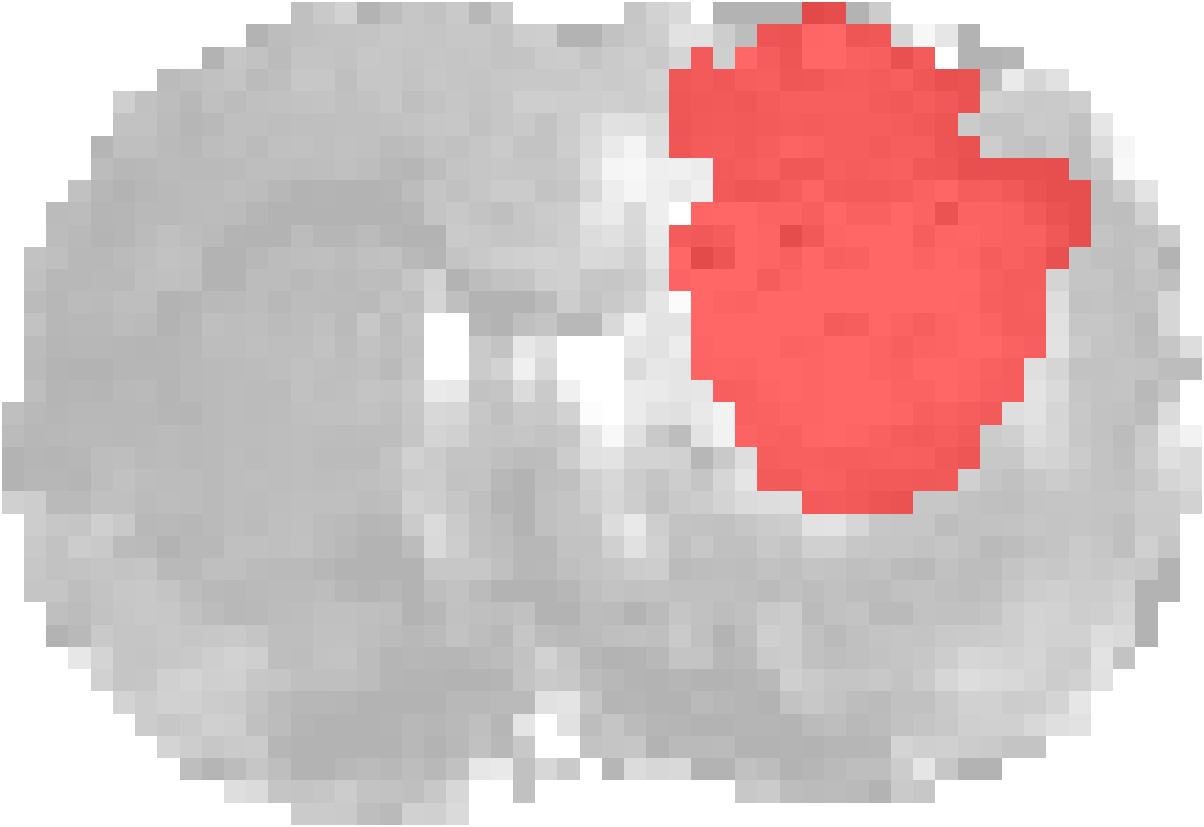} & 
            \includegraphics[width=0.15\textwidth]{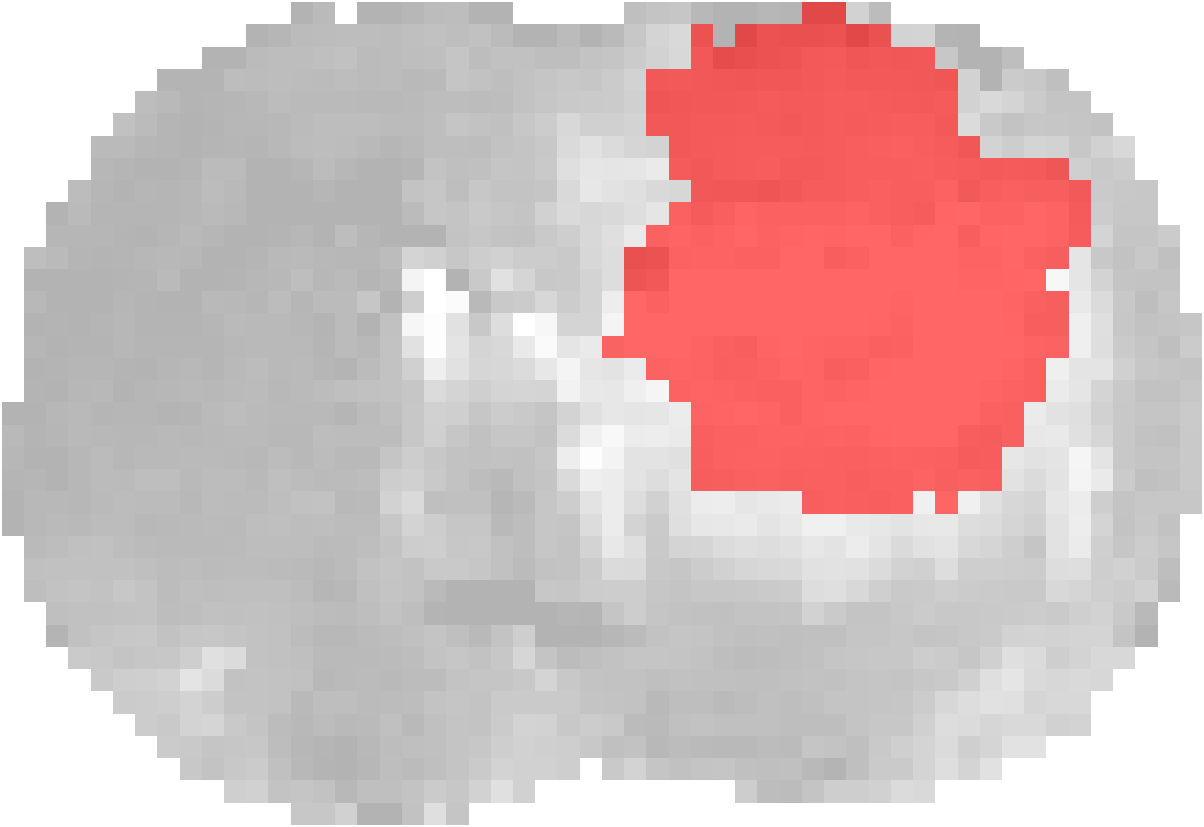} & 
            \includegraphics[width=0.15\textwidth]{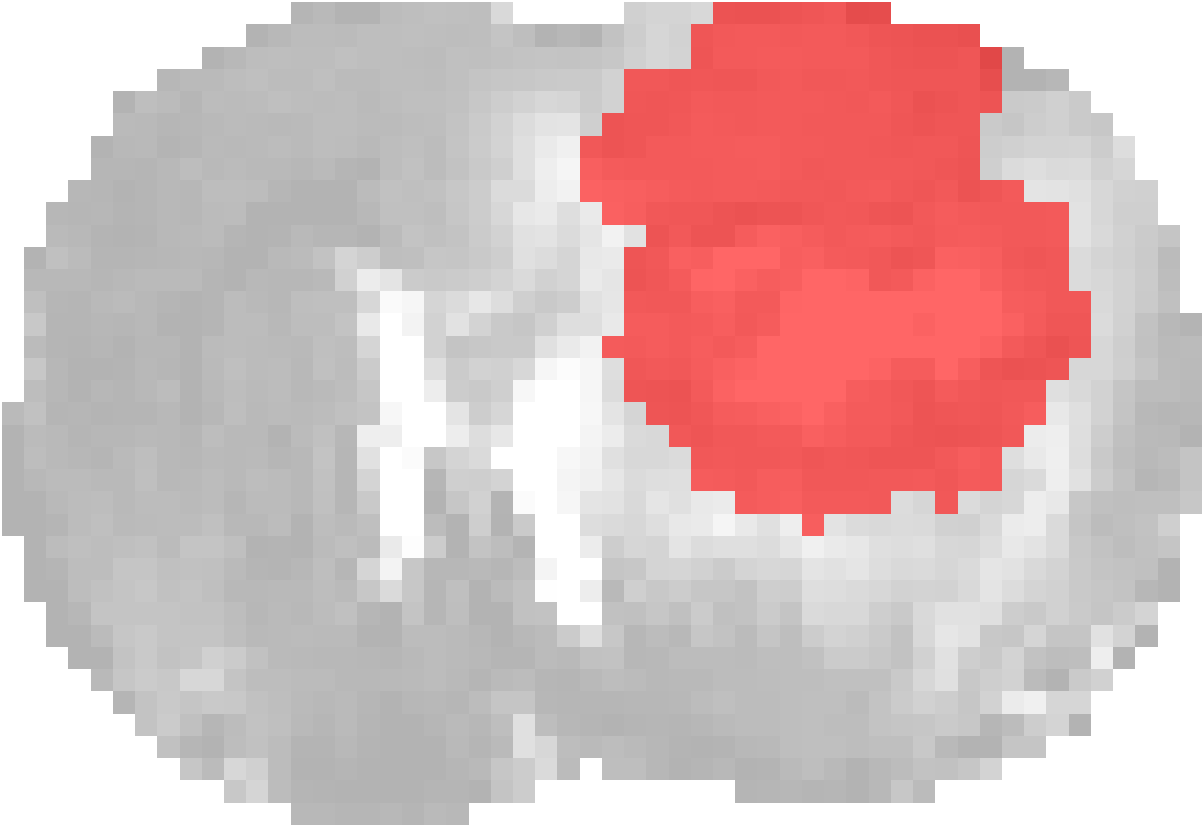} & 
            \includegraphics[width=0.15\textwidth]{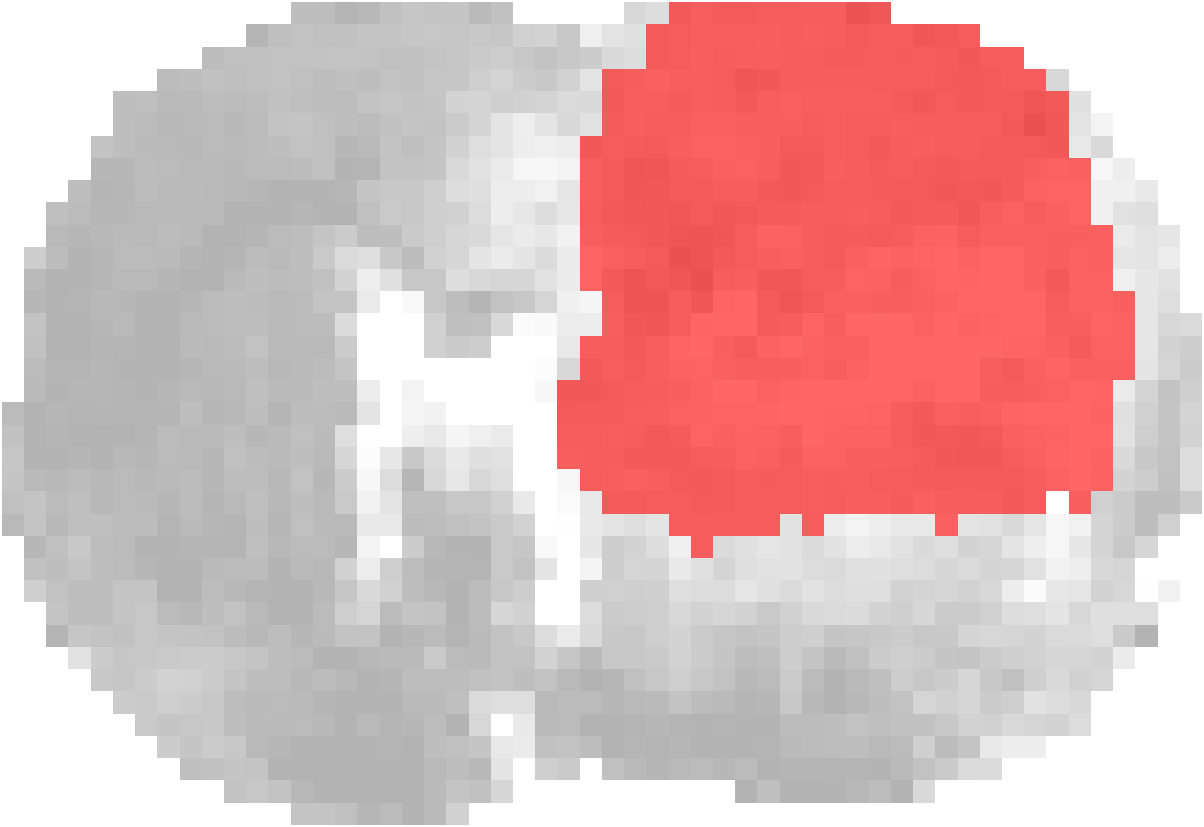} \\ [4pt] 
            
            \includegraphics[width=0.152\textwidth]{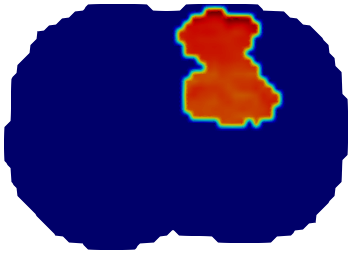} & 
            \includegraphics[width=0.152\textwidth]{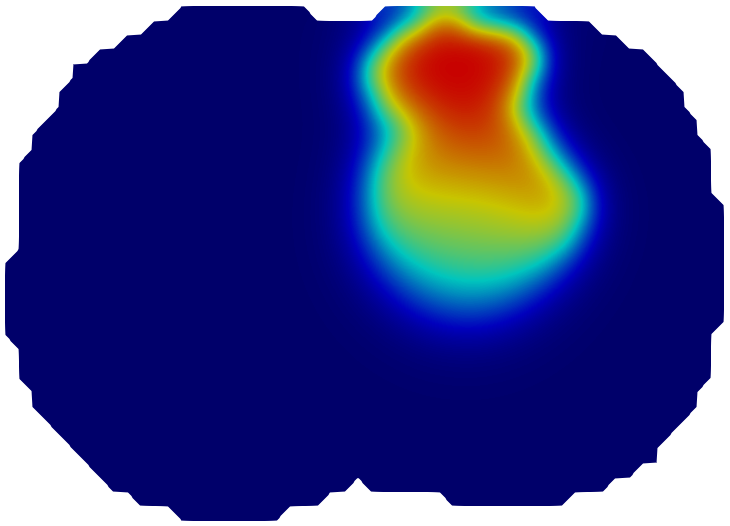} & 
            {\fboxrule=1.5pt\fcolorbox{green!60}{white}{\includegraphics[width=0.152\textwidth]{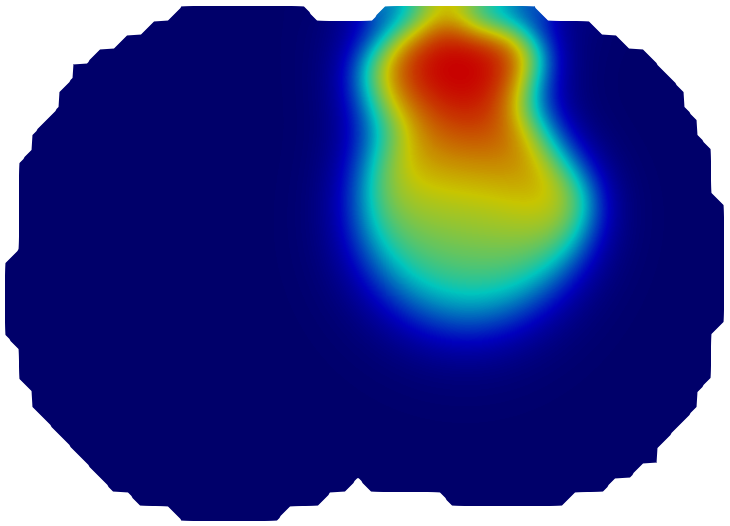}}} & 
            & & \\ [0pt]
            
            \includegraphics[width=0.152\textwidth]{Figures/Pred/sim-rat-I-Day12.png} & 
            \includegraphics[width=0.152\textwidth]{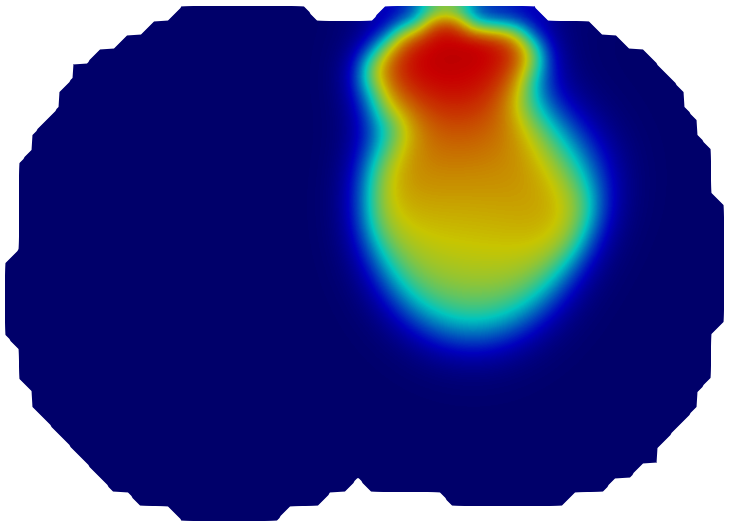} & 
            \includegraphics[width=0.152\textwidth]{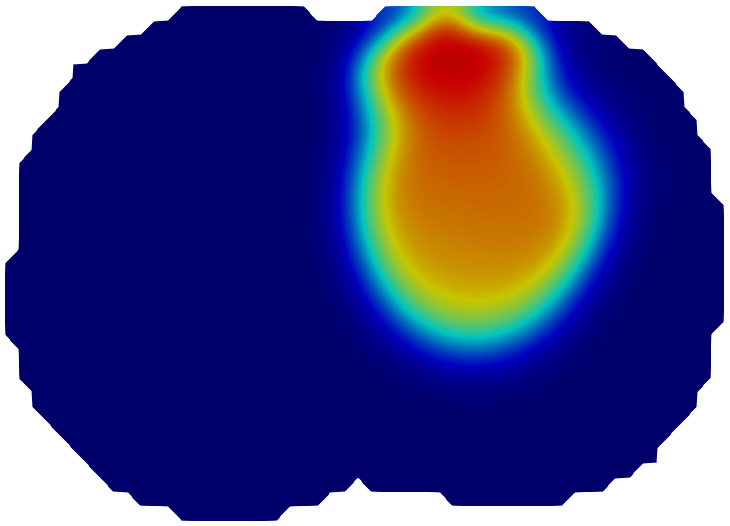} & 
            {\fboxrule=1.5pt\fcolorbox{green!60}{white}{\includegraphics[width=0.152\textwidth]{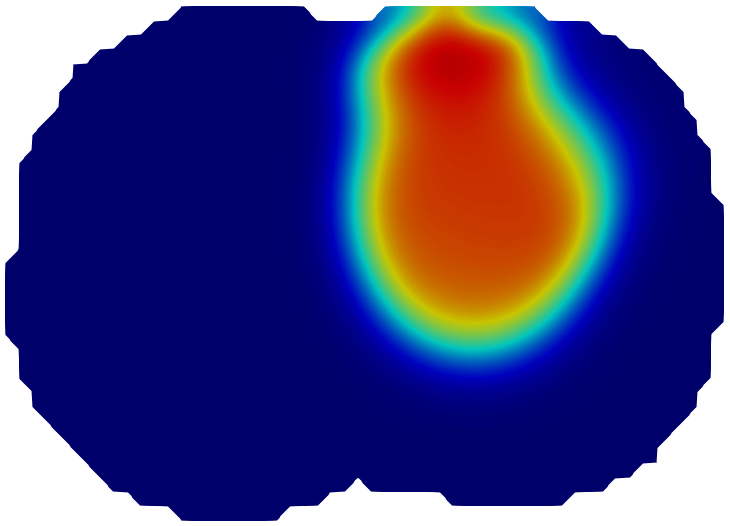}}} & 
            & \\ [0pt]
            
            \includegraphics[width=0.152\textwidth]{Figures/Pred/sim-rat-I-Day12.png} & 
            \includegraphics[width=0.152\textwidth]{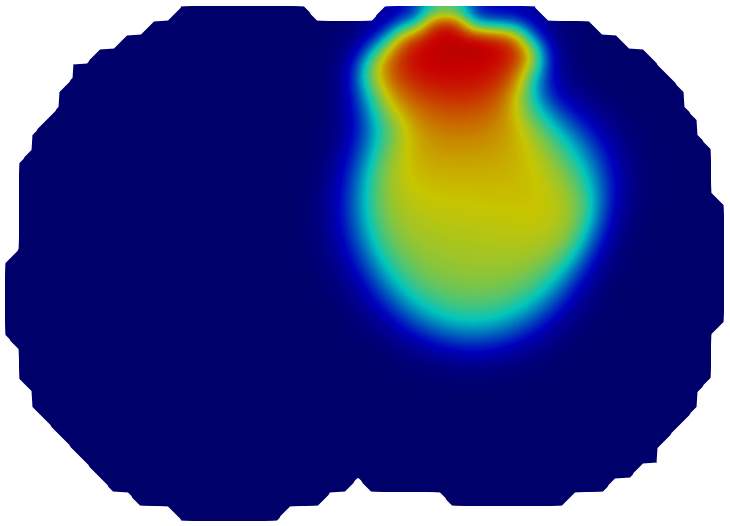} & 
            \includegraphics[width=0.152\textwidth]{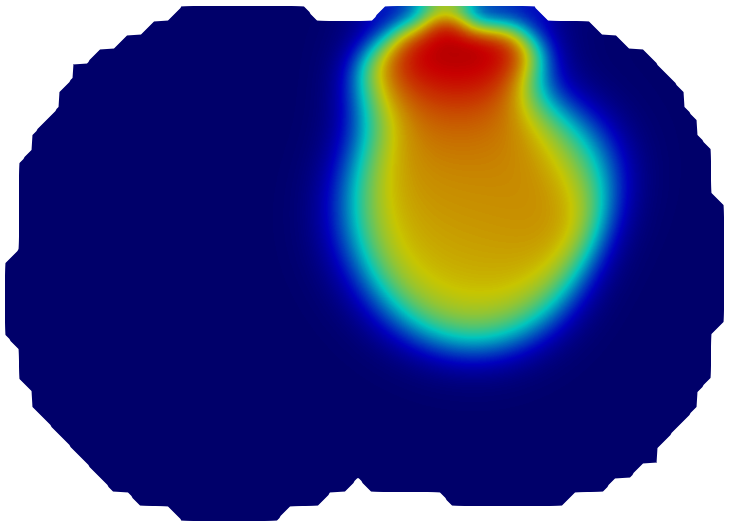} & 
            \includegraphics[width=0.152\textwidth]{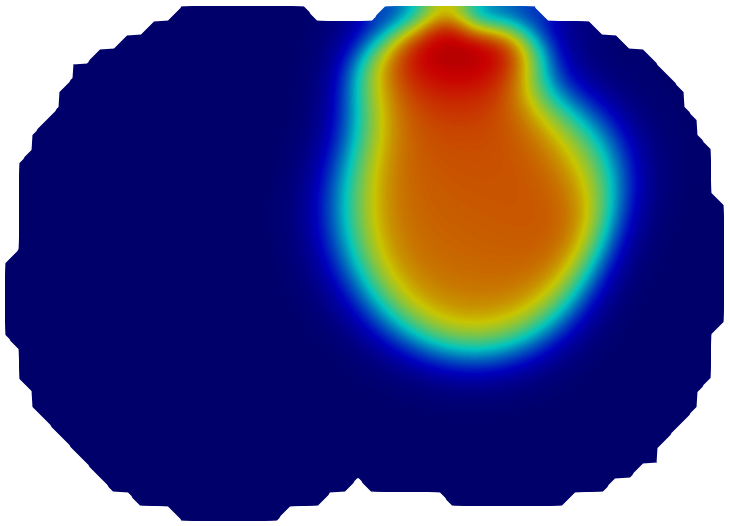} & 
            {\fboxrule=1.5pt\fcolorbox{green!60}{white}{\includegraphics[width=0.152\textwidth]{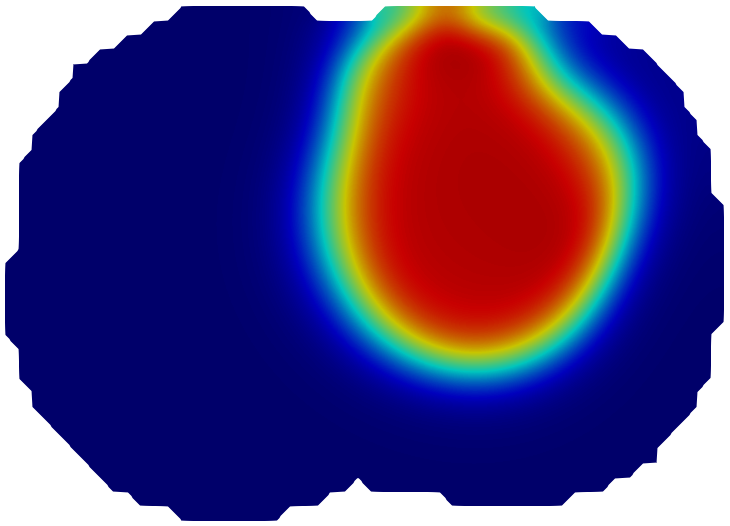}}} & \\ [0pt]
            
            \includegraphics[width=0.152\textwidth]{Figures/Pred/sim-rat-I-Day12.png} & 
            \includegraphics[width=0.152\textwidth]{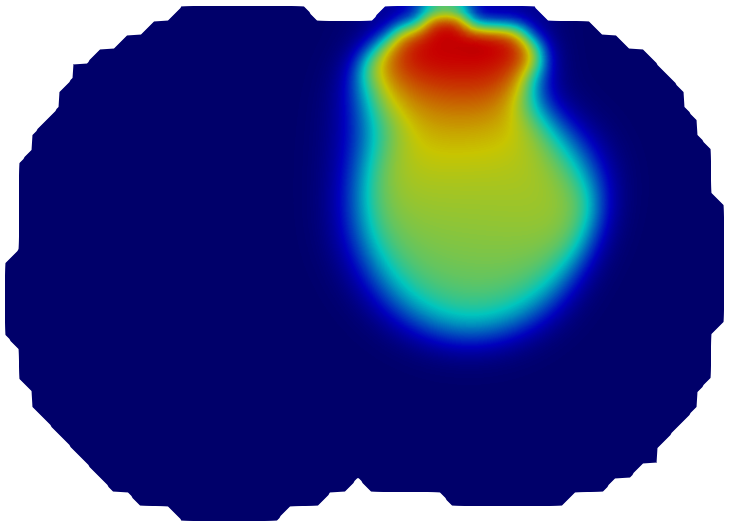} & 
            \includegraphics[width=0.152\textwidth]{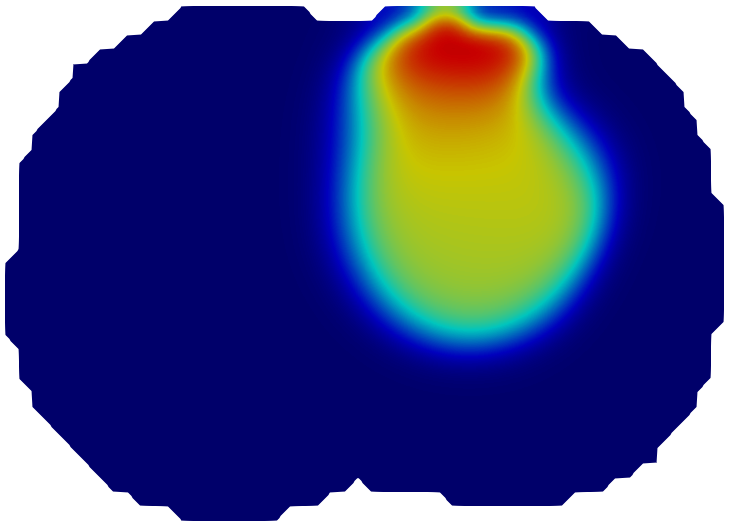} & 
            \includegraphics[width=0.152\textwidth]{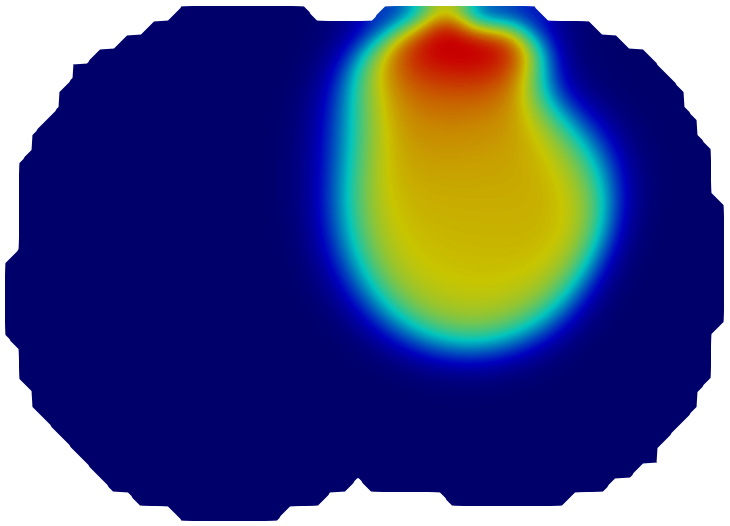} & 
            \includegraphics[width=0.152\textwidth]{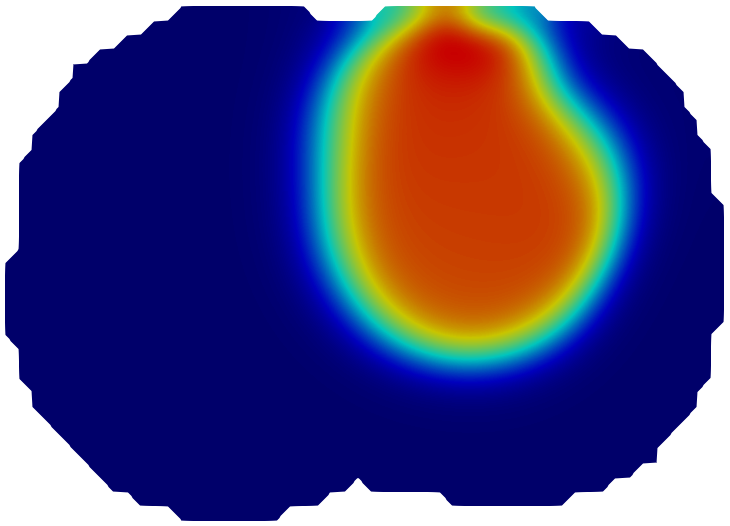} & 
            {\fboxrule=1.5pt\fcolorbox{green!60}{white}{\includegraphics[width=0.152\textwidth]{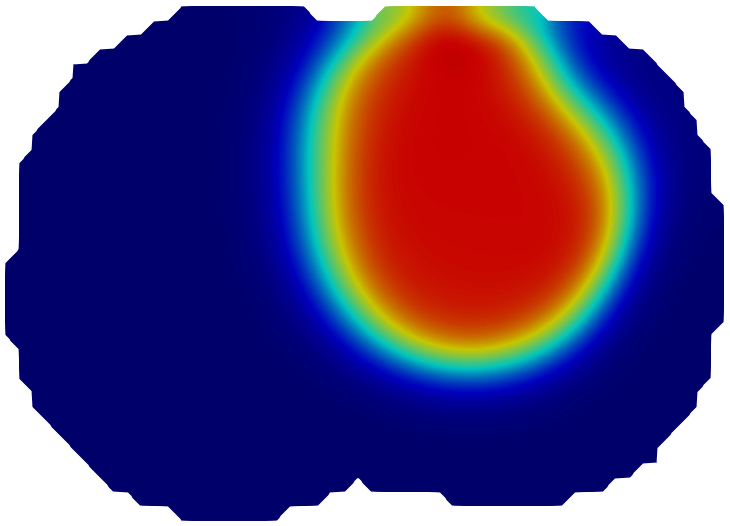}}} \\
        \end{tabular} & 
        
        \makebox[0pt][l]{
            \hspace{0.1pt}
            \raisebox{-4.6cm}{
                \includegraphics[width=.95cm, height=6.8cm, keepaspectratio=false]{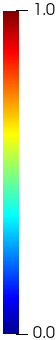}
            }
            \hspace{-20pt}
            \raisebox{-3cm}{
                \rotatebox{90}{\small Tumor Volume Fraction}
            }}
        
    \end{tabular}
    \vspace{-0.1in}
    \caption{
    Sequential one-scan-ahead prediction for Rat~I at MAP parameters
    \blue{
    using the hyperelastic mechanically coupled tumor growth model defined in \eqref{eq:rd_PDE_REF} and \eqref{eq:mech_PDE_REF}.
    }
    Top row: MRI-derived tumor volume fraction.
    Bottom rows: model predictions obtained by assimilating data up to~$t_k$ and predicting the subsequent scan at~$t_{k+1}$
    (boxed panels). Dice values between
    the predicted and MRI-derived tumor at Day~15, Day~16, Day~19 and Day~21 are 0.85, 0.91, 0.9, and 0.92 respectively.}
    \label{fig:ratI_MRI-MAP}
    \vspace{-0.2in}
\end{figure}

\subsection{Sequential Bayesian inference of biomechanical tumor growth model}\label{sec:results_hyperelastic}
Figure~\ref{fig:eigen_analysis} demonstrates the scalability of the
LA solution of the high-dimensional spatial inference
in Rat~III MRI. 
The rapid and mesh-independent decay of the
prior-preconditioned data misfit Hessian spectrum indicates that the
effective dimension of the inference problem is governed by the
information content of the imaging data rather than mesh resolution.
Leading eigenvectors localize in regions influenced by tumor growth,
confirming that data primarily inform parameters in mechanically active
domains. 
\blue{Based on the truncation error in spectral decay indicated in equation (S.5) of the \textit{Supplementary Information}, the leading 50 Hessian eigenmodes were retained in the low-rank posterior approximation for all animals. A rank-sensitivity check using 60 modes is reported in the \textit{Supplementary Information} and showed no change in posterior model plausibility to the reported precision.}

\begin{figure}[!ht]
    \centering
    \raggedright
    \renewcommand{\arraystretch}{0}
    \setlength{\fboxsep}{0pt} 
    \setlength{\tabcolsep}{1pt} 
    
    \begin{tabular}{@{}l@{}l@{}}
        
        \begin{tabular}{@{} *{6}{>{\centering\arraybackslash}p{0.148\textwidth}} @{}} 
            \colorbox{black}{\makebox[0.148\textwidth][c]{\vrule width 0pt height 0.4cm \textcolor{white}{\small Day 10}}} & 
            \colorbox{black}{\makebox[0.148\textwidth][c]{\vrule width 0pt height 0.4cm \textcolor{white}{\small Day 12}}} & 
            \colorbox{black}{\makebox[0.148\textwidth][c]{\vrule width 0pt height 0.4cm \textcolor{white}{\small Day 14}}} & 
            \colorbox{black}{\makebox[0.148\textwidth][c]{\vrule width 0pt height 0.4cm \textcolor{white}{\small Day 15}}} & 
            \colorbox{black}{\makebox[0.148\textwidth][c]{\vrule width 0pt height 0.4cm \textcolor{white}{\small Day 16}}} & 
            \colorbox{black}{\makebox[0.148\textwidth][c]{\vrule width 0pt height 0.4cm \textcolor{white}{\small Day 18}}} \\ [3pt]
            
            \includegraphics[width=0.15\textwidth]{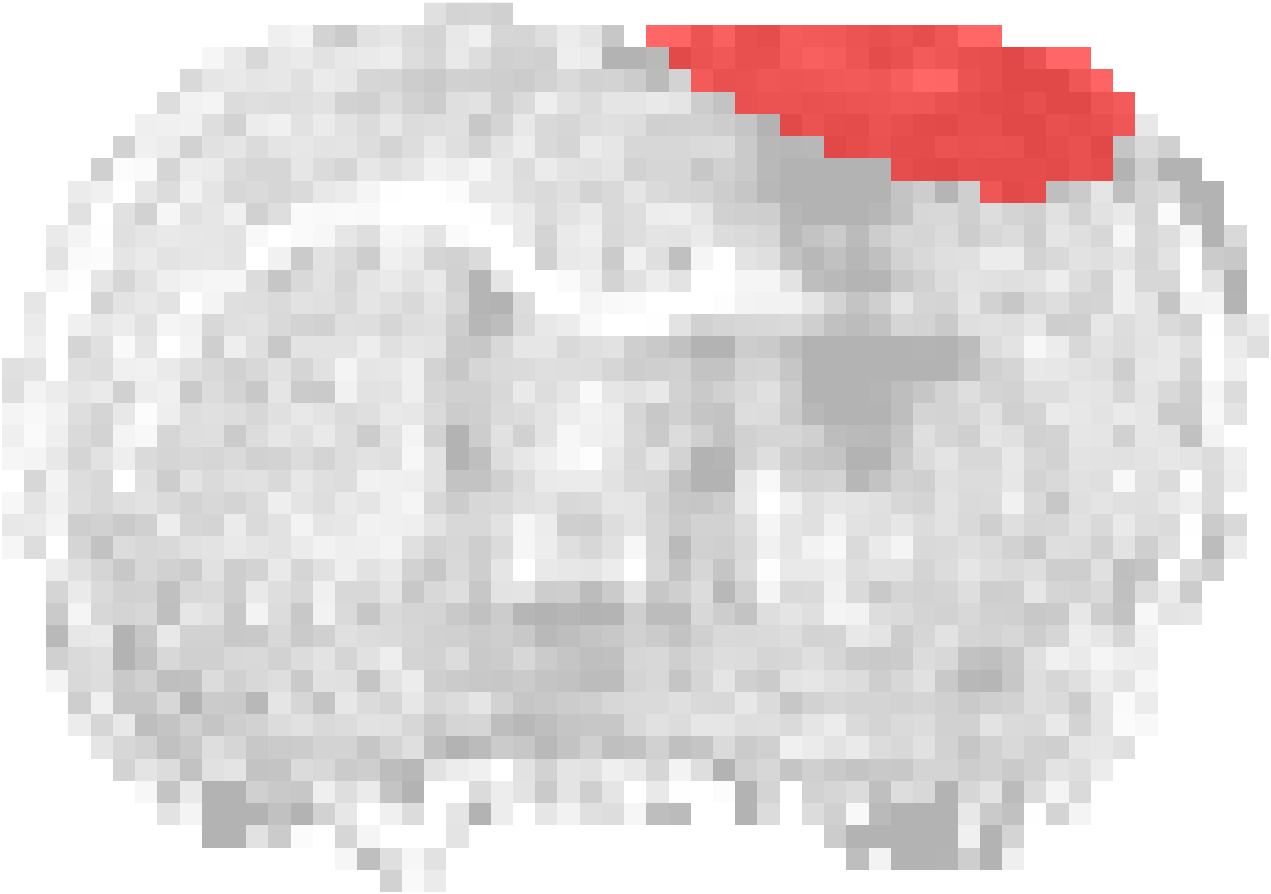} & 
            \includegraphics[width=0.15\textwidth]{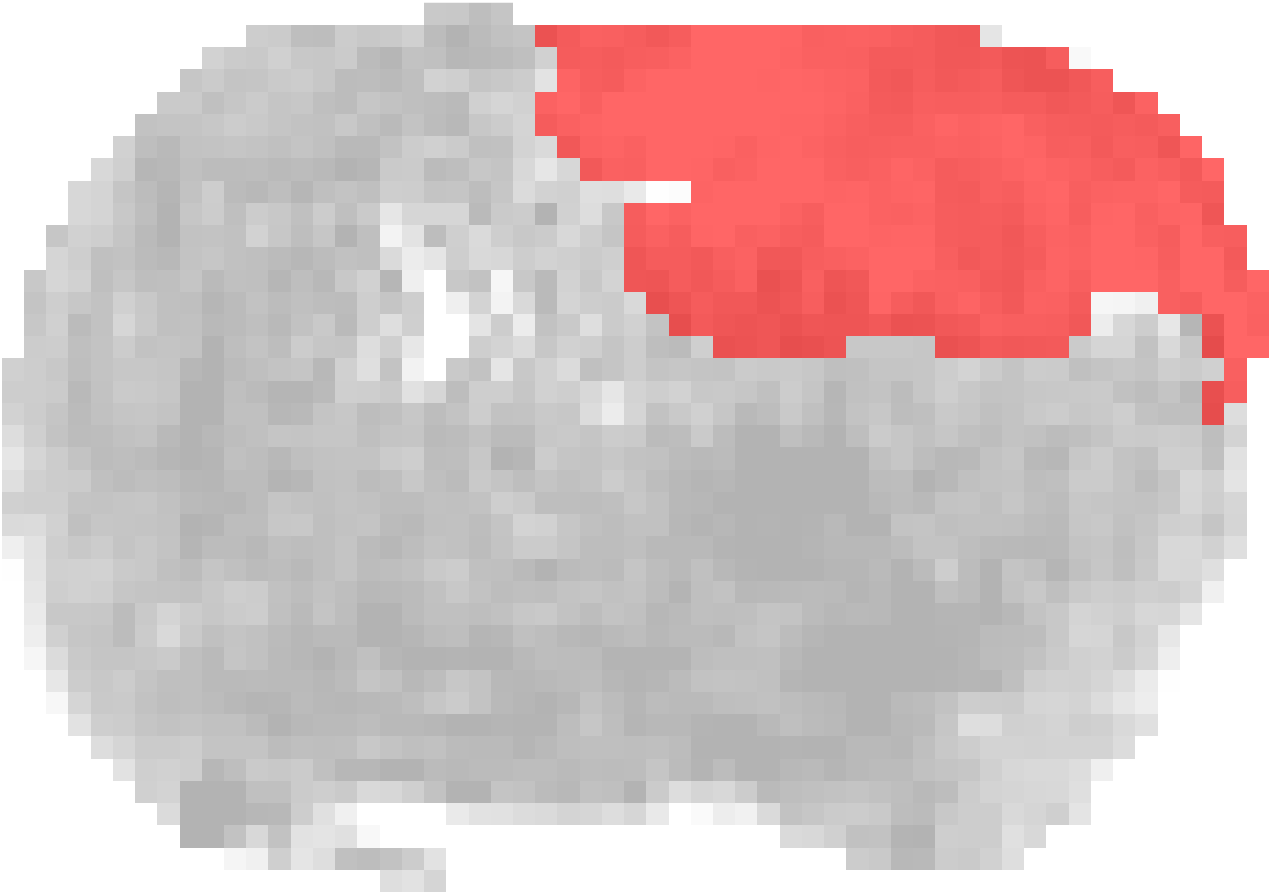} & 
            \includegraphics[width=0.15\textwidth]{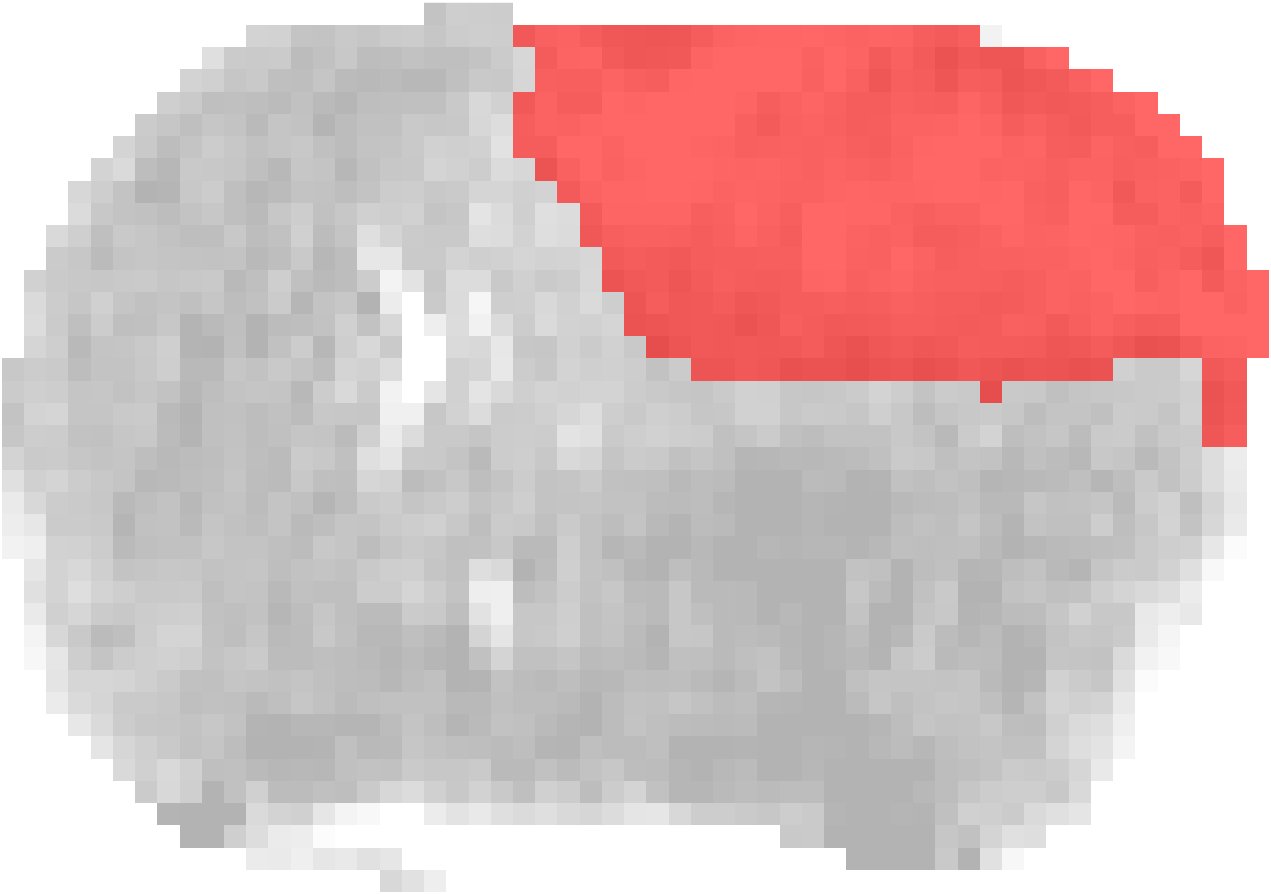} & 
            \includegraphics[width=0.15\textwidth]{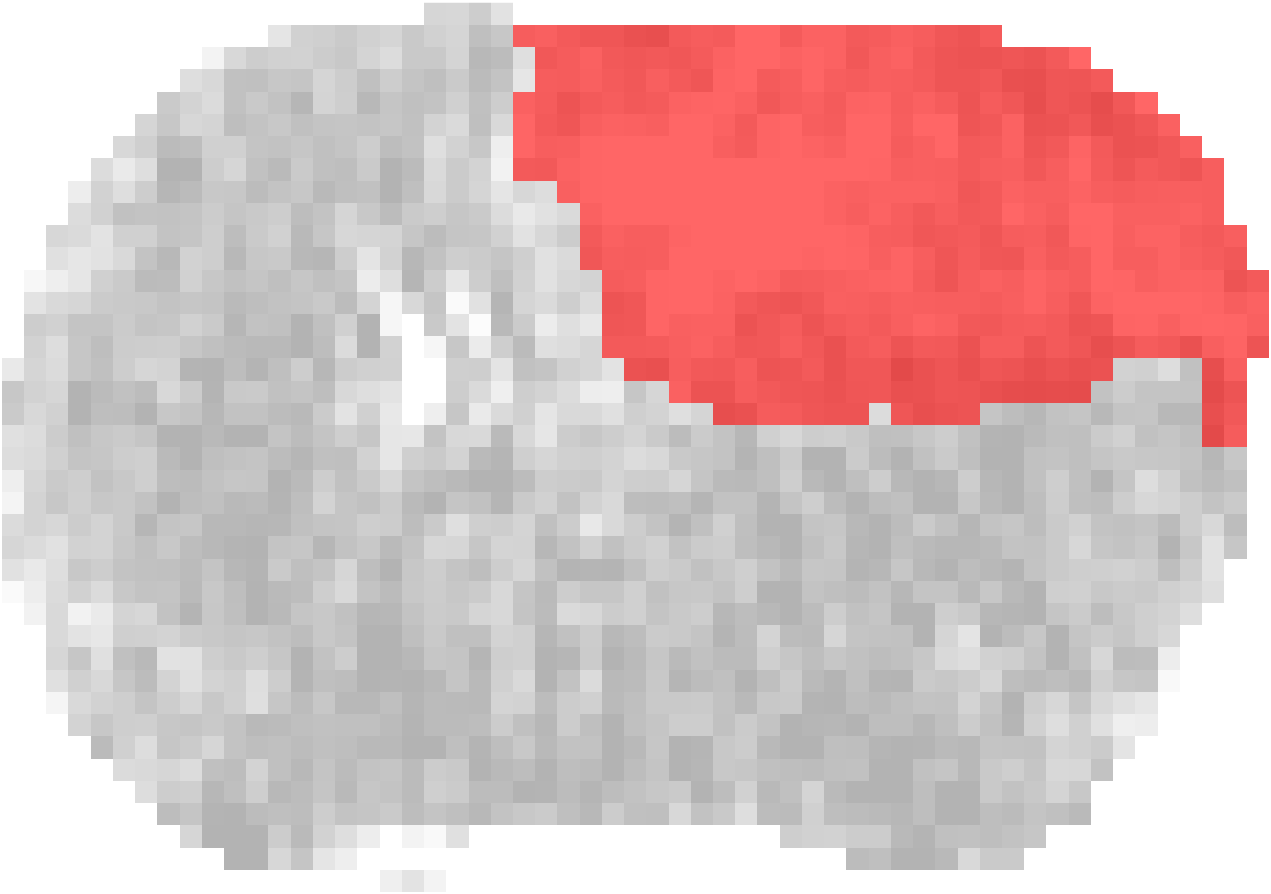} & 
            \includegraphics[width=0.15\textwidth]{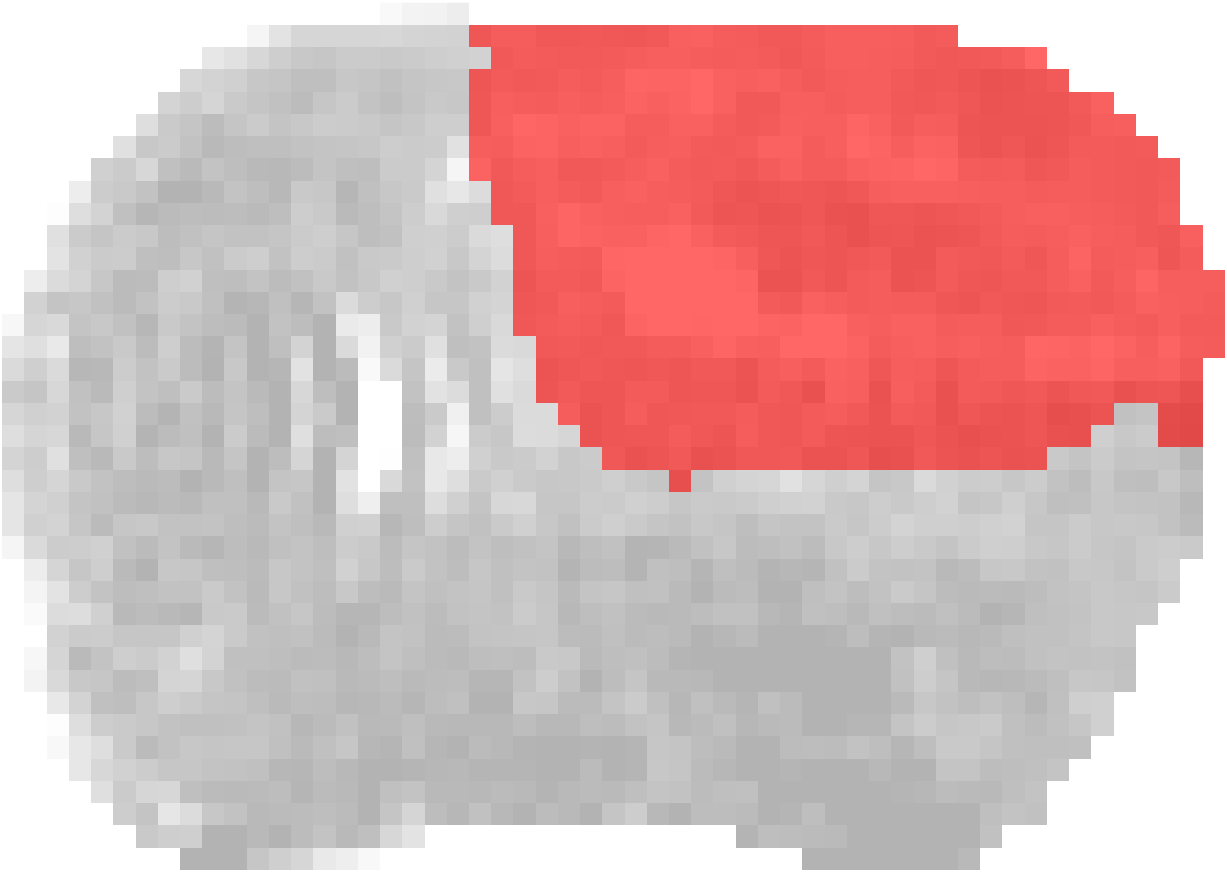} & 
            \includegraphics[width=0.15\textwidth]{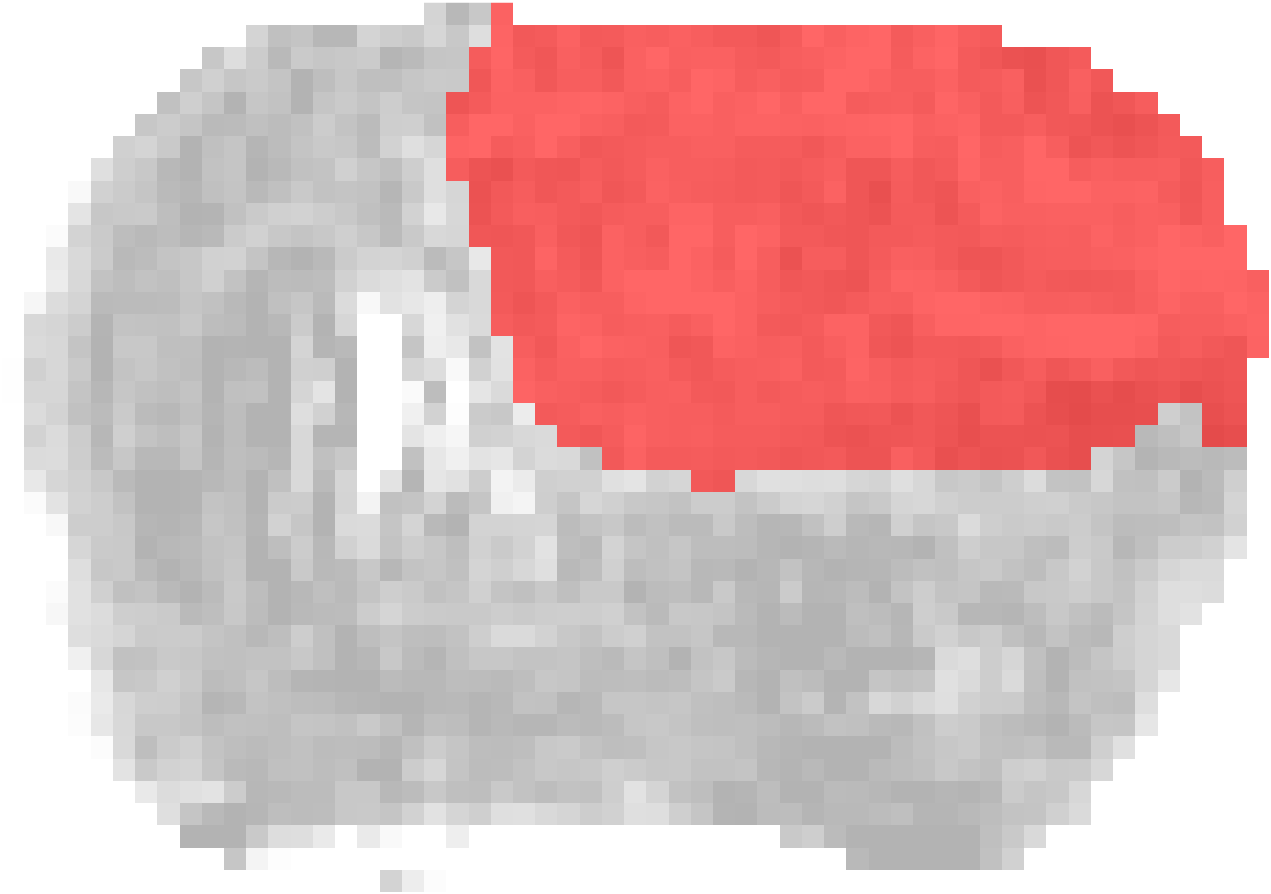} \\ [4pt] 
            
            \includegraphics[width=0.152\textwidth]{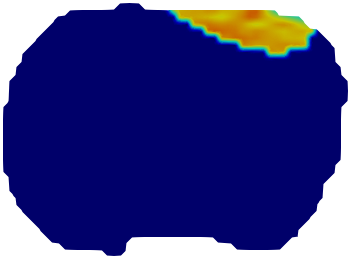} & 
            \includegraphics[width=0.152\textwidth]{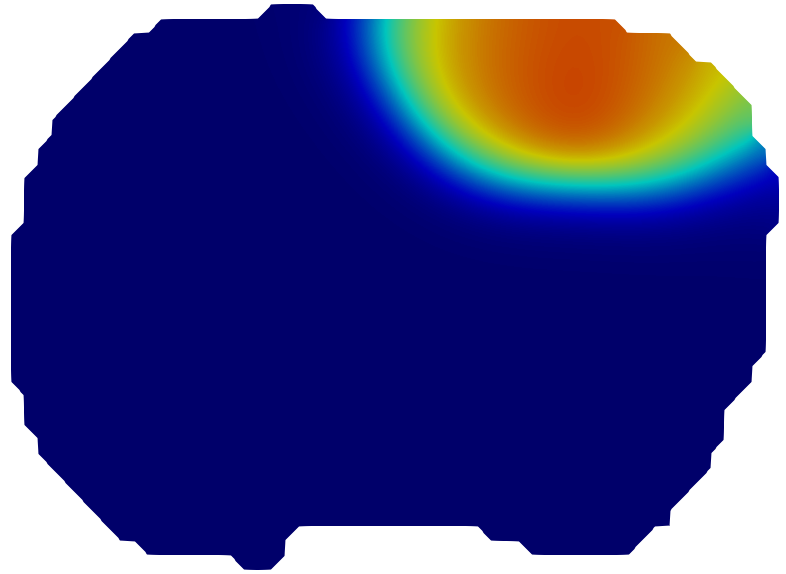} & 
            {\fboxrule=1.5pt\fcolorbox{green!60}{white}{\includegraphics[width=0.152\textwidth]{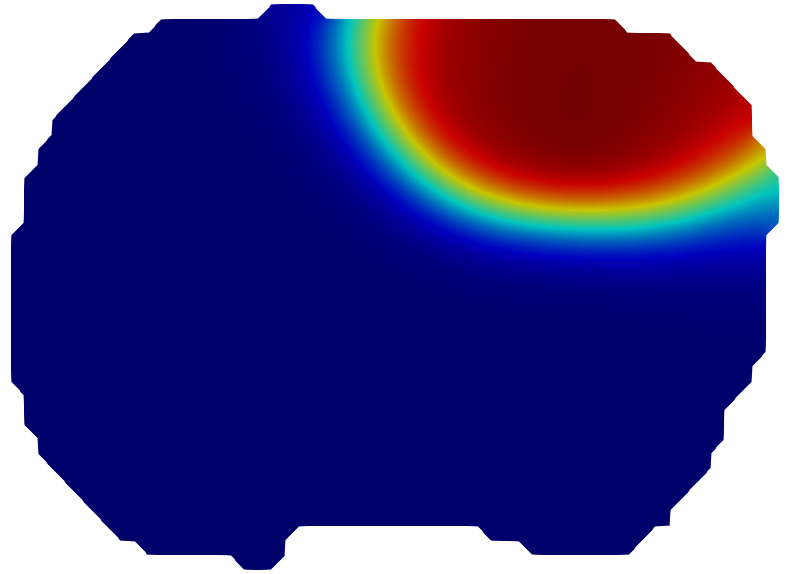}}} & 
            & & \\ [0pt]
            
            \includegraphics[width=0.152\textwidth]{Figures/Pred/sim-rat-II-Day10.png} & 
            \includegraphics[width=0.152\textwidth]{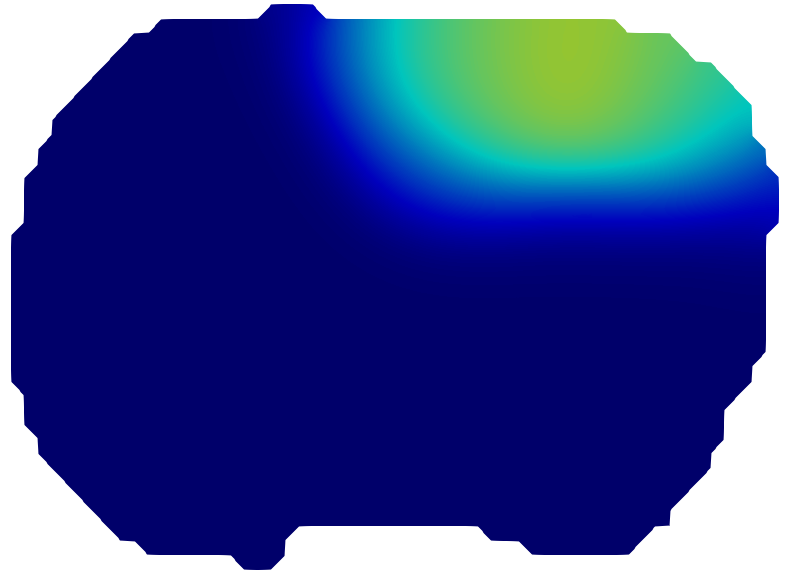} & 
            \includegraphics[width=0.152\textwidth]{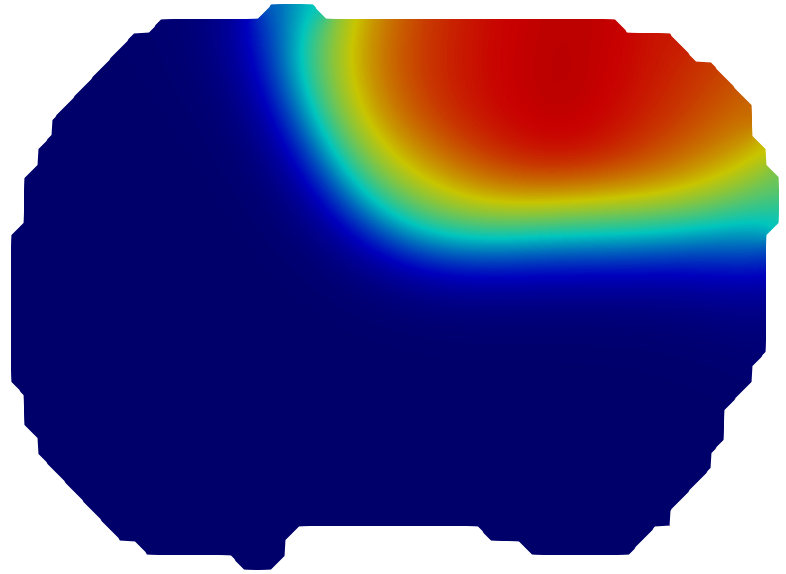} & 
            {\fboxrule=1.5pt\fcolorbox{green!60}{white}{\includegraphics[width=0.152\textwidth]{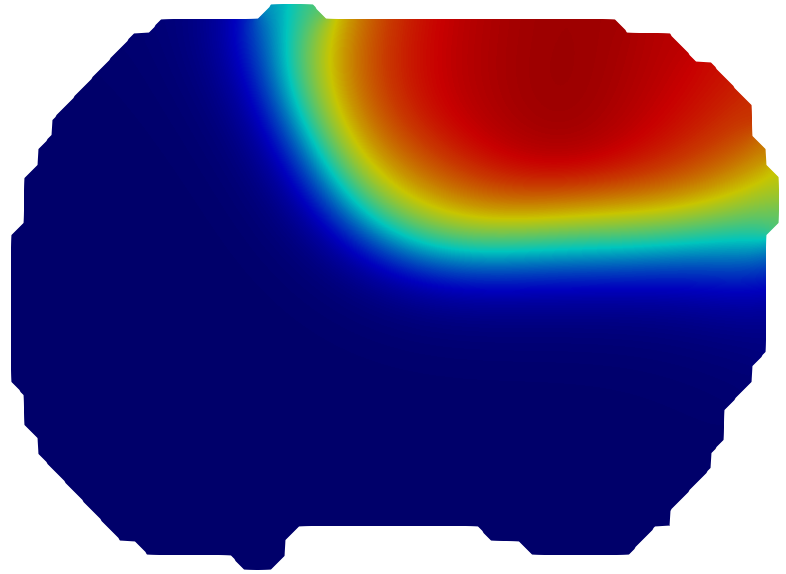}}} & 
            & \\ [0pt]
            
            \includegraphics[width=0.152\textwidth]{Figures/Pred/sim-rat-II-Day10.png} & 
            \includegraphics[width=0.152\textwidth]{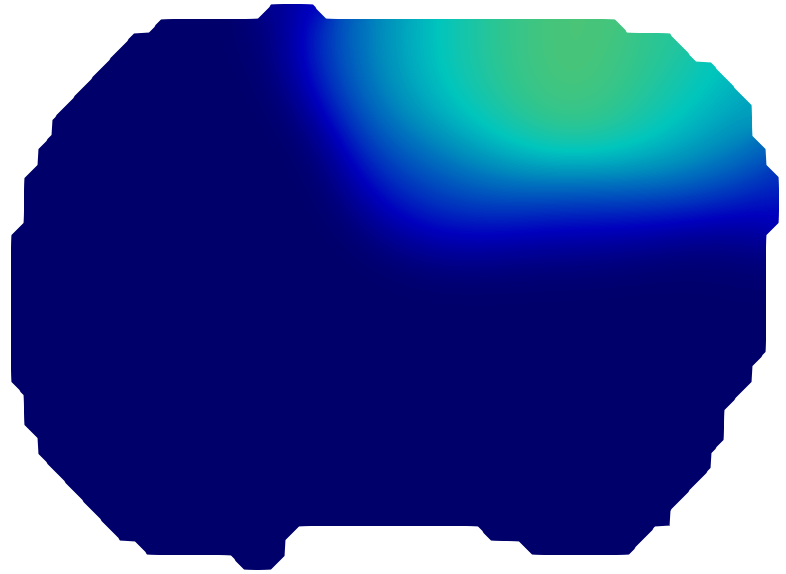} & 
            \includegraphics[width=0.152\textwidth]{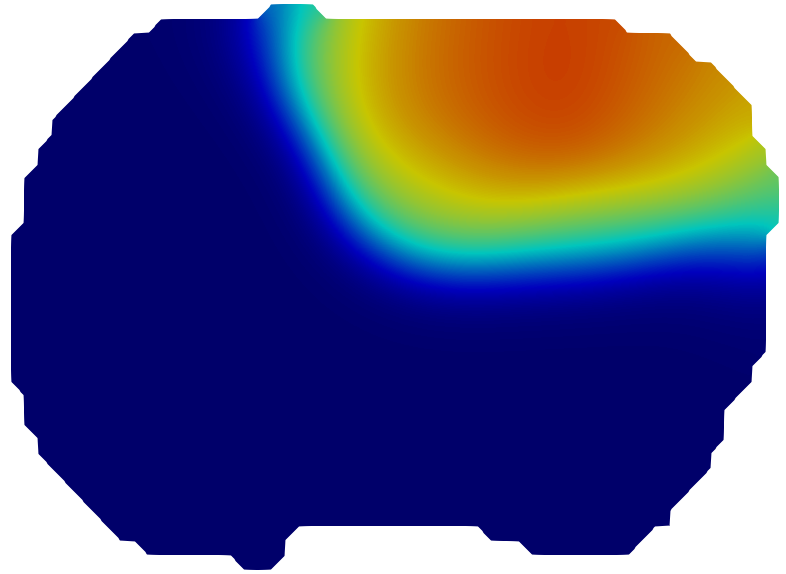} & 
            \includegraphics[width=0.152\textwidth]{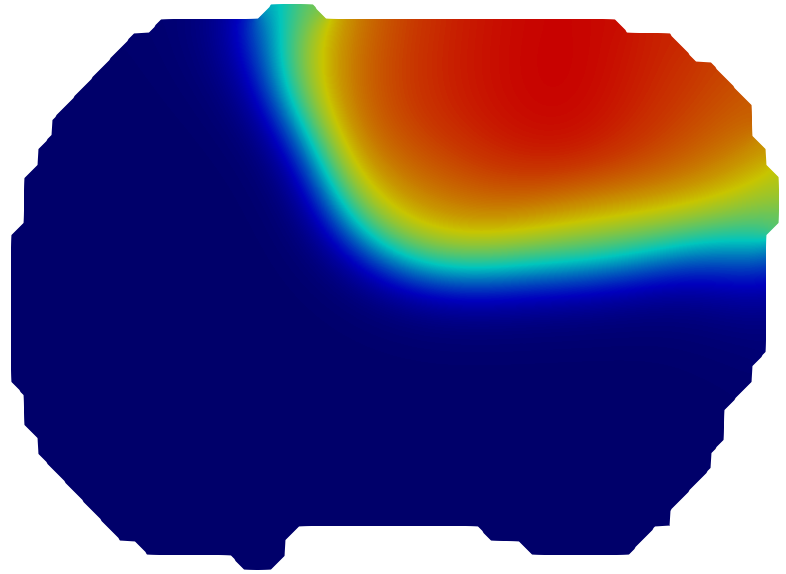} & 
            {\fboxrule=1.5pt\fcolorbox{green!60}{white}{\includegraphics[width=0.152\textwidth]{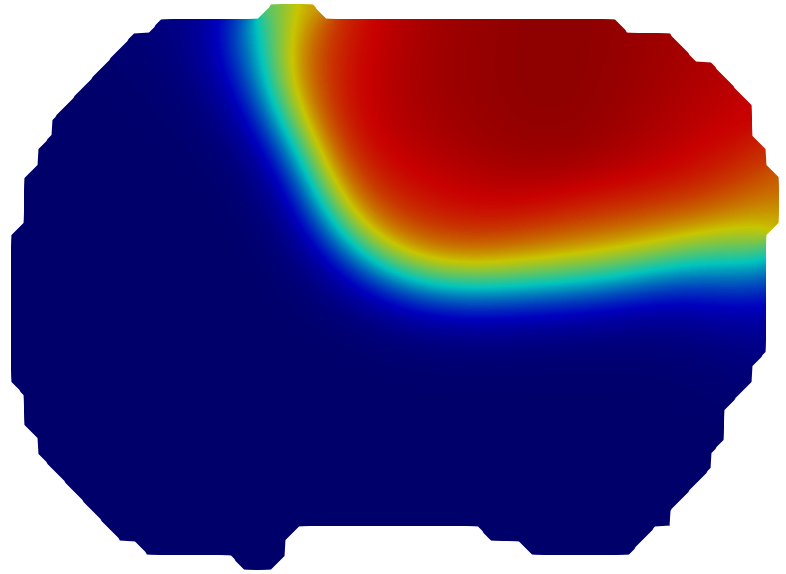}}} & \\ [0pt]
            
            \includegraphics[width=0.152\textwidth]{Figures/Pred/sim-rat-II-Day10.png} & 
            \includegraphics[width=0.152\textwidth]{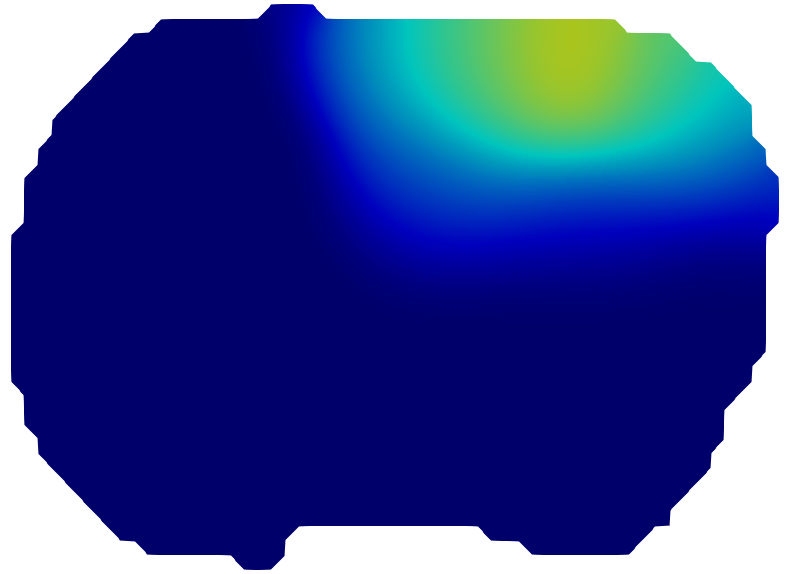} & 
            \includegraphics[width=0.152\textwidth]{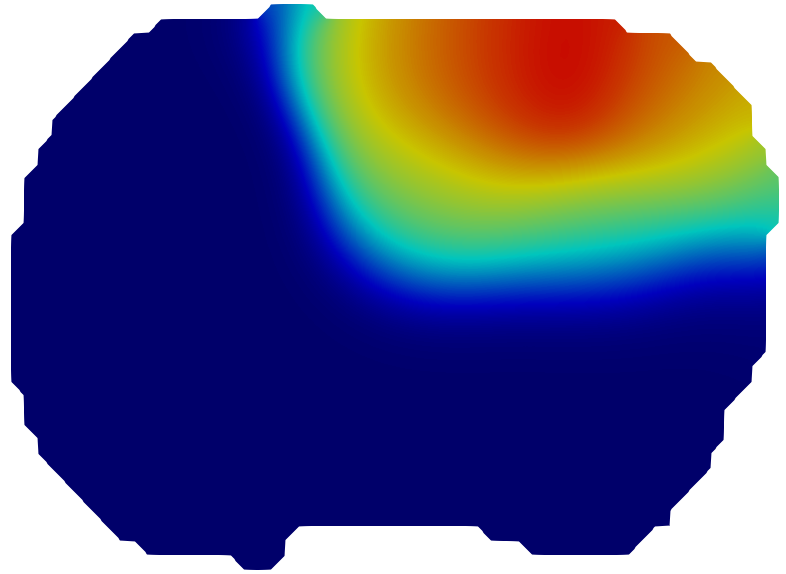} & 
            \includegraphics[width=0.152\textwidth]{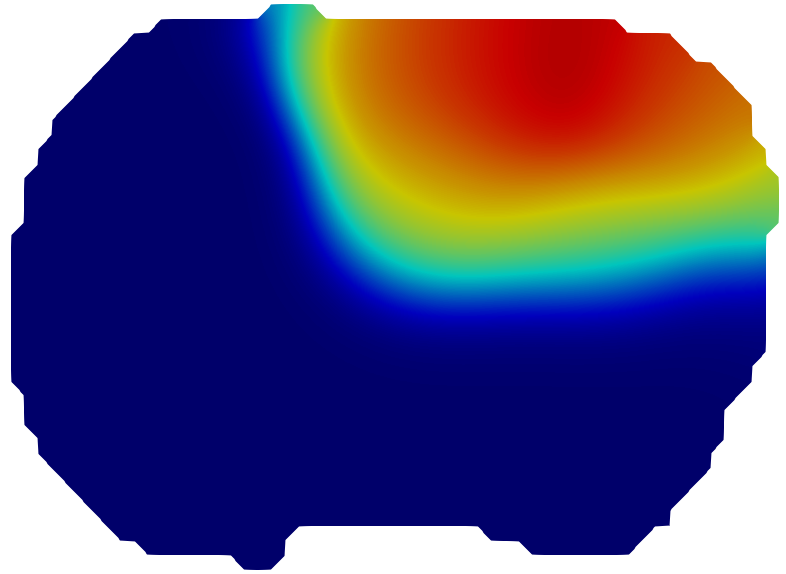} & 
            \includegraphics[width=0.152\textwidth]{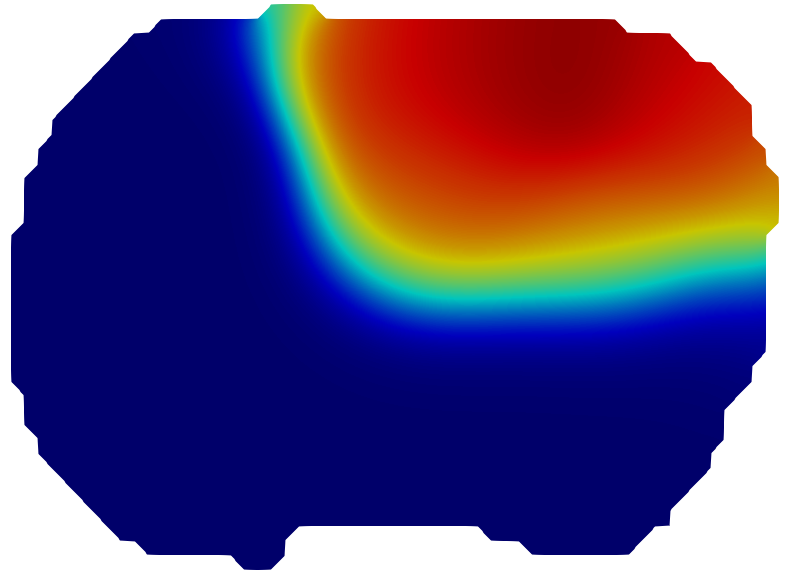} & 
            {\fboxrule=1.5pt\fcolorbox{green!60}{white}{\includegraphics[width=0.152\textwidth]{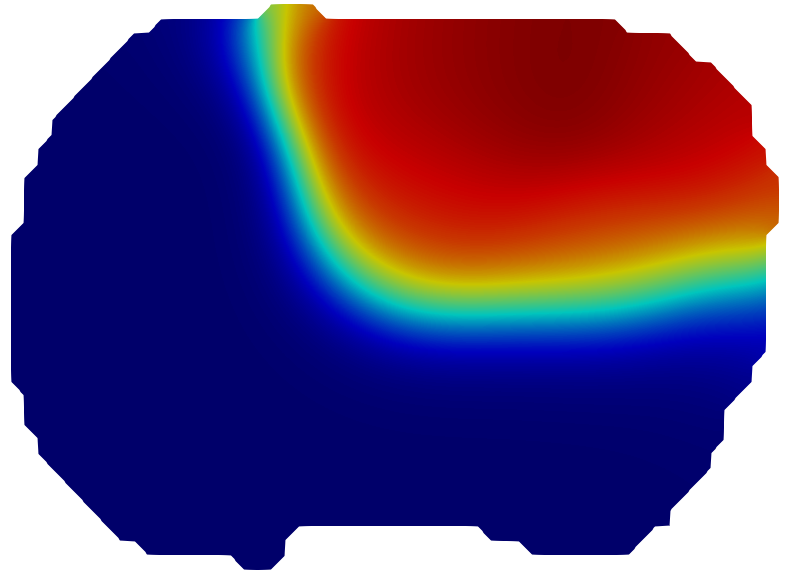}}} \\
        \end{tabular} & 
        
        \makebox[0pt][l]{
            \hspace{.1pt}
            \raisebox{-4.6cm}{
                \includegraphics[width=.95cm, height=6.8cm, keepaspectratio=false]{Figures/Pred/color-bar.png}
            }
            \hspace{-20pt}
            \raisebox{-3cm}{
                \rotatebox{90}{\small Tumor Volume Fraction}
            }}
        
    \end{tabular}
    \vspace{-0.1in}
    \caption{
    Sequential one-scan-ahead prediction for Rat-II at MAP parameters
    \blue{
    using the hyperelastic mechanically coupled tumor growth model defined in \eqref{eq:rd_PDE_REF} and \eqref{eq:mech_PDE_REF}.
    }.
    Top row: MRI-derived tumor volume fraction.
    Bottom rows: model predictions obtained by assimilating data up to
    ~$t_k$ and predicting the subsequent scan at ~$t_{k+1}$
    (boxed panels). Dice values between
    the predicted and MRI-derived tumor at Day~14, Day~15, Day~16 and Day~18 are 0.88, 0.96, 0.95, and 0.95 respectively.}
    \label{fig:ratII_MRI-MAP}
    \vspace{-0.2in}
\end{figure}

\begin{figure}[!ht]
    \centering
    \raggedright
    \renewcommand{\arraystretch}{0}
    \setlength{\fboxsep}{0pt} 
    \setlength{\tabcolsep}{2pt} 
    
    \begin{tabular}{@{}l@{}l@{}}
        
        \begin{tabular}{@{} *{6}{>{\centering\arraybackslash}p{0.144\textwidth}} @{}} 
            \colorbox{black}{\makebox[0.146\textwidth][c]{\vrule width 0pt height 0.4cm \textcolor{white}{\small Day 10}}} & 
            \colorbox{black}{\makebox[0.146\textwidth][c]{\vrule width 0pt height 0.4cm \textcolor{white}{\small Day 12}}} & 
            \colorbox{black}{\makebox[0.146\textwidth][c]{\vrule width 0pt height 0.4cm \textcolor{white}{\small Day 14}}} & 
            \colorbox{black}{\makebox[0.146\textwidth][c]{\vrule width 0pt height 0.4cm \textcolor{white}{\small Day 15}}} & 
            \colorbox{black}{\makebox[0.146\textwidth][c]{\vrule width 0pt height 0.4cm \textcolor{white}{\small Day 16}}} & 
            \colorbox{black}{\makebox[0.146\textwidth][c]{\vrule width 0pt height 0.4cm \textcolor{white}{\small Day 19}}} \\ [3pt]
            
            \includegraphics[width=0.148\textwidth]{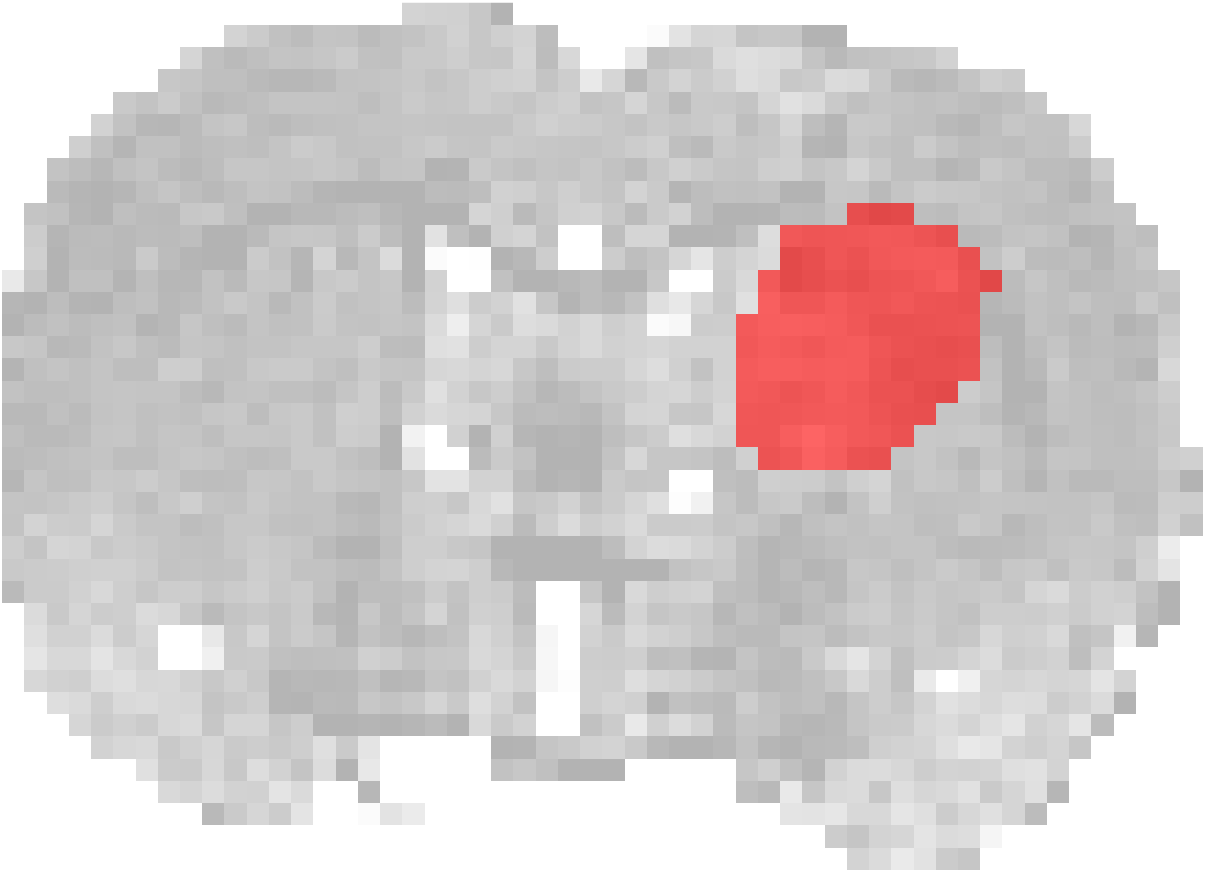} & 
            \includegraphics[width=0.148\textwidth]{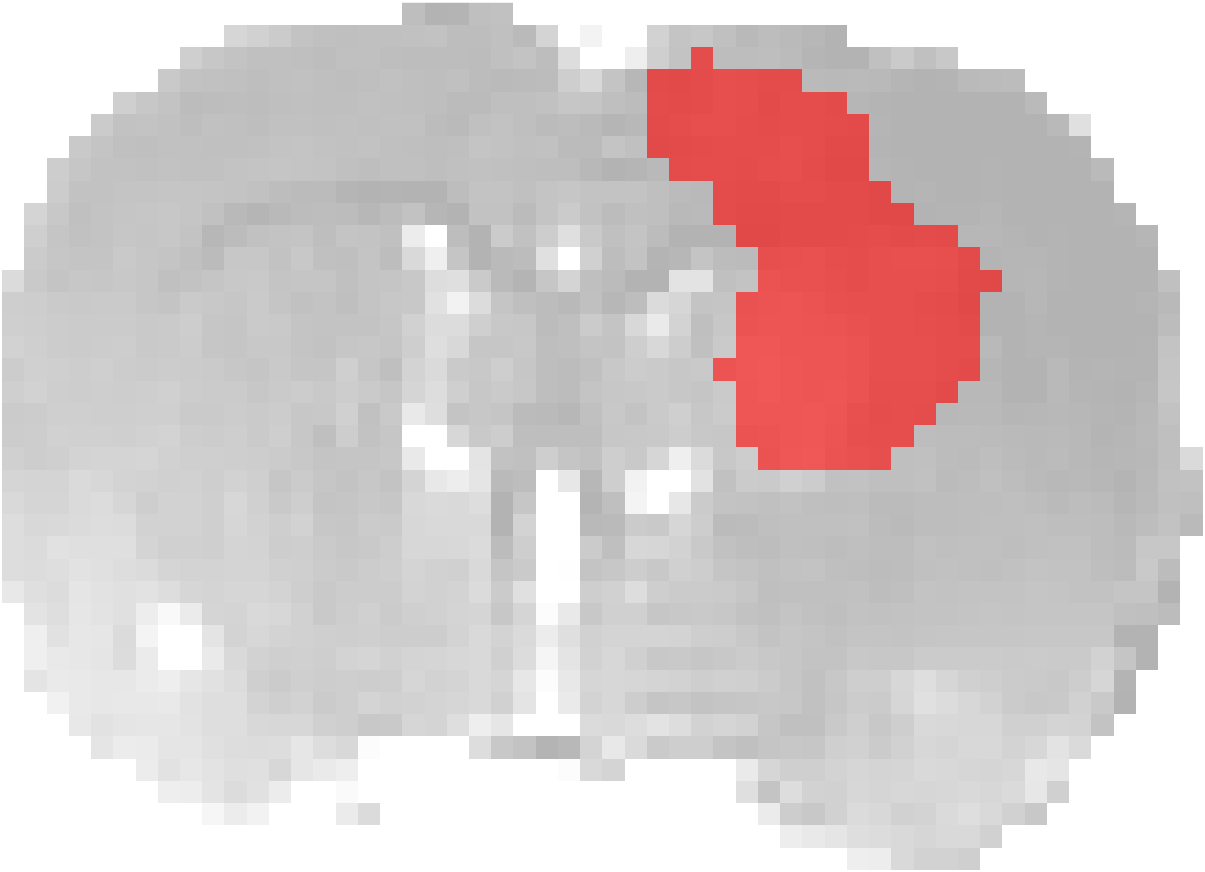} & 
            \includegraphics[width=0.148\textwidth]{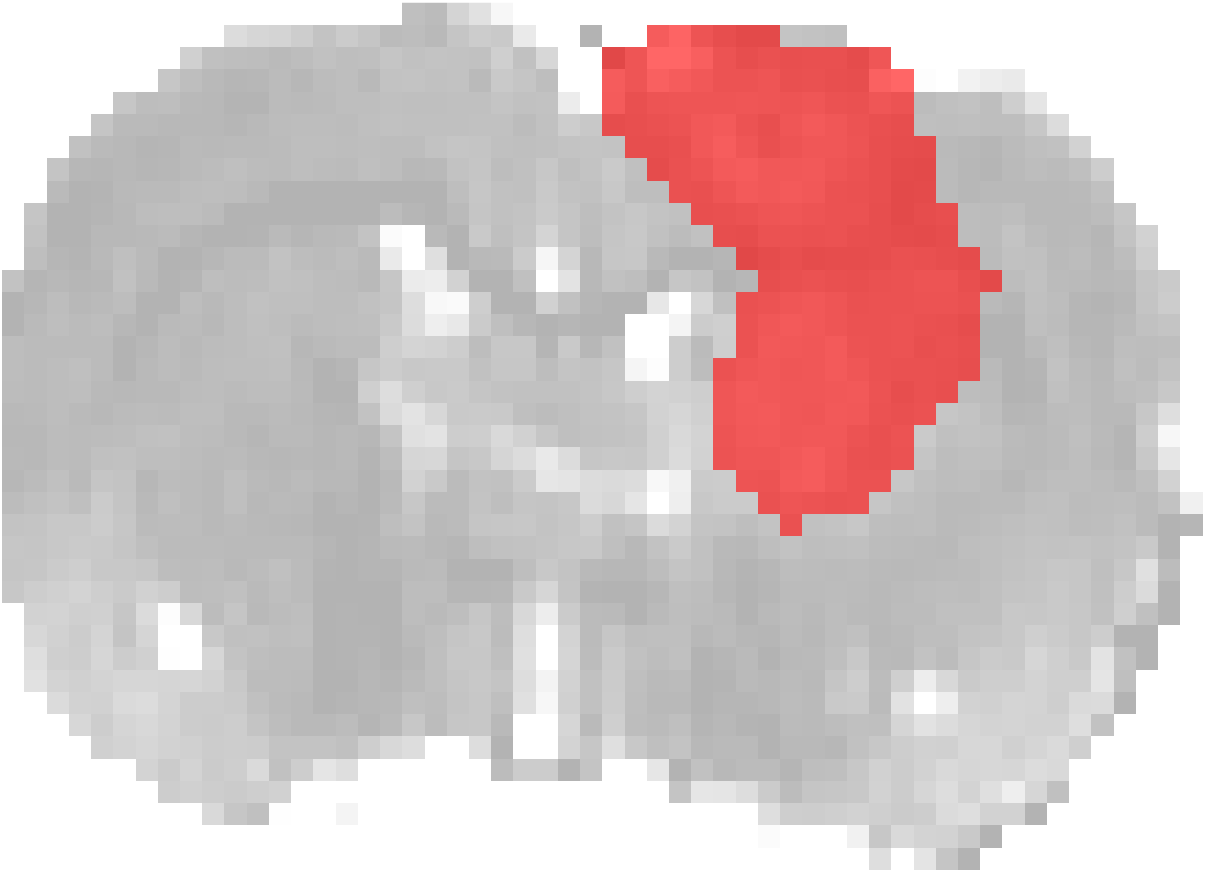} & 
            \includegraphics[width=0.148\textwidth]{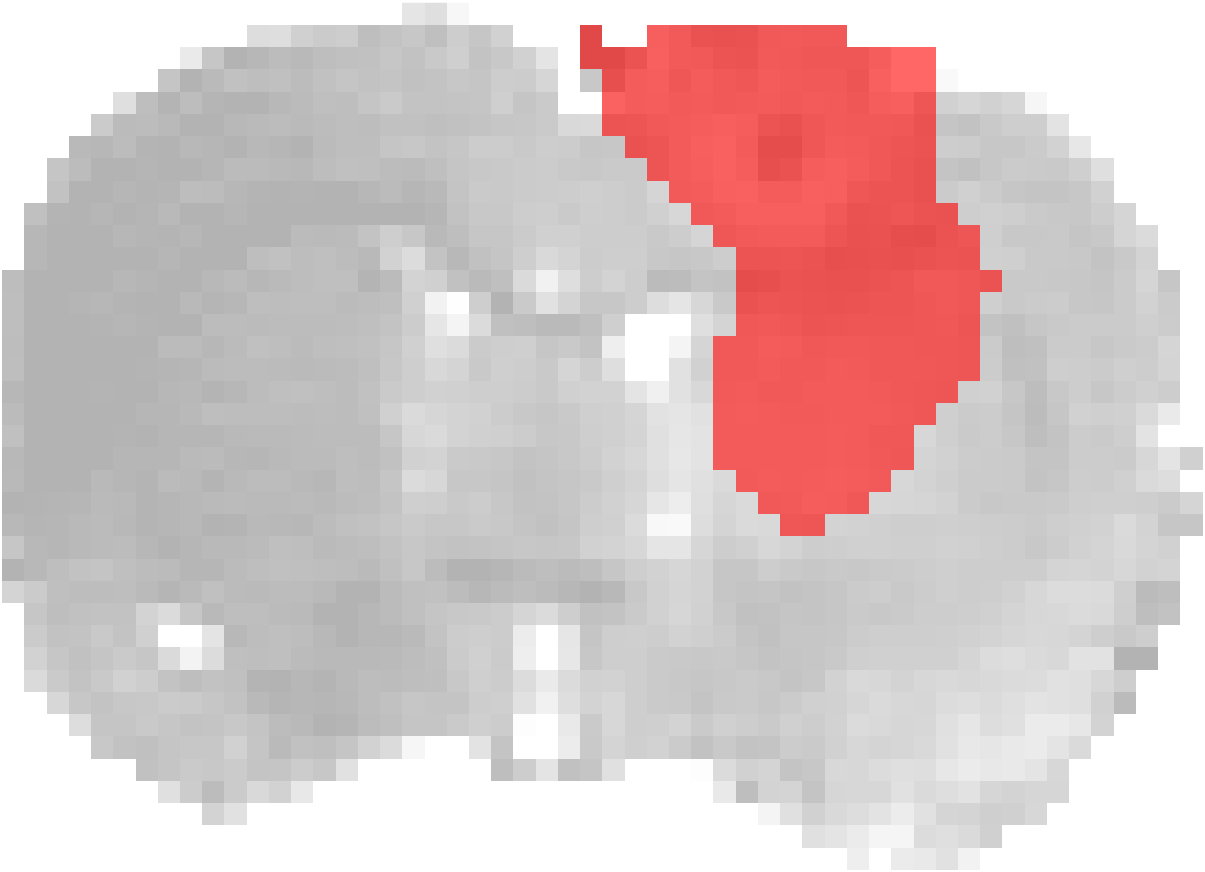} & 
            \includegraphics[width=0.148\textwidth]{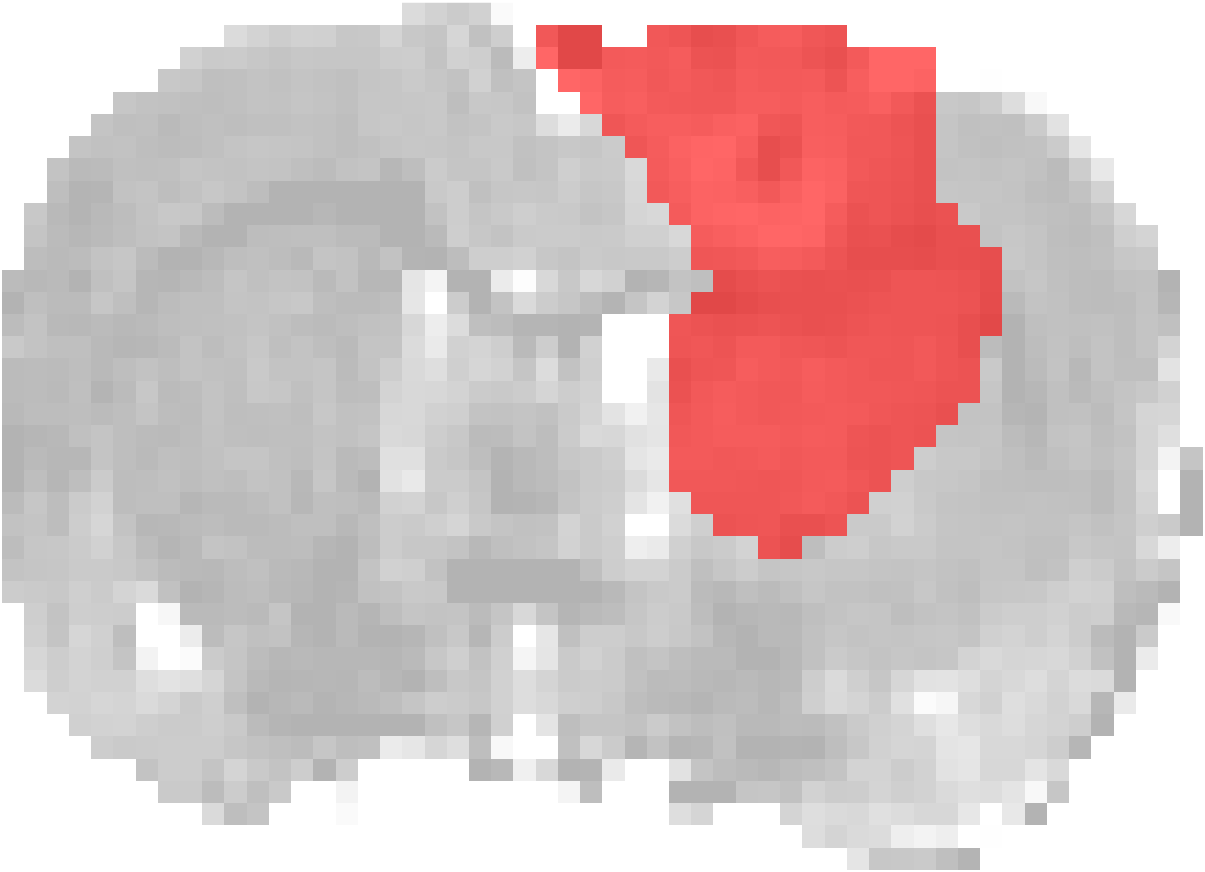} & 
            \includegraphics[width=0.148\textwidth]{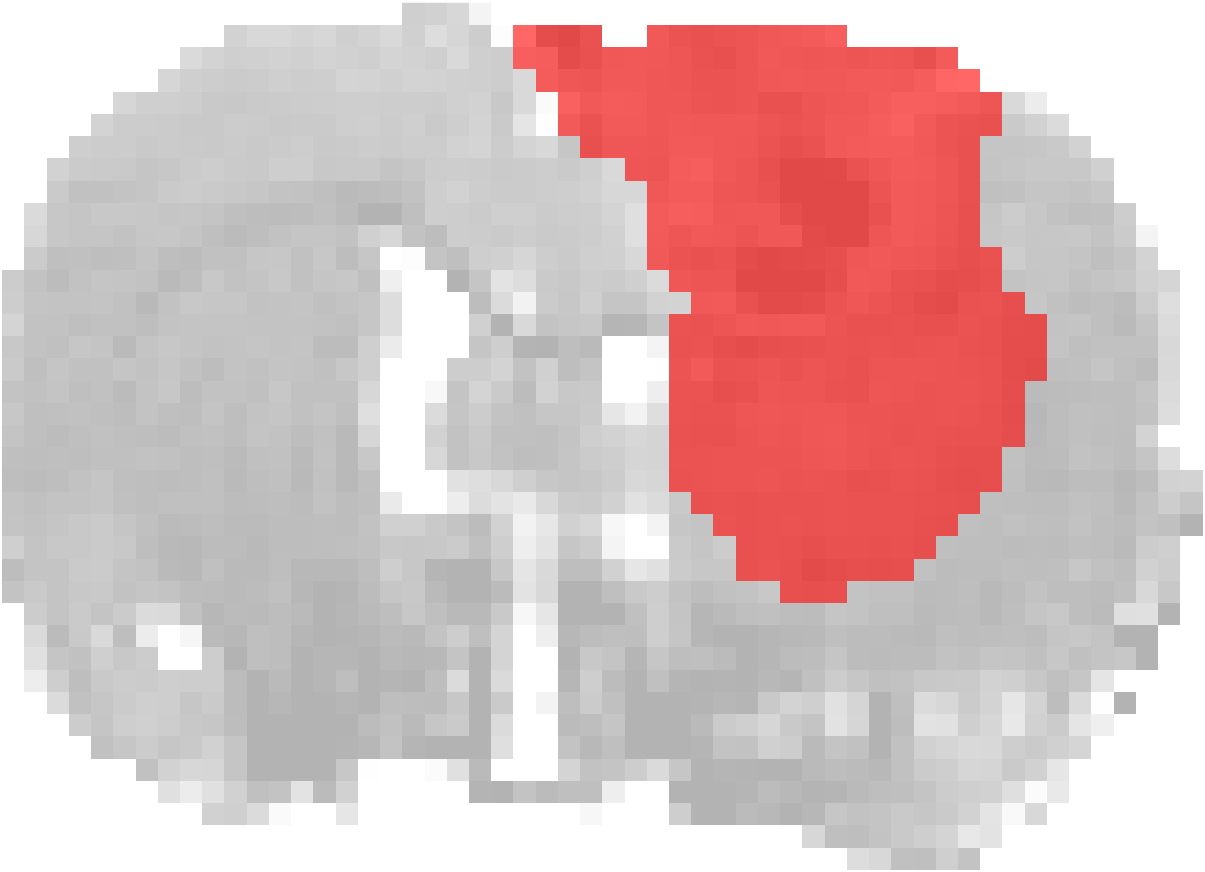} \\ [4pt]
            
            \includegraphics[width=0.15\textwidth]{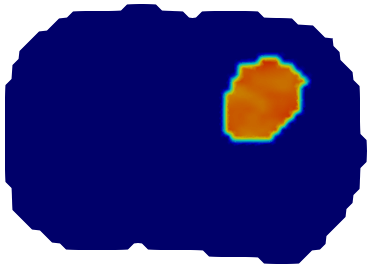} & 
            \includegraphics[width=0.15\textwidth]{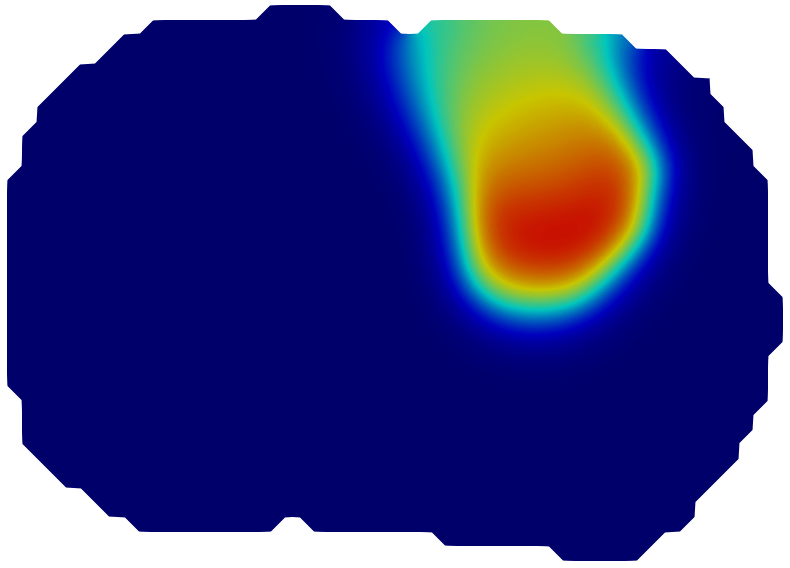} & 
            {\fboxrule=1.5pt\fcolorbox{green!60}{white}{\includegraphics[width=0.15\textwidth]{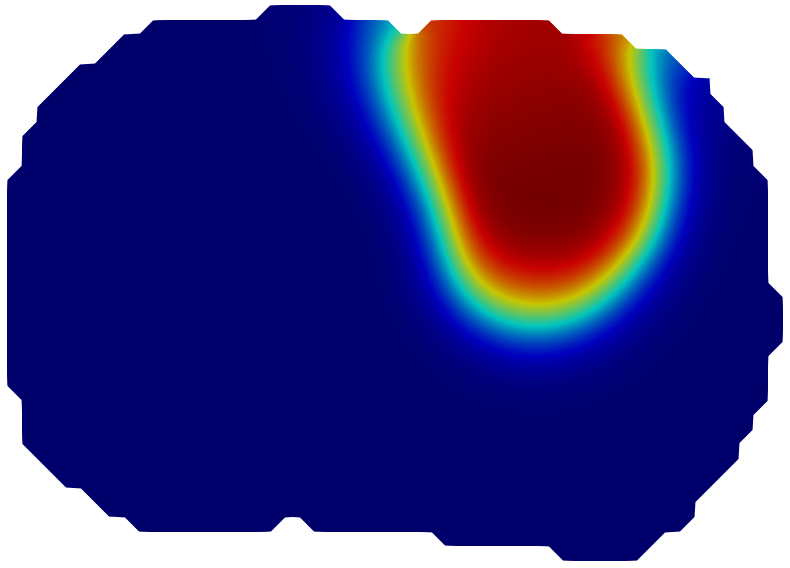}}} & 
            & & \\ [0pt]
            
            \includegraphics[width=0.15\textwidth]{Figures/Pred/sim-rat-III-Day10.png} & 
            \includegraphics[width=0.15\textwidth]{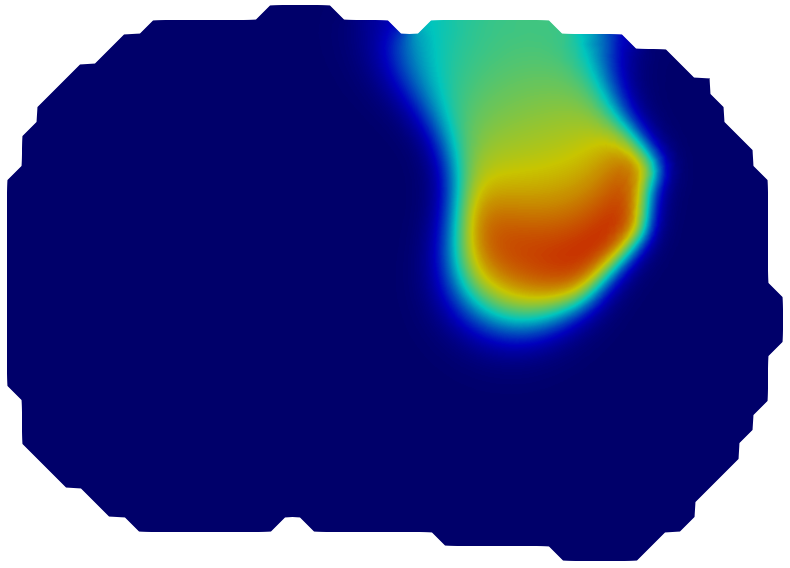} & 
            \includegraphics[width=0.15\textwidth]{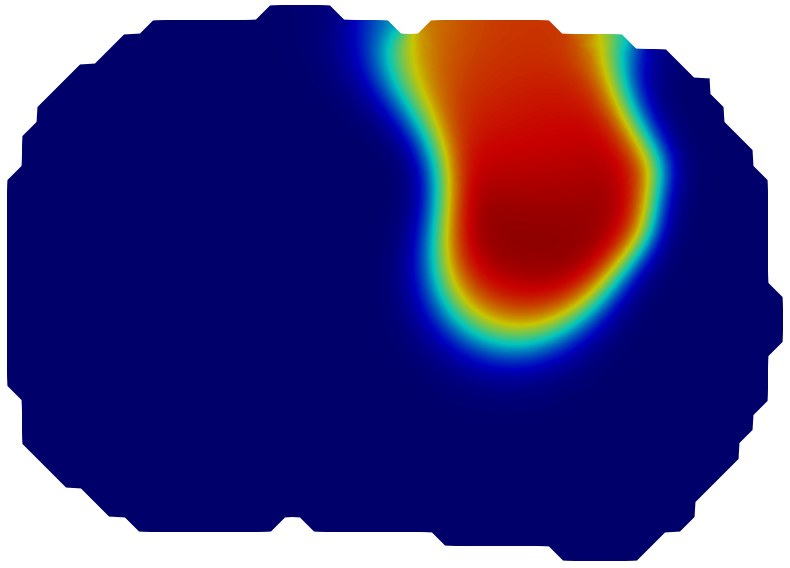} & 
            {\fboxrule=1.5pt\fcolorbox{green!60}{white}{\includegraphics[width=0.15\textwidth]{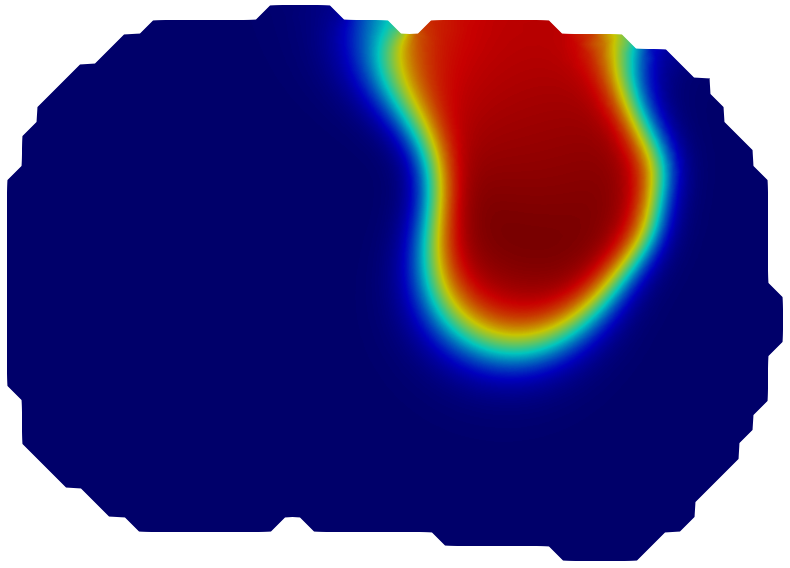}}} & 
            & \\ [0pt]
            
            \includegraphics[width=0.15\textwidth]{Figures/Pred/sim-rat-III-Day10.png} & 
            \includegraphics[width=0.15\textwidth]{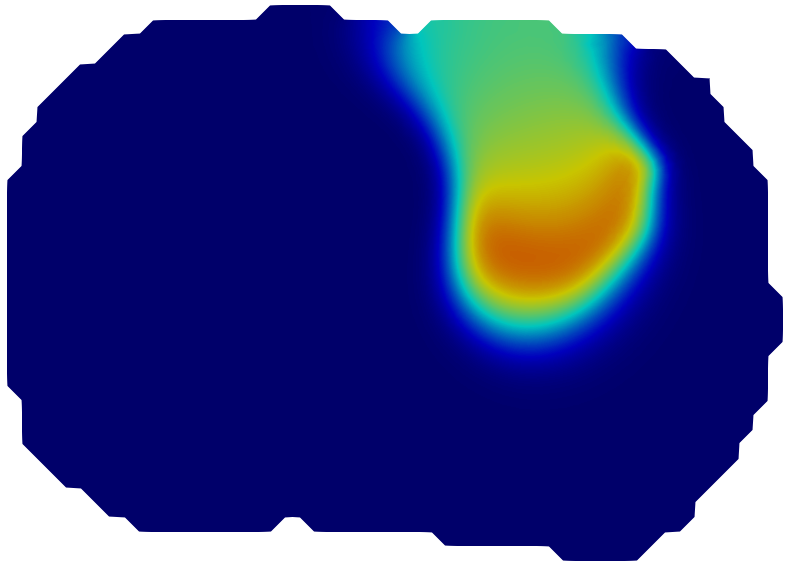} & 
            \includegraphics[width=0.15\textwidth]{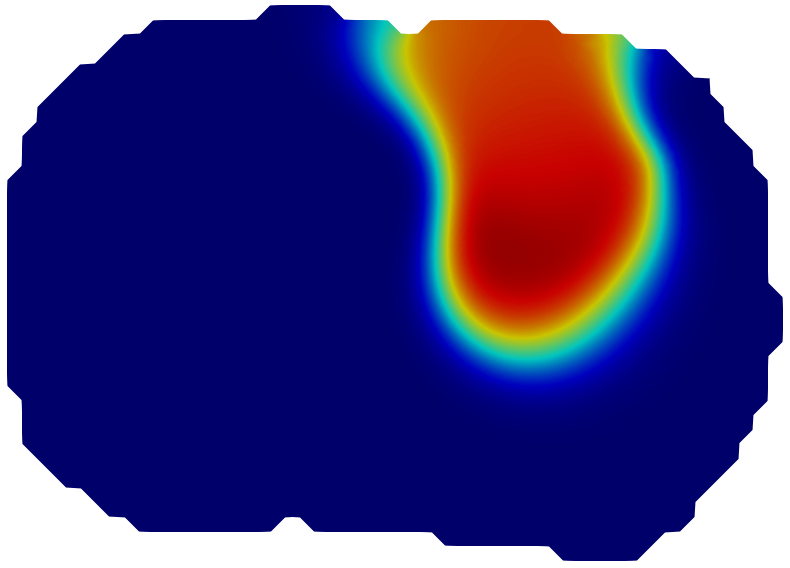} & 
            \includegraphics[width=0.15\textwidth]{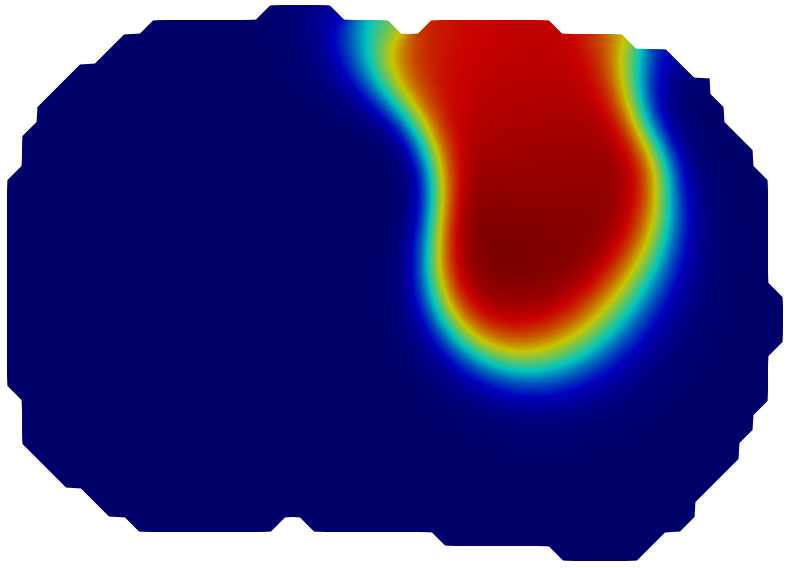} & 
            {\fboxrule=1.5pt\fcolorbox{green!60}{white}{\includegraphics[width=0.15\textwidth]{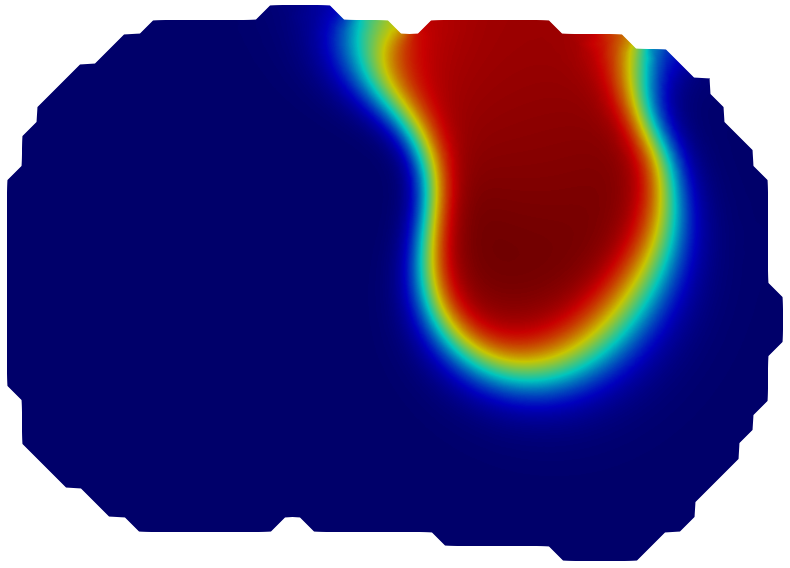}}} & \\ [0pt]
            
            \includegraphics[width=0.15\textwidth]{Figures/Pred/sim-rat-III-Day10.png} & 
            \includegraphics[width=0.15\textwidth]{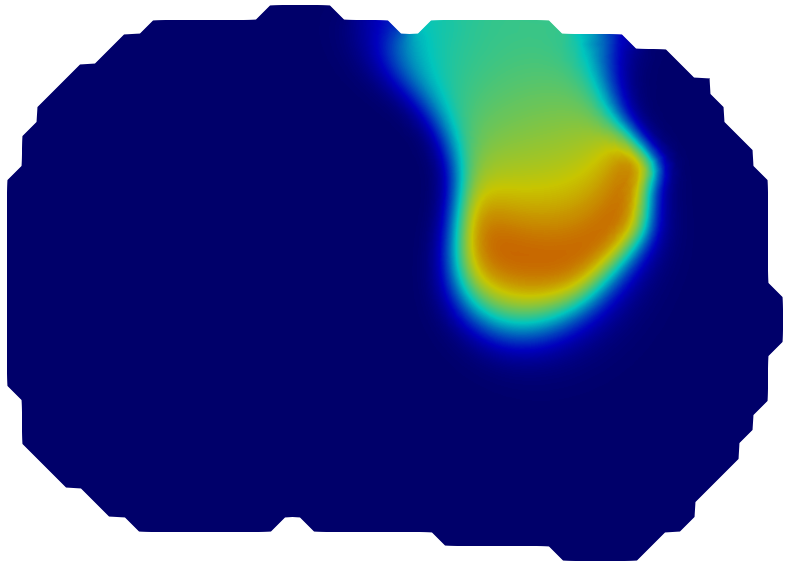} & 
            \includegraphics[width=0.15\textwidth]{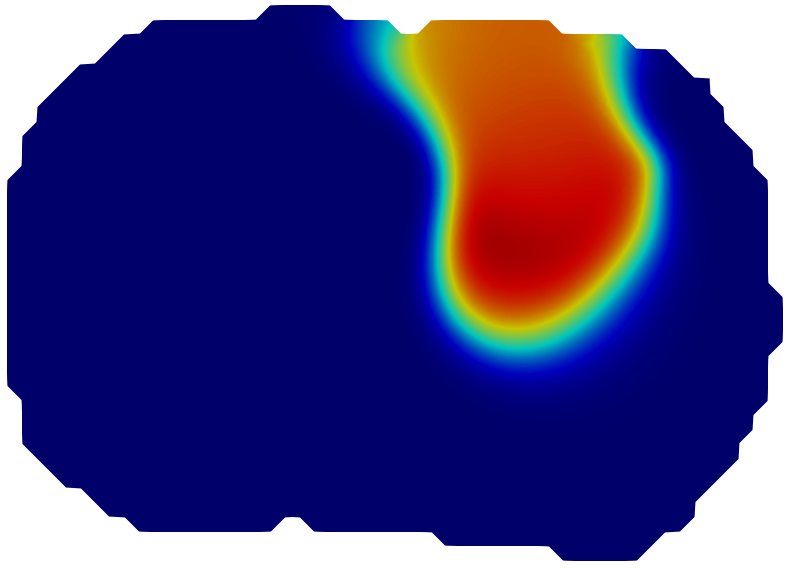} & 
            \includegraphics[width=0.15\textwidth]{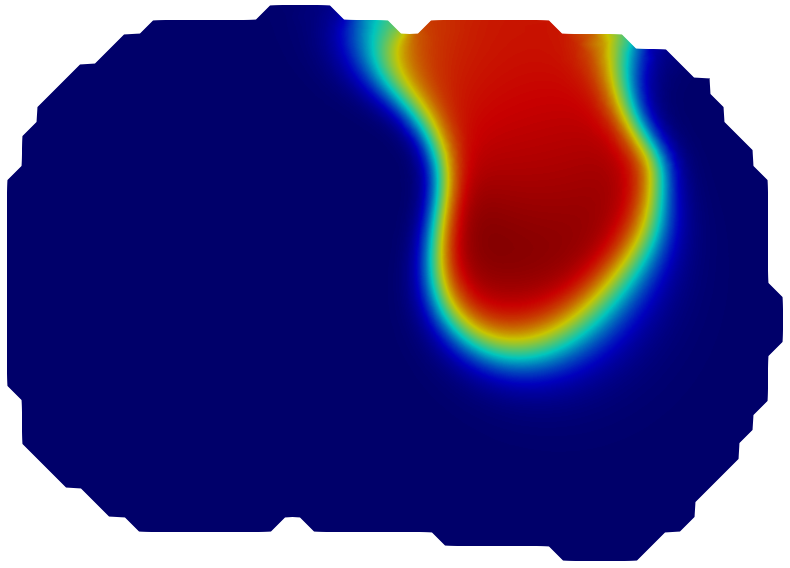} & 
            \includegraphics[width=0.15\textwidth]{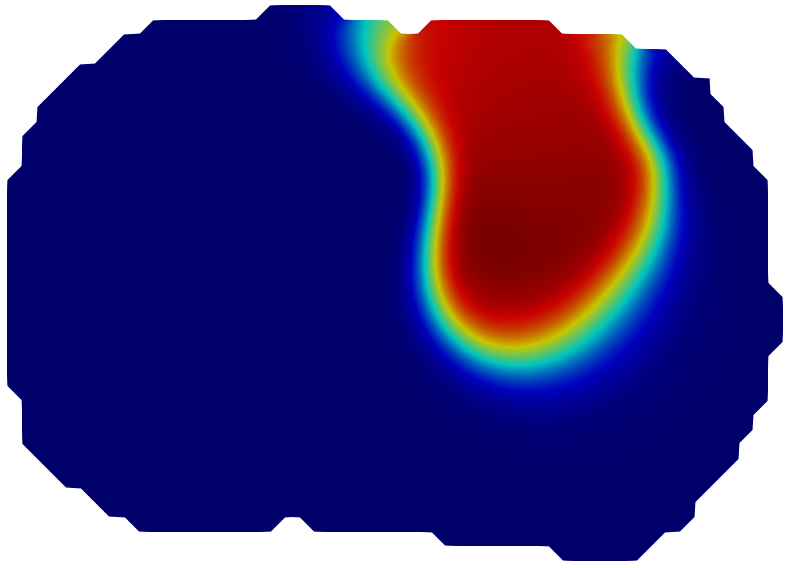} & 
            {\fboxrule=1.5pt\fcolorbox{green!60}{white}{\includegraphics[width=0.15\textwidth]{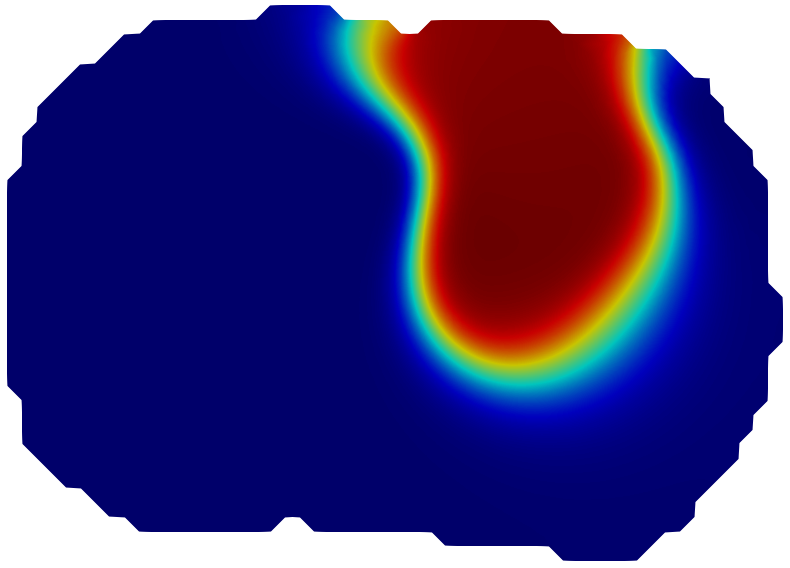}}} \\
        \end{tabular} & 
        
        \makebox[0pt][l]{
            \hspace{.1pt}
            \raisebox{-4.6cm}{
                \includegraphics[width=.95cm, height=6.8cm, keepaspectratio=false]{Figures/Pred/color-bar.png}
            }
            \hspace{-22pt}
            \raisebox{-3cm}{
                \rotatebox{90}{\small Tumor Volume Fraction}
            }}
        
    \end{tabular}
    \vspace{-0.1in}
    \caption{
    Sequential one-scan-ahead prediction for Rat-III at MAP parameters
    \blue{
    using the hyperelastic mechanically coupled tumor growth model defined in \eqref{eq:rd_PDE_REF} and \eqref{eq:mech_PDE_REF}.
    }.
    Top row: MRI-derived tumor volume fraction.
    Bottom rows: model predictions obtained by assimilating data up to
    ~$t_k$ and predicting the subsequent scan at ~$t_{k+1}$
    (boxed panels). Dice values between
    the predicted and MRI-derived tumor at Day~14, Day~15, Day~16 and Day~19 are 0.86, 0.9, 0.82, and 0.92 respectively.}    
    \label{fig:ratIII_MRI-MAP}
    \vspace{-0.1in}
\end{figure}

\begin{figure}[!ht]
    \centering
    \raggedright
    \renewcommand{\arraystretch}{0}
    \setlength{\fboxsep}{0pt} 
    \setlength{\tabcolsep}{2pt} 
    
    \begin{tabular}{@{}l@{}l@{}}
        
        \begin{tabular}{@{} *{6}{>{\centering\arraybackslash}p{0.144\textwidth}} @{}} 
            \colorbox{black}{\makebox[0.146\textwidth][c]{\vrule width 0pt height 0.4cm \textcolor{white}{\small Day 10}}} & 
            \colorbox{black}{\makebox[0.146\textwidth][c]{\vrule width 0pt height 0.4cm \textcolor{white}{\small Day 12}}} & 
            \colorbox{black}{\makebox[0.146\textwidth][c]{\vrule width 0pt height 0.4cm \textcolor{white}{\small Day 14}}} & 
            \colorbox{black}{\makebox[0.146\textwidth][c]{\vrule width 0pt height 0.4cm \textcolor{white}{\small Day 16}}} & 
            \colorbox{black}{\makebox[0.146\textwidth][c]{\vrule width 0pt height 0.4cm \textcolor{white}{\small Day 18}}} & 
            \colorbox{black}{\makebox[0.146\textwidth][c]{\vrule width 0pt height 0.4cm \textcolor{white}{\small Day 20}}} \\ [3pt]
            
            \includegraphics[width=0.15\textwidth]{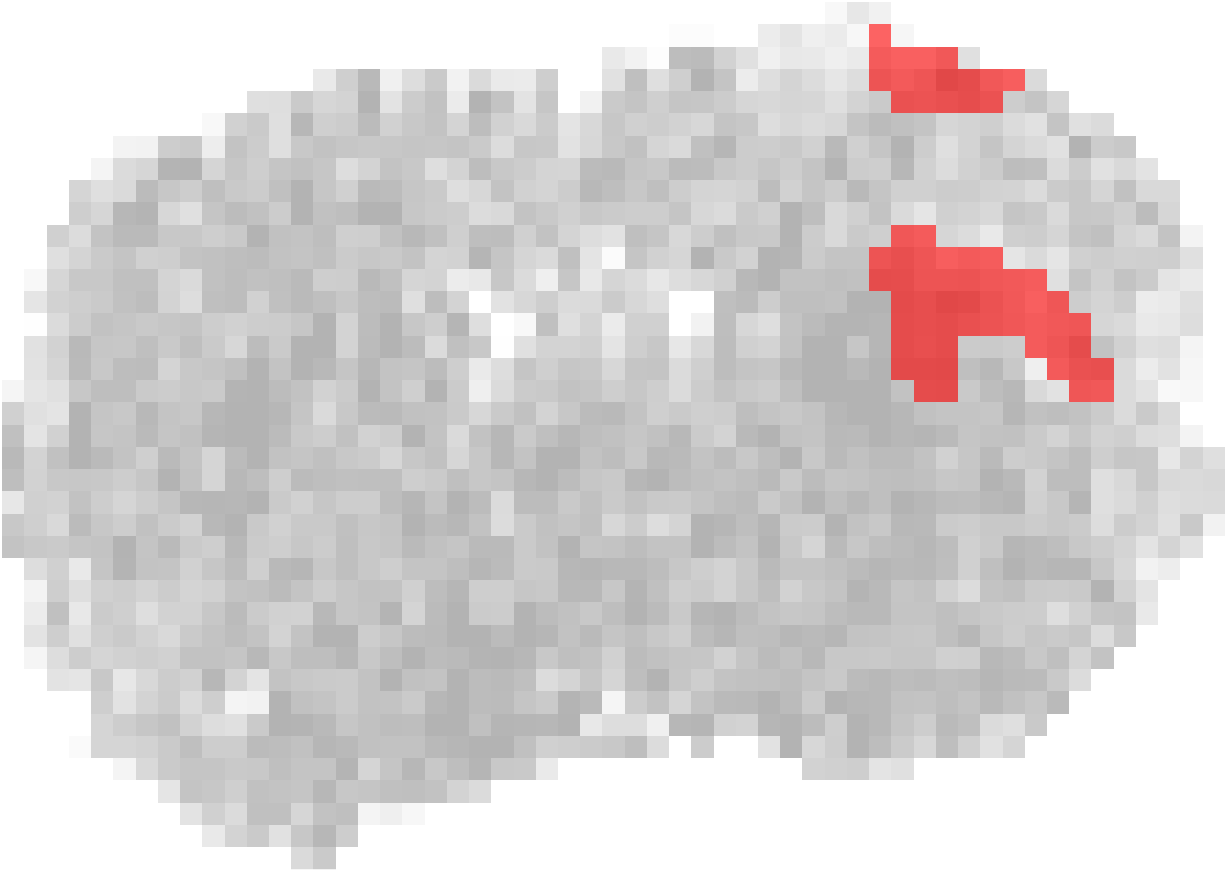} & 
            \includegraphics[width=0.15\textwidth]{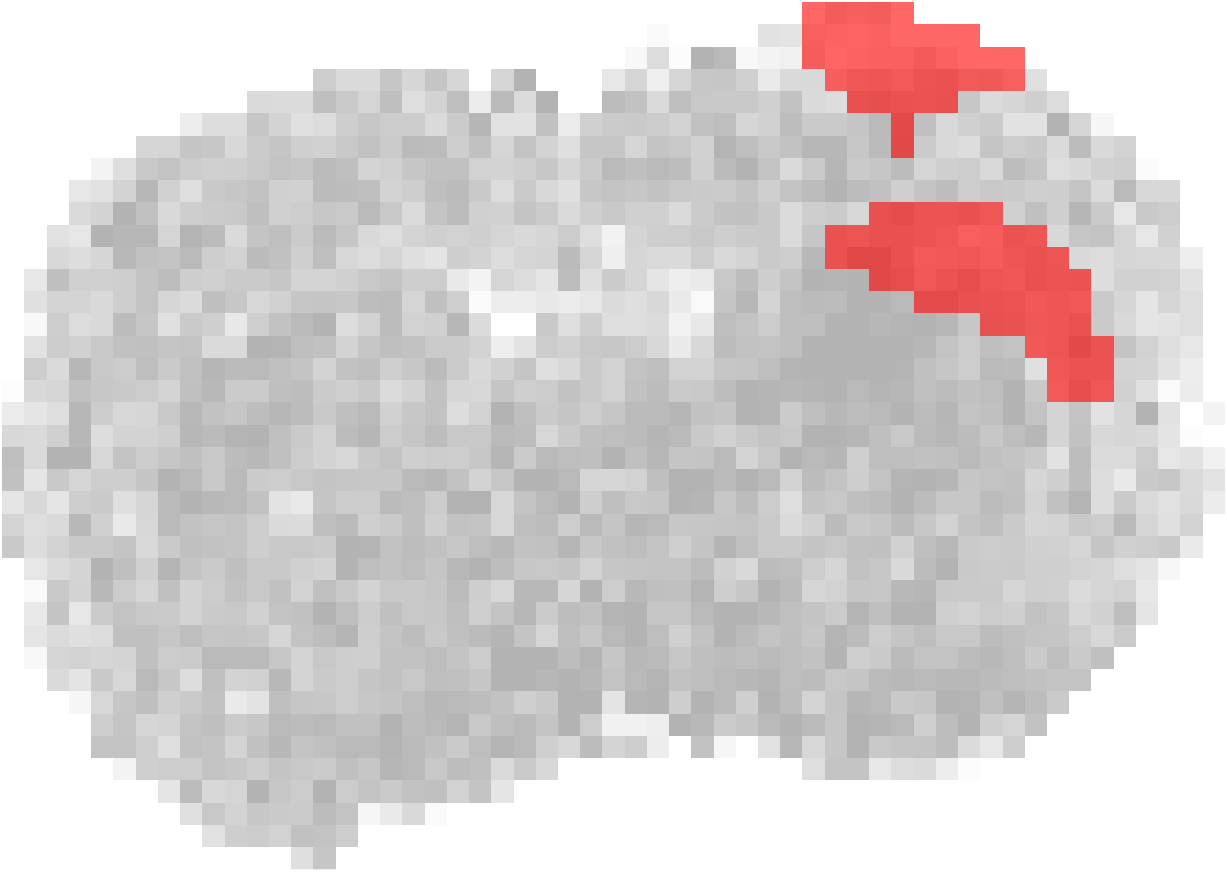} & 
            \includegraphics[width=0.15\textwidth]{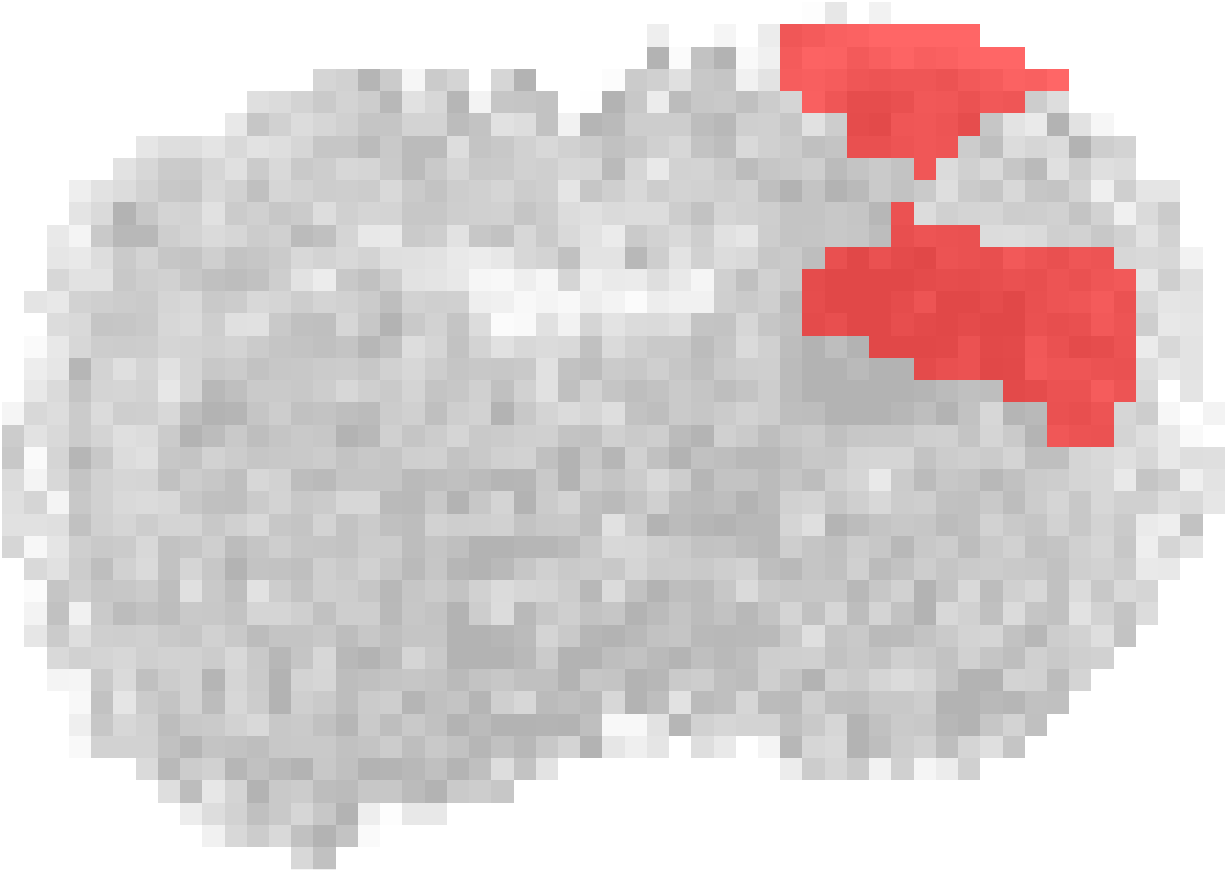} & 
            \includegraphics[width=0.15\textwidth]{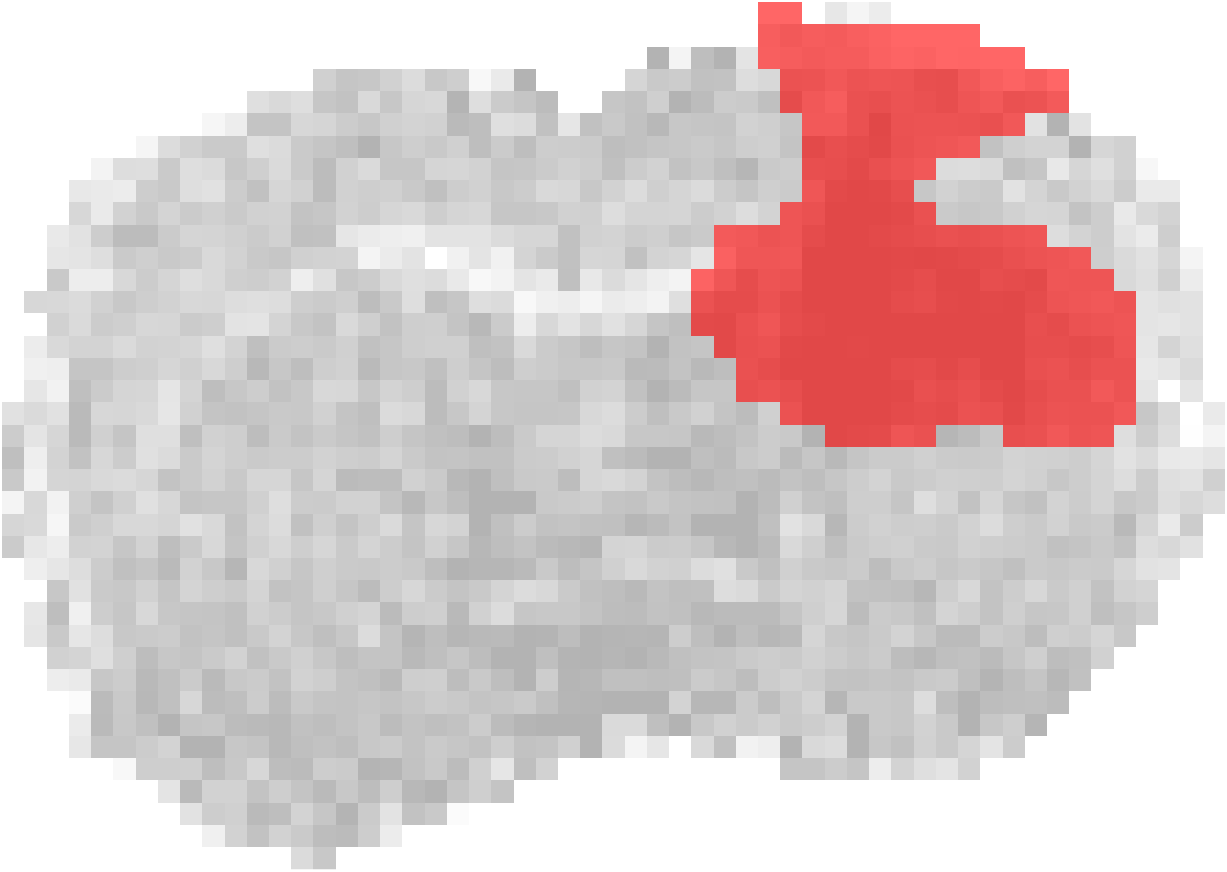} & 
            \includegraphics[width=0.15\textwidth]{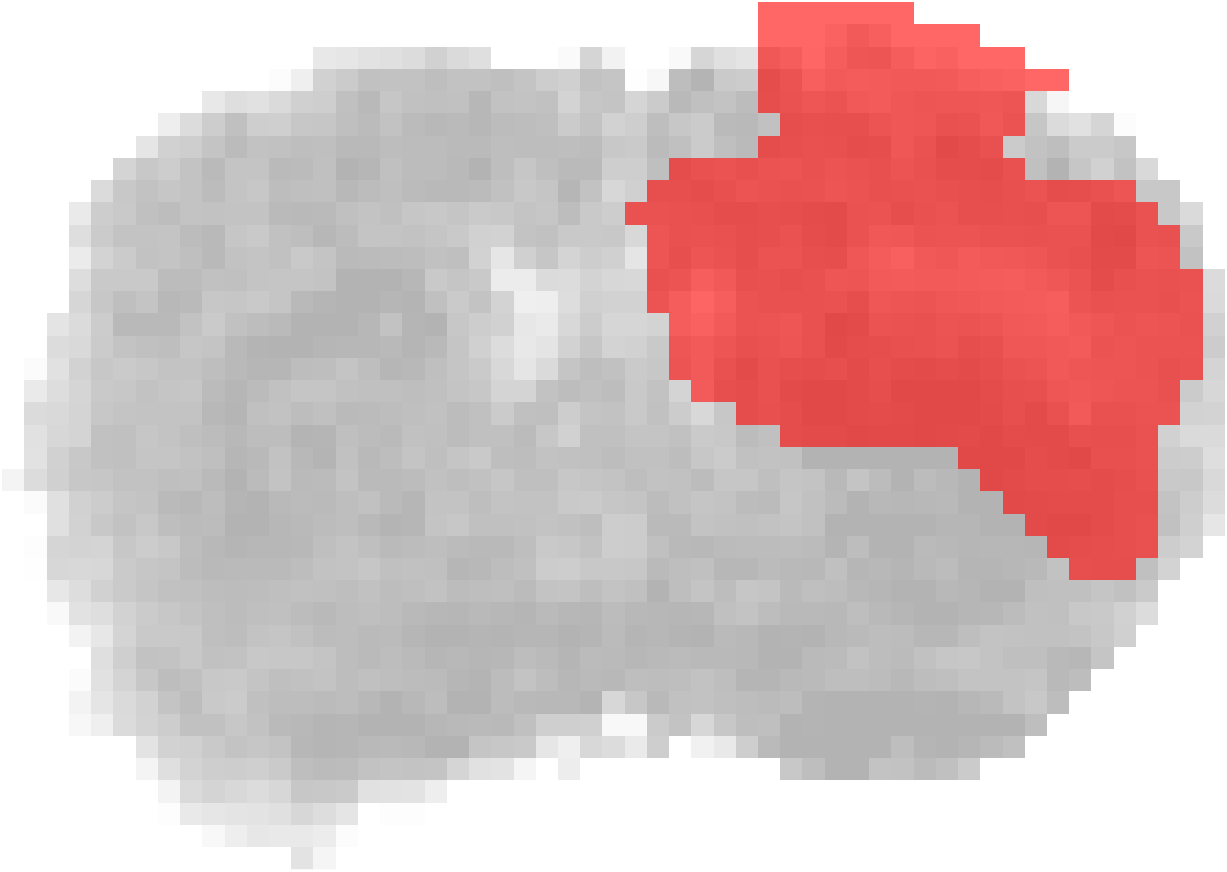} & 
            \includegraphics[width=0.15\textwidth]{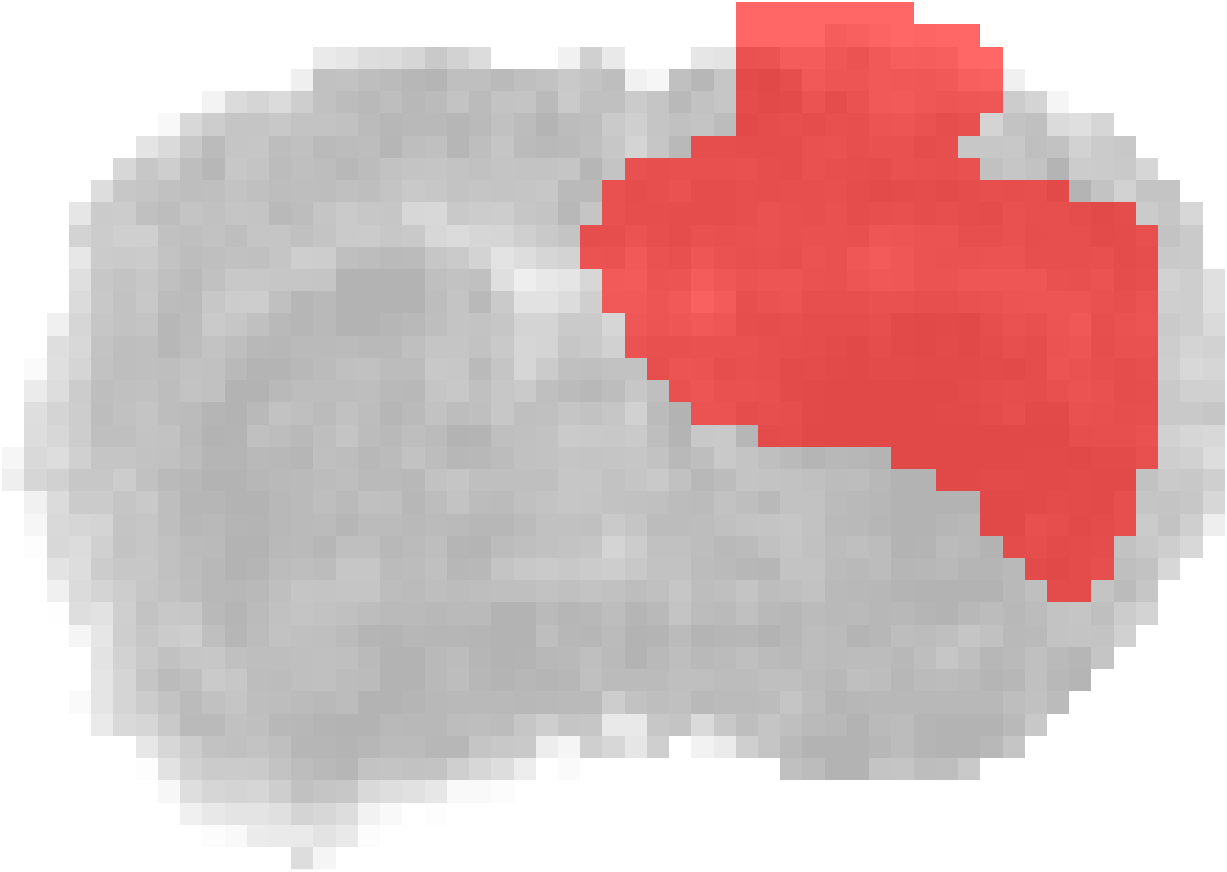} \\ [4pt]
            
            \includegraphics[width=0.151\textwidth]{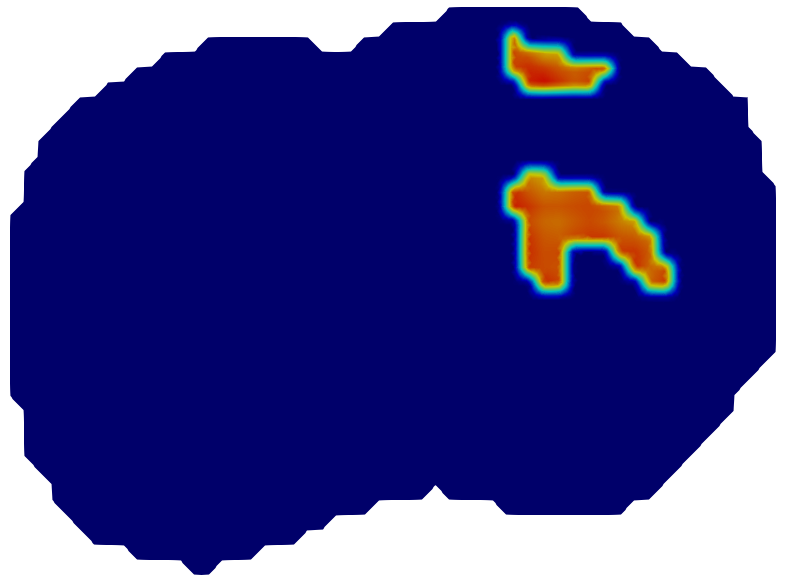} & 
            \includegraphics[width=0.151\textwidth]{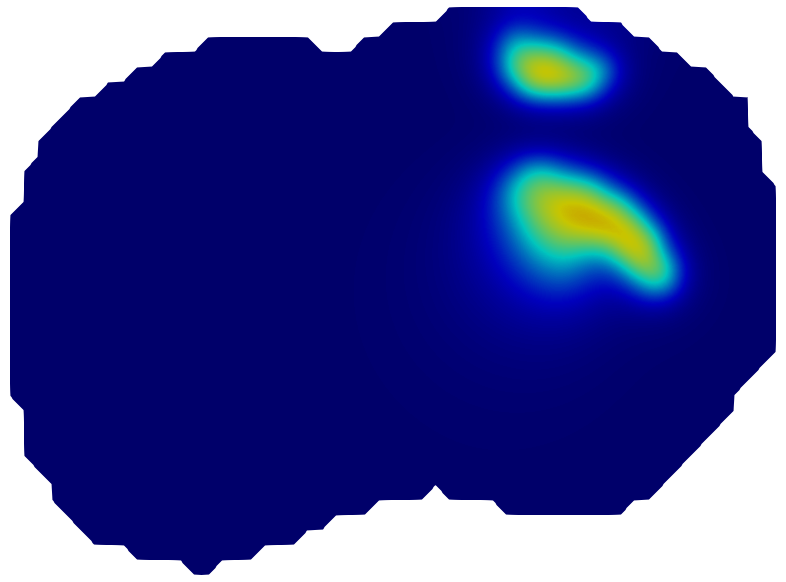} & 
            {\fboxrule=1.5pt\fcolorbox{green!60}{white}{\includegraphics[width=0.151\textwidth]{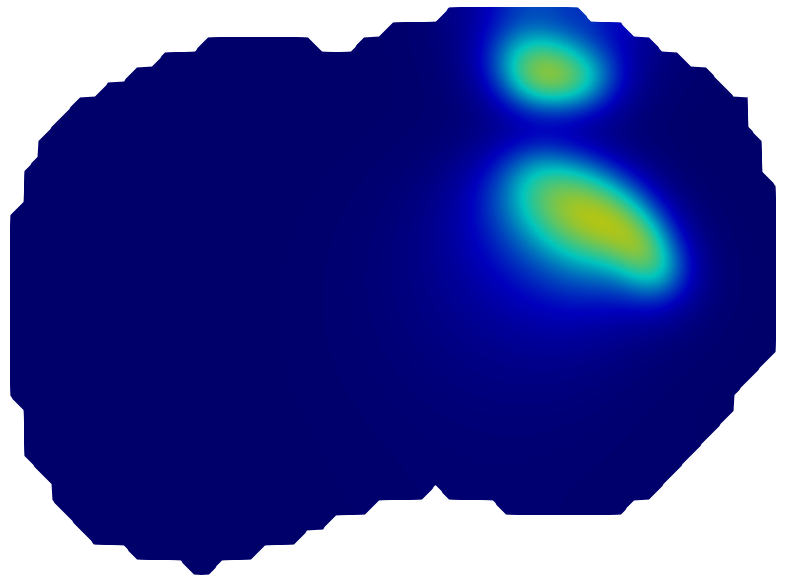}}} & 
            & & \\ [0pt]
            
            \includegraphics[width=0.151\textwidth]{Figures/Pred/Data4-sim-rat-IV-Day10.png} & 
            \includegraphics[width=0.151\textwidth]{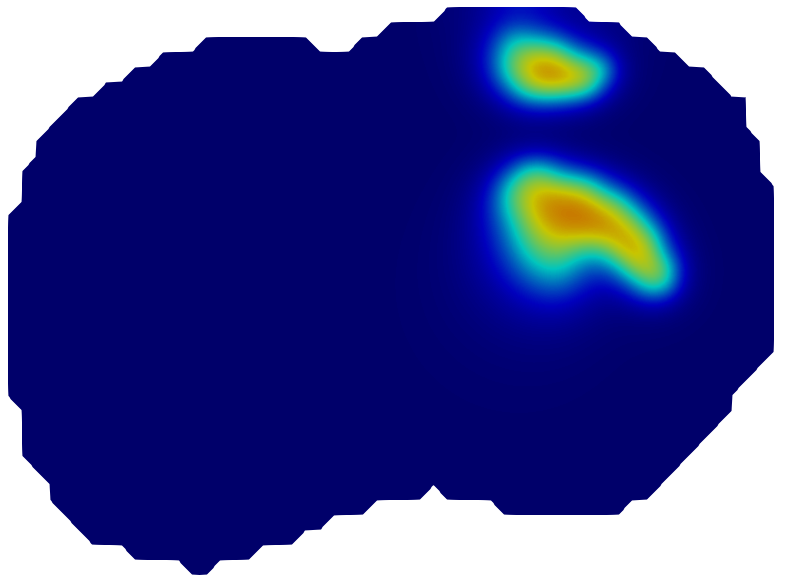} & 
            \includegraphics[width=0.151\textwidth]{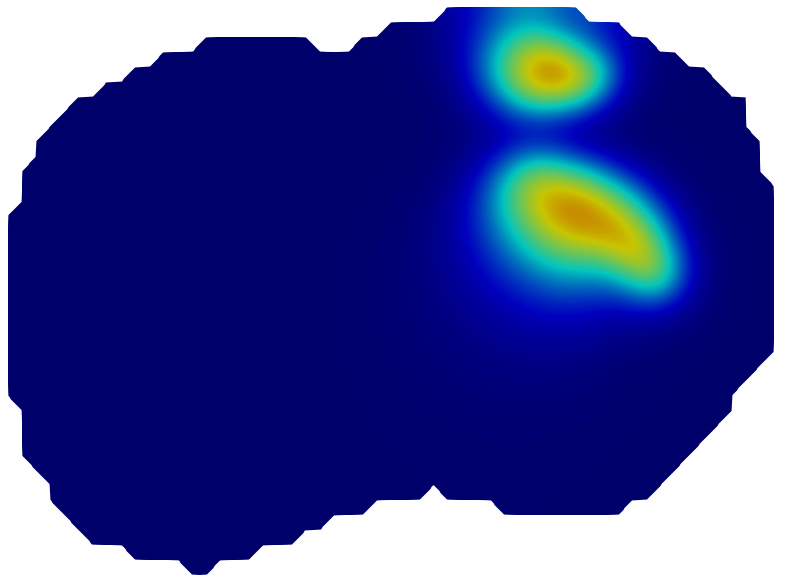} & 
            {\fboxrule=1.5pt\fcolorbox{green!60}{white}{\includegraphics[width=0.151\textwidth]{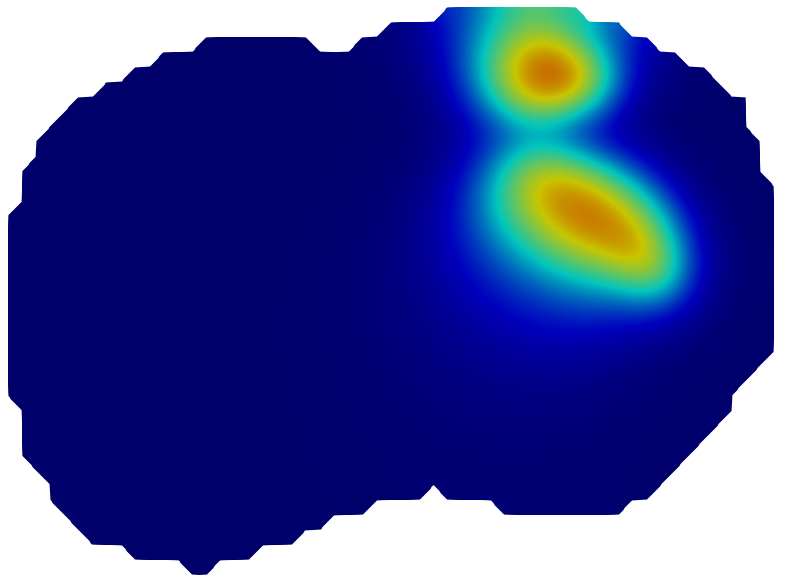}}} & 
            & \\ [0pt]
            
            \includegraphics[width=0.151\textwidth]{Figures/Pred/Data4-sim-rat-IV-Day10.png} & 
            \includegraphics[width=0.151\textwidth]{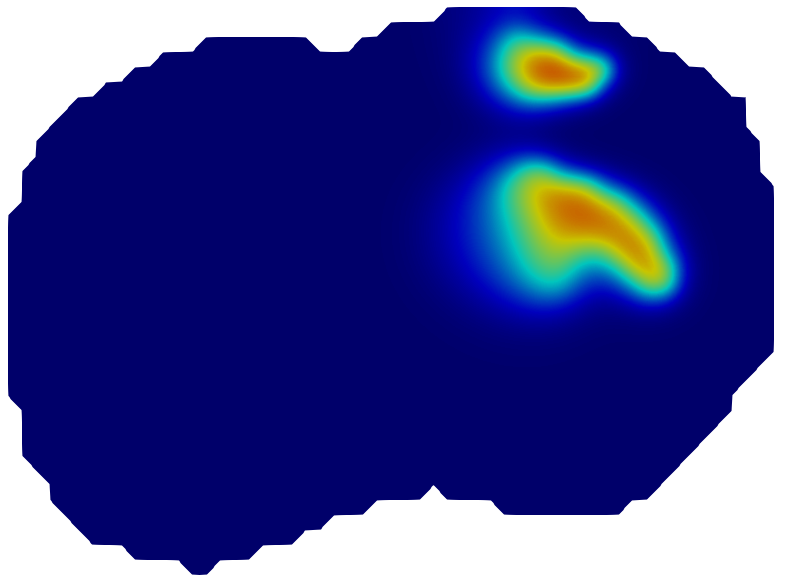} & 
            \includegraphics[width=0.151\textwidth]{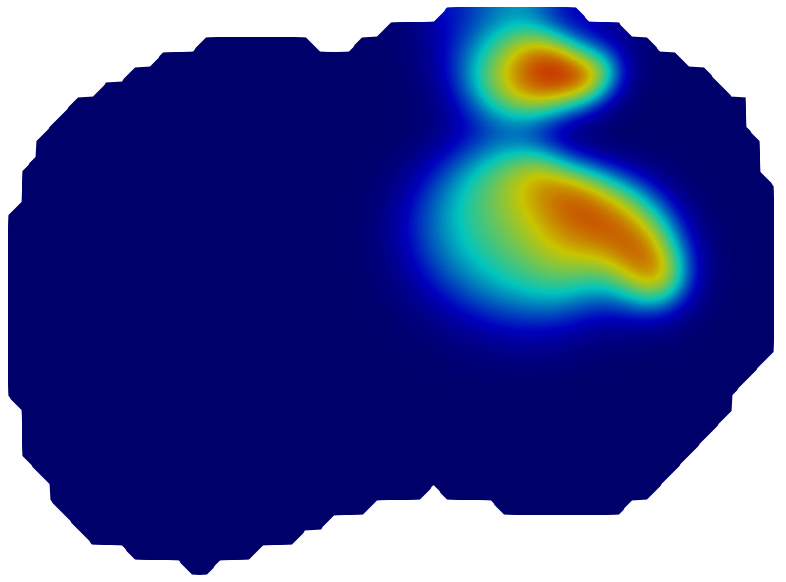} & 
            \includegraphics[width=0.151\textwidth]{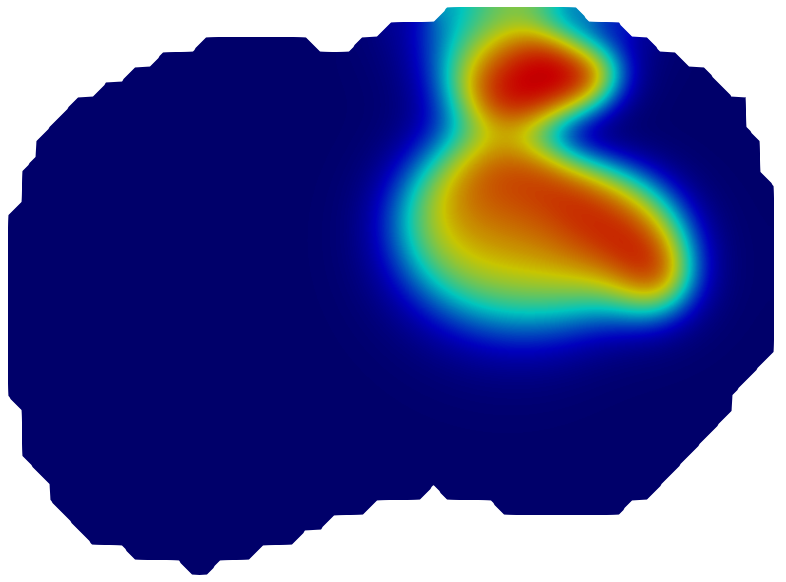} & 
            {\fboxrule=1.5pt\fcolorbox{green!60}{white}{\includegraphics[width=0.151\textwidth]{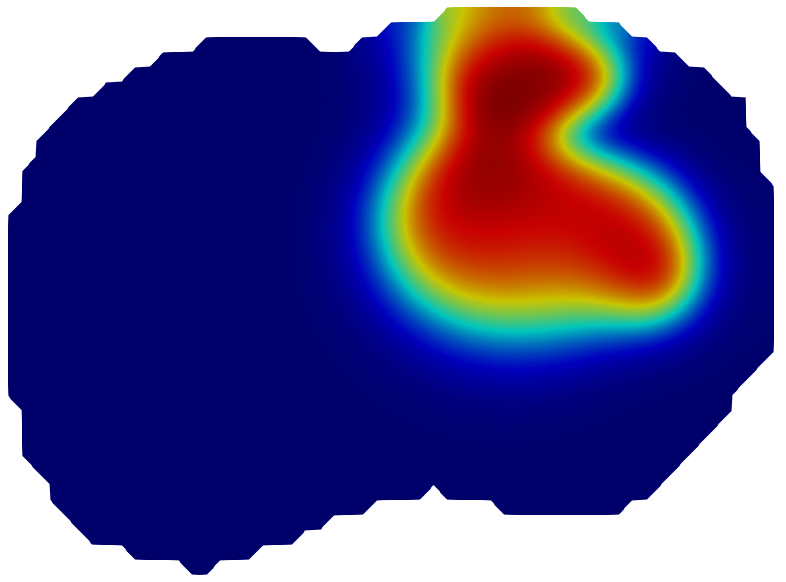}}} & \\ [0pt]
            
            \includegraphics[width=0.151\textwidth]{Figures/Pred/Data4-sim-rat-IV-Day10.png} & 
            \includegraphics[width=0.151\textwidth]{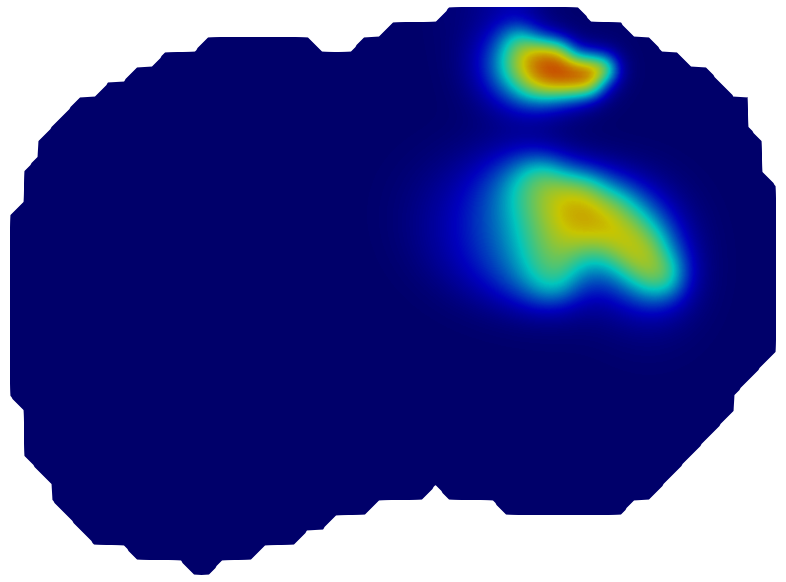} & 
            \includegraphics[width=0.151\textwidth]{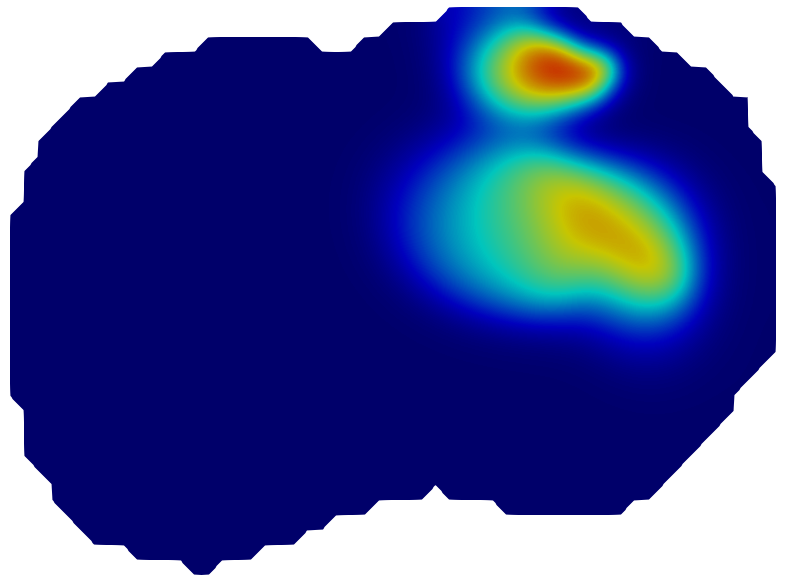} & 
            \includegraphics[width=0.151\textwidth]{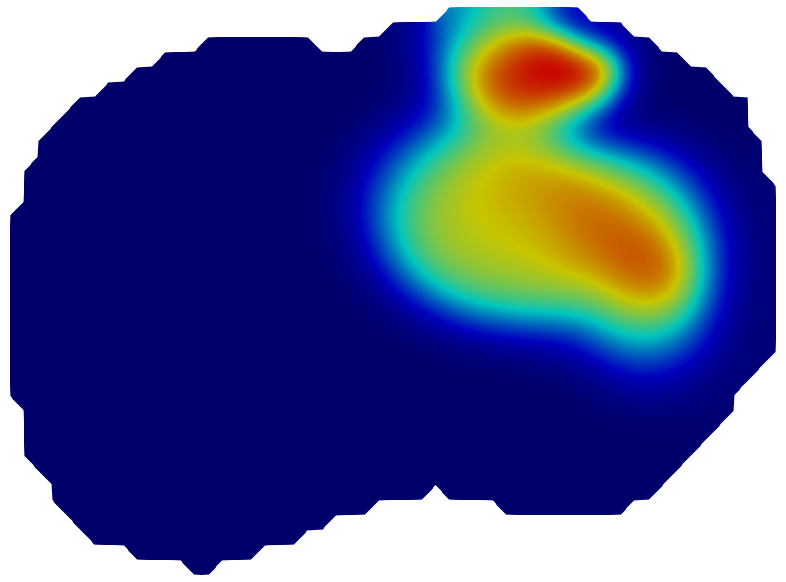} & 
            \includegraphics[width=0.151\textwidth]{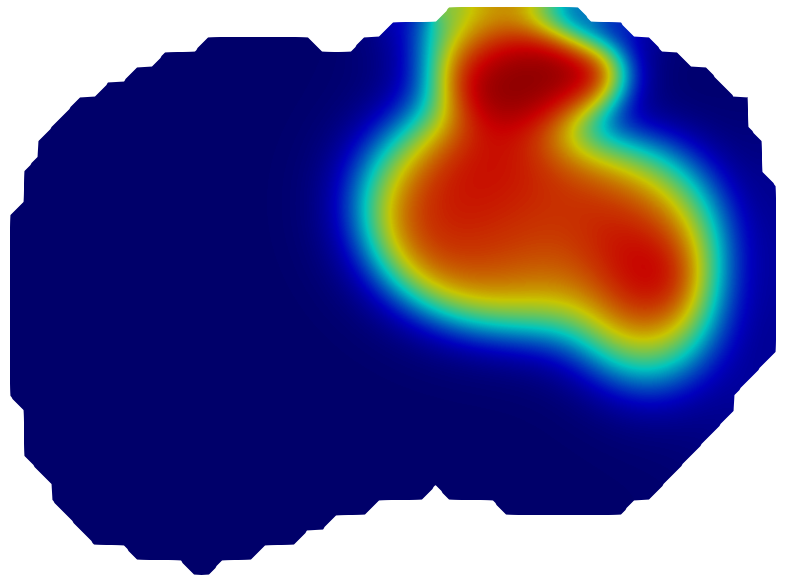} & 
            {\fboxrule=1.5pt\fcolorbox{green!60}{white}{\includegraphics[width=0.151\textwidth]{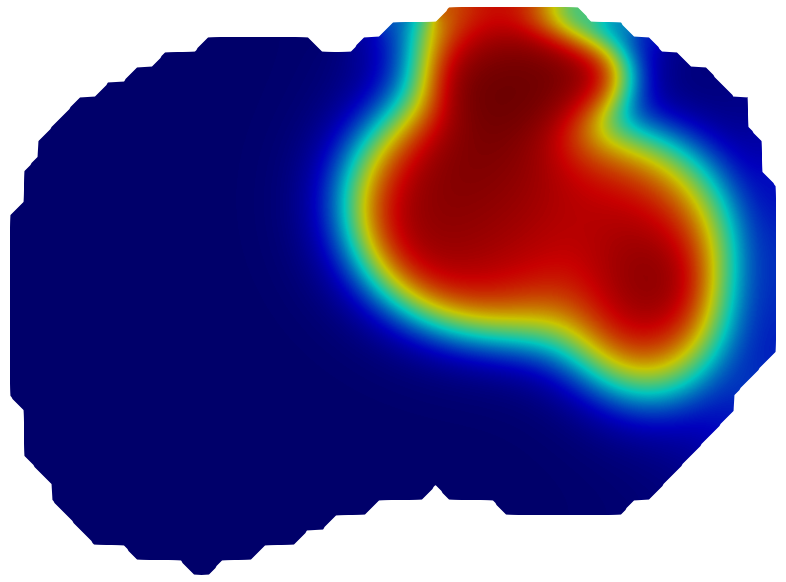}}} \\
        \end{tabular} & 
        
        \makebox[0pt][l]{
            \hspace{.1pt}
            \raisebox{-4.6cm}{
                \includegraphics[width=.95cm, height=6.8cm, keepaspectratio=false]{Figures/Pred/color-bar.png}
            }
            \hspace{-22pt}
            \raisebox{-3cm}{
                \rotatebox{90}{\small Tumor Volume Fraction}
            }}
        
    \end{tabular}
    \vspace{-0.1in}
    \caption{\blue{
    Sequential one-scan-ahead prediction for Rat~IV at MAP parameters using the hyperelastic mechanically coupled tumor growth model defined in \eqref{eq:rd_PDE_REF} and \eqref{eq:mech_PDE_REF}.
    Top row: MRI-derived tumor volume fraction.
    Bottom rows: model predictions obtained by assimilating data up to
    ~$t_k$ and predicting the subsequent scan at ~$t_{k+1}$
    (boxed panels). Dice values between
    the predicted and MRI-derived tumor at Day~14, Day~16, Day~18 and Day~20 are 0.72, 0.71, 0.85, and 0.89 respectively.}}    
    \label{fig:ratIV_MRI-MAP}
    \vspace{-0.0in}
\end{figure}

\blue{Figures~\ref{fig:ratI_MRI-MAP}--\ref{fig:ratIV_MRI-MAP}} compare MRI-derived tumor
volume fractions with sequential predictions obtained from the \blue{hyperelastic mechanically coupled tumor growth model
\eqref{eq:rd_PDE_REF} and \eqref{eq:mech_PDE_REF},} evaluated at the maximum a posteriori (MAP) parameters for \blue{four}
rats. For each subject, the model is sequentially informed by imaging data available up to a
given time point and used to predict tumor evolution at the subsequent imaging
time.
We refer to this as \textit{one-scan-ahead}
prediction to the next, unseen MRI acquisition time.
\blue{The results show that subject-specific Bayesian filter of spatially varying calibration parameters captures the overall anisotropic tumor spread and irregular growth patterns across animals. 
Prediction accuracy improves as additional scans are assimilated including subject-to-subject variability in predictive performance. A multi-step prediction experiment without intermediate data assimilation is reported in the \textit{Supplementary Information} and shows the expected degradation of prediction accuracy over longer forecast horizons.}

\begin{figure}[!htbp]
\centering
\renewcommand{\arraystretch}{0} 
\setlength{\tabcolsep}{-0.5pt}
\setlength{\fboxsep}{0pt}

\newcommand{\dayhead}[1]{
    \colorbox{black}{
        \makebox[2.9cm][c]{
            \vrule width 0pt height 0.32cm depth 0.2cm
            \textcolor{white}{\textbf{\small #1}}
        }}}

\newcommand{\blackimg}[1]{
    \includegraphics[width=3cm, trim=3 3 3 3, clip]{#1}
}

\begin{tabular}{ccccc}
\dayhead{Day~14} & \dayhead{Day~15} & \dayhead{Day~16} & \dayhead{Day~19} & \\ [10pt]

\blackimg{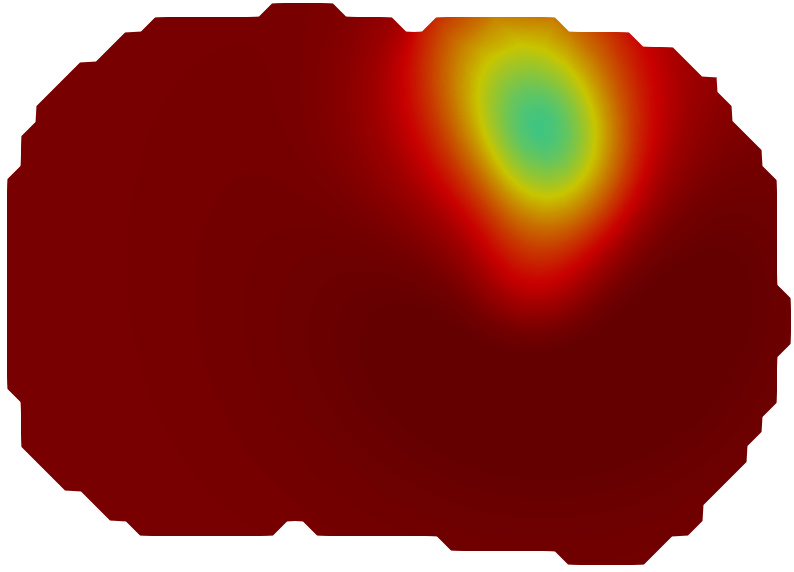} & 
\blackimg{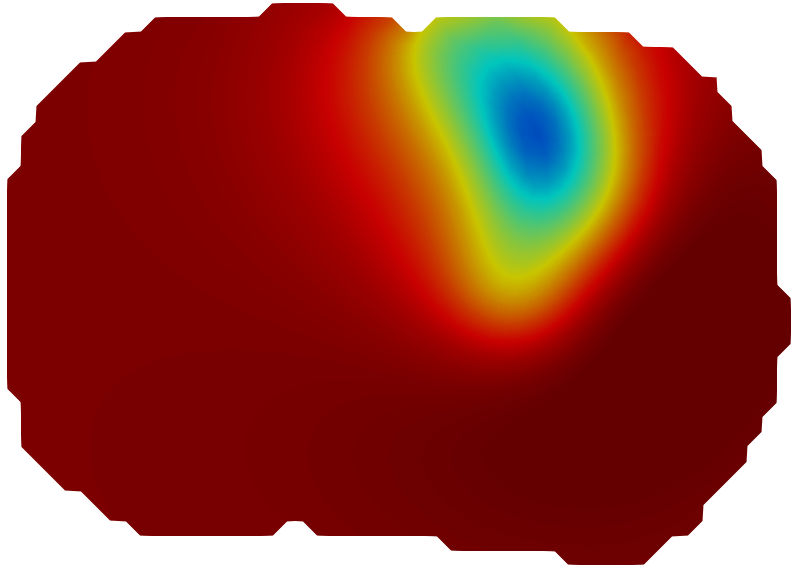} & 
\blackimg{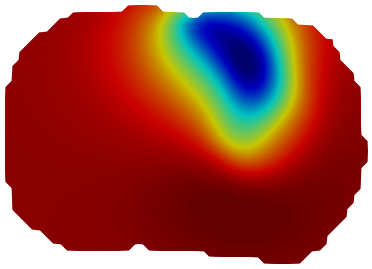} & 
\blackimg{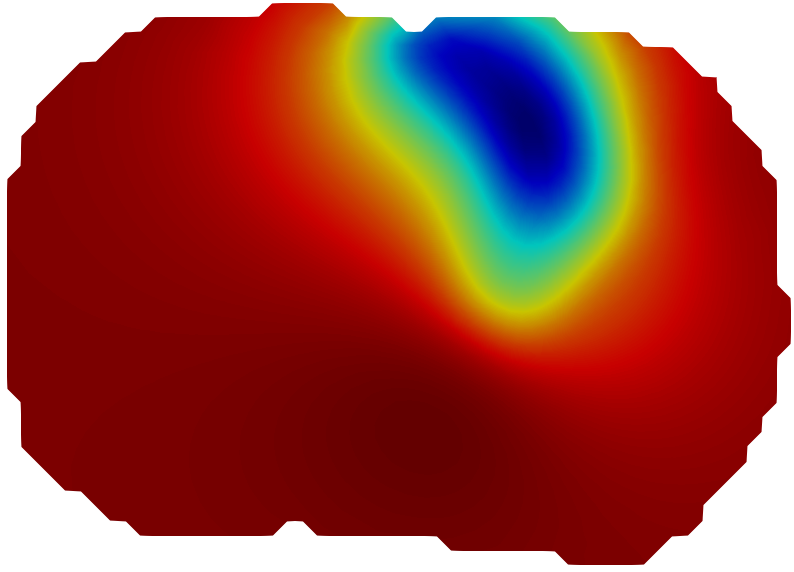} & 
\begin{tabular}{cl} 
\vspace{-2.1cm} \\
{\includegraphics[width=0.8cm, height=2.2cm, keepaspectratio=false]{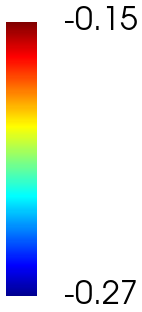}} &
\hspace{-1pt}\rotatebox{90}{
    \raisebox{10pt}{
        \begin{tabular}{@{}c@{}} \textbf{\tiny log(E)} \\[5pt] \textbf{\tiny [log(KPa)]} \end{tabular}
    }
}
\end{tabular} \\ [6pt]

\blackimg{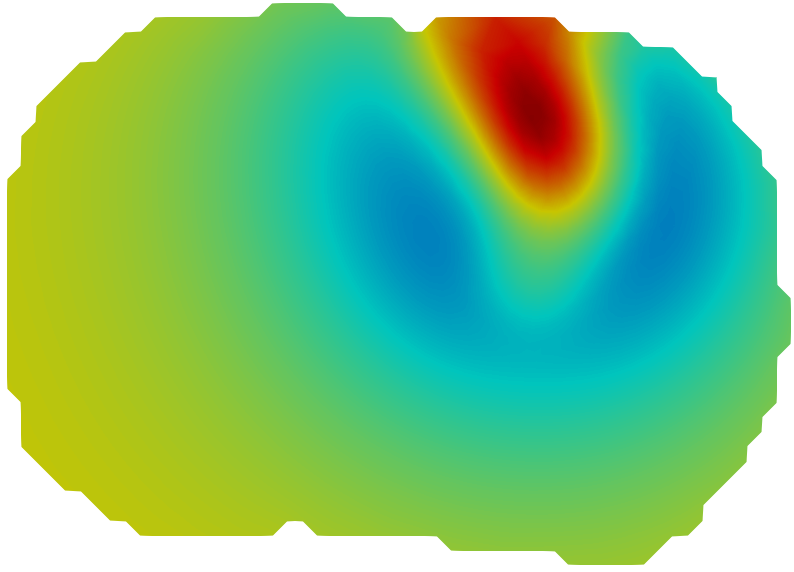} & 
\blackimg{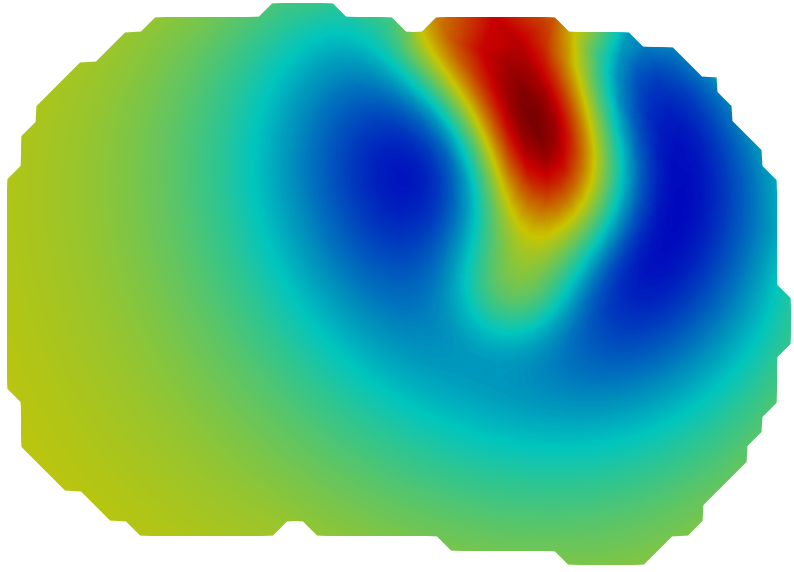} & 
\blackimg{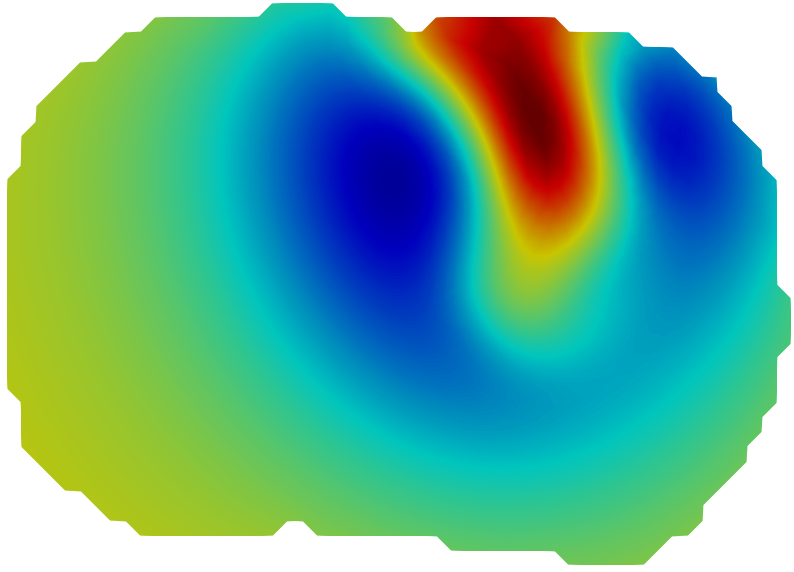} & 
\blackimg{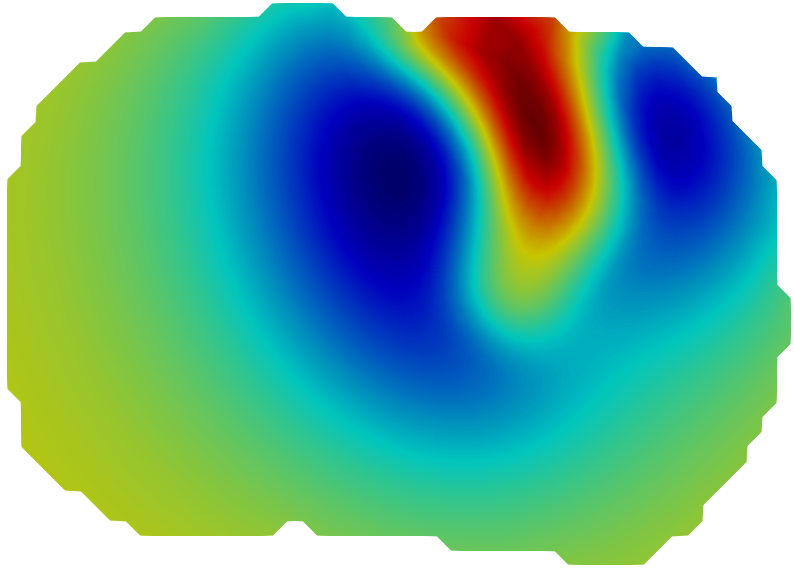} & 
\begin{tabular}{cl} 
\vspace{-2.2cm} \\
{\includegraphics[width=0.8cm, height=2.2cm, keepaspectratio=false]{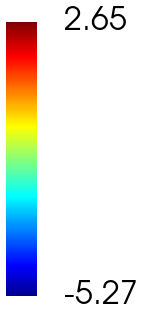}} &
\hspace{-1pt}\rotatebox{90}{
    \raisebox{10pt}{
        \begin{tabular}{@{}c@{}} \textbf{\tiny log(D)} \\[5pt] \textbf{\tiny [log(mm$^2$/day)]} \end{tabular}
    }
}
\end{tabular} \\ [6pt] 

\blackimg{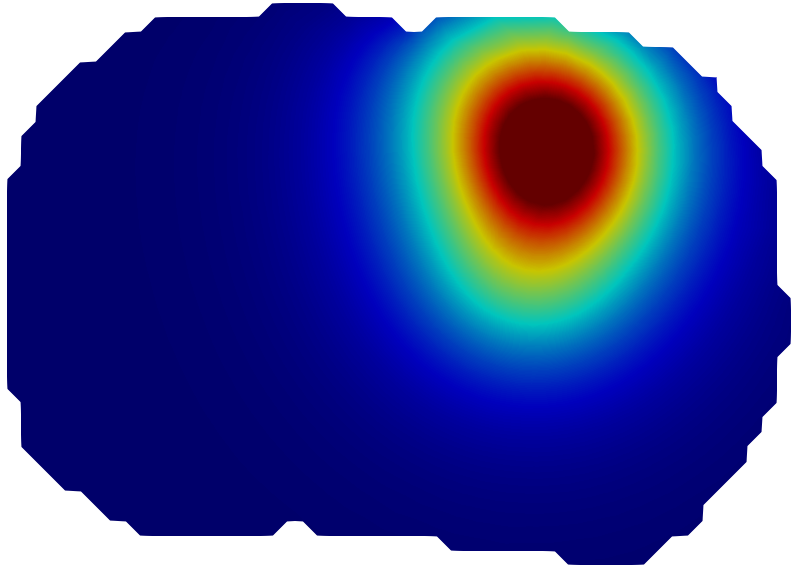} & 
\blackimg{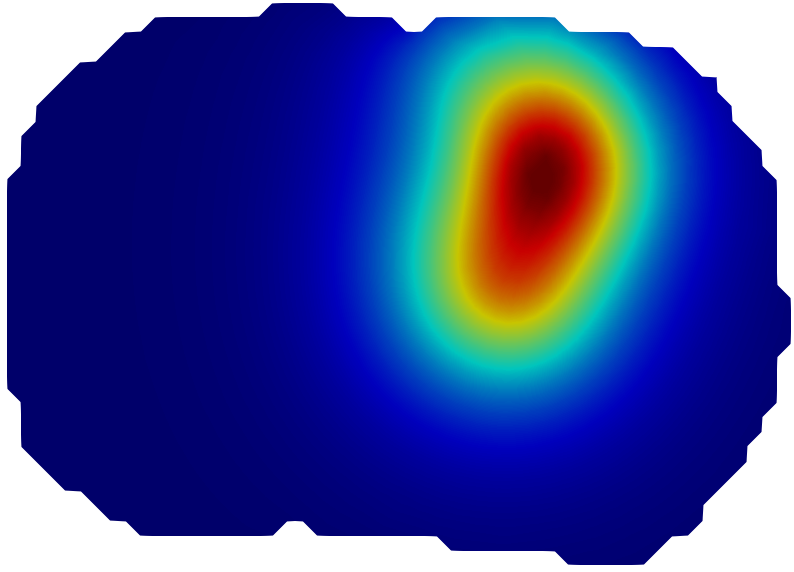} & 
\blackimg{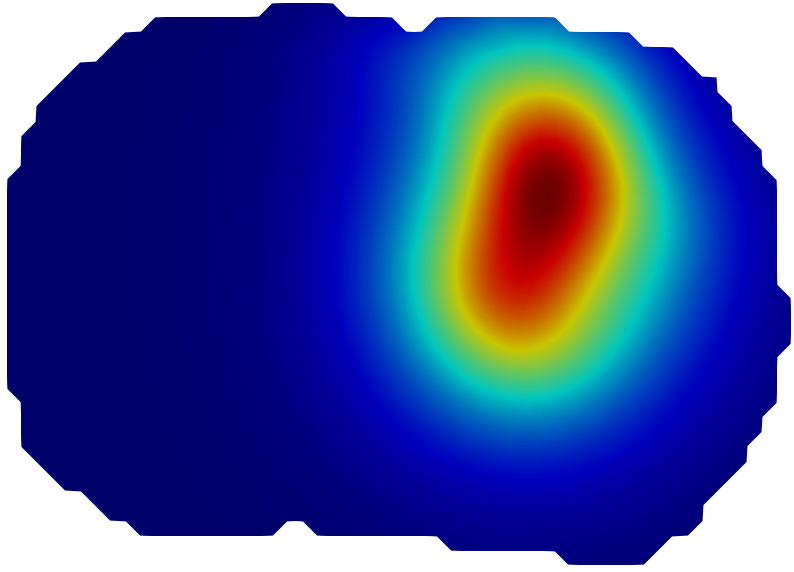} & 
\blackimg{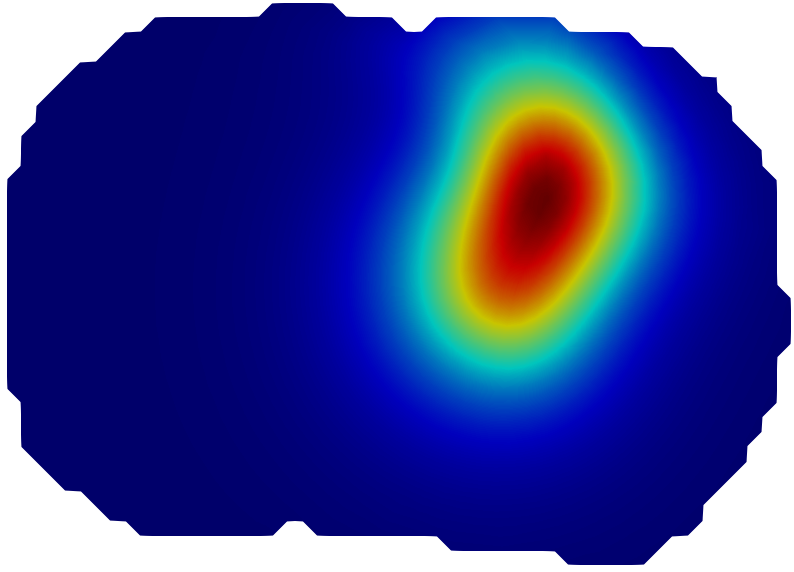} & 
\begin{tabular}{cl} 
\vspace{-2.1cm} \\
{\includegraphics[width=0.8cm, height=2.2cm, keepaspectratio=false]{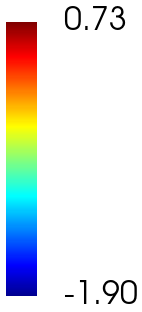}} &
\hspace{-1pt}\rotatebox{90}{
    \raisebox{10pt}{
        \begin{tabular}{@{}c@{}} \textbf{\tiny log(G)} \\[5pt] \textbf{\tiny [log(1/day)]} \end{tabular}
    }
}
\end{tabular} \\

\end{tabular}
\vspace{-0.1in}
\caption{
Evolution of MAP estimates of spatially varying elastic modulus $E(\mathbf{X})$, tumor
diffusivity $D(\mathbf{X})$, and proliferation rate $G(\mathbf{X})$ for Rat-III during
sequential Bayesian inference.
    }
\label{fig:RatIII-map_evol}
\vspace{-0.1in}
\end{figure}

Figure~\ref{fig:RatIII-map_evol} shows the evolution of the MAP estimates of the
spatially varying tumor diffusivity $D$, proliferation rate $G$, and elastic
modulus $E$ for Rat~III during sequential Bayesian inference. As additional MRI
data are assimilated, the inferred parameter fields retain coherent spatial
structure while becoming progressively refined. 
\blue{These fields provide one estimate of the spatial heterogeneity, while the degree to which each parameter is informed by the data is assessed below using posterior uncertainty reduction.}
Moreover, the spatial correlation imposed by the priors enables information from
the tumor region to propagate into the surrounding tissue, supporting inference
of heterogeneous growth patterns and accurate prediction of the tumor shapes. 
In particular, the inferred elastic modulus $E$ exhibits spatial contrast between the tumor region and adjacent tissue, with lower $E$ in regions of high tumor burden. Through the constitutive relation defining the first Piola--Kirchhoff stress $\mathbf{P}$, this stiffness heterogeneity modulates the mechanical response in \eqref{eq:mech_PDE_REF}, thereby shaping the stress field that feeds back into tumor expansion via the stress-dependent diffusivity term in \eqref{eq:rd_PDE_REF}. The inferred diffusivity $D$ exhibits spatial heterogeneity aligned with the observed directions of tumor spread, while the proliferation rate $G$ remains more localized, with elevated values confined to the actively growing tumor region. \blue{The inferred baseline diffusivity should not be interpreted as a pointwise map of invasion intensity at the apparent tumor boundary; in the reaction--diffusion model, invasion is governed by the diffusive flux, which depends on both $D$ and the spatial gradient of $\phi$, and in the mechanically coupled model also on stress-mediated modulation.}

\begin{figure}[!htbp]
\centering
\hspace*{-0.4cm}

\renewcommand{\arraystretch}{0}
\setlength{\tabcolsep}{1pt}
\setlength{\fboxsep}{0pt}

\newcommand{\casehead}[1]{
    \colorbox{black}{
        \makebox[4.3cm][c]{
            \vrule width 0pt height 0.38cm depth 0.22cm
            \textcolor{white}{\textbf{\small #1}}
        }}}

\newcommand{\sensimg}[1]{
    \includegraphics[width=4.5cm, trim=3 3 3 3, clip]{#1}
}

\begin{tabular}{ccc}

\casehead{log(E)} &
\casehead{log(D)} &
\casehead{log(G)} \\ [12pt]

\sensimg{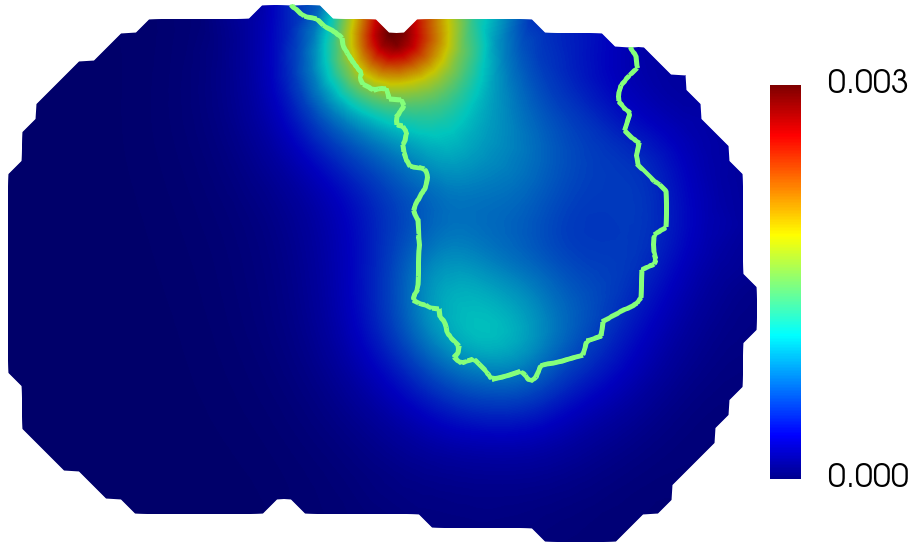} &
\sensimg{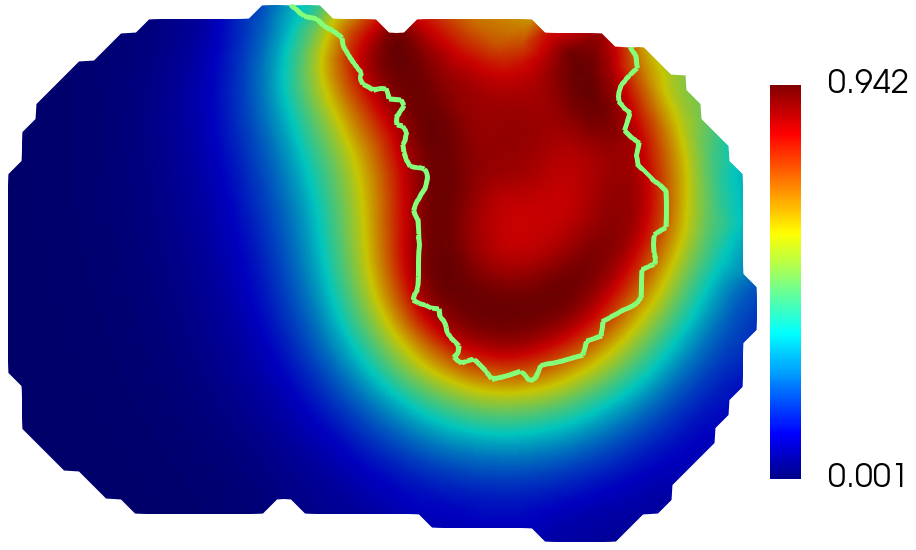} &
\sensimg{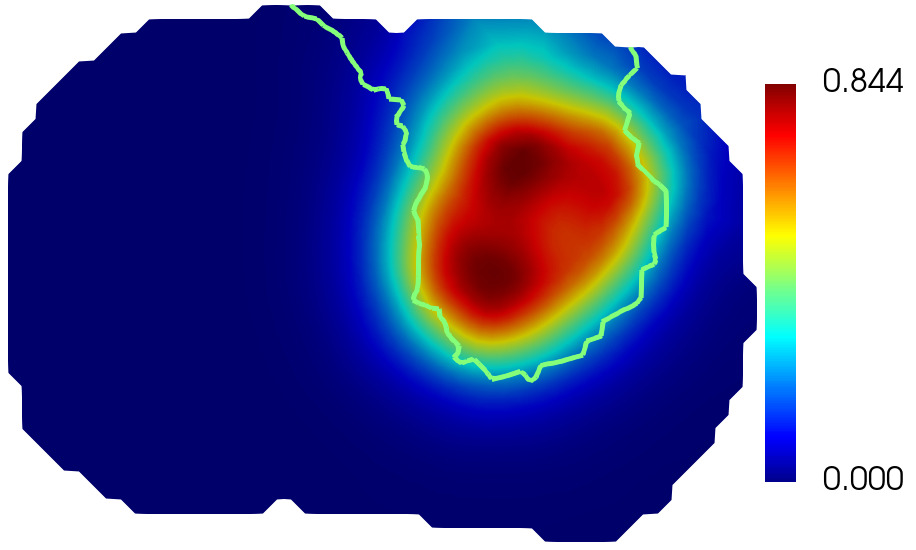} \\

\end{tabular}

\vspace{-0.05in}
\caption{\blue{
Spatial maps of the local posterior uncertainty reduction index $\mathcal R_\theta$ for the inferred parameter fields for Rat~III on Day~19 including the overlay of the model-predicted tumor boundary. Higher values correspond to regions where the longitudinal MRI data provide stronger local constraints on the parameter field, while lower values indicate regions that remain weakly informed by the data.
}}
\label{fig:uncertainty_red}
\vspace{-0.1in}
\end{figure}

\blue{
To assess the how much the MRI data informs the inferred spatial parameter fields, we quantified the reduction of posterior uncertainty relative to the prior. For each inferred parameter field $\theta \in {\log D,\log G,\log E}$, we computed the local posterior uncertainty reduction index,
\begin{equation}
\mathcal{R}_\theta=
1-
\frac{
\sqrt{\operatorname{Var}_{\pi_{\mathrm{post}}}(\theta)}
}{
\sqrt{\operatorname{Var}_{\pi_{\mathrm{pr}}}(\theta)}
}.
\end{equation}
Here, $\mathcal R_\theta$ close to zero indicates small local reduction in uncertainty relative to the prior, whereas larger values indicate regions where the MRI data more strongly constrain the parameter field. The resulting maps for Rat~III at Day~19 are shown in Figure~\ref{fig:uncertainty_red}. The uncertainty reduction is concentrated in the tumor-bearing and peritumoral regions, with stronger reduction for diffusivity and proliferation than for stiffness. This trend is expected because $D$ and $G$ directly control tumor volume fraction evolution, whereas $E$ affects the observed tumor volume fraction only indirectly through the mechanical coupling.
}

\begin{figure}[!ht]
    \centering
    {\setlength{\fboxrule}{1pt}
    \setlength{\fboxsep}{1pt}
    {\includegraphics[width=0.25\linewidth]{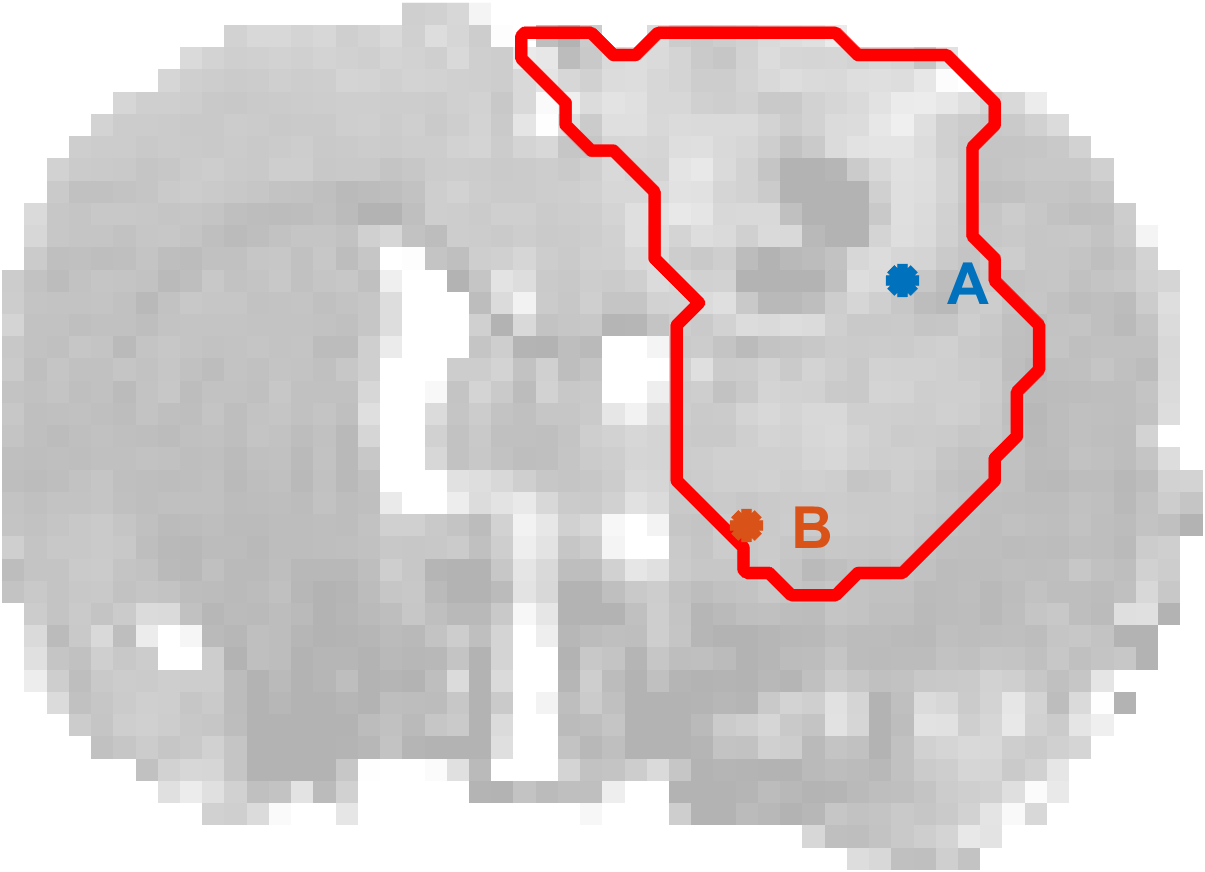}}}
    \\
    \includegraphics[width=0.25\linewidth]{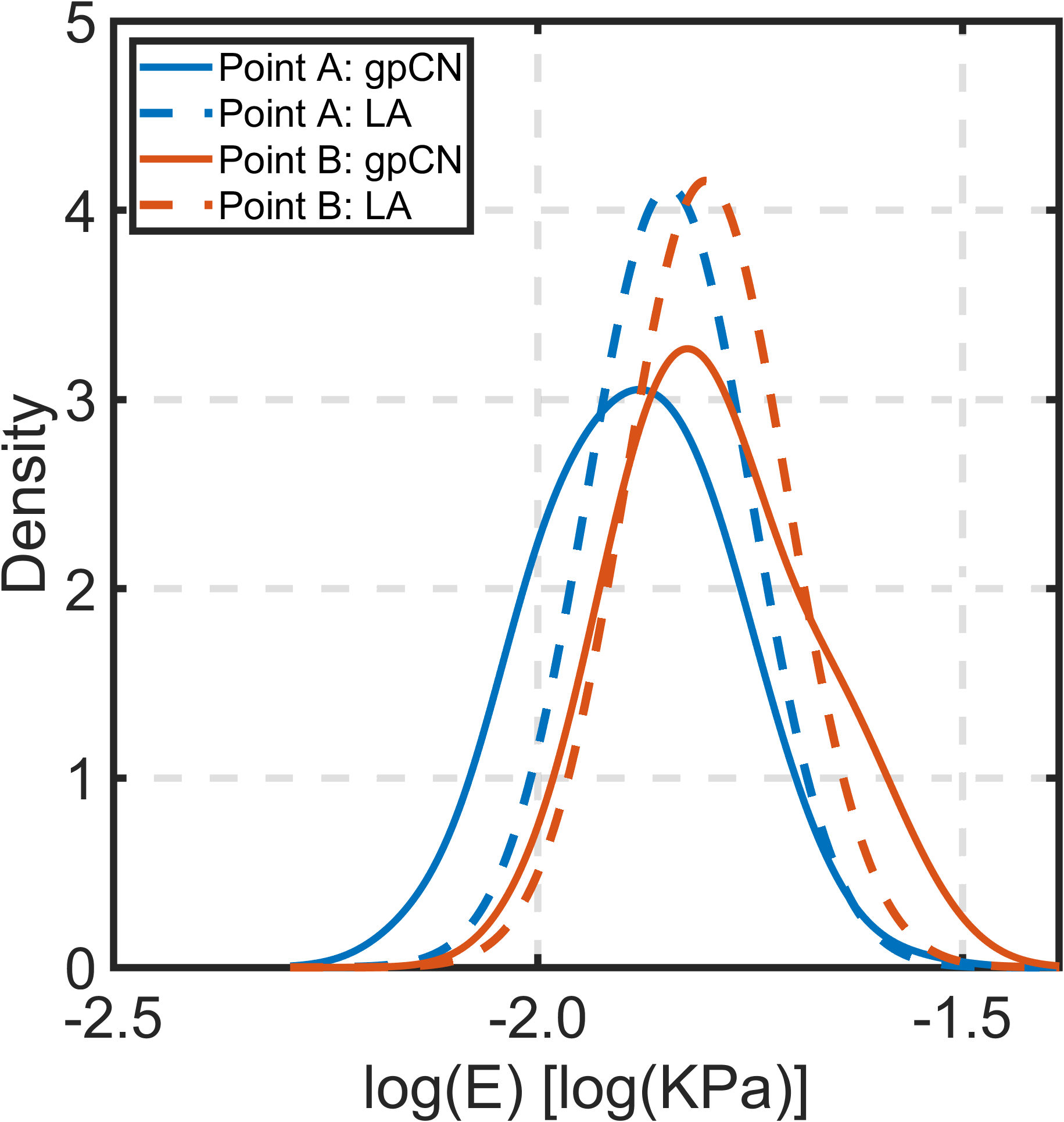}
    ~
    \includegraphics[width=0.27\linewidth]{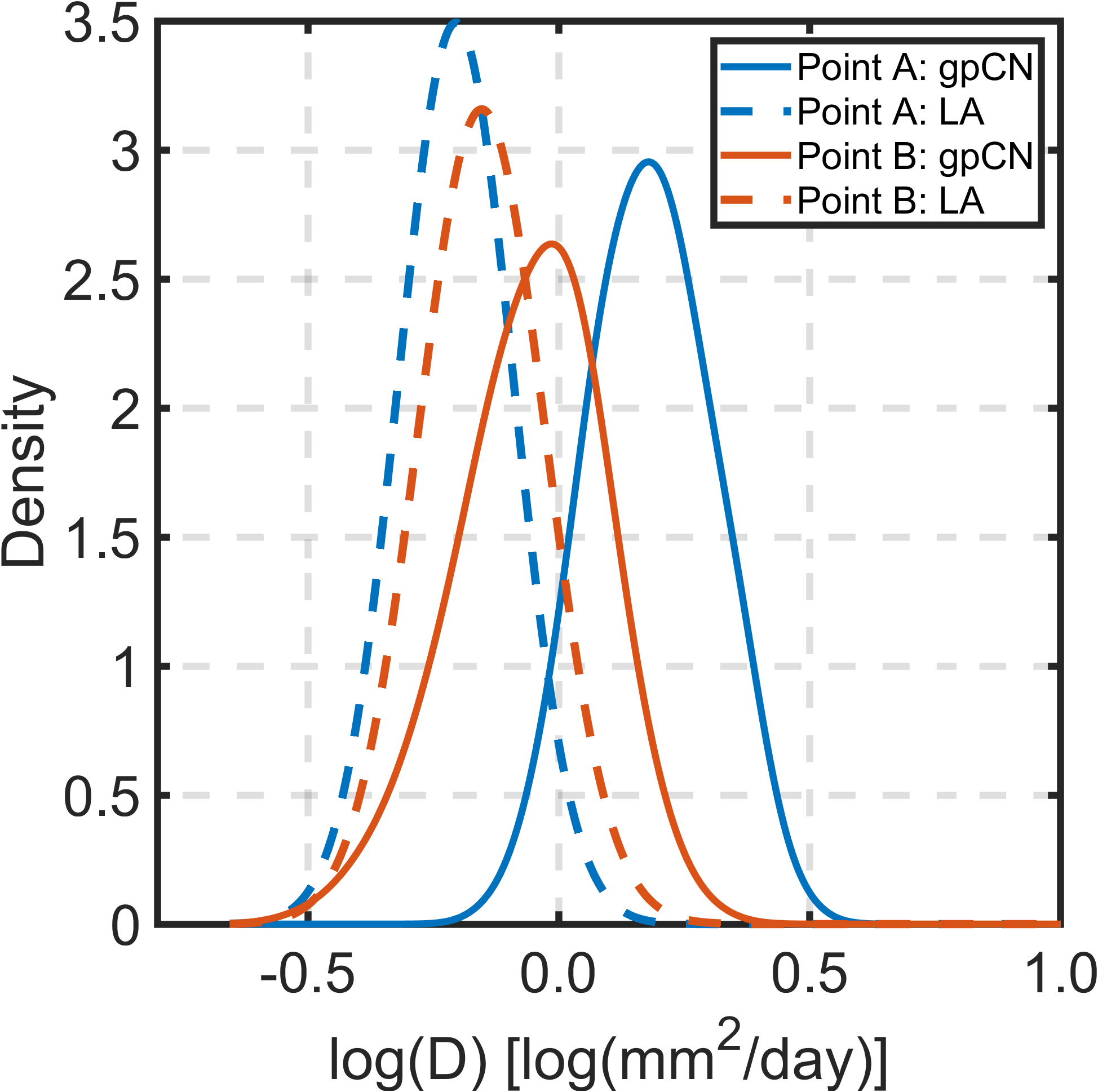}
    ~
    \includegraphics[width=0.28\linewidth]{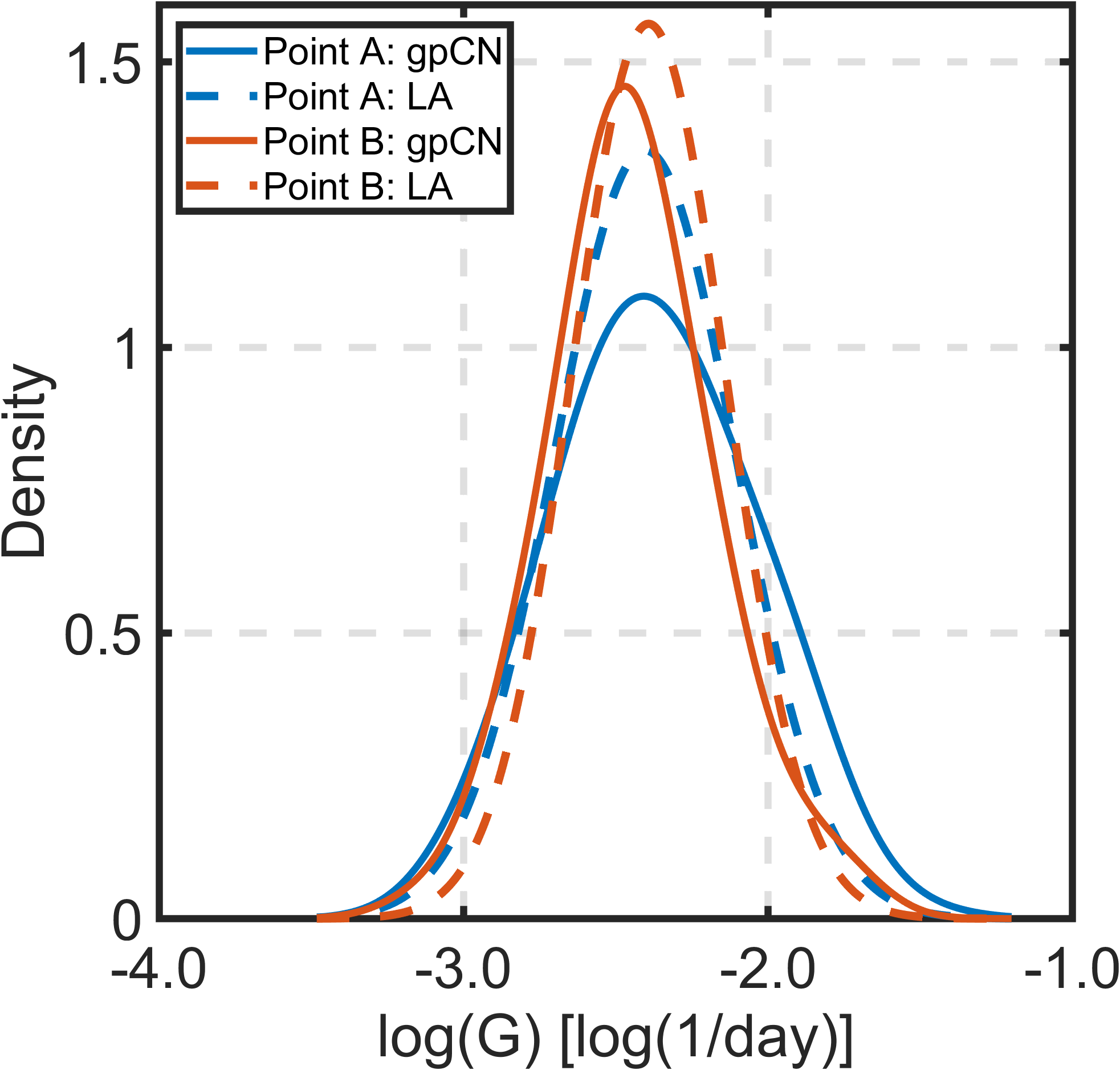}
\caption{Comparison of LA and gpCN posterior estimates: point-wise kernel density estimates (KDEs) of inference parameters at two locations within the tumor in Rat~III.
}
\vspace{-0.2in}
    \label{fig:la_gpcn_kde}
\end{figure}
\begin{figure}[!ht]
\centering
\makebox[\textwidth][c]{
    \renewcommand{\arraystretch}{0} 
    \setlength{\tabcolsep}{6pt}
    \setlength{\fboxsep}{0pt}

    \newcommand{\panelimg}[2]{
        \begin{tabular}{@{}l@{}}
            \textbf{(#1)} \\[-2pt]
            \includegraphics[width=4.6cm, trim=3 3 3 3, clip]{#2}
        \end{tabular}
    }

    \begin{tabular}{cc}
    \panelimg{a}{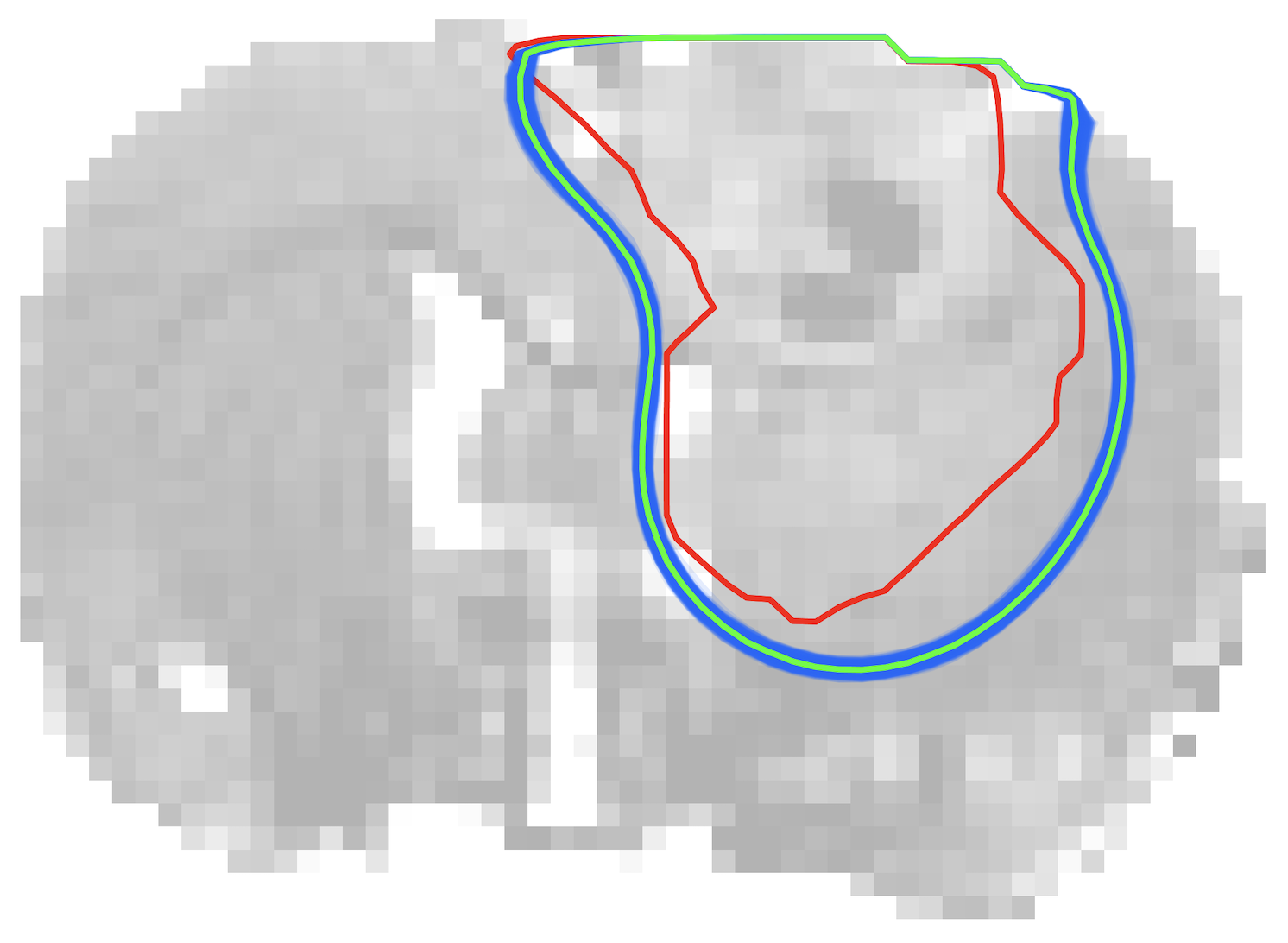} & \panelimg{b}{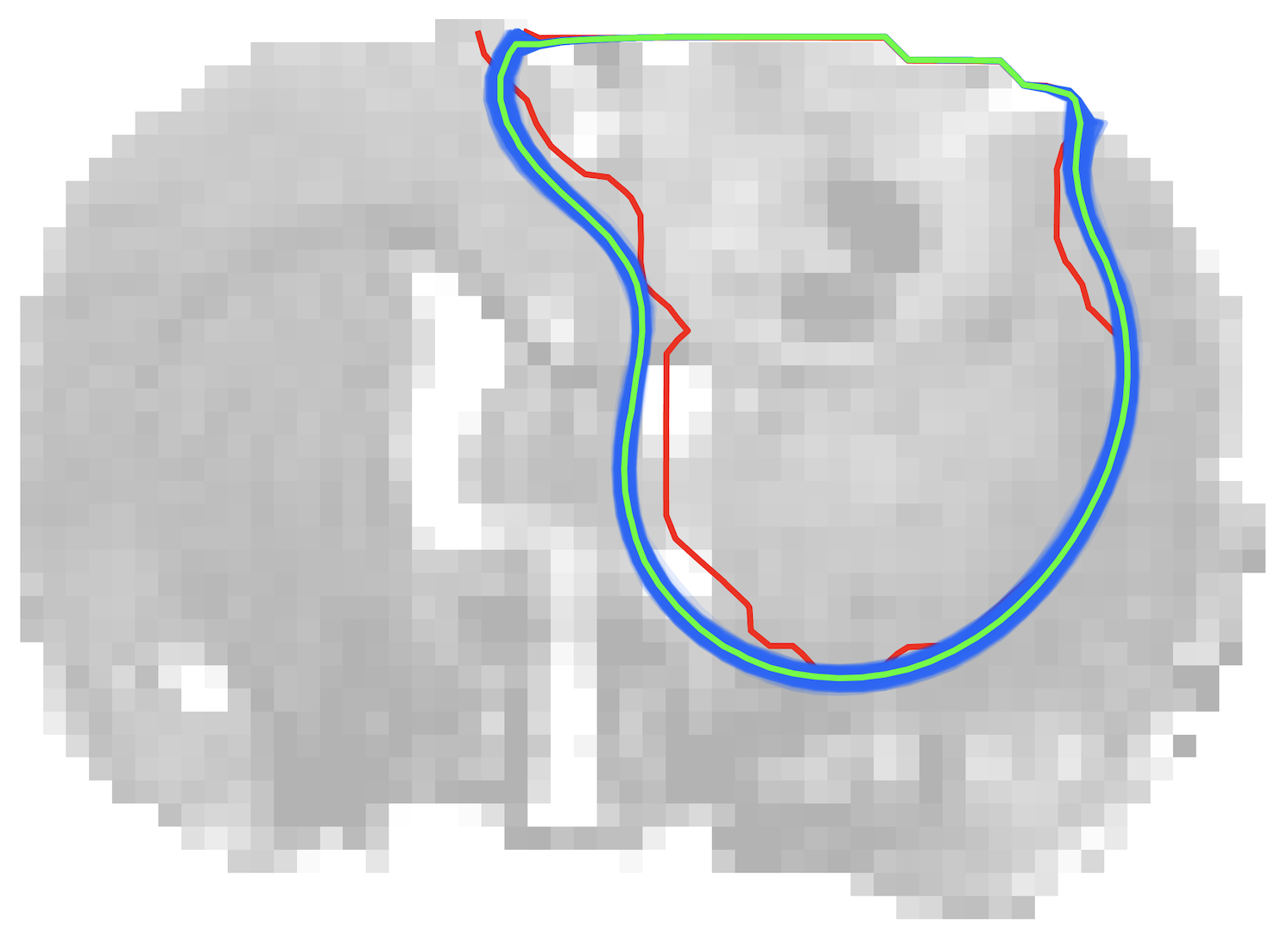} \\ [10pt]

    \panelimg{c}{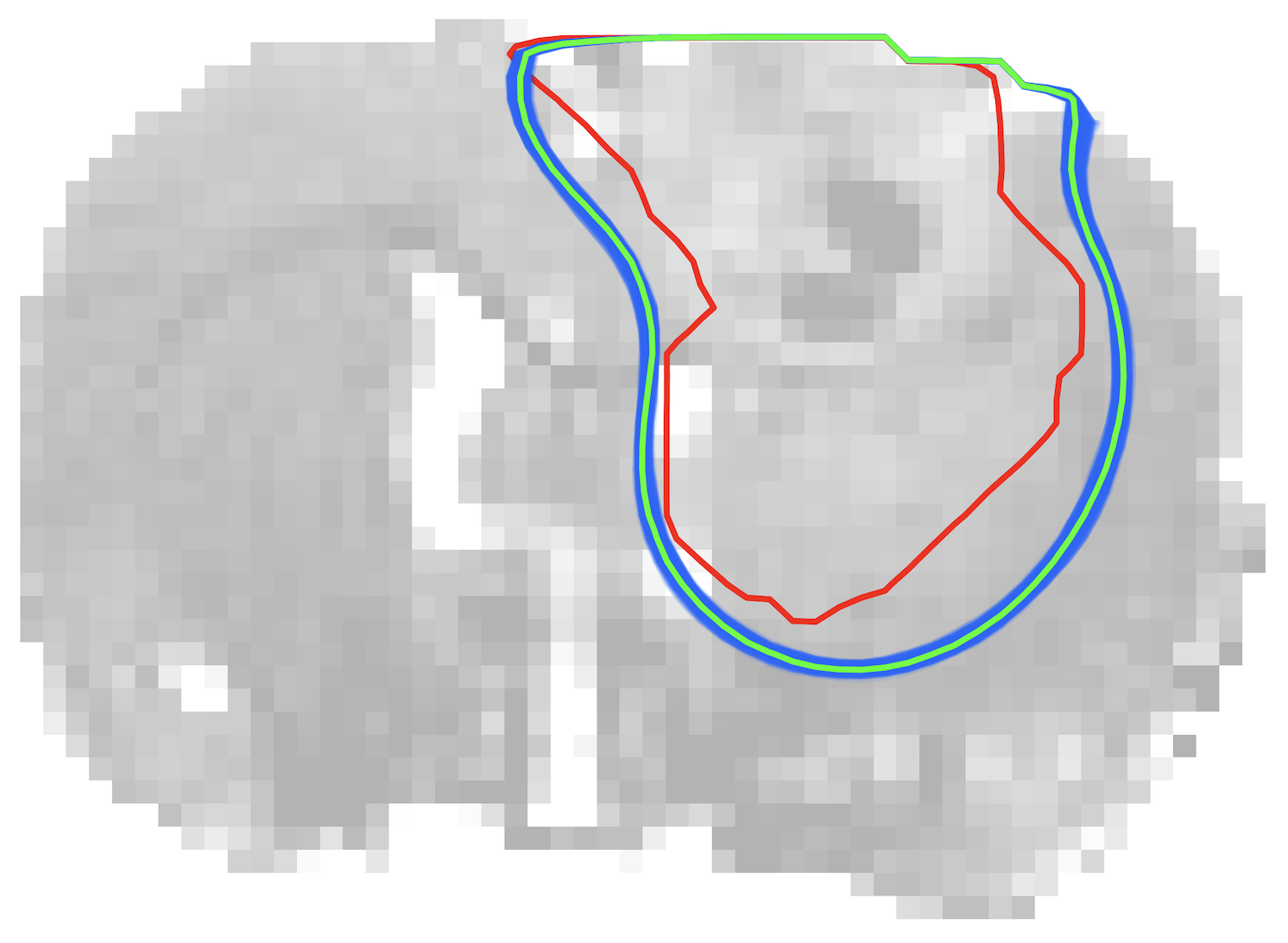} & \panelimg{d}{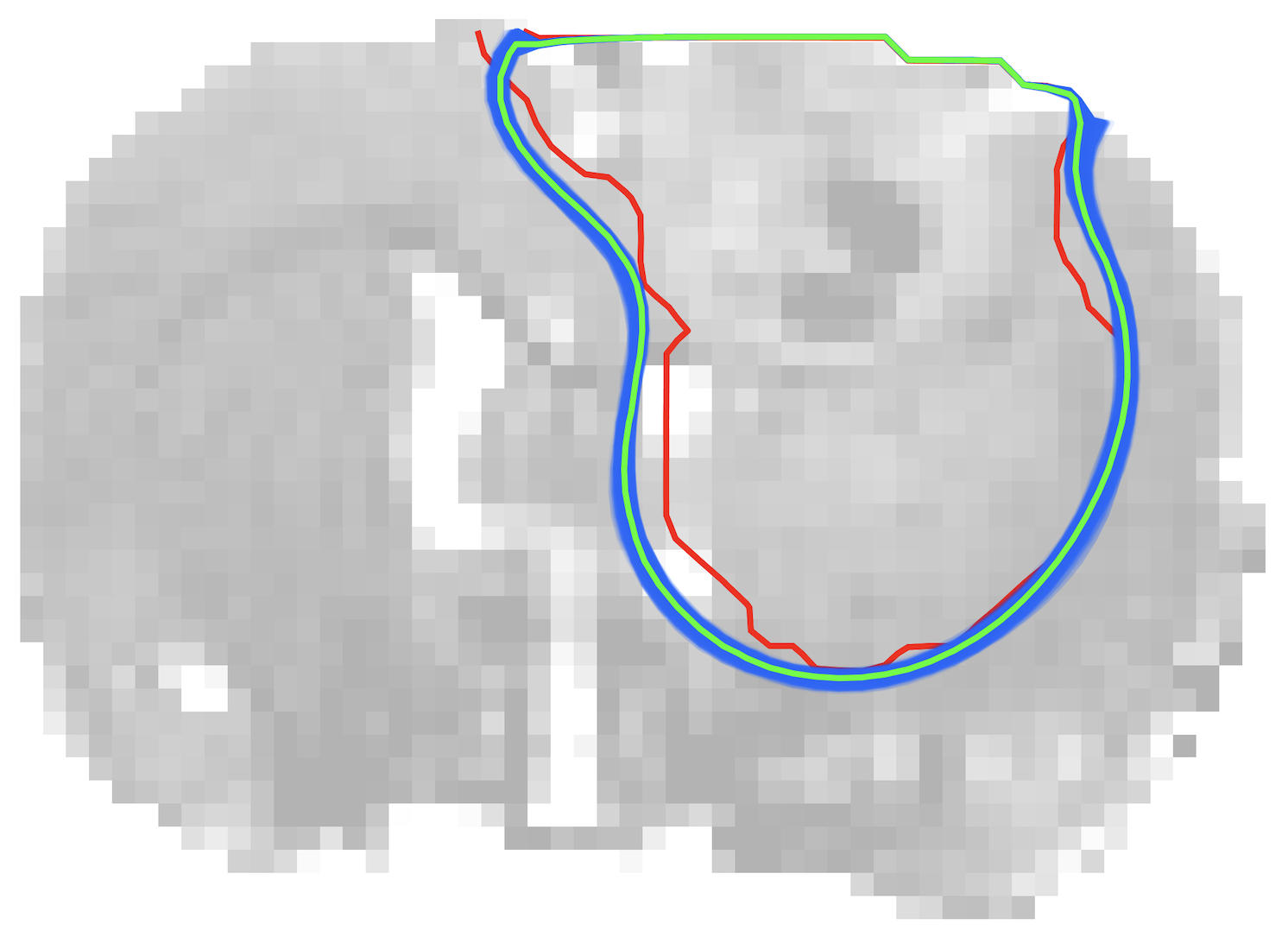} \\ [15pt]

    \multicolumn{2}{c}{
        \includegraphics[width=7cm]{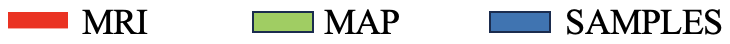}
    } \\
    \end{tabular}
}
\vspace{-1pt}
\caption{Comparison of tumor boundary predictions from LA and gpCN posterior samples (blue) with MRI-derived boundaries (red) and MAP predictions (green) in Rat~III:  (a,b) LA samples and (c,d) gpCN samples. Panels (a,c) correspond to Day 16 and (b,d) to Day 19 predictions. A total of 1000 samples are used for each method.}
\label{fig:la_gpcn_boundsamples}
\end{figure}
\begin{figure}[!htbp]
\centering
\hspace*{-0.05\textwidth}
\makebox[\textwidth][c]{
    \renewcommand{\arraystretch}{0}
    \setlength{\tabcolsep}{2pt}
    \setlength{\fboxsep}{0pt}

    \begin{minipage}[t]{0.18\textwidth}
        \centering

        \vspace{0pt} 
        \textbf{\small MAP} \\ 
        \vspace{35pt} 
        \includegraphics[width=1.0\textwidth, trim=3 3 3 3, clip]{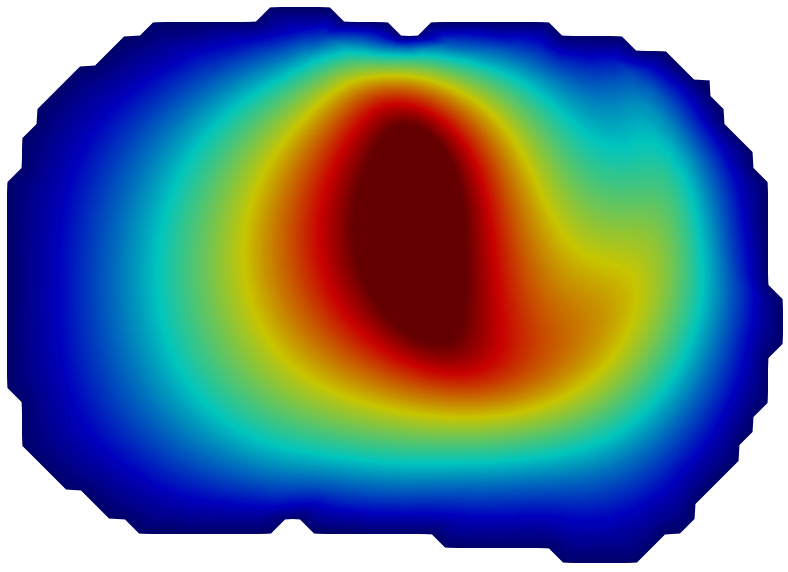} \\ 

        \vspace{62pt} 
        \includegraphics[width=1.0\textwidth, trim=3 3 3 3, clip]{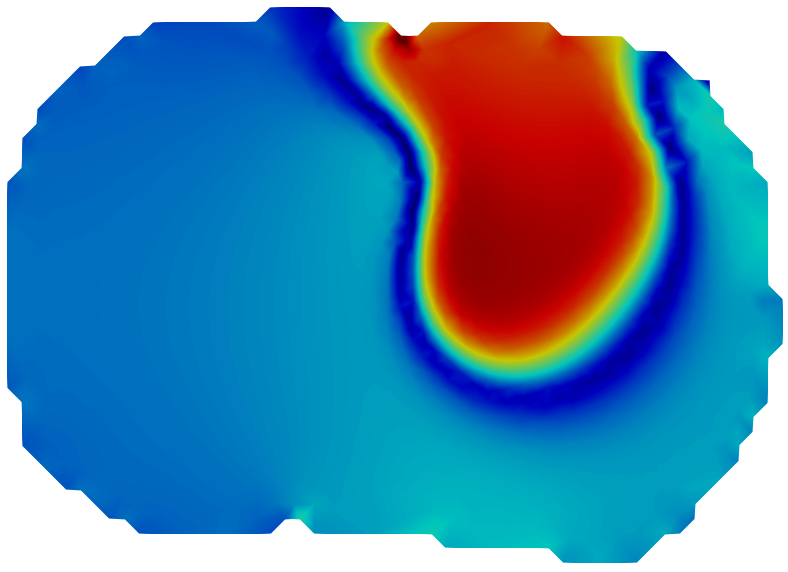}
    \end{minipage}

    \hspace{1.5pt}
    {\color{gray}\vrule width 1pt} 
    \hspace{1pt}

    \begin{minipage}[t]{0.78\textwidth}
        \centering

        \vspace{0pt} 
        \textbf{\small Samples} \\[10pt]
        \begin{tabular}{m{0.03\textwidth}*{4}{c} c @{\hspace{-25pt}} m{0.05\textwidth}}
            
            \raisebox{18pt}[0pt][0pt]{\rotatebox{90}{\textbf{\footnotesize LA}}} &
            \includegraphics[width=0.21\textwidth]{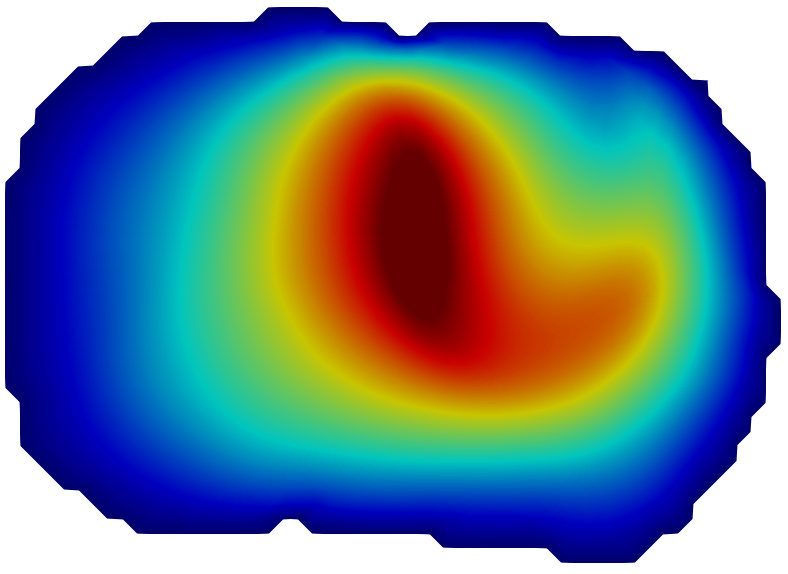} &
            \includegraphics[width=0.21\textwidth]{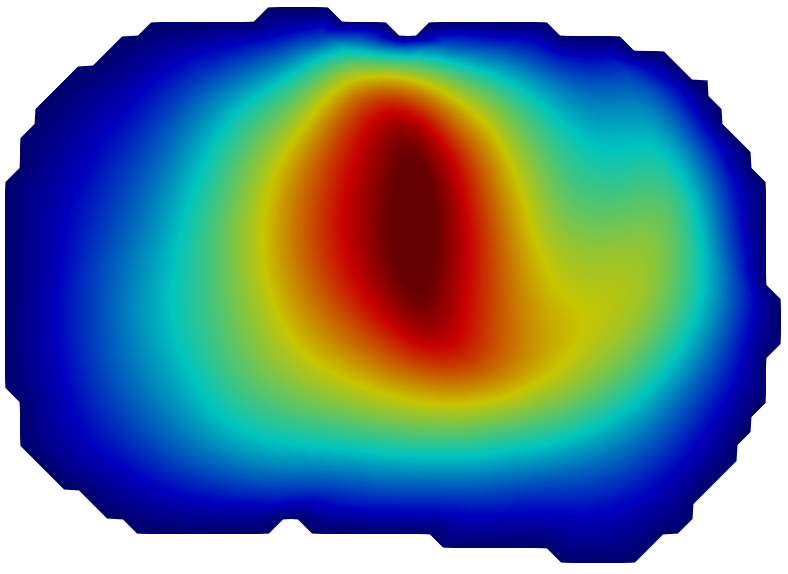} &
            \includegraphics[width=0.21\textwidth]{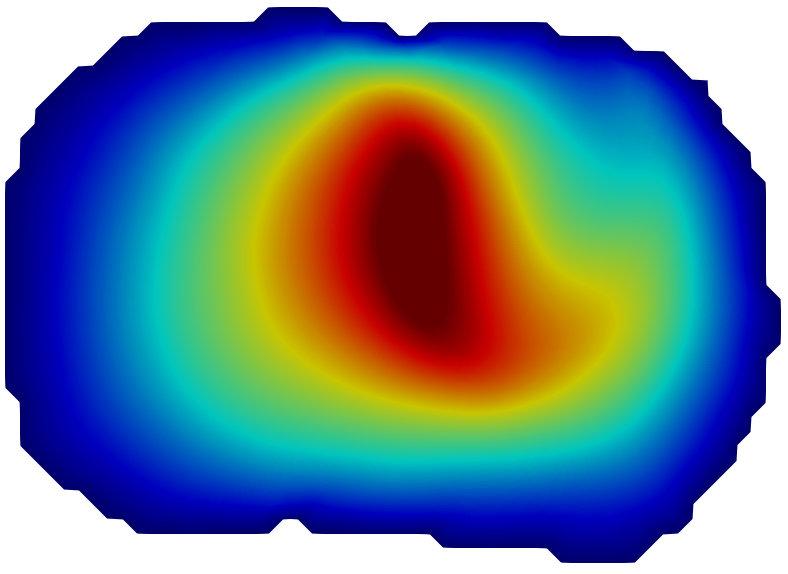} &
            \includegraphics[width=0.21\textwidth]{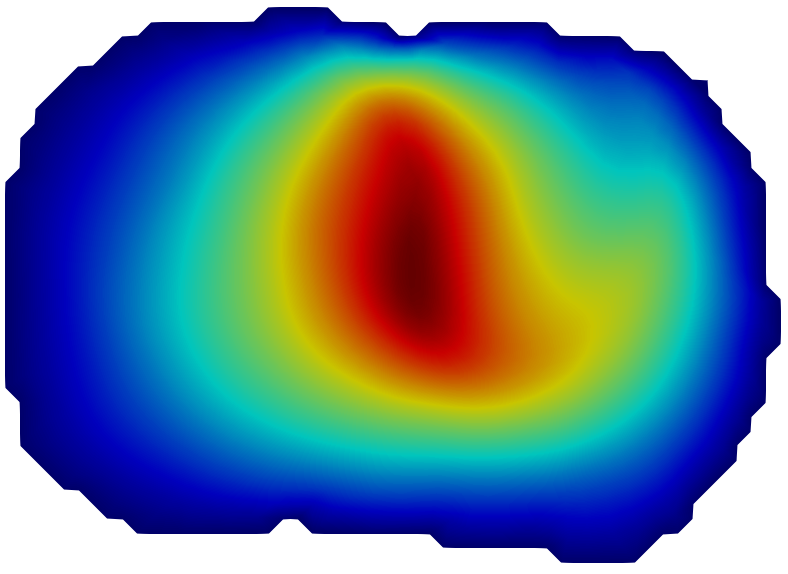} &
            
            \multirow[t]{2}{*}{
                \raisebox{-1.5cm}{
                \includegraphics[width=0.8cm, height=3cm, keepaspectratio=false]{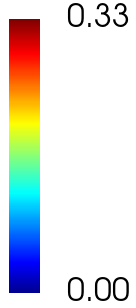}
                }
            } & 

            \multirow[t]{2}{*}{
                \raisebox{-10pt}[0pt][0pt]{
                    \rotatebox{90}{\begin{tabular}{@{}c@{}} \textbf{\tiny ||u||} \\[.12cm] \textbf{\tiny [mm]} \end{tabular}}
                }
            } \\ [2pt]
            
            \raisebox{10pt}[0pt][0pt]{\rotatebox{90}{\textbf{\footnotesize gpCN}}} &
            \includegraphics[width=0.21\textwidth]{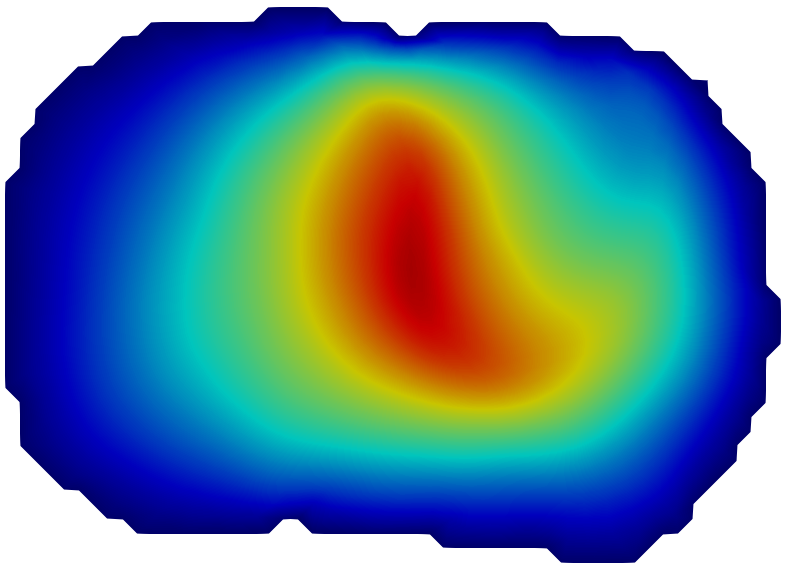} &
            \includegraphics[width=0.21\textwidth]{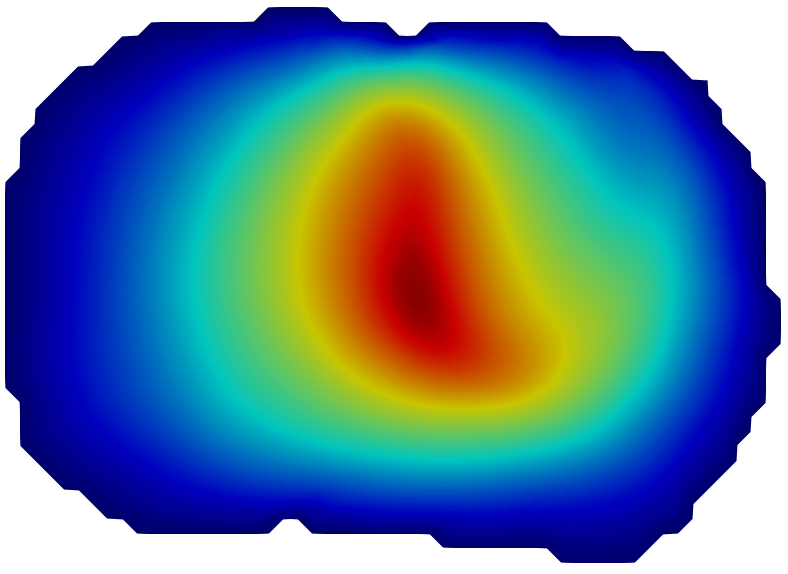} &
            \includegraphics[width=0.21\textwidth]{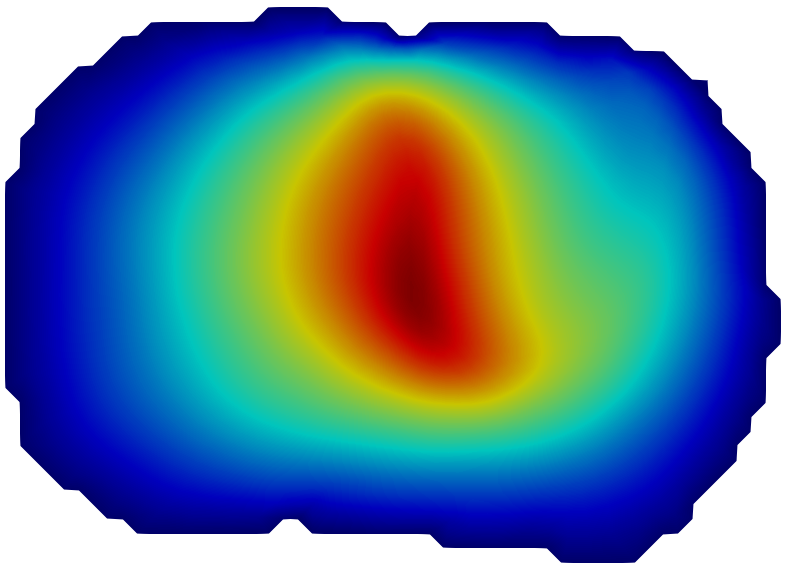} &
            \includegraphics[width=0.21\textwidth]{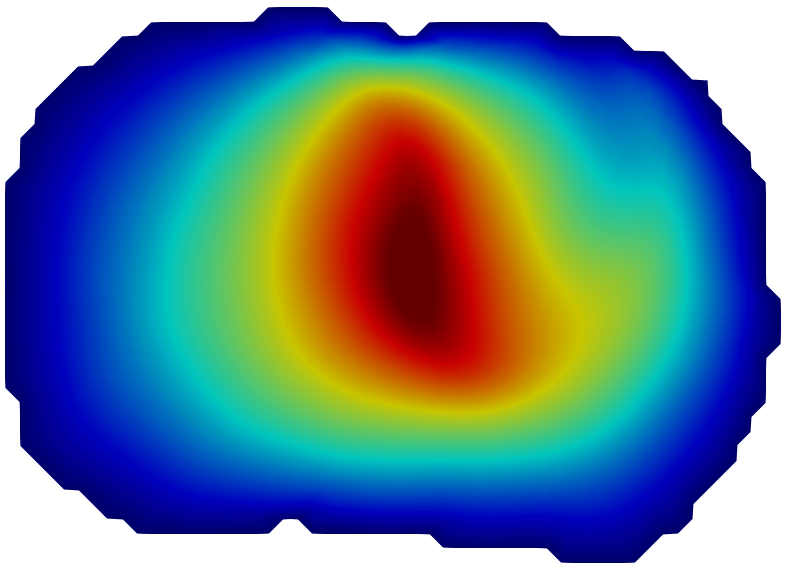} & & \\ [18pt] 
            
            \raisebox{18pt}[0pt][0pt]{\rotatebox{90}{\textbf{\footnotesize LA}}} &
            \includegraphics[width=0.21\textwidth]{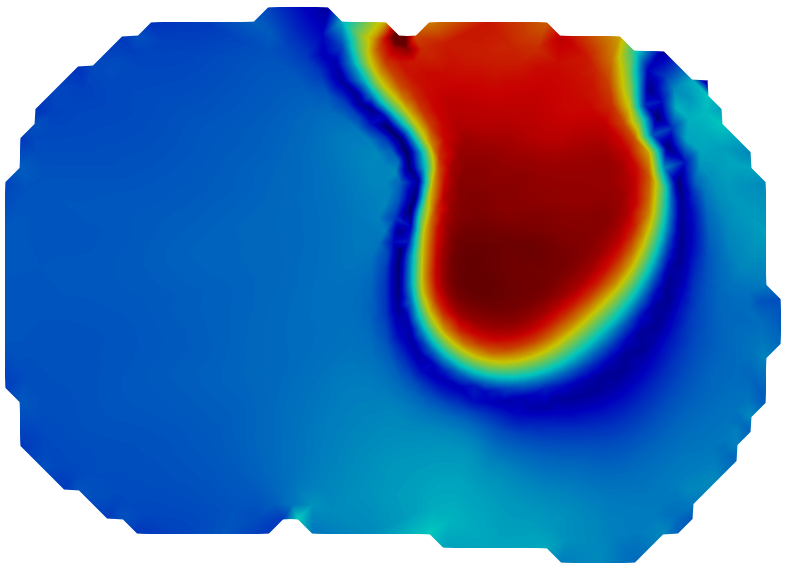} &
            \includegraphics[width=0.21\textwidth]{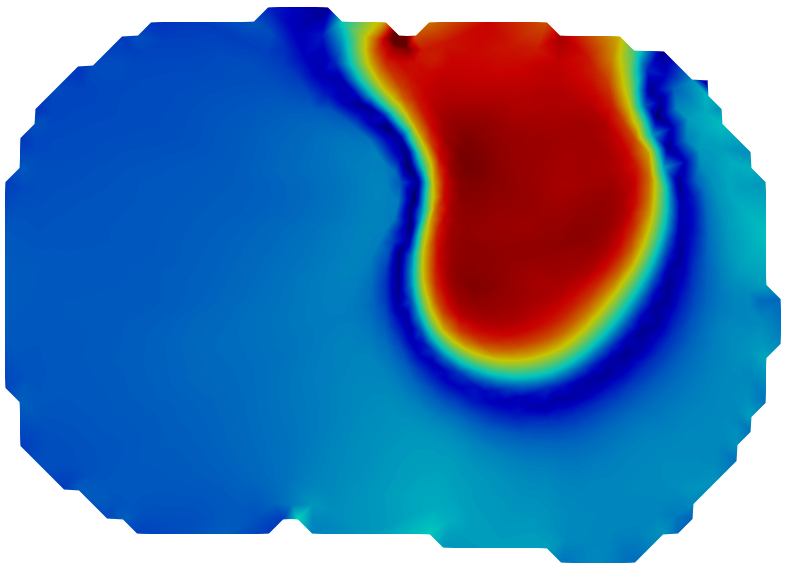} &
            \includegraphics[width=0.21\textwidth]{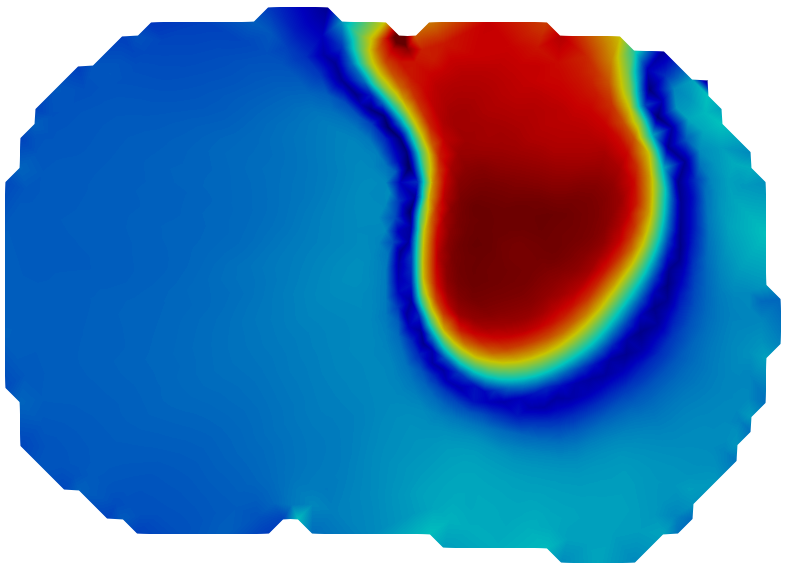} &
            \includegraphics[width=0.21\textwidth]{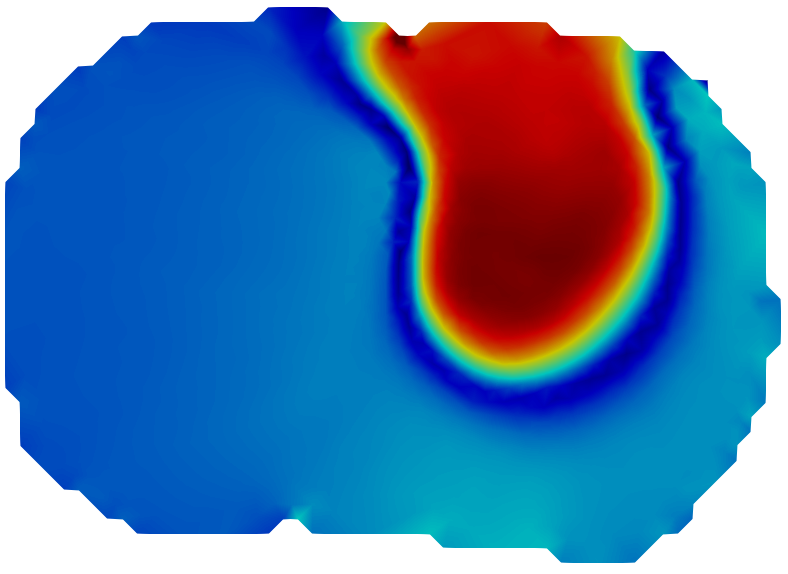} &

            \multirow[t]{2}{*}{
                \raisebox{-1.5cm}{
                \includegraphics[width=0.8cm, height=3cm, keepaspectratio=false]{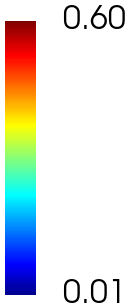}
                }
            } & 

            \multirow[t]{2}{*}{
                \raisebox{-10pt}[0pt][0pt]{
                    \rotatebox{90}{\begin{tabular}{@{}c@{}} \textbf{\tiny ||P||} \\ [.12cm]\textbf{\tiny [KPa]} \end{tabular}}
                }

            } \\ [2pt]

            \raisebox{10pt}[0pt][0pt]{\rotatebox{90}{\textbf{\footnotesize gpCN}}} &
            \includegraphics[width=0.21\textwidth]{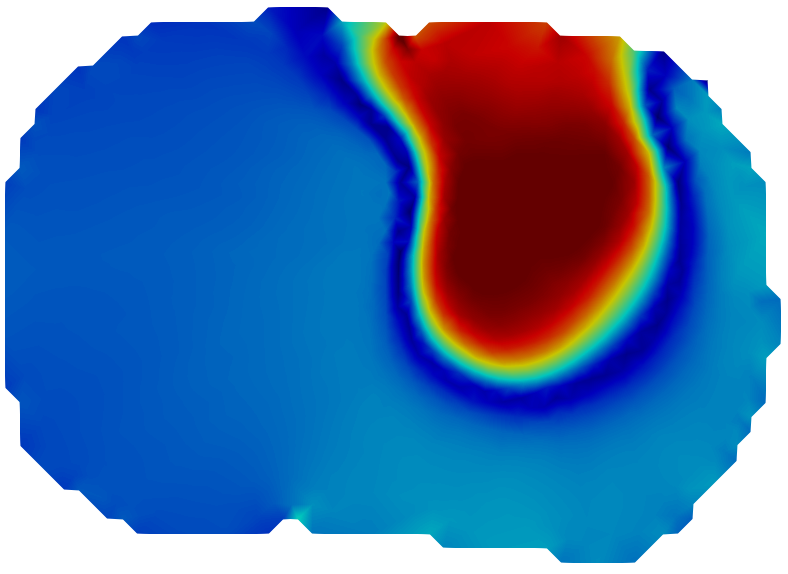} &
           \includegraphics[width=0.21\textwidth]{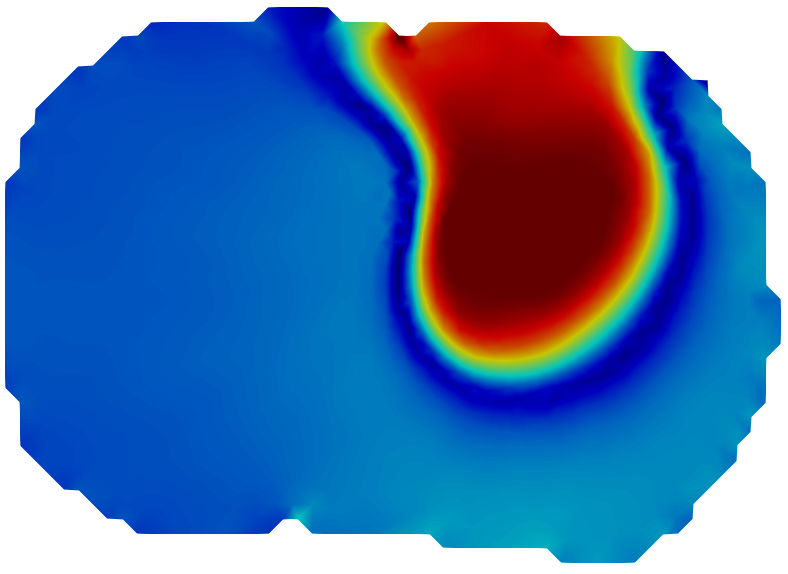} &
           \includegraphics[width=0.21\textwidth]{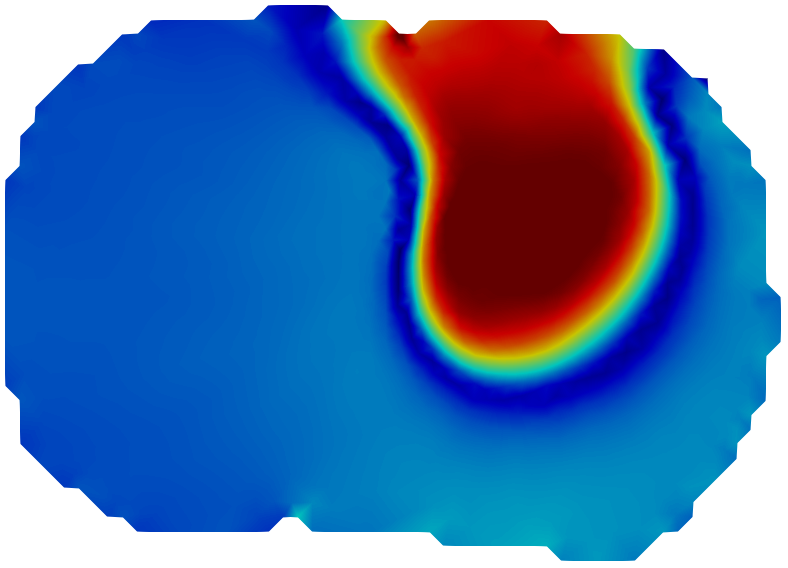} &
           \includegraphics[width=0.21\textwidth]{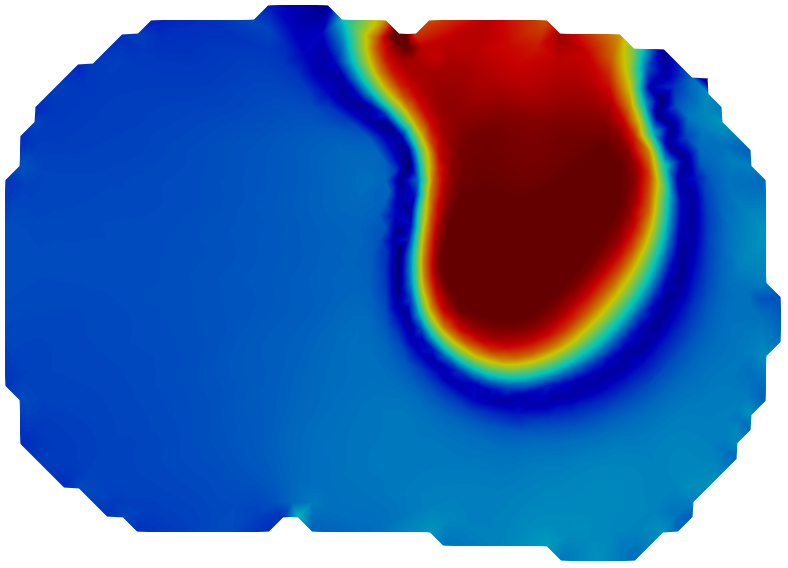} & & 

        \end{tabular}
    \end{minipage}
}
\vspace{0pt}
\caption{Comparison of displacement and stress fields for Rat~III: MAP estimates and posterior samples obtained from LA and gpCN for the Day 19 prediction.}
\label{fig:la_gpcn_stress_disp}
\end{figure}

%
Figures~\ref{fig:la_gpcn_kde}---\ref{fig:la_gpcn_stress_disp} 
compare the low-rank LA with the sampling-based gpCN algorithm for Rat~III to assess the accuracy of the LA posterior approximation. \blue{The gpCN sampler is used here as an independent sampling-based reference rather than as an exact ground truth, since both approaches approximate the same Bayesian posterior using different computational strategies. As discussed in \textit{Supplementary Information}, both methods are applicable to high-dimensional parameter spaces, although gpCN is substantially more computationally demanding because it requires many posterior samples and repeated finite element solves.}
In this study, gpCN was run using four chains of length 5000, with each chain requiring approximately 56.4 hours of computation, whereas the corresponding LA solution required about 4 hours on the same processor.
Figure \ref{fig:la_gpcn_kde}
compares LA and gpCN estimates of the marginal posterior distributions of inference parameters at two locations within the tumor in Rat~III, using MRI data from Day~10 to Day~16. 
Overall, LA provides a reasonable approximation of the local posterior, with the best agreement observed for $\log(G)$ and $\log(E)$ and the largest discrepancy occurring for $\log(D)$. 
\blue{As an additional sampling-based validation of the LA model evidence approximation, (S.14) in the \textit{Supplementary Information}, we also estimated the relative model plausibility from the gpCN posterior samples for Rat~III at Day~19. This comparison was restricted to the two mechanically coupled formulations and therefore the reported plausibility values were normalized over the linear elastic and hyperelastic coupled reaction--diffusion models (see Section \ref{sec:results_dynselec} for more details). 
The LA and gpCN estimates yielded nearly identical plausibility values, preserving the same model ranking. When normalized over the two mechanically coupled formulations, both approaches assigned approximately $49\%$ plausibility to the linear elastic model and $51\%$ plausibility to the hyperelastic model.
This agreement indicates that the low-rank LA evidence approximation provides a reliable estimate of relative model plausibility for the dynamic model-selection framework.} 
Figure~\ref {fig:la_gpcn_boundsamples} compares MRI-derived tumor boundaries with the MAP prediction and boundary ensembles generated from posterior samples obtained by LA and gpCN. The two methods produce tumor boundary envelopes of similar width at both Day~16 and Day~19, indicating comparable uncertainty in the predicted tumor morphology.
Figure~\ref{fig:la_gpcn_stress_disp} compares the displacement and stress fields at the MAP estimate and across posterior samples for the Day~19 prediction. While LA and gpCN produce qualitatively consistent spatial patterns, gpCN samples exhibit slightly smaller displacement and stress magnitudes. This difference is consistent with the diffusivity discrepancies observed in Figure~\ref{fig:la_gpcn_kde}, consistent with the slightly larger posterior diffusivity inferred by gpCN.

\blue{Overall, these comparisons show that LA reproduces the main characteristics of the sampling-based posterior relevant to this study, including marginal parameter distributions, tumor-boundary uncertainty, mechanically induced displacement and stress patterns, and relative model plausibility. Although modest differences remain, particularly in the posterior of $\log(D)$, these differences do not alter the predicted tumor morphology or the qualitative ordering of competing biomechanical models. Owing to its substantially lower computational cost, LA is adopted in the remainder of this work for sequential Bayesian inference and dynamic model selection.}

\subsection{Dynamic selection of biomechanical tumor models}\label{sec:results_dynselec}
While spatially varying parameter inference captures heterogeneity in tumor growth and tissue mechanics within a prescribed model, it does not by itself determine whether the assumed biomechanical formulation remains the most plausible description as additional MRI data are acquired. Glioma growth is a dynamic process in which the relative importance of diffusion, proliferation, mass effect, and stress-mediated feedback may evolve as the tumor expands within the confined brain. 
\blue{Therefore, rather than assuming a single fixed mechanical formulation a priori, we evaluate competing biomechanical models sequentially for each animal. At each assimilation time, the candidate models are recalibrated using the accumulated MRI data from that subject, their posterior model plausibilities are recomputed, and the resulting model ranking can be used to select the formulation for subject-specific one-scan-ahead prediction. To this end, \textit{dynamic selection} refers to time-updated, subject-specific posterior model comparison, not automatic discovery of new model equations.}

%

%
\begin{figure}[H]
    \centering
    \includegraphics[width=0.249\linewidth]{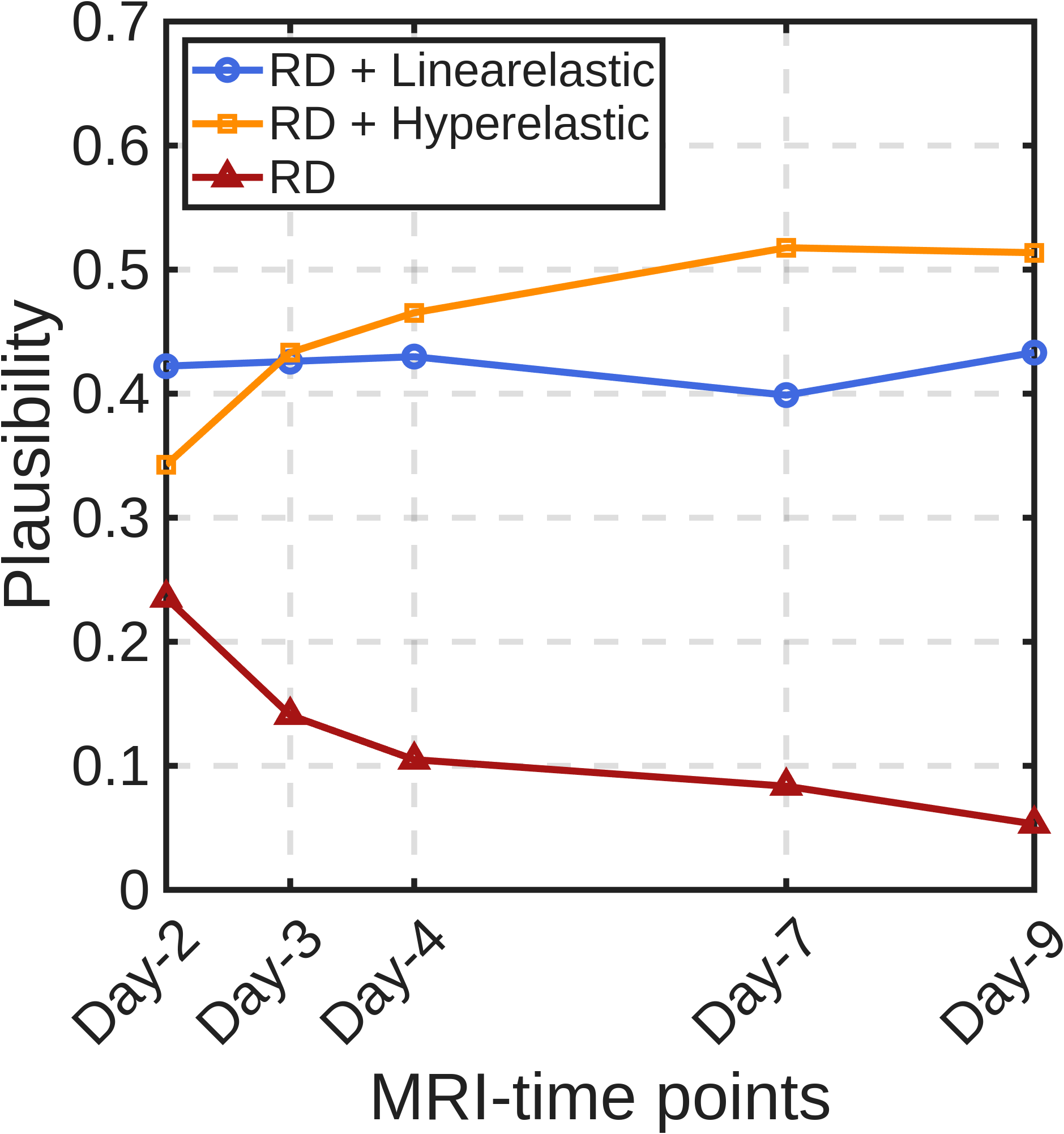}
    ~
    \includegraphics[width=0.255\linewidth]{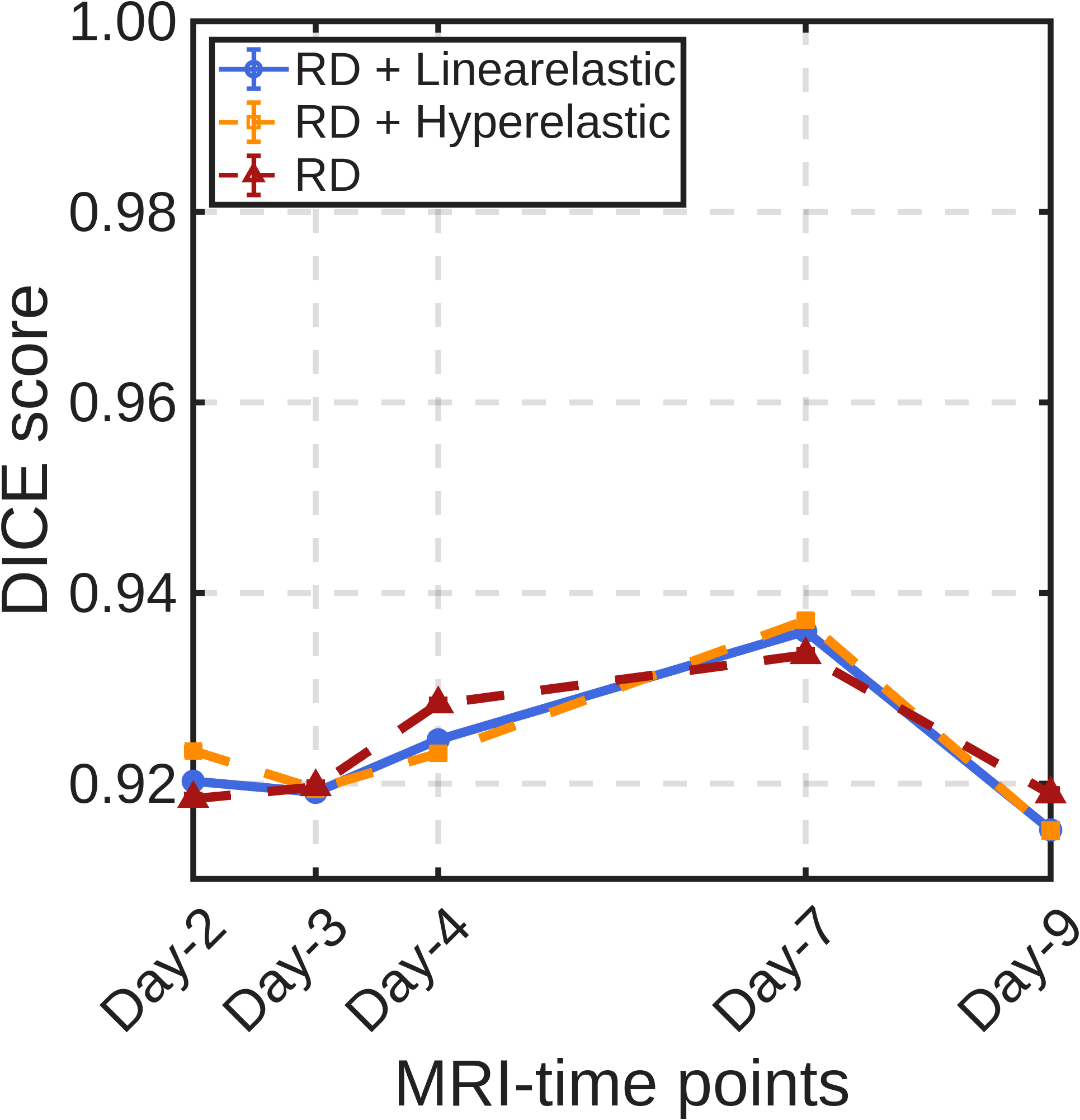}
    ~
    \includegraphics[width=0.255\linewidth]{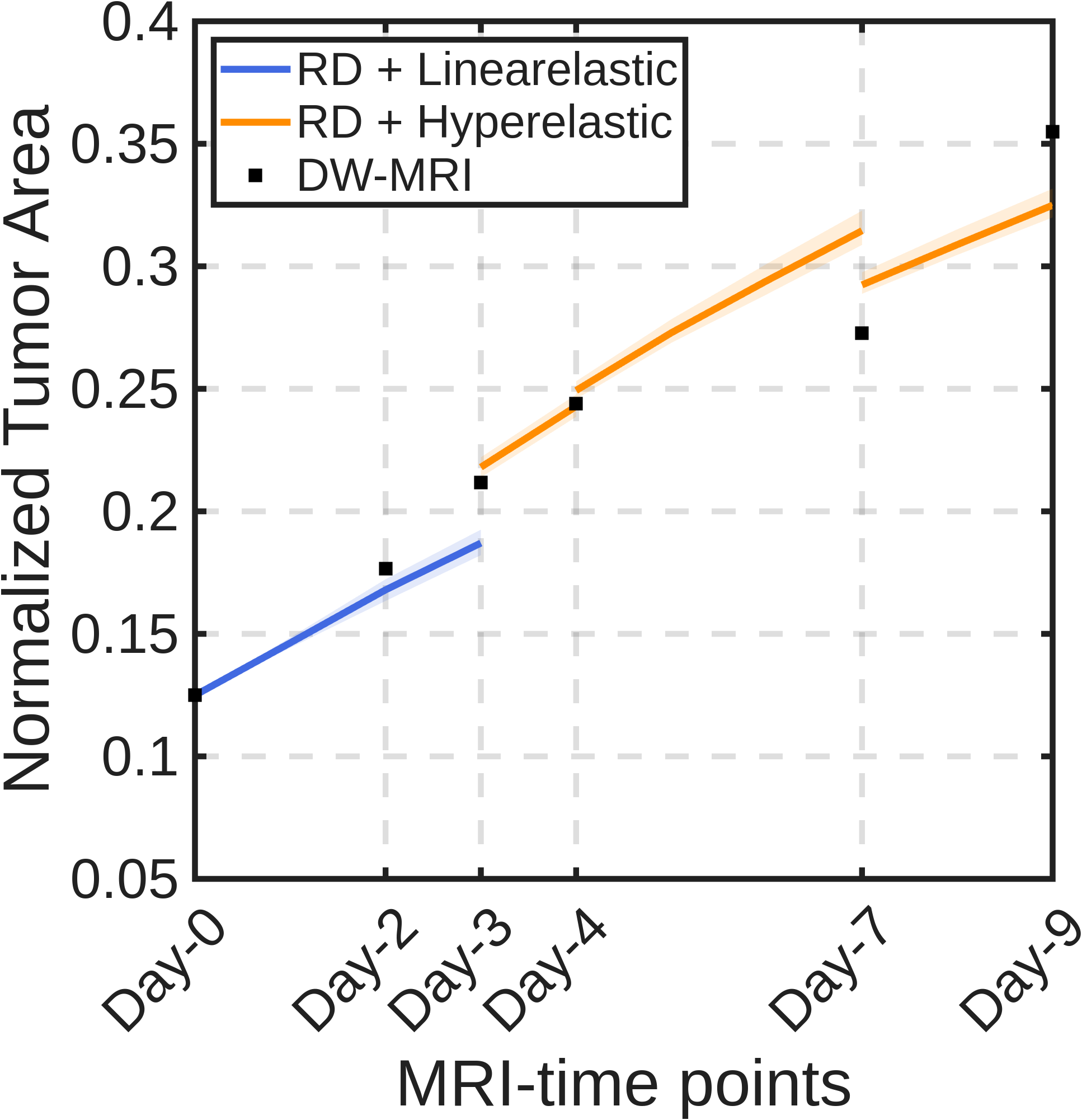}
    \\ (a) \\
    \includegraphics[width=0.251\linewidth]{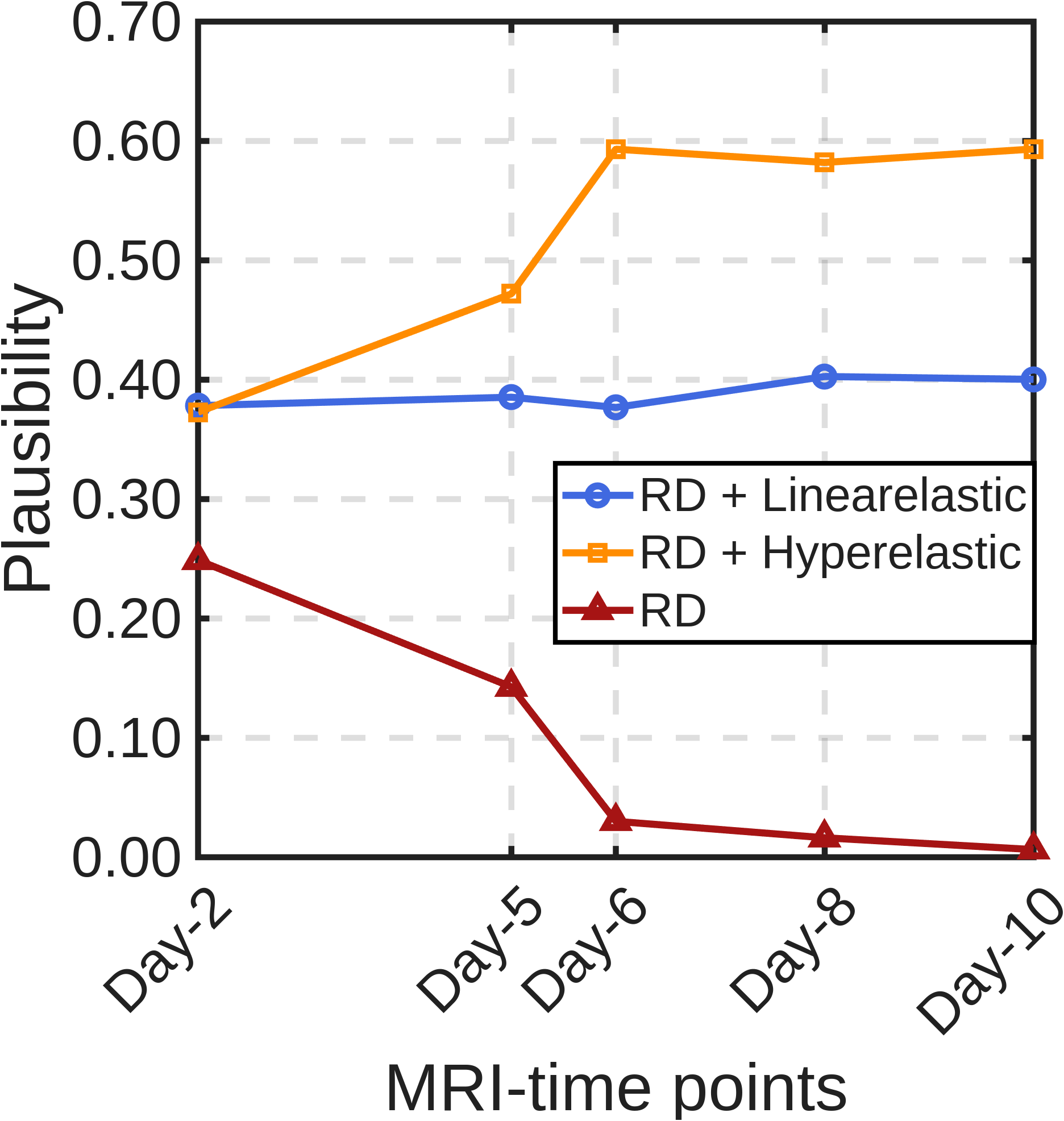}
    ~
    \includegraphics[width=0.252\linewidth]{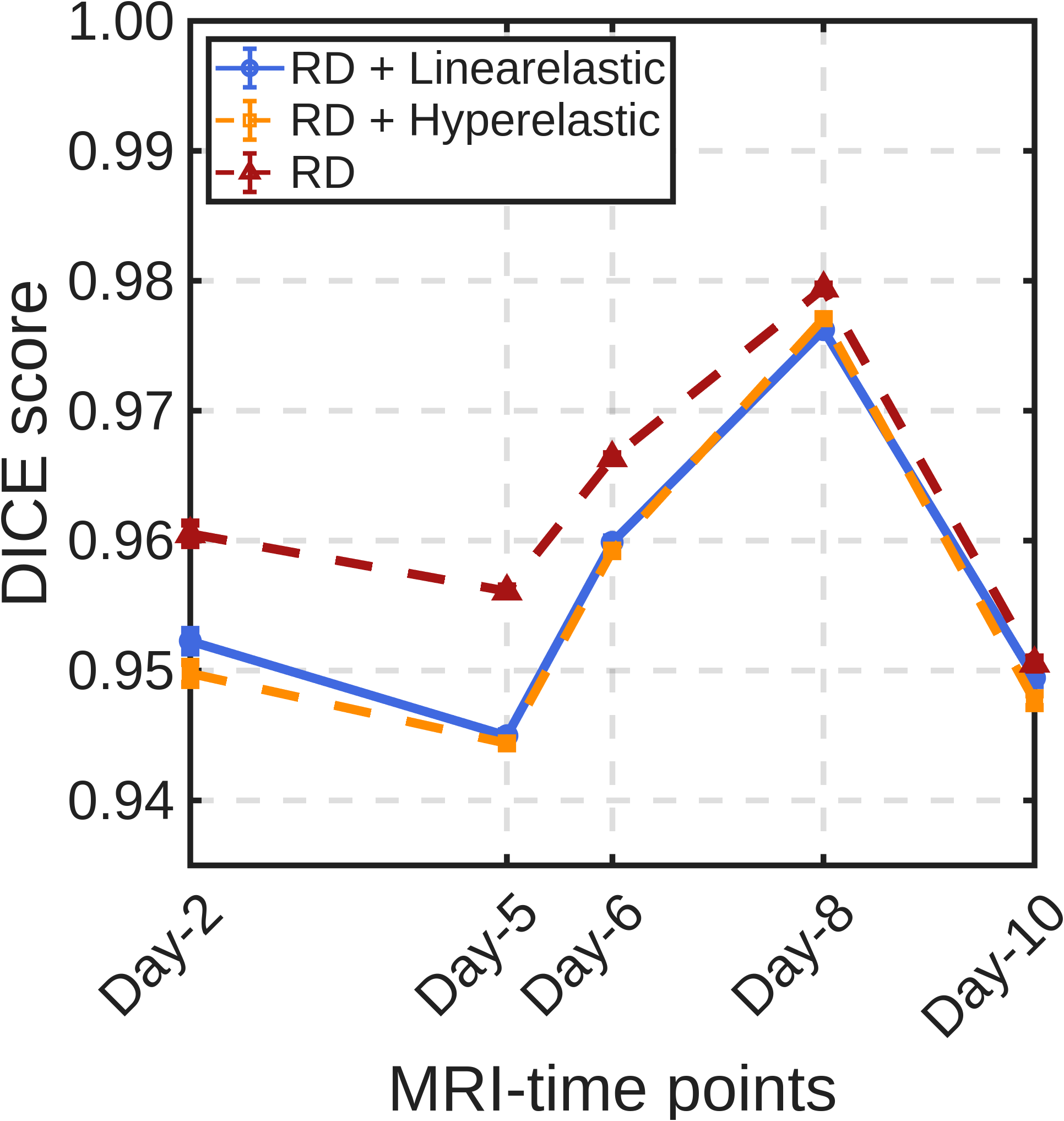}
    ~
    \includegraphics[width=0.252\linewidth]{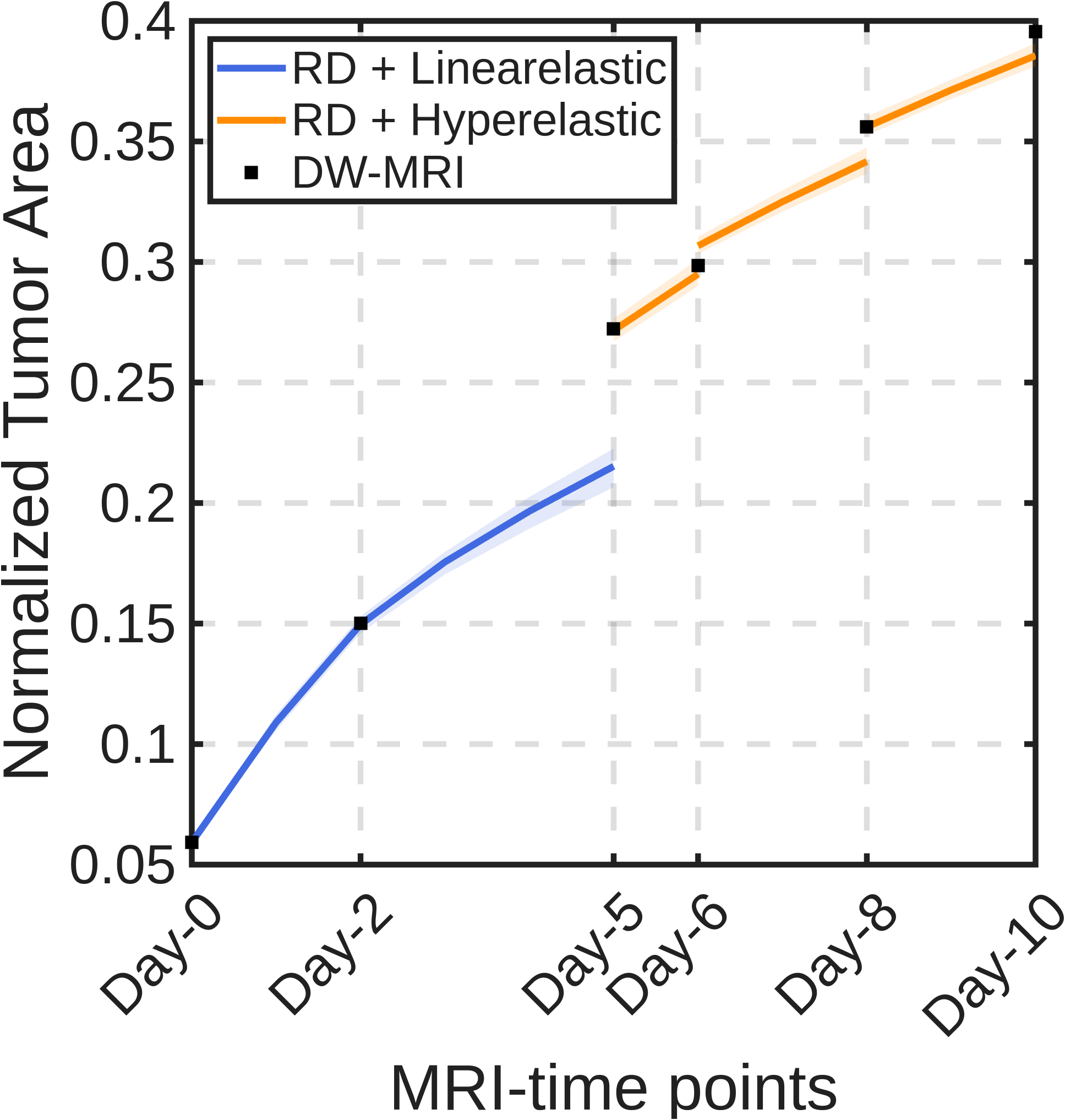}
    \\ (b) \\
    \includegraphics[width=0.246\linewidth]{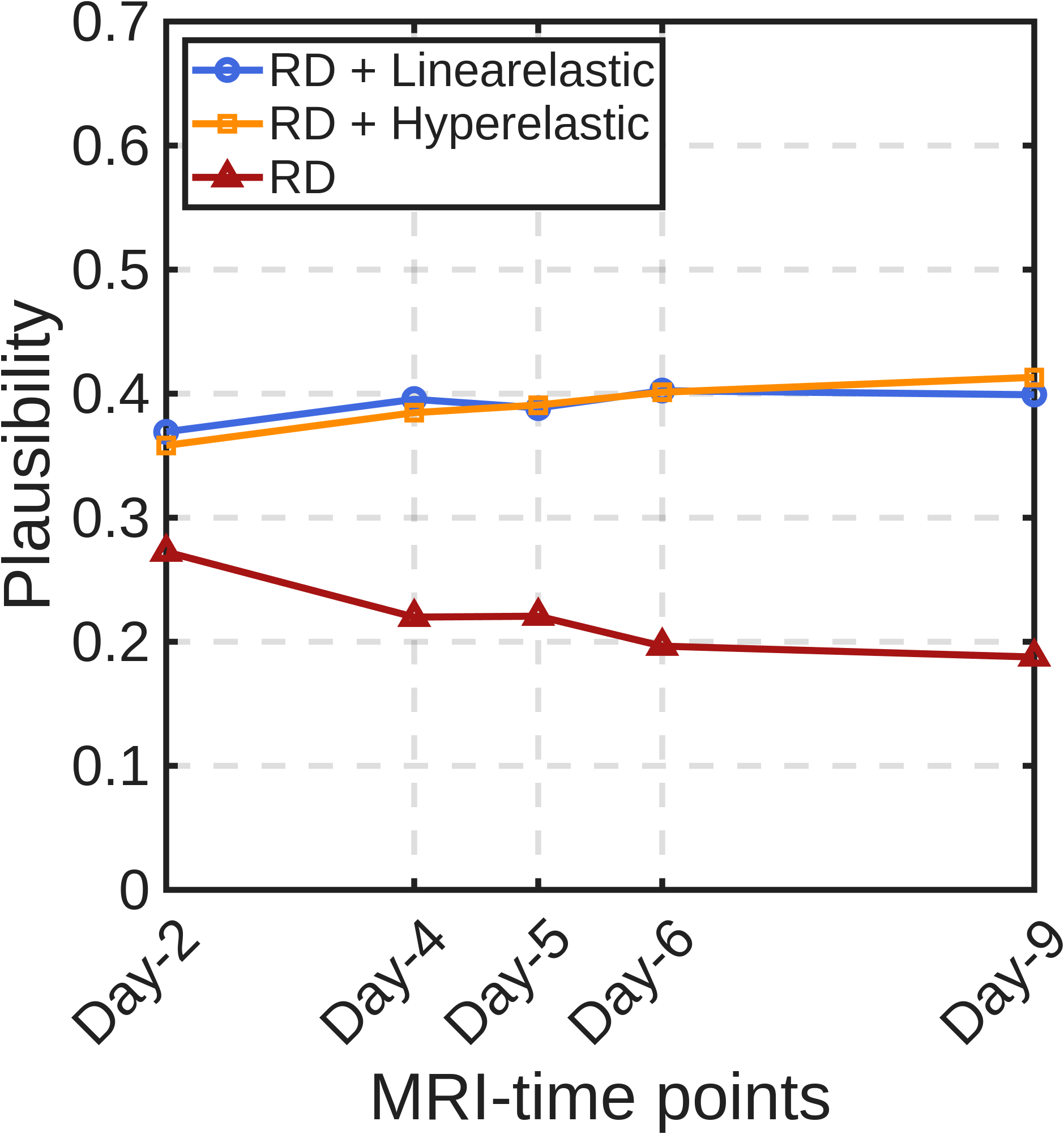}
    ~
    \includegraphics[width=0.252\linewidth]{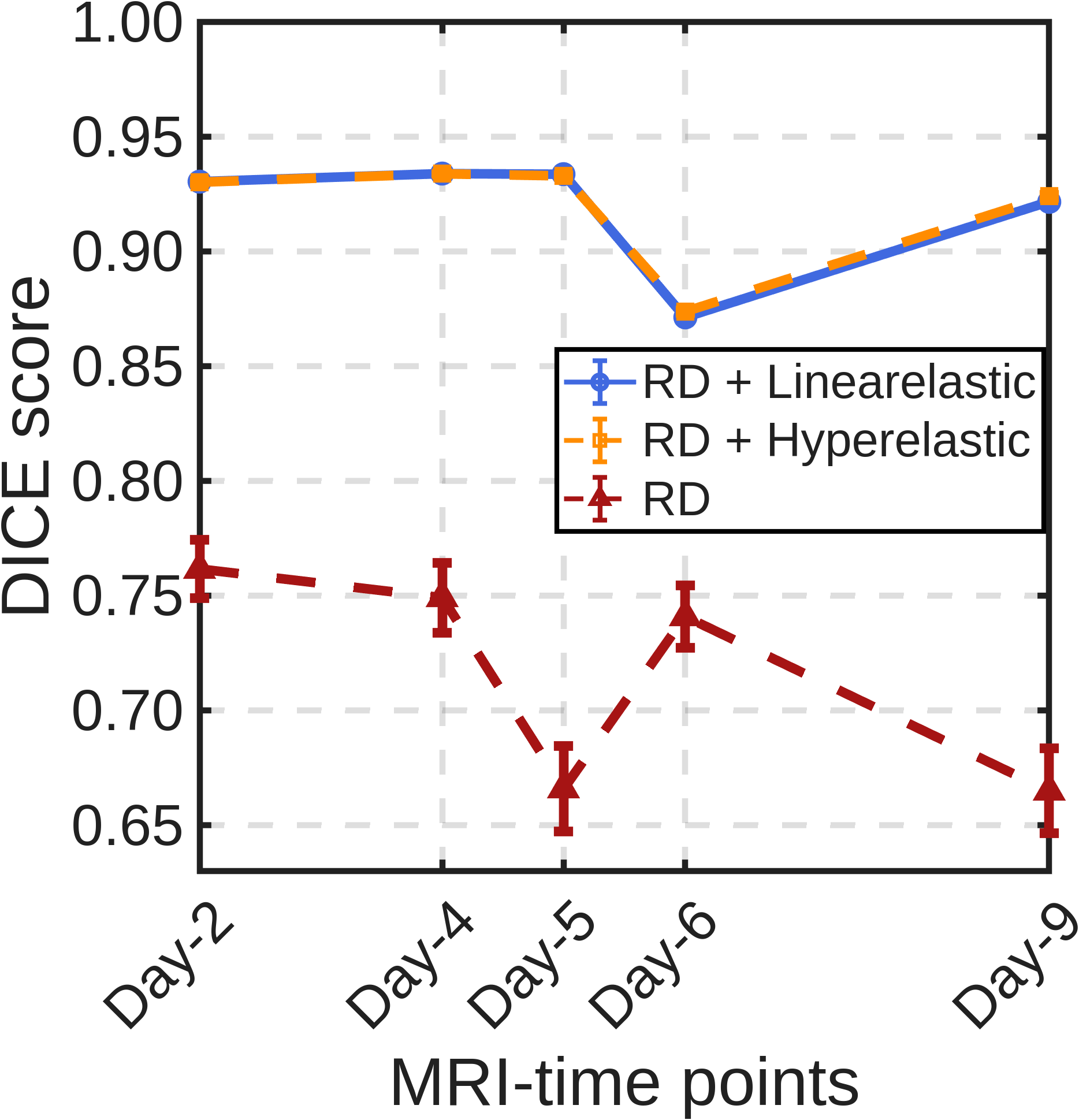}
    ~
    \includegraphics[width=0.252\linewidth]{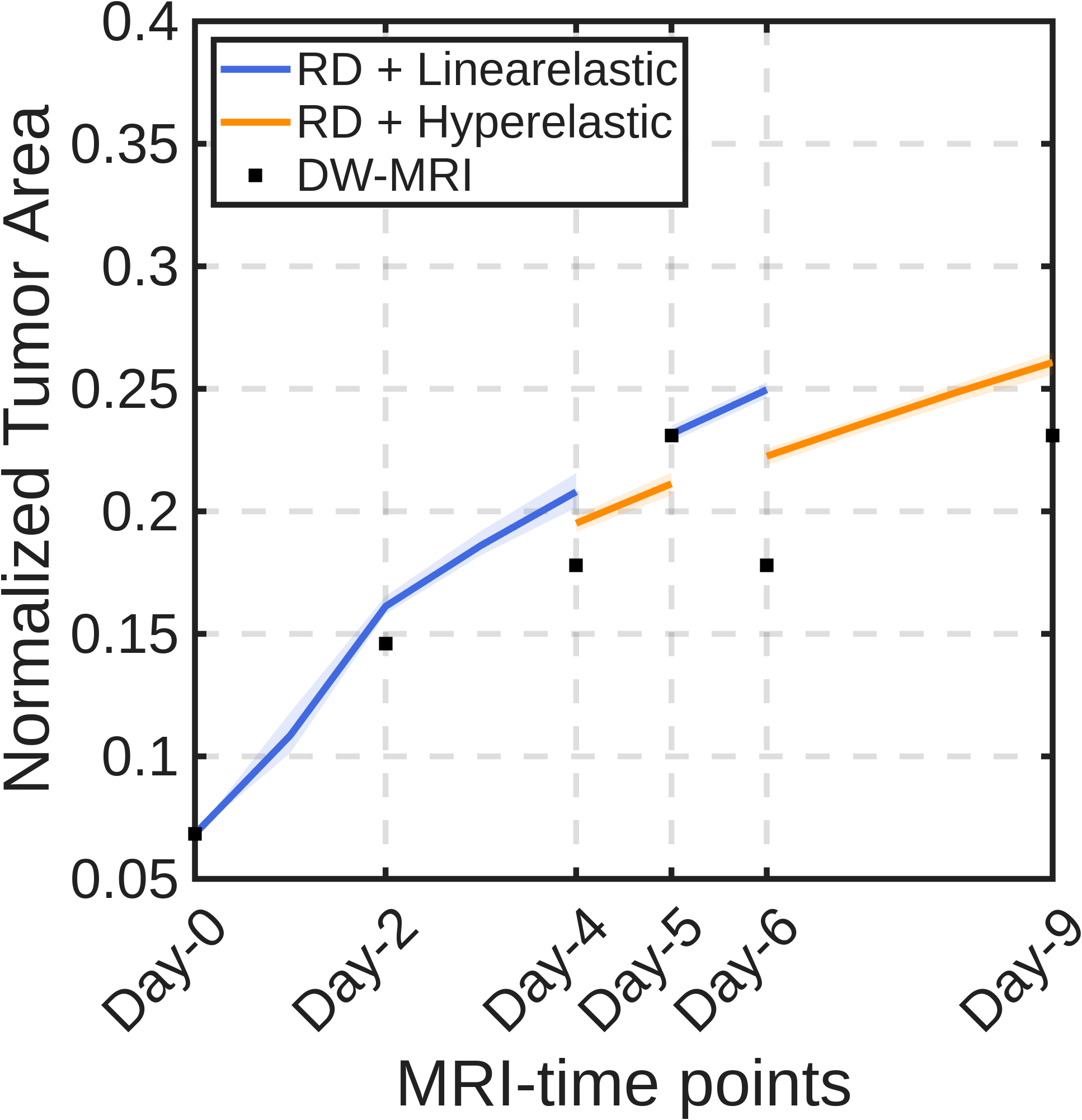}
    \\ (c) \\
    \includegraphics[width=0.248\linewidth]{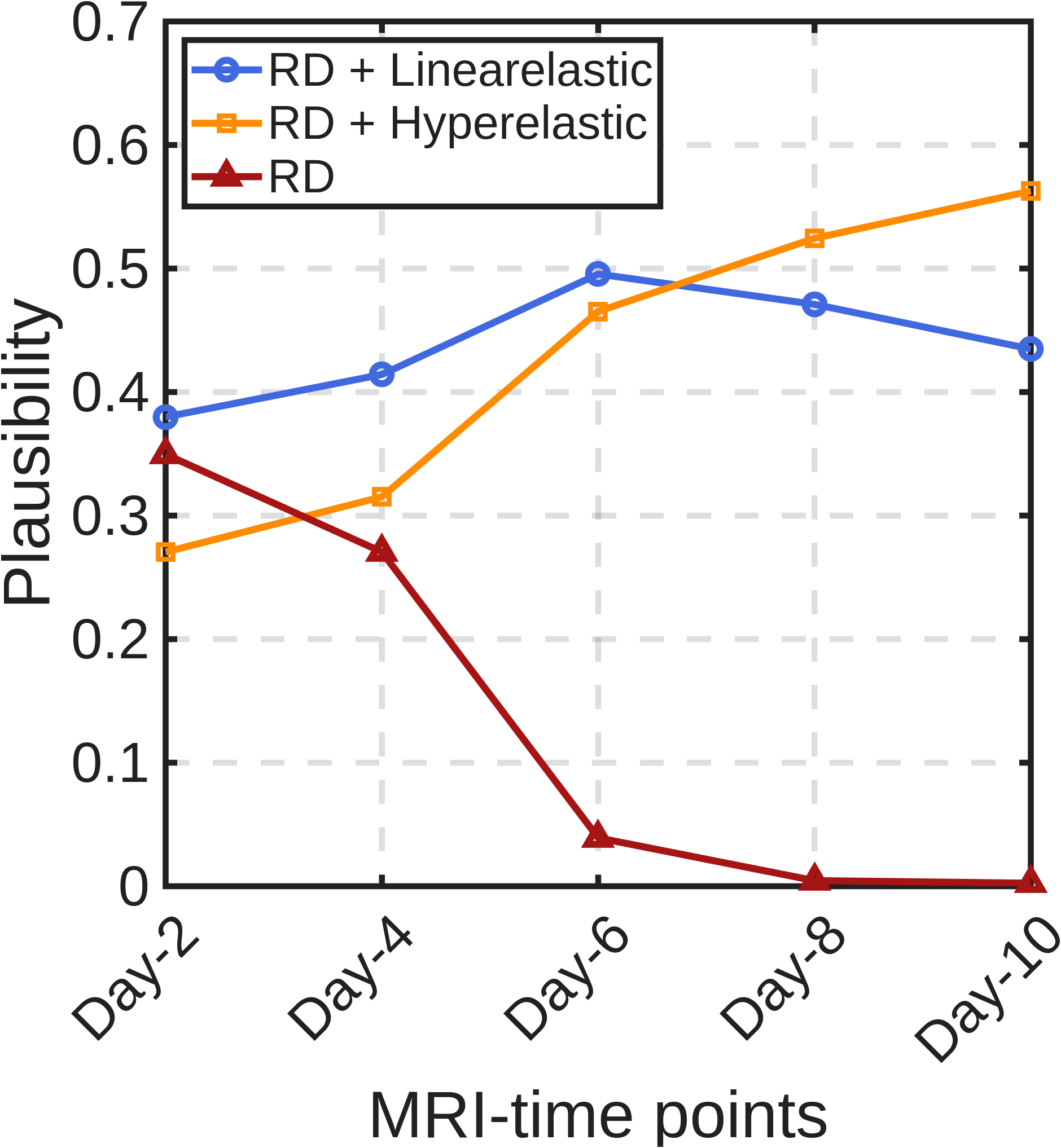}
    ~
    \includegraphics[width=0.252\linewidth]{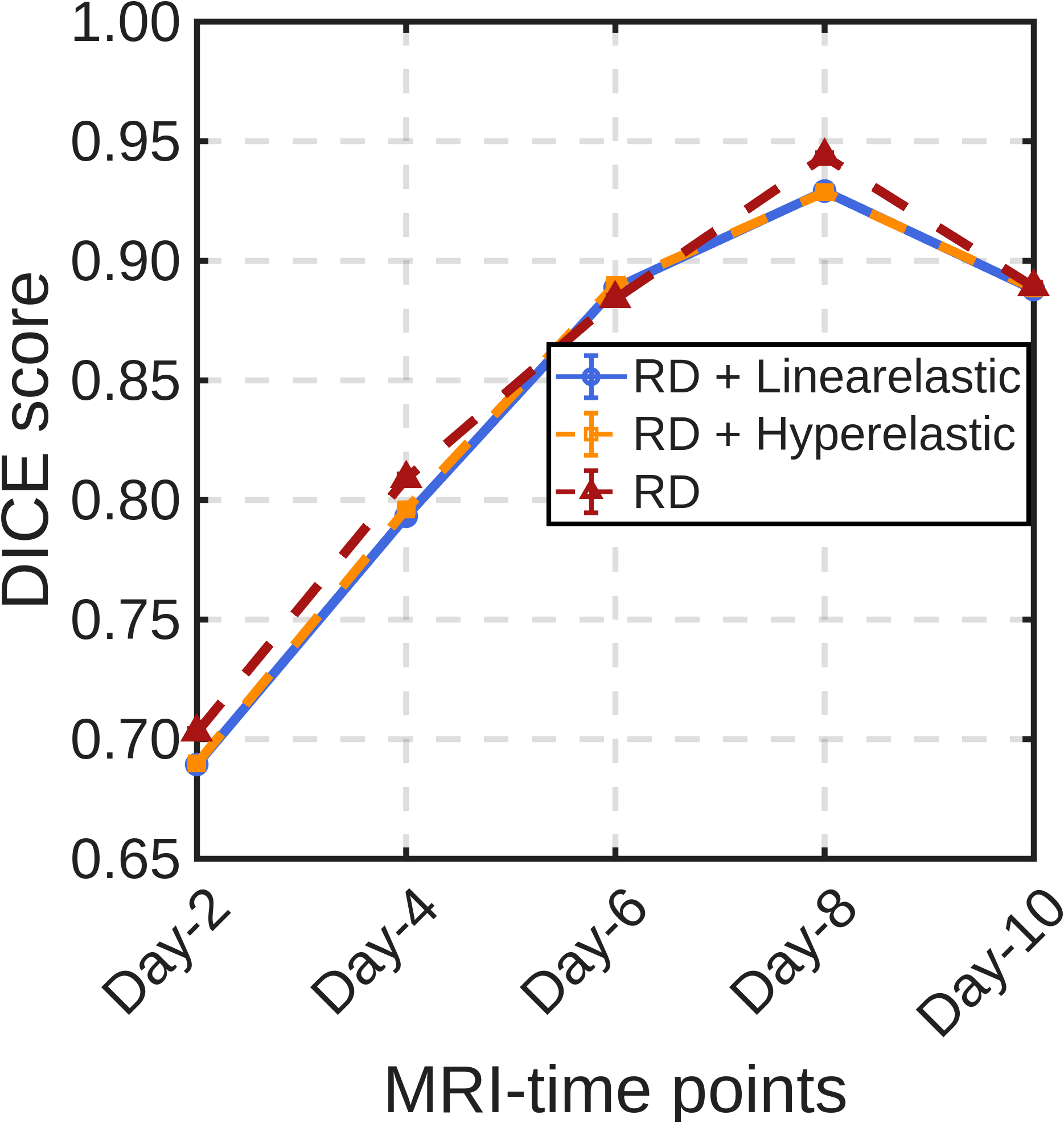}
    ~
    \includegraphics[width=0.252\linewidth]{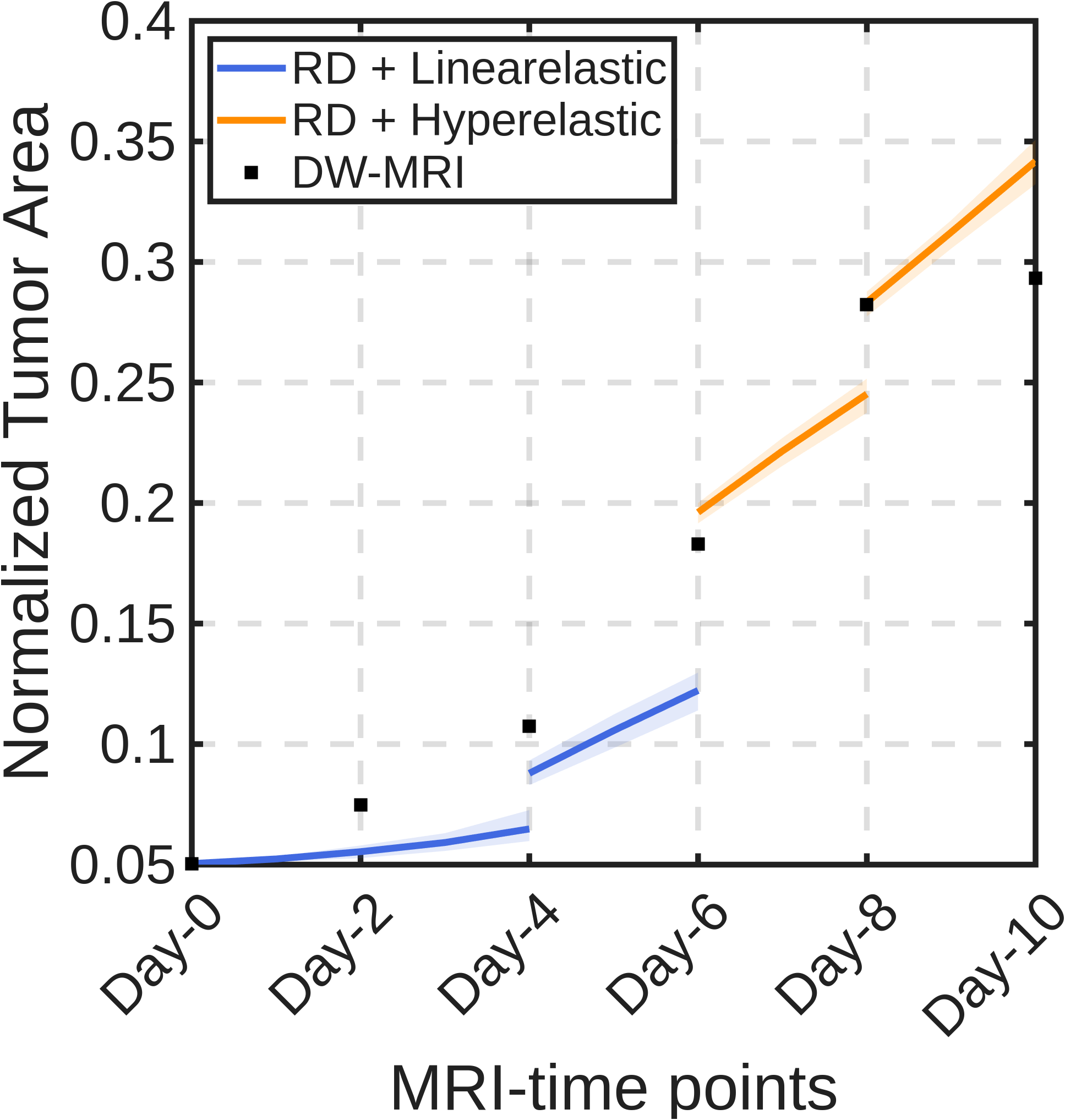}
    \\ \blue{(d)}
    \vspace{-0.1in}
    \caption{
Temporal evolution of posterior model plausibility, one-scan-ahead predictive Dice score, and normalized tumor area (NTA) in (a) Rat~I, (b) Rat~II, (c) Rat~III, and \blue{(d) Rat~IV}. Posterior plausibility is shown for the three candidate models: reaction--diffusion without mechanics (\textit{RD}), reaction--diffusion coupled to linear elasticity (\textit{RD+Linearelastic}), and reaction--diffusion coupled to hyperelastic mechanics (\textit{RD+Hyperelastic}). \blue{The NTA panel shows the prediction from the most plausible model at each assimilation time, for each subject.} Error bars in Dice and shaded bounds in NTA denote 95\% credible intervals propagated from the posterior parameter distributions.
}
    \label{fig:plausibility_dice_rats}
    \vspace{-0.2in}
\end{figure}
\blue{In this work, we demonstrate this approach by comparing} three competing tumor growth
models: a reaction-diffusion model with hyperelastic mechanics (\textit{RD+Hyperelastic}) as detailed in Section \ref{sec:model}, a
reaction-diffusion model with linear elastic mechanics (\textit{RD+Linearelastic}), and a
reaction-diffusion model without mechanical deformation (\textit{RD}).
For the \textit{RD+Linearelastic} model, the governing equations
\eqref{eq:rd_PDE_REF}--\eqref{eq:mech_PDE_REF} are simplified by assuming
infinitesimal deformations. Under this assumption,
growth-induced volume change is neglected and the
first Piola--Kirchhoff stress in \eqref{eq:PK1} is replaced by the linear elastic
stress, parameterized by the elastic modulus $E$ and Poisson’s ratio $\nu$,
\begin{equation}
\label{eq:stress_linear}
\boldsymbol{P}
=
\frac{E}{(1+\nu)(1-2\nu)}
\left[
\nu\,\mathrm{tr}(\boldsymbol{\epsilon})\,\mathbf{I}
+
(1-2\nu)\boldsymbol{\epsilon}
\right],
\qquad
\boldsymbol{\epsilon}
=
\frac{1}{2}\left(\nabla \mathbf{u} + \nabla \mathbf{u}^T\right),
\end{equation}
where $\boldsymbol{\epsilon}$ is the strain tensor. In the
small-strain limit, the first Piola--Kirchhoff stress is equivalent to the Cauchy
stress, and distinctions between reference and deformed configurations are
neglected. The \textit{RD} model omits mechanical coupling by removing the
mechanical PDE \eqref{eq:mech_PDE_REF} and setting the stress-dependent term
$e^{-H p_{\boldsymbol{\tau}}}=1$ in \eqref{eq:rd_PDE_REF}.

\blue{
Figure~\ref{fig:plausibility_dice_rats} 
summarize the subject-specific dynamic model selection results for the four rats by tracking posterior model plausibility together with one-scan-ahead predictive Dice. For each animal, posterior plausibility is recomputed at each assimilation time using the MRI data accumulated for that animal up to that point. \blue{The NTA panel reports the prediction obtained from the model with the highest posterior plausibility at each assimilation time, rather than a fixed model used throughout the full time series. Thus, the NTA trajectory represents the dynamically selected subject-specific prediction, with the selected model switching between \textit{RD+Linearelastic} and \textit{RD+Hyperelastic} at some intermediate imaging times.}
The \textit{RD} model is generally the least plausible, particularly at later imaging times, indicating that diffusion--proliferation alone is less supported by the longitudinal data than mechanically coupled  formulation. Thus, within the present murine dataset, the inclusion of mechanical coupling through mass effect and stress-mediated feedback, results in more plausible computational models.}
Across rats, the competition is primarily between \textit{RD+Linearelastic} and \textit{RD+Hyperelastic}.
\blue{The posterior model plausibility for these two mechanically coupled models evolves with the available MRI data and varies by subject, with the hyperelastic formulation often becoming more plausible at later imaging times.} 
This trend is consistent with increasing relevance of finite-deformation mechanics as tumor burden and growth-induced deformation increase. Importantly, the Dice and NTA trends in this figure do not reliably discriminate between \textit{RD+Linearelastic} and \textit{RD+Hyperelastic}, because these metrics primarily measure tumor shape agreement and are only weakly sensitive to the underlying mechanical state. Posterior model plausibility therefore provides a complementary, more
mechanistically meaningful criterion criterion for comparing whether each formulation consistently explains the observed spatiotemporal tumor evolution under its own biomechanical assumptions.

\begin{figure}[!ht]
\centering
\renewcommand{\arraystretch}{.6}
\setlength{\fboxsep}{0pt}
\setlength{\tabcolsep}{.1pt}
\setlength{\arrayrulewidth}{0.2pt}

\makebox[\textwidth][c]{
\begin{tabular}{m{0.04\textwidth}*{5}{>{\centering\arraybackslash}m{0.18\textwidth}}@{\hspace{-10pt}}>{\centering\arraybackslash}m{0.14\textwidth}@{\hspace{-15pt}}m{0.1\textwidth}}

\makebox[0.04\textwidth][c]{\vrule width 0pt height 0.45cm} &
\colorbox{black}{\makebox[0.177\textwidth][c]{\vrule width 0pt height 0.4cm \textcolor{white}{\small Day 14}}} &
\colorbox{black}{\makebox[0.177\textwidth][c]{\vrule width 0pt height 0.4cm \textcolor{white}{\small Day 15}}} &
\colorbox{black}{\makebox[0.177\textwidth][c]{\vrule width 0pt height 0.4cm \textcolor{white}{\small Day 16}}} &
\colorbox{black}{\makebox[0.177\textwidth][c]{\vrule width 0pt height 0.4cm \textcolor{white}{\small Day 19}}} &
\colorbox{black}{\makebox[0.177\textwidth][c]{\vrule width 0pt height 0.4cm \textcolor{white}{\small Day 21}}} &
\makebox[0.1\textwidth][c]{\vrule width 0pt height 0.45cm} \\ [4pt]

\centering\rotatebox{90}{
    \begin{tabular}{@{}c@{}} \textbf{\footnotesize RD+hyper} \\ \textbf{\footnotesize elastic} \end{tabular}
}
& \includegraphics[width=0.176\textwidth]{Figures/Pred/Data4-sim-rat-I-Day14.png}
& \includegraphics[width=0.176\textwidth]{Figures/Pred/Data4-sim-rat-I-Day15.png}
& \includegraphics[width=0.176\textwidth]{Figures/Pred/Data4-sim-rat-I-Day16.png}
& \includegraphics[width=0.176\textwidth]{Figures/Pred/Data4-sim-rat-I-Day19.png}
& \includegraphics[width=0.176\textwidth]{Figures/Pred/Data4-sim-rat-I-Day21.png}
& \multirow[t]{2}{*}{ 
    \raisebox{-2.8cm}{
        \includegraphics[width=0.95cm,height=3.1cm,keepaspectratio=false]{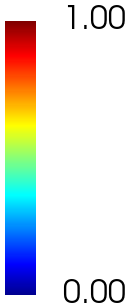}
        \hspace{-15pt}
        \raisebox{15pt}{\rotatebox{90}{\textbf{\tiny Volume fraction}}}
    }
} \\

\centering\rotatebox{90}{
    \begin{tabular}{@{}c@{}} \textbf{\footnotesize RD+linear} \\ \textbf{\footnotesize elastic} \end{tabular}
}
& \includegraphics[width=0.176\textwidth]{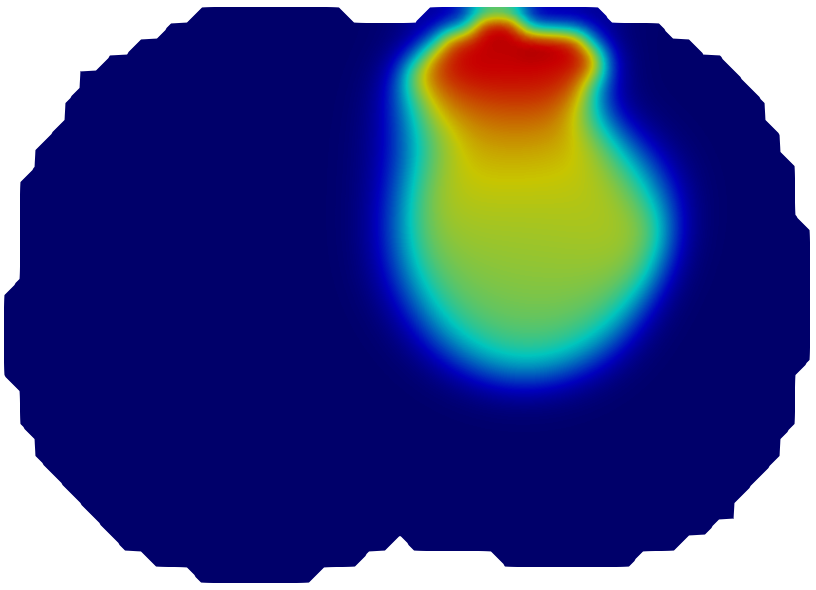}
& \includegraphics[width=0.176\textwidth]{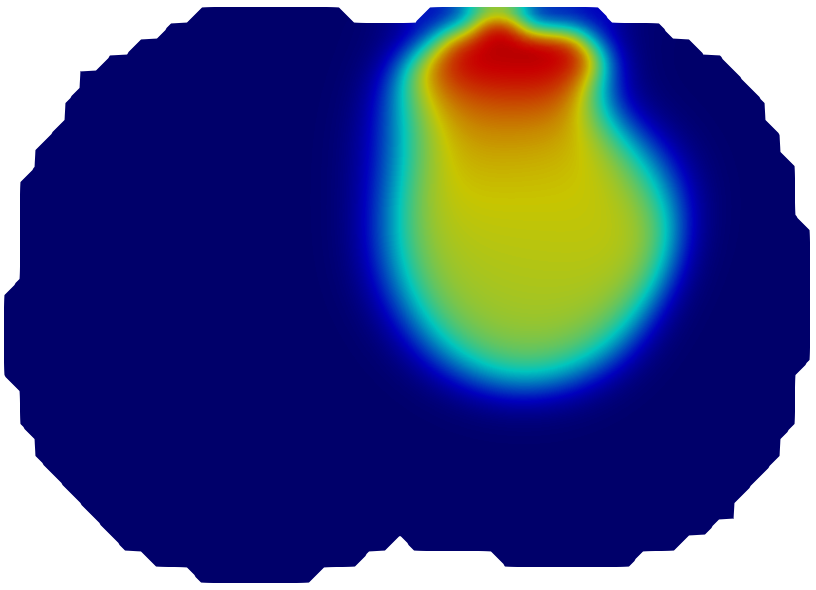}
& \includegraphics[width=0.176\textwidth]{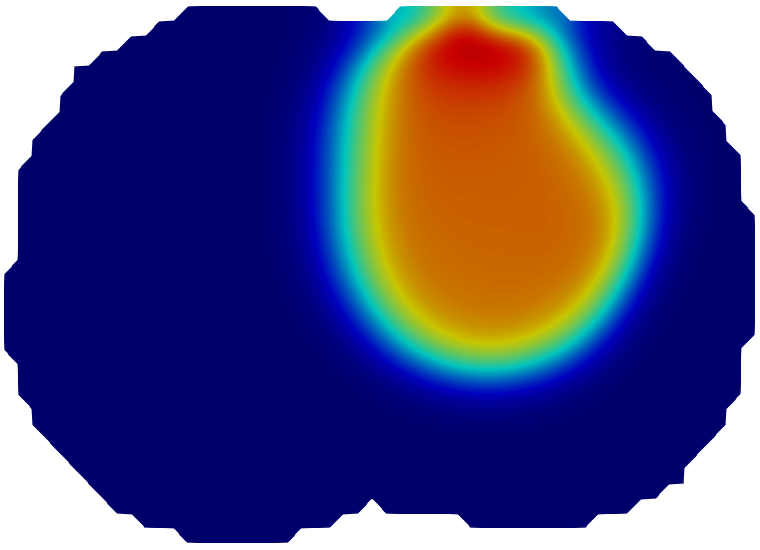}
& \includegraphics[width=0.176\textwidth]{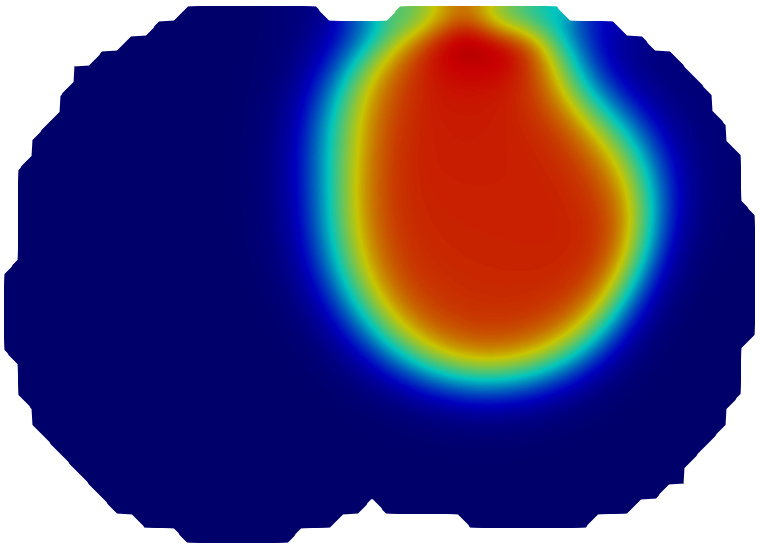}
& \includegraphics[width=0.176\textwidth]{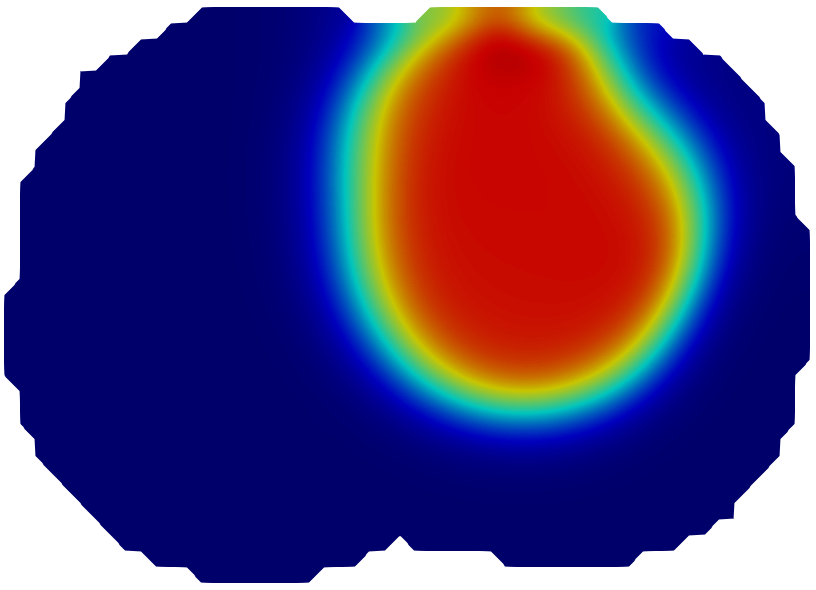}
& \\ [15pt] 

\centering\rotatebox{90}{
    \begin{tabular}{@{}c@{}} \textbf{\footnotesize RD+hyper} \\ \textbf{\footnotesize elastic} \end{tabular}
}
& \includegraphics[width=0.176\textwidth]{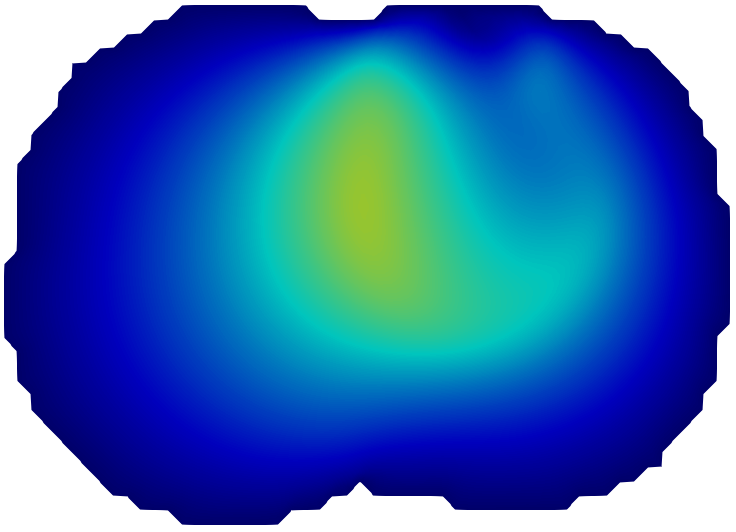}
& \includegraphics[width=0.176\textwidth]{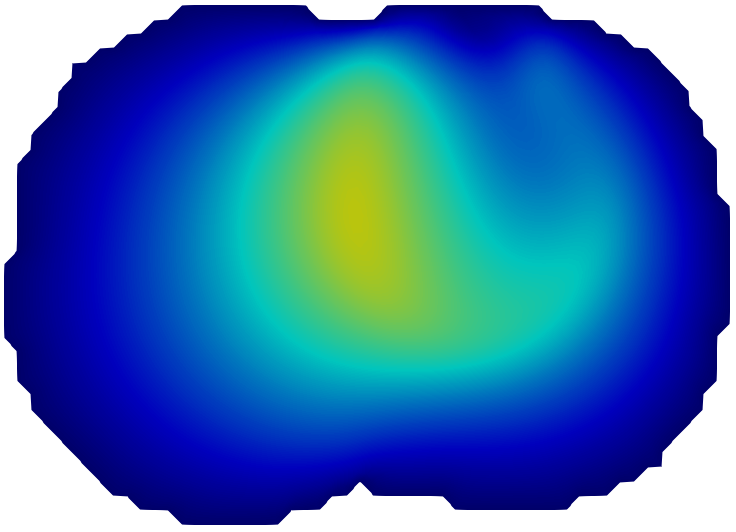}
& \includegraphics[width=0.176\textwidth]{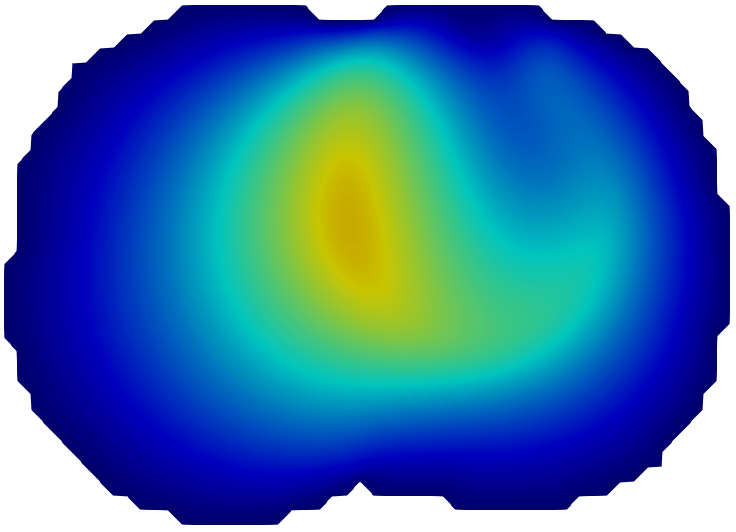}
& \includegraphics[width=0.176\textwidth]{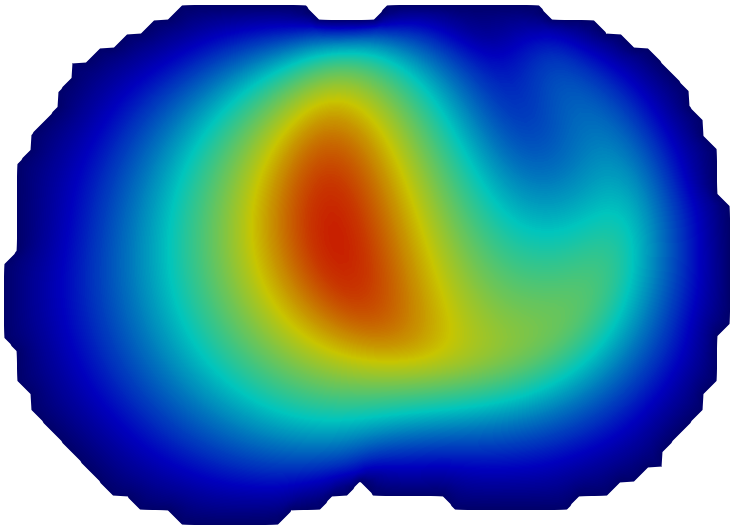}
& \includegraphics[width=0.176\textwidth]{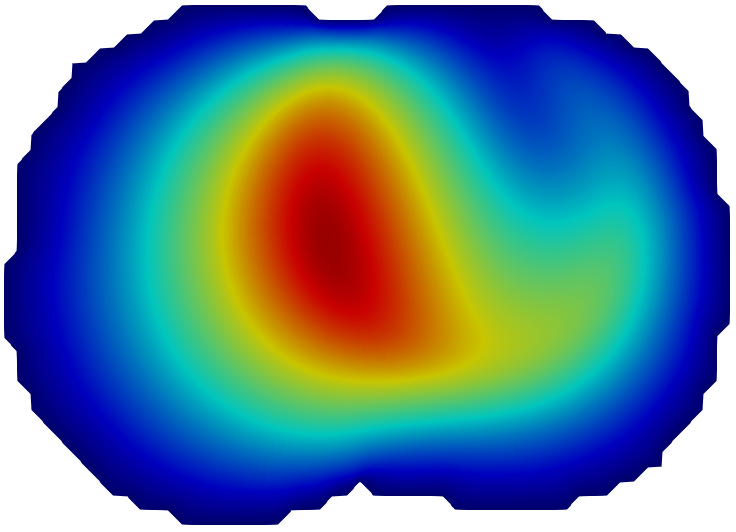}
& \multirow[t]{2}{*}{
    \raisebox{-2.7cm}{
        \includegraphics[width=0.95cm,height=3.2cm,keepaspectratio=false]{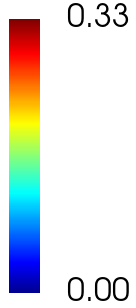}
        \hspace{-22pt}
        \raisebox{35pt}{\rotatebox{90}{\begin{tabular}{@{}c@{}} \textbf{\tiny ||u||} \\[3pt] \textbf{\tiny [mm]} \end{tabular}}}
    }
} \\

\centering\rotatebox{90}{
    \begin{tabular}{@{}c@{}} \textbf{\footnotesize RD+linear} \\ \textbf{\footnotesize elastic} \end{tabular}
}
& \includegraphics[width=0.176\textwidth]{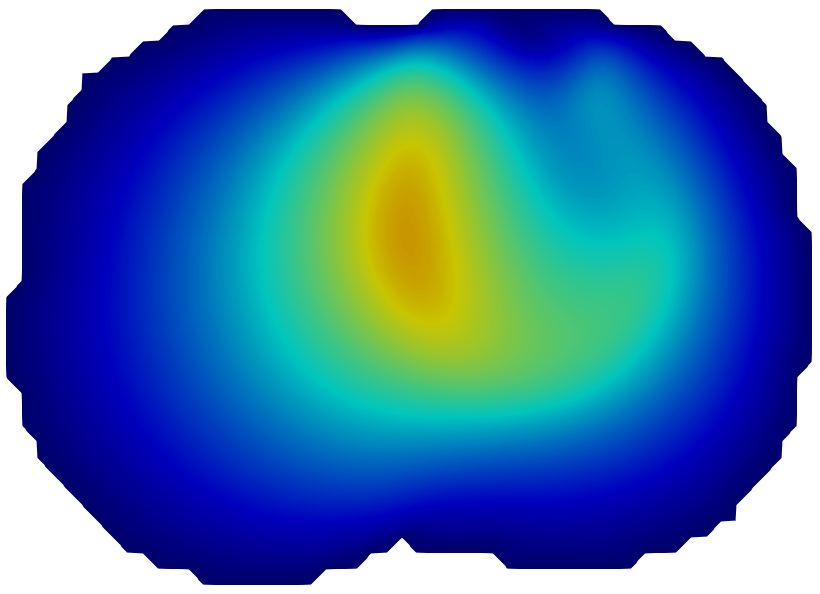}
& \includegraphics[width=0.176\textwidth]{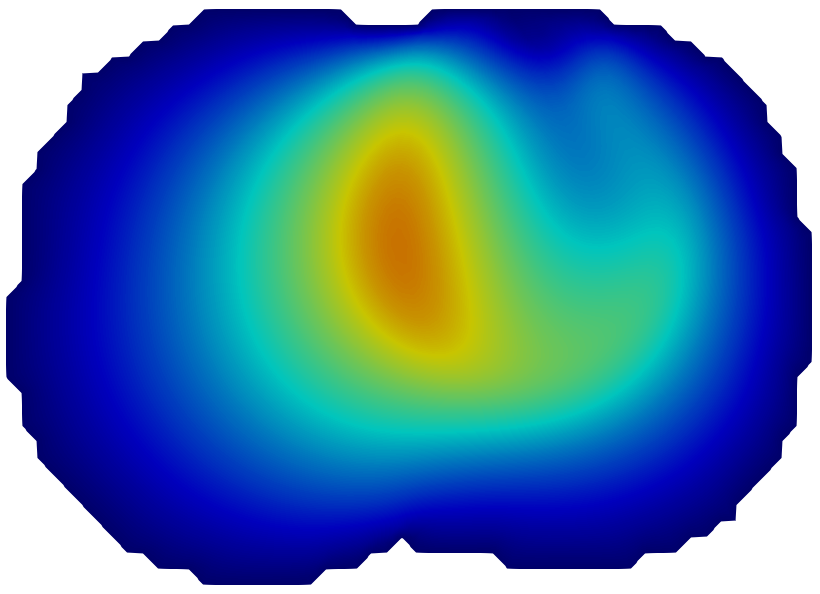}
& \includegraphics[width=0.176\textwidth]{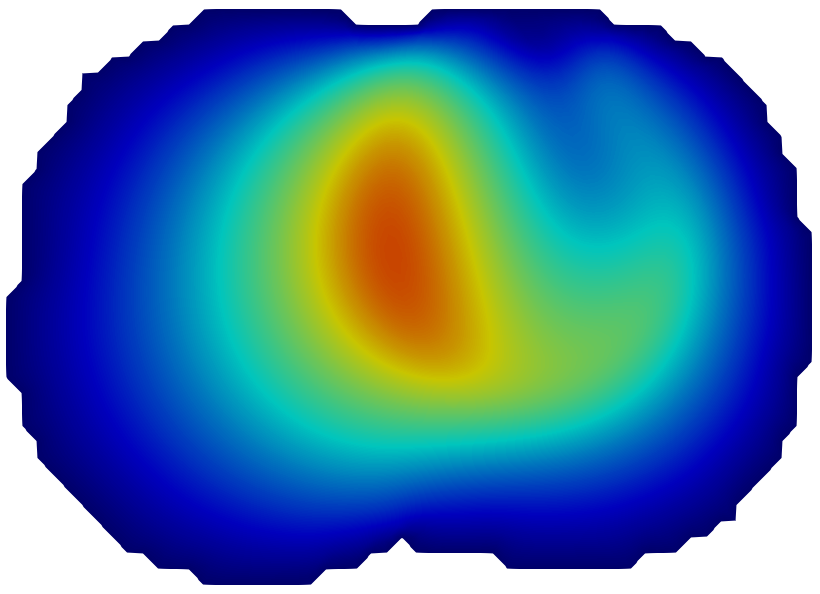}
& \includegraphics[width=0.176\textwidth]{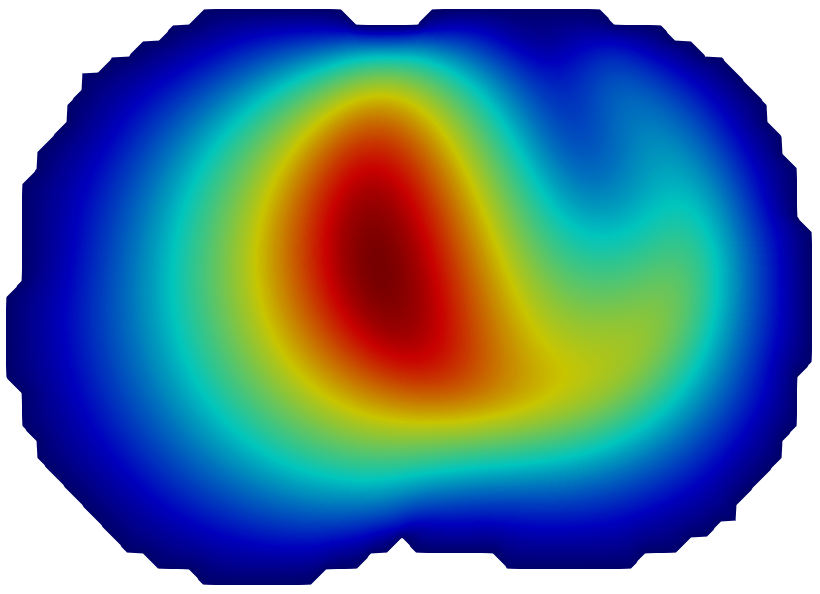}
& \includegraphics[width=0.176\textwidth]{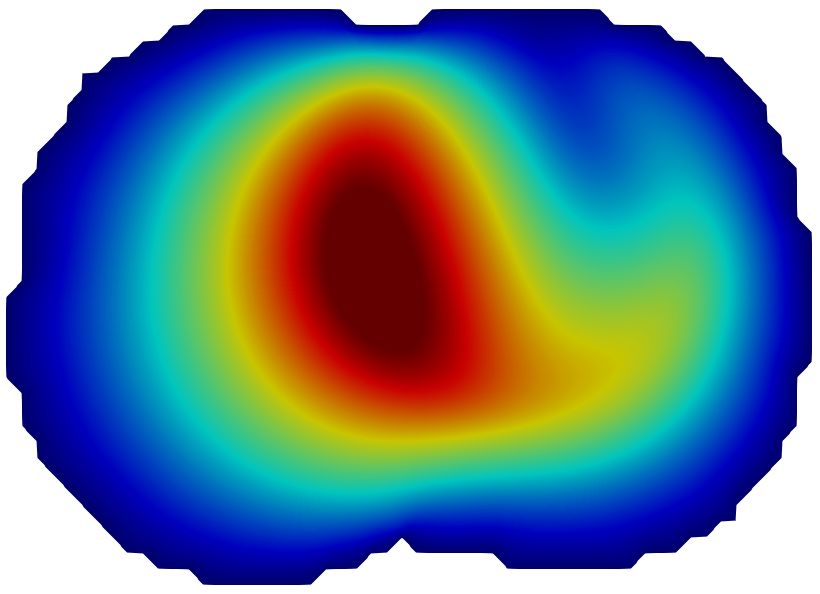}
& \\ [15pt]

\centering\rotatebox{90}{
    \begin{tabular}{@{}c@{}} \textbf{\footnotesize RD+hyper} \\ \textbf{\footnotesize elastic} \end{tabular}
}
& \includegraphics[width=0.176\textwidth]{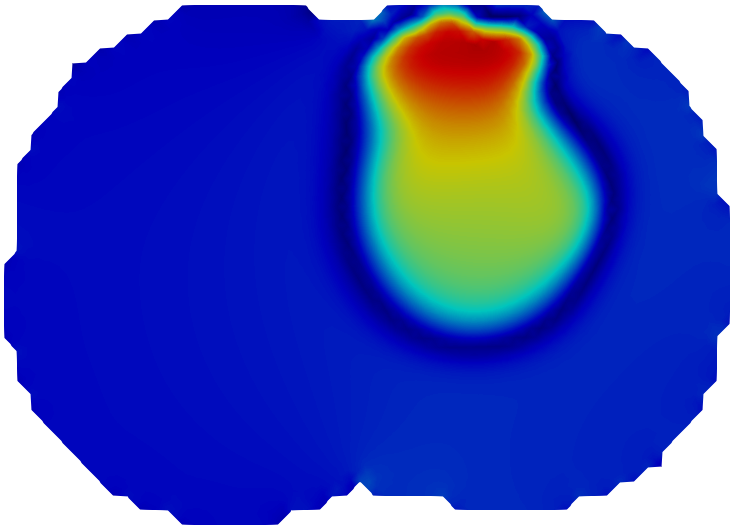}
& \includegraphics[width=0.176\textwidth]{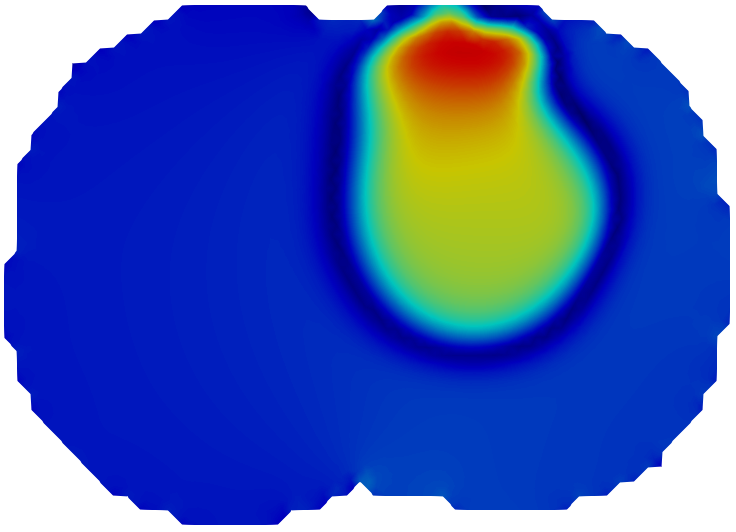}
& \includegraphics[width=0.176\textwidth]{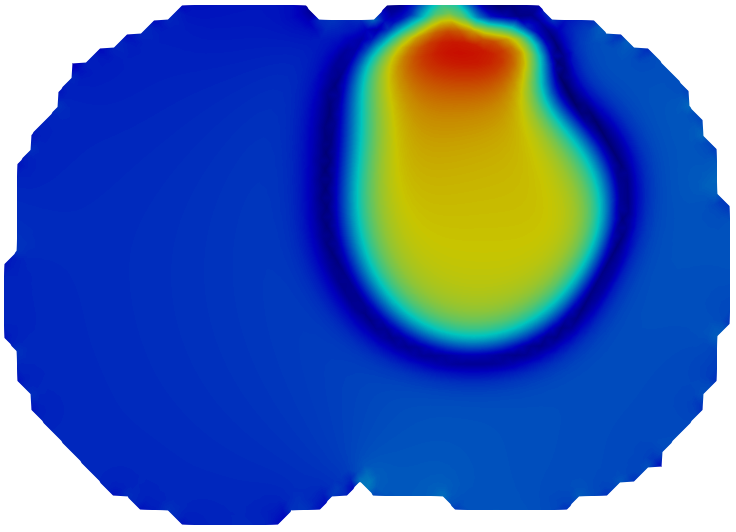}
& \includegraphics[width=0.176\textwidth]{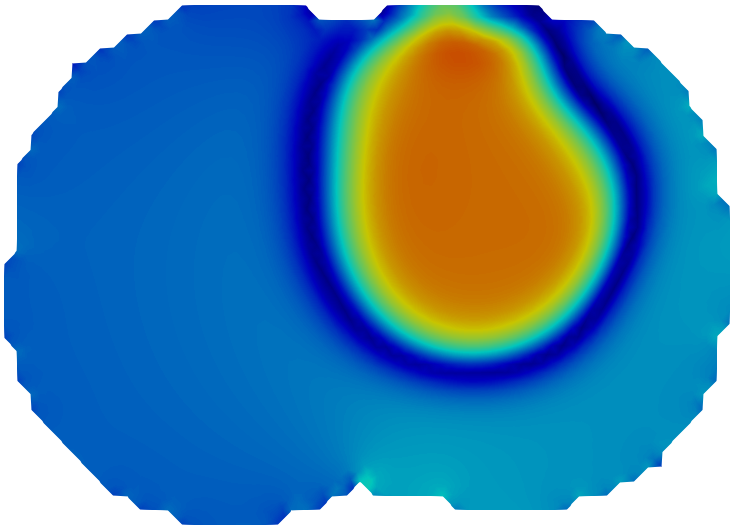}
& \includegraphics[width=0.176\textwidth]{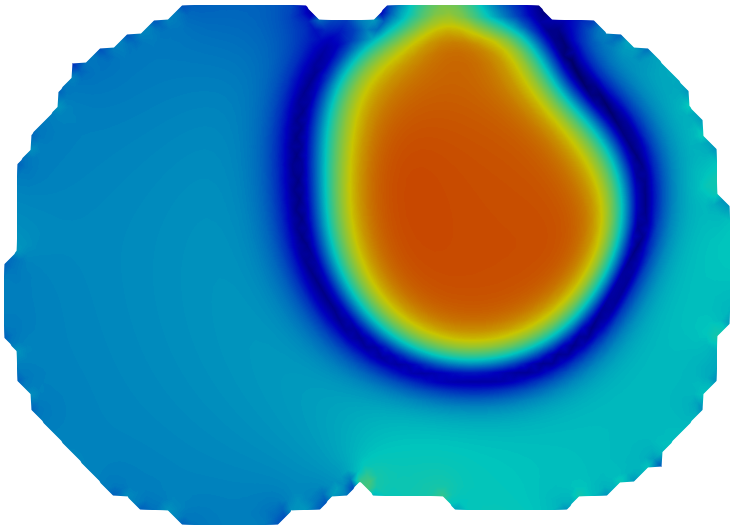}
& \multirow[t]{2}{*}{
    \raisebox{-2.7cm}{
        \includegraphics[width=0.95cm,height=3cm,keepaspectratio=false]{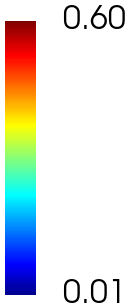}
        \hspace{-22pt}
        \raisebox{35pt}{\rotatebox{90}{\begin{tabular}{@{}c@{}} \textbf{\tiny ||\textbf{P}||} \\[3pt] \textbf{\tiny [KPa]} \end{tabular}}}
    }
} \\

\centering\rotatebox{90}{
    \begin{tabular}{@{}c@{}} \textbf{\footnotesize RD+linear} \\ \textbf{\footnotesize elastic} \end{tabular}
}
& \includegraphics[width=0.176\textwidth]{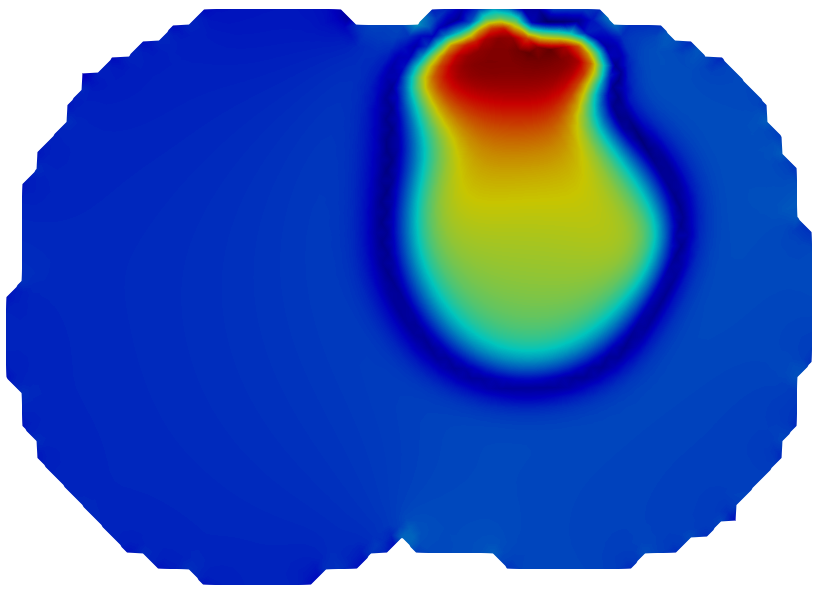}
& \includegraphics[width=0.176\textwidth]{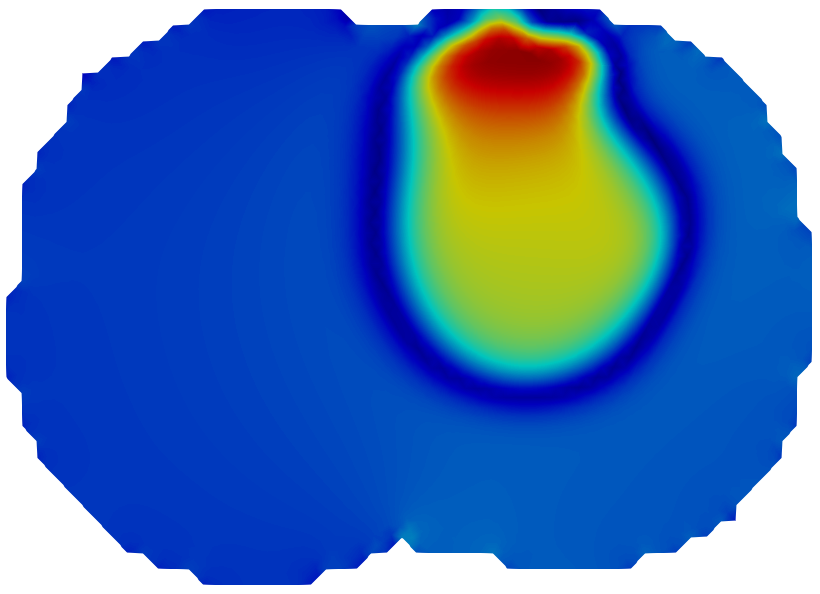}
& \includegraphics[width=0.176\textwidth]{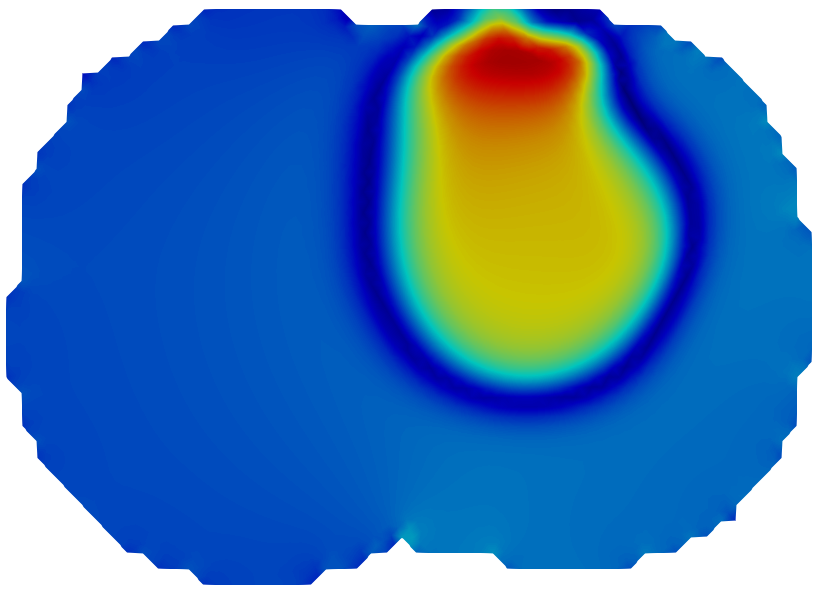}
& \includegraphics[width=0.176\textwidth]{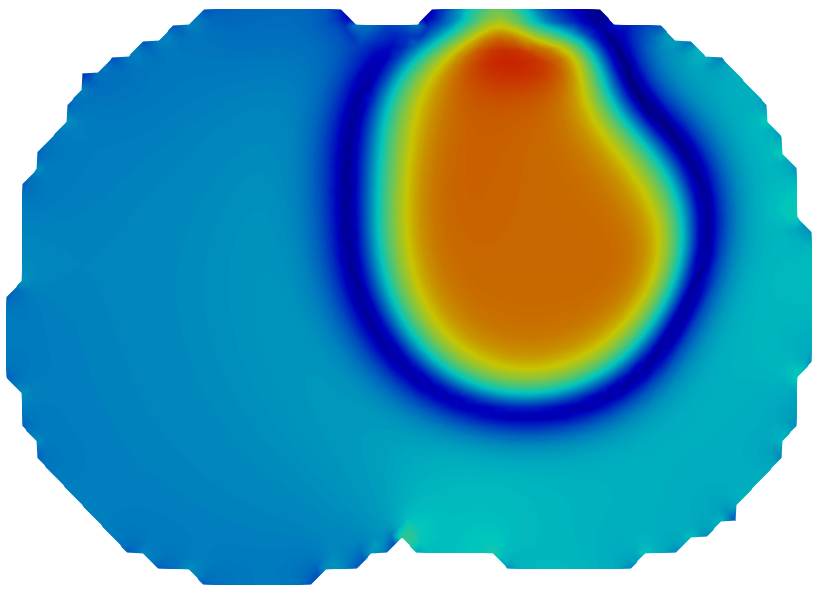}
& \includegraphics[width=0.176\textwidth]{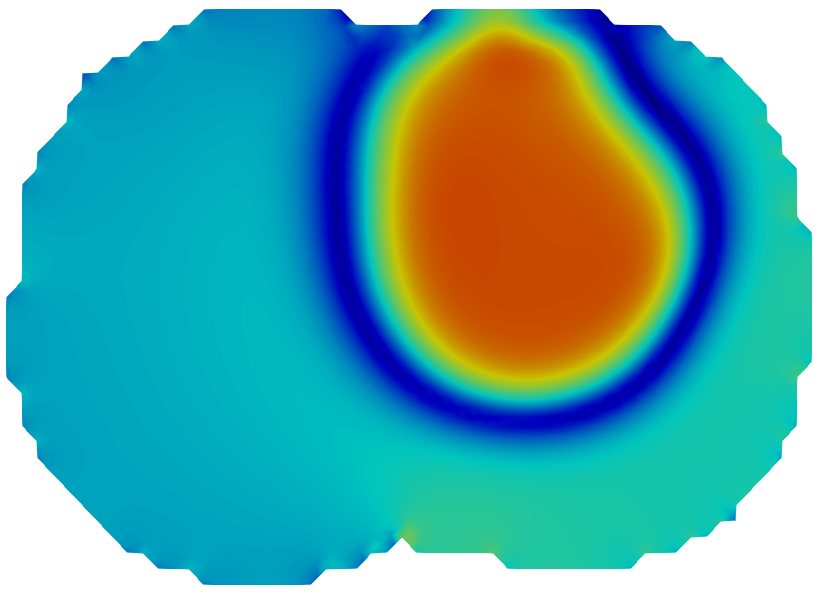}
& \\

\end{tabular}}
\vspace{-0.1in}
\caption{Comparison of one-scan-ahead predictions for Rat-I under 
\textit{RD+Hyperelastic} and \textit{RD+Linearelastic} formulations. 
Tumor volume fraction $\phi(\mathbf{X})$, displacement magnitude $\|\mathbf{u}(\mathbf{X})\|$, 
and first Piola--Kirchhoff stress magnitude $\|\mathbf{P}(\mathbf{X})\|$ are shown 
at sequential imaging times.
}
\label{fig:disp_stress_hyperlinear_ratI}
\vspace{-0.1in}
\end{figure}

\begin{figure}[!h]
    \centering
    \includegraphics[width=0.31\linewidth]{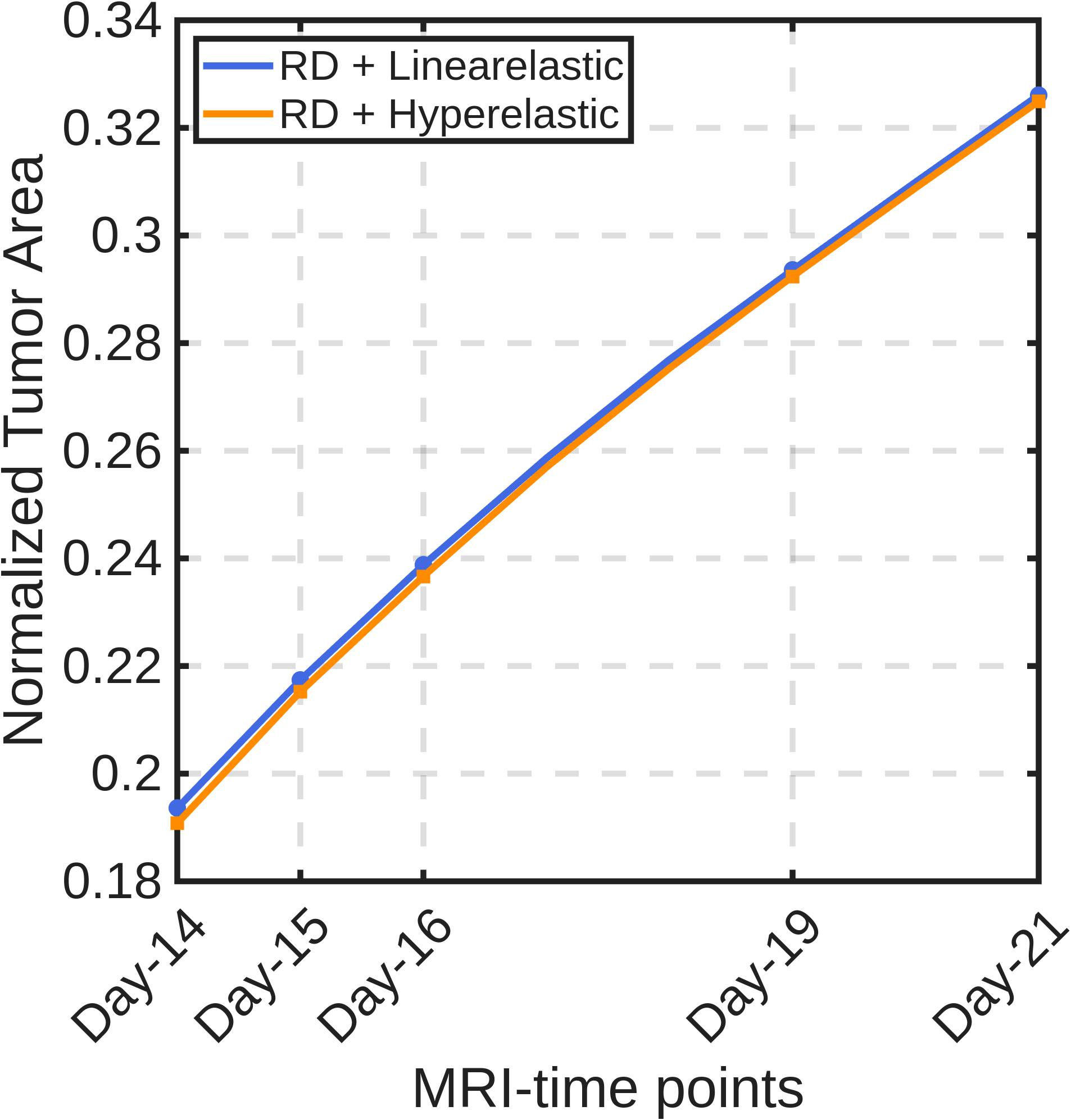}
    ~
    \includegraphics[width=0.31\linewidth]{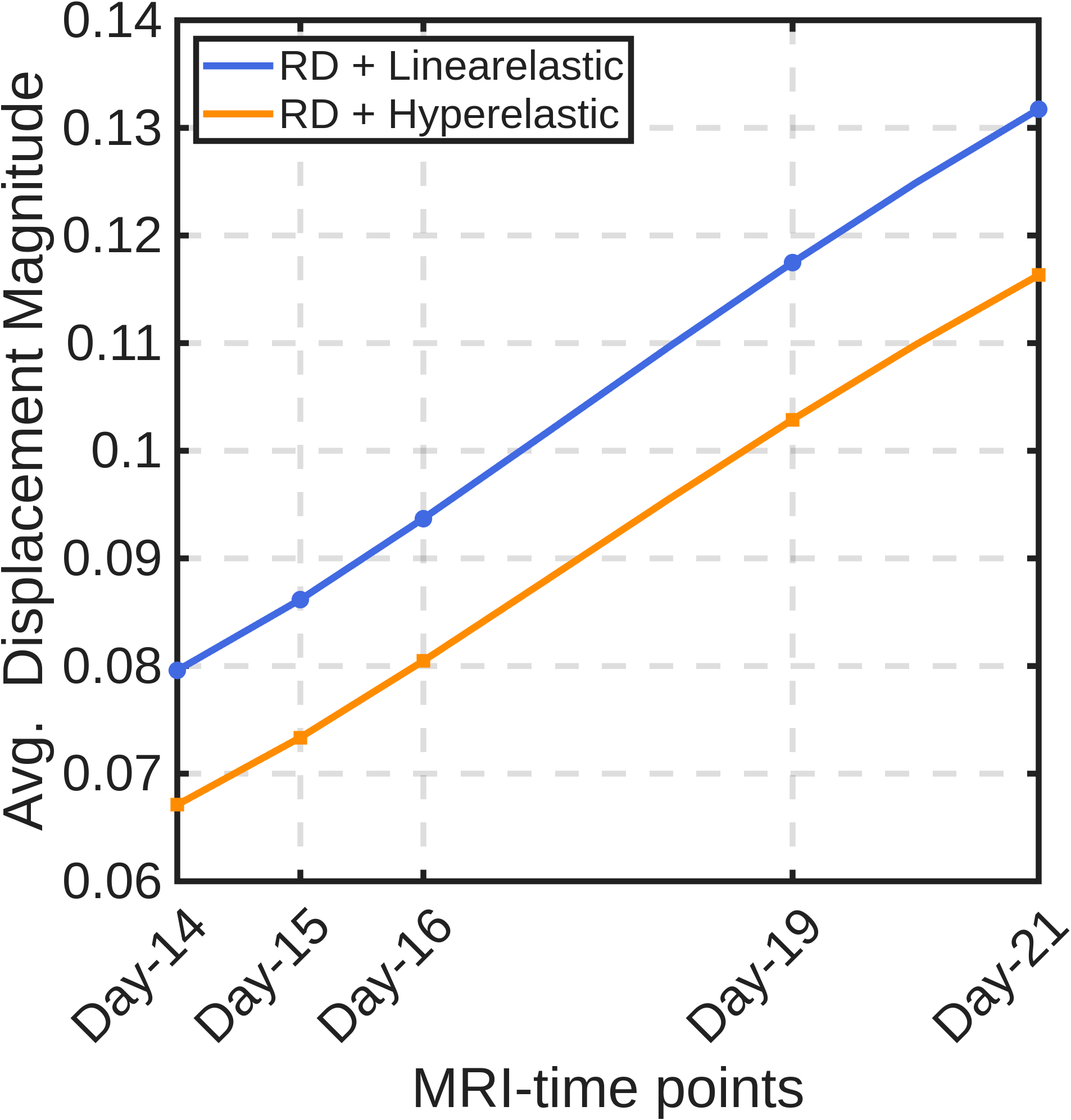}
    ~
     \includegraphics[width=0.31\linewidth]{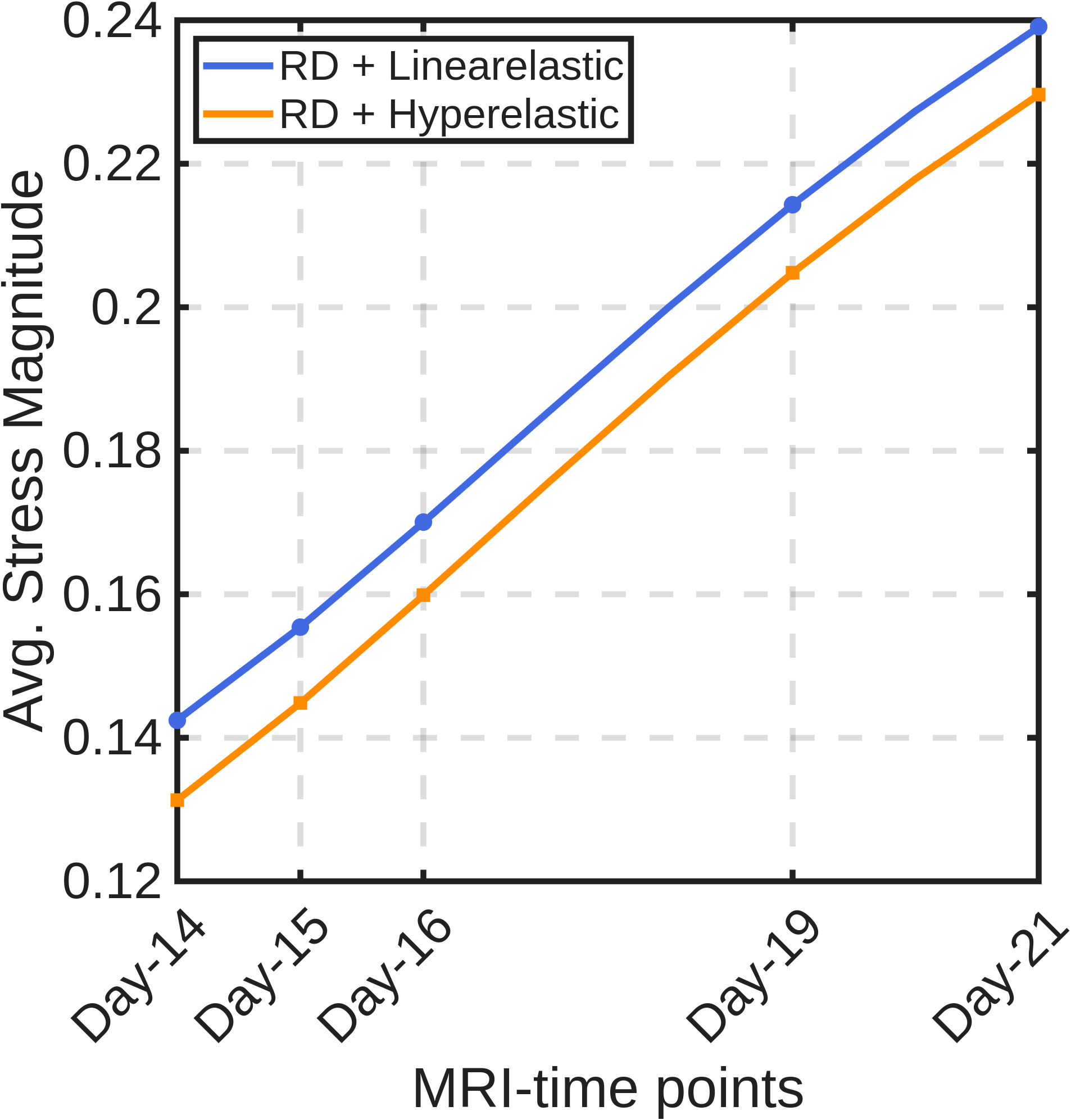}
    \vspace{-0.1in}
    \caption{\blue{
    Comparison of \textit{RD+Hyperelastic} and \textit{RD+Linearelastic} predictions for Rat~I in terms of normalized tumor area (NTA), domain-averaged displacement magnitude, and domain-averaged first Piola--Kirchhoff stress magnitude over time.
    }}
    \label{fig:numerical_value_ratI}
    \vspace{-0.2in}
\end{figure}

\begin{figure}[!ht]
\centering
\renewcommand{\arraystretch}{.6}
\setlength{\fboxsep}{0pt}
\setlength{\tabcolsep}{-3pt}
\setlength{\arrayrulewidth}{0.2pt}
\begin{tabular}{m{0.05\textwidth}*{4}{>{\centering\arraybackslash}m{0.21\textwidth}}m{0.1\textwidth}}

\makebox[0.05\textwidth][c]{\vrule width 0pt height 0.45cm} &
\colorbox{black}{\makebox[0.19\textwidth][c]{\vrule width 0pt height 0.4cm \textcolor{white}{\small Day 15}}} &
\colorbox{black}{\makebox[0.19\textwidth][c]{\vrule width 0pt height 0.4cm \textcolor{white}{\small Day 16}}} &
\colorbox{black}{\makebox[0.19\textwidth][c]{\vrule width 0pt height 0.4cm \textcolor{white}{\small Day 19}}} &
\colorbox{black}{\makebox[0.19\textwidth][c]{\vrule width 0pt height 0.4cm \textcolor{white}{\small Day 21}}} &
\makebox[0.1\textwidth][c]{\vrule width 0pt height 0.45cm} \\ [4pt] 

\centering\rotatebox{90}{
    \begin{tabular}{@{}c@{}} \textbf{\footnotesize RD+hyper} \\ \textbf{\footnotesize elastic} \end{tabular}
}
& \includegraphics[width=0.19\textwidth]{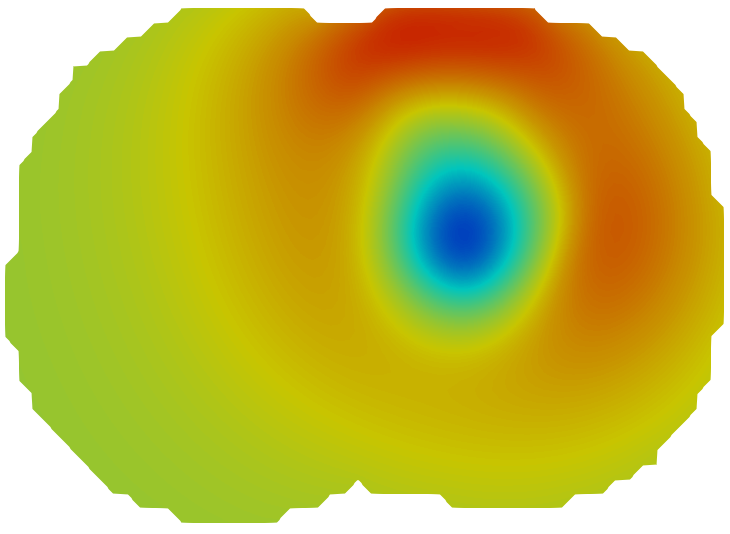} 
& \includegraphics[width=0.19\textwidth]{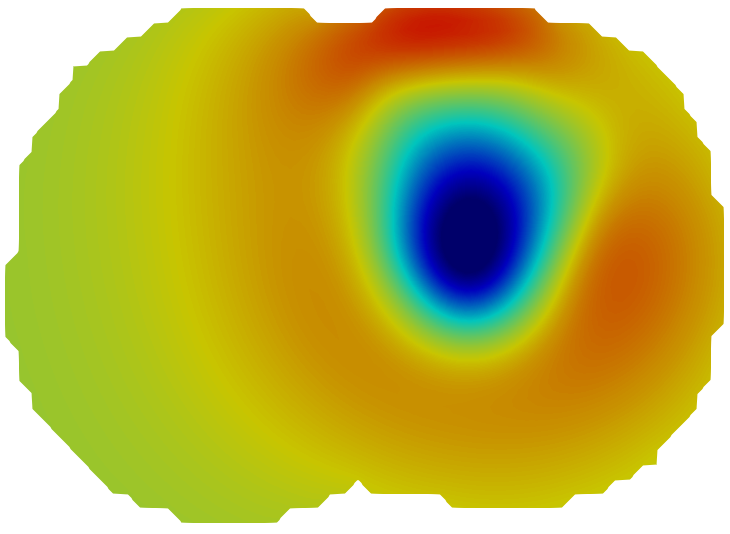} 
& \includegraphics[width=0.19\textwidth]{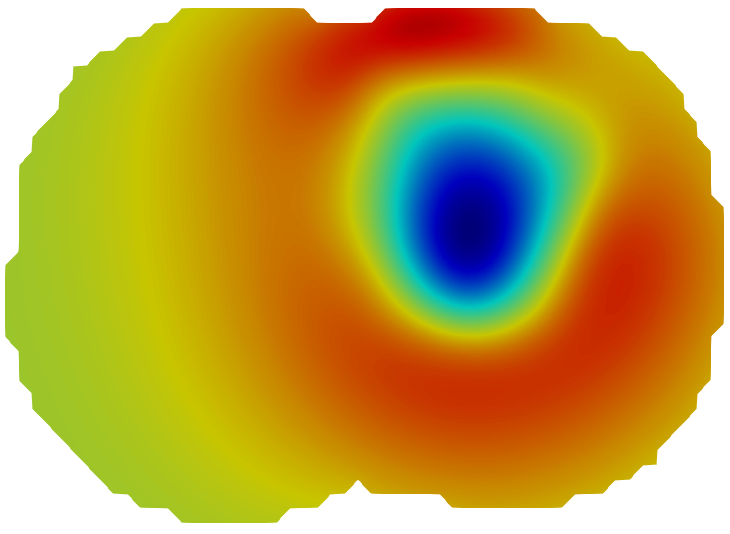} 
& \includegraphics[width=0.19\textwidth]{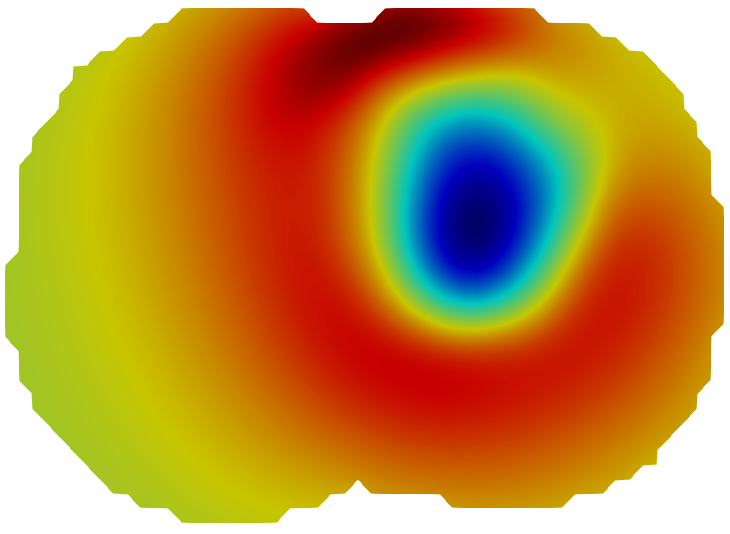} 
&
\multirow[t]{2}{*}{ 
    \raisebox{-3cm}{
        \includegraphics[width=1.0cm,height=3.6cm,keepaspectratio=false]{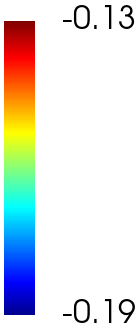} 
        \hspace{-30pt}
        \raisebox{28pt}{
            \rotatebox{90}{
                \begin{tabular}{@{}c@{}} \textbf{\tiny log(E)} \\[3pt] \textbf{\tiny [log(KPa)]} \end{tabular}}}
    }
}\\

\centering\rotatebox{90}{
    \begin{tabular}{@{}c@{}} \textbf{\footnotesize RD+linear} \\ \textbf{\footnotesize elastic} \end{tabular}
}
& \includegraphics[width=0.19\textwidth]{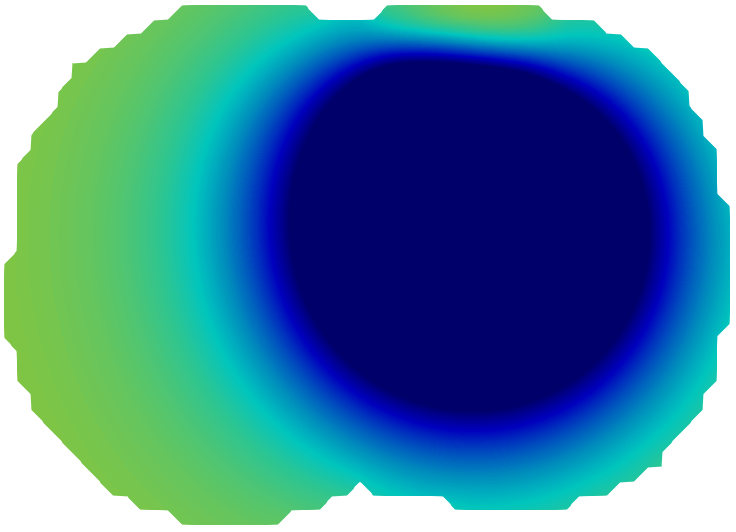} 
& \includegraphics[width=0.19\textwidth]{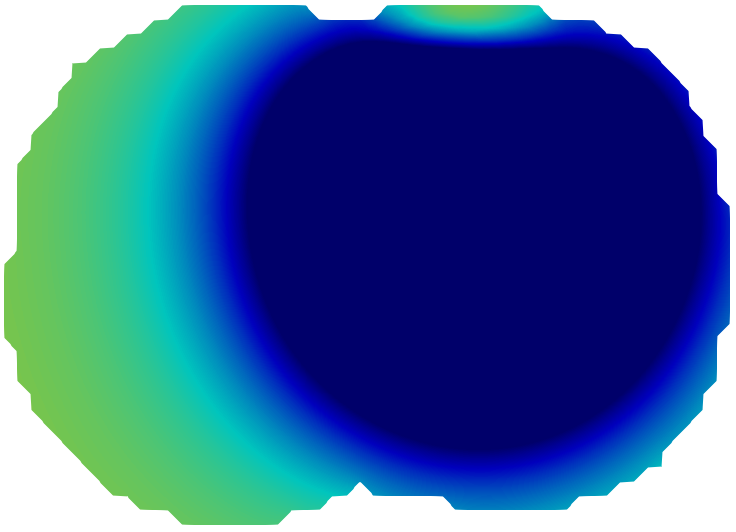} 
& \includegraphics[width=0.19\textwidth]{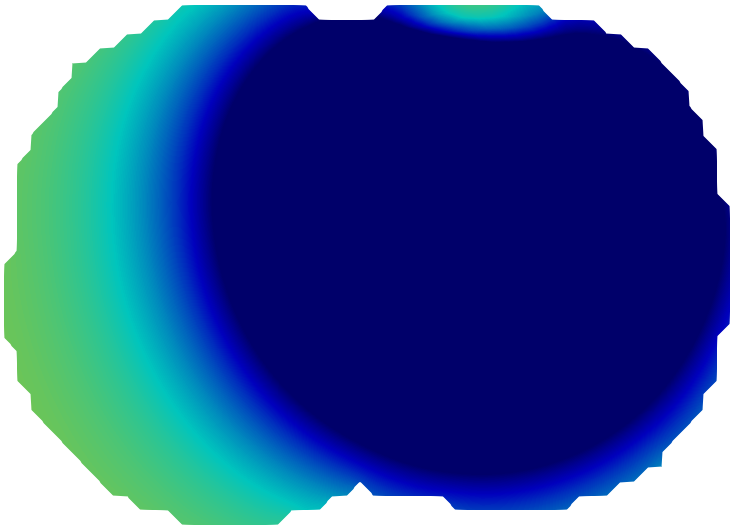} 
& \includegraphics[width=0.19\textwidth]{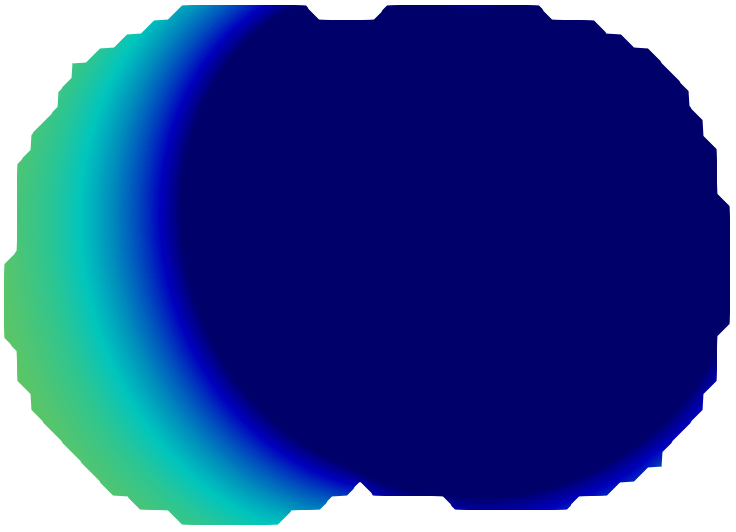} 
& \\ [10pt]

\centering\rotatebox{90}{
    \begin{tabular}{@{}c@{}} \textbf{\footnotesize RD+hyper} \\ \textbf{\footnotesize elastic} \end{tabular}
}
& \includegraphics[width=0.19\textwidth]{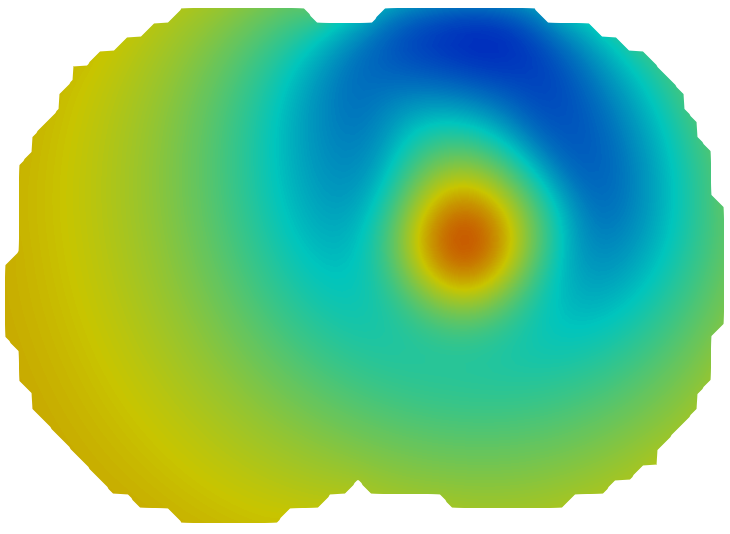} 
& \includegraphics[width=0.19\textwidth]{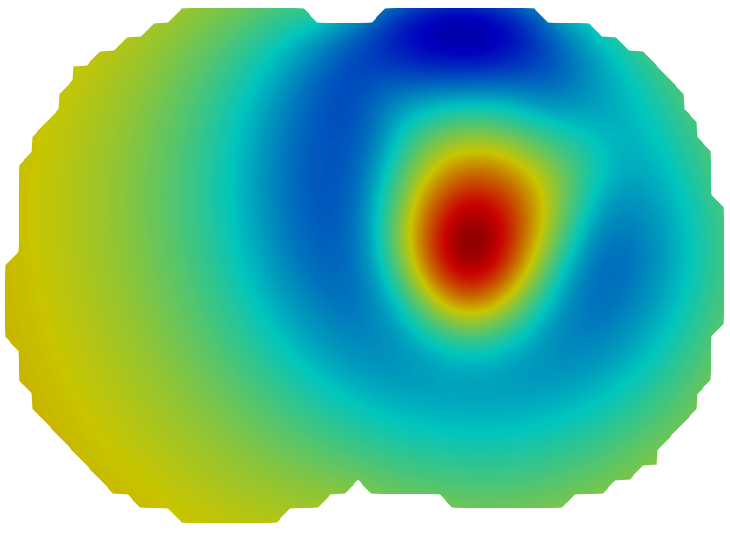} 
& \includegraphics[width=0.19\textwidth]{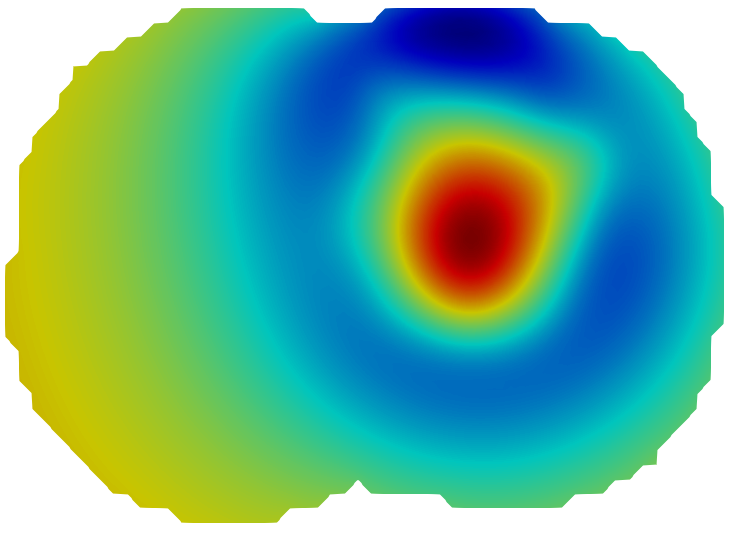} 
& \includegraphics[width=0.19\textwidth]{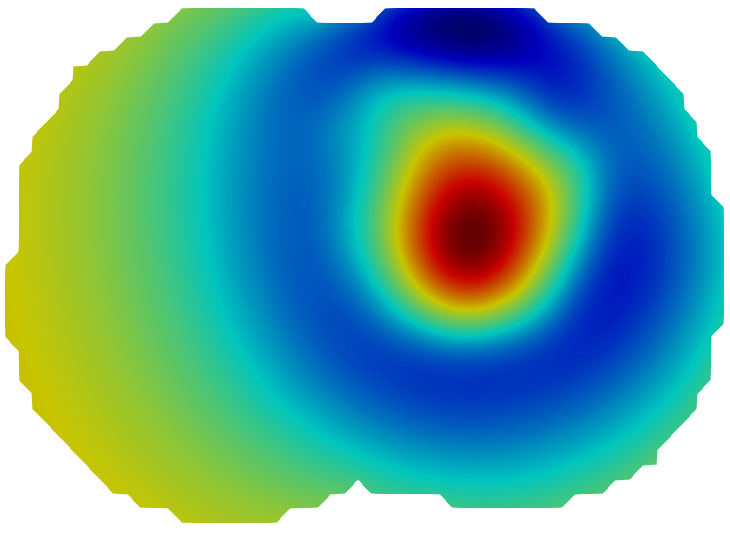} 
&
\multirow[t]{2}{*}{
    \raisebox{-3cm}{
        \includegraphics[width=1cm,height=3.6cm,keepaspectratio=false]{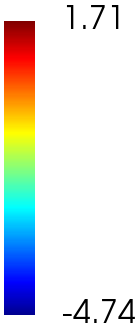} 
        \hspace{-30pt}
        \raisebox{18pt}{
            \rotatebox{90}{
                \begin{tabular}{@{}c@{}} \textbf{\tiny log(D)} \\[3pt] \textbf{\tiny [log(mm$^2$/day)]} \end{tabular}}}
    }
}\\

\centering\rotatebox{90}{
    \begin{tabular}{@{}c@{}} \textbf{\footnotesize RD+linear} \\ \textbf{\footnotesize elastic} \end{tabular}
}
& \includegraphics[width=0.19\textwidth]{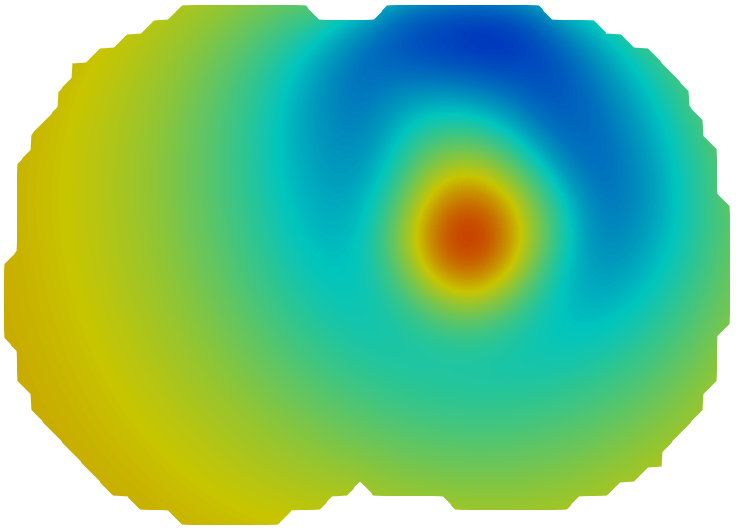}
& \includegraphics[width=0.19\textwidth]{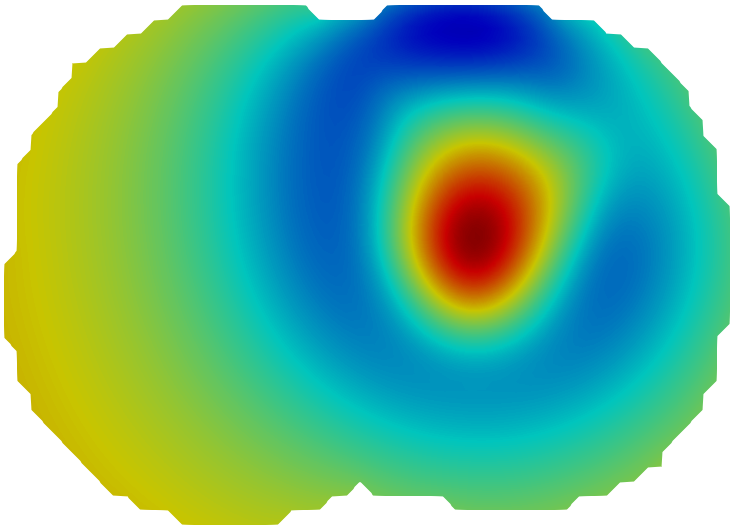}
& \includegraphics[width=0.19\textwidth]{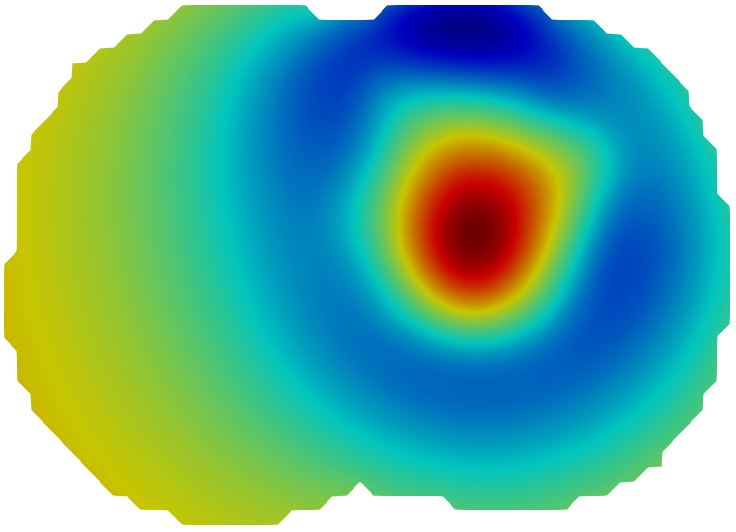}
& \includegraphics[width=0.19\textwidth]{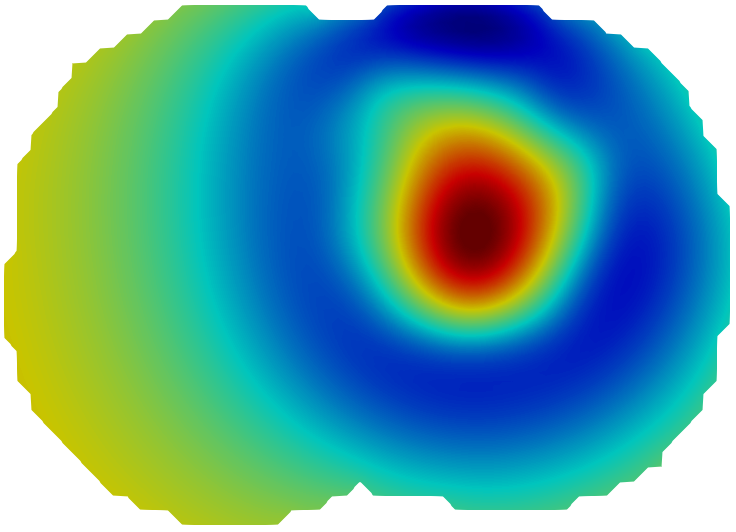}
& \\ [10pt]

\centering\rotatebox{90}{
    \begin{tabular}{@{}c@{}} \textbf{\footnotesize RD+hyper} \\ \textbf{\footnotesize elastic} \end{tabular}
}
& \includegraphics[width=0.19\textwidth]{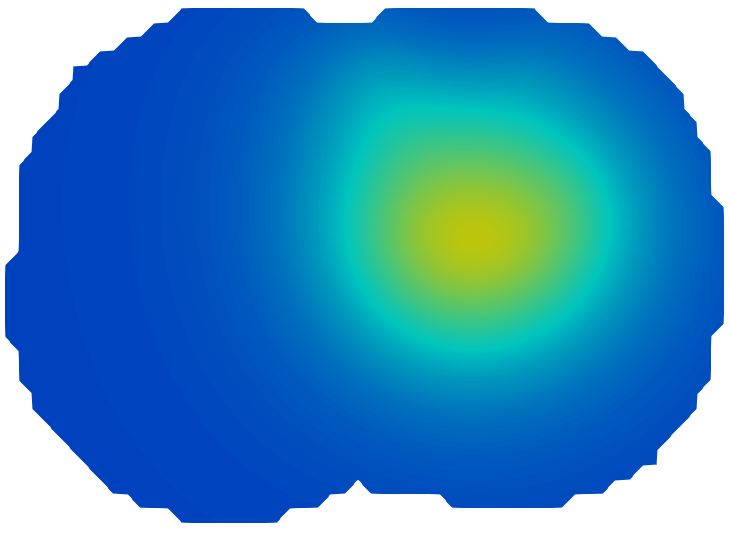} 
& \includegraphics[width=0.19\textwidth]{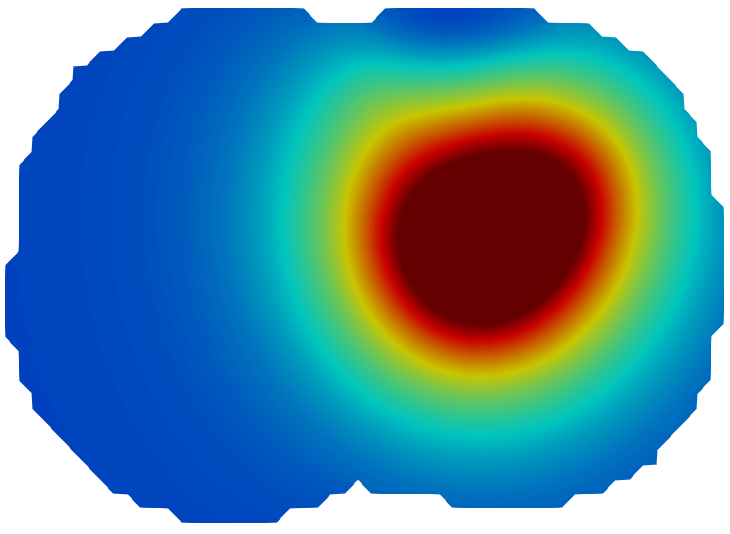} 
& \includegraphics[width=0.19\textwidth]{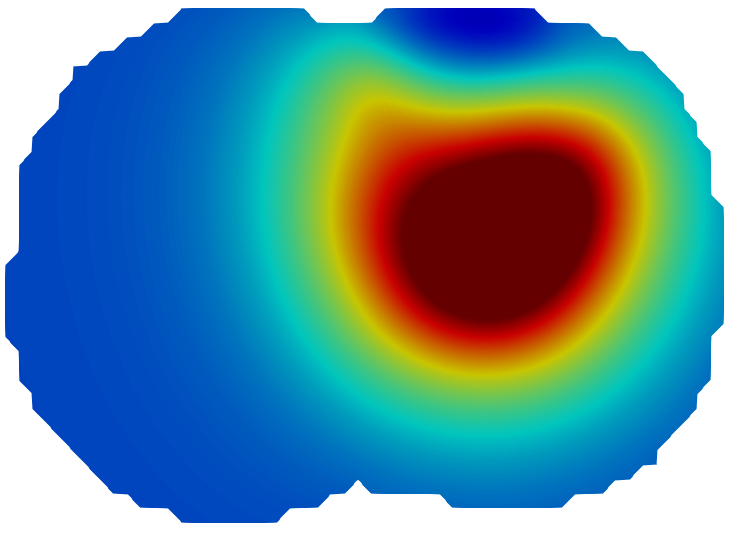} 
& \includegraphics[width=0.19\textwidth]{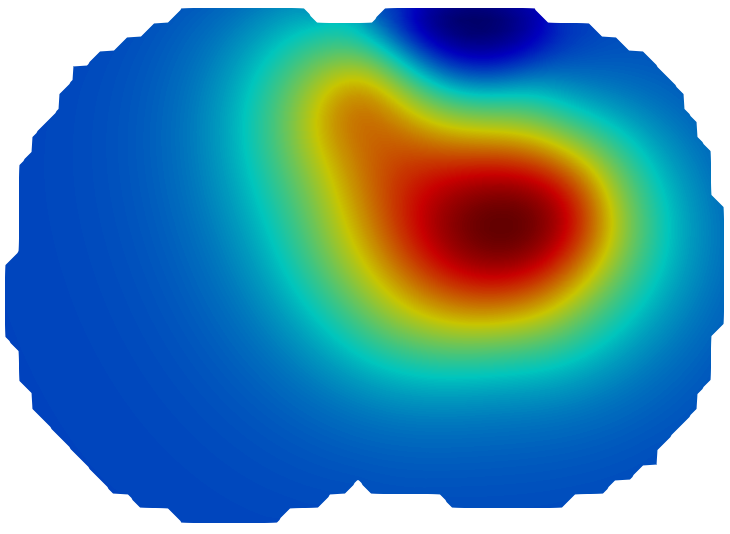} 
&
\multirow[t]{2}{*}{
    \raisebox{-3cm}{
        \includegraphics[width=1cm,height=3.6cm,keepaspectratio=false]{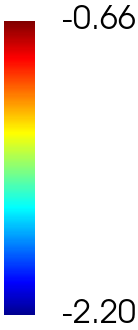} 
        \hspace{-30pt}
        \raisebox{23pt}{
            \rotatebox{90}{
                \begin{tabular}{@{}c@{}} \textbf{\tiny log(G)} \\[3pt] \textbf{\tiny [log(1/day)]} \end{tabular}}}
    }
}\\

\centering\rotatebox{90}{
    \begin{tabular}{@{}c@{}} \textbf{\footnotesize RD+linear} \\ \textbf{\footnotesize elastic} \end{tabular}
}
& \includegraphics[width=0.19\textwidth]{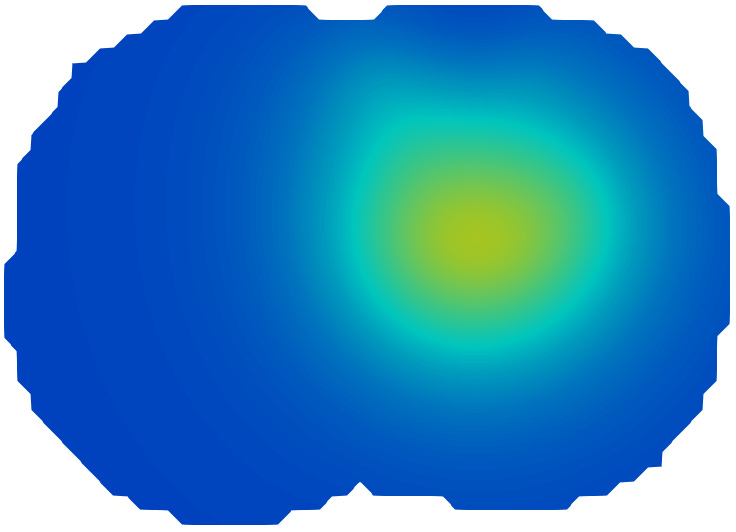}
& \includegraphics[width=0.19\textwidth]{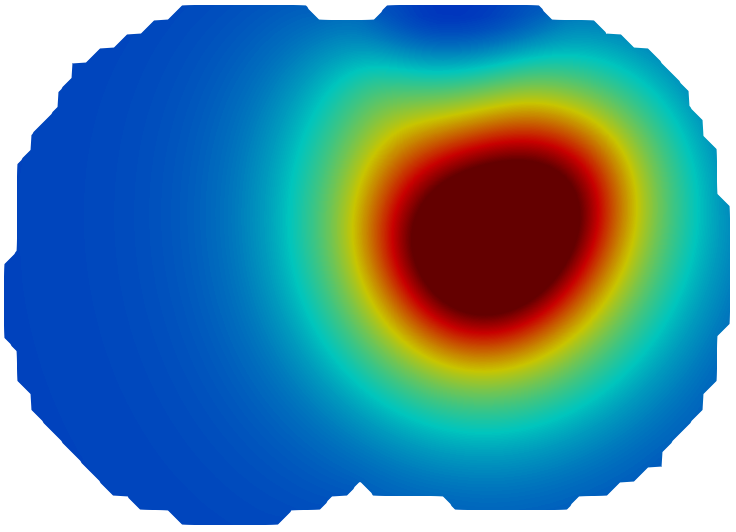}
& \includegraphics[width=0.19\textwidth]{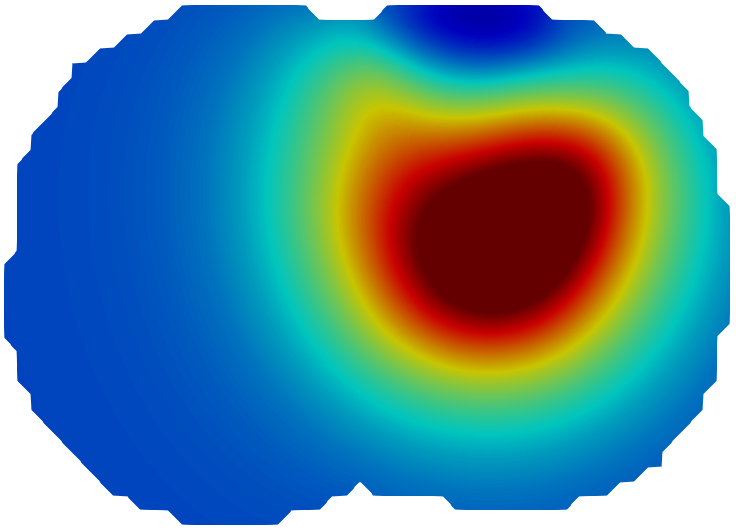}
& \includegraphics[width=0.19\textwidth]{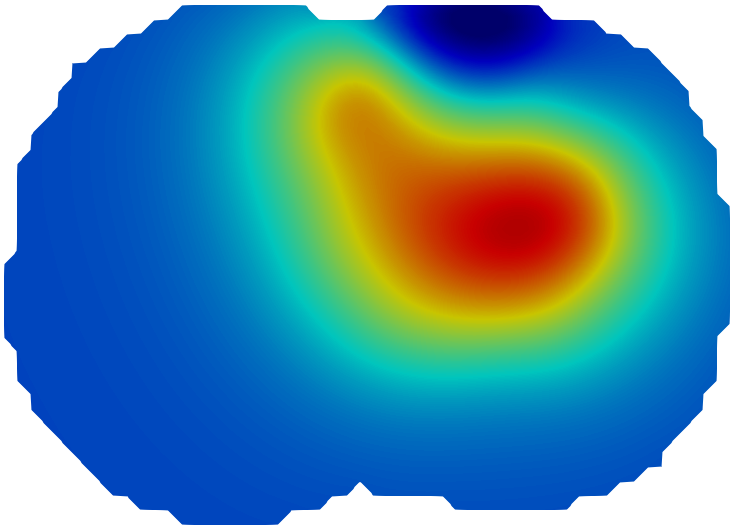}
& \\

\end{tabular}
\caption{Evolution of MAP estimates of spatially varying 
elastic modulus $E(\mathbf{X})$, tumor diffusivity $D(\mathbf{X})$, 
and proliferation rate $G(\mathbf{X})$ for Rat~I under 
\textit{RD+Hyperelastic} and \textit{RD+Linearelastic} models.}
\label{fig:params_hyperlinear_ratI}
\vspace{-0.1in}
\end{figure}

To directly connect model plausibility to biomechanics, Figures
\ref{fig:disp_stress_hyperlinear_ratI}--\ref{fig:params_hyperlinear_ratI} compare the \textit{RD+Linearelastic} and \textit{RD+Hyperelastic} predictions and inferred parameter fields for Rat~I. Although the predicted tumor volume fraction fields $\phi$ are visually similar between the two models, the displacement and stress fields differ and diverge further at later times
(Figure~\ref{fig:disp_stress_hyperlinear_ratI}), when \textit{RD+Hyperelastic} receives higher posterior plausibility (Figure~\ref{fig:plausibility_dice_rats}). 
This demonstrates that comparable tumor
shape agreement does not imply comparable biomechanics: constitutive assumptions
can produce similar tumor morphology while inducing substantially different
mechanical states. \blue{The quantitative comparison in Figure~\ref{fig:numerical_value_ratI} supports this interpretation that NTA evolves similarly for the two models, whereas the domain-averaged displacement and stress magnitudes show larger differences over time.}

The inferred parameter evolution clarifies the mechanical origin of the model
preference. While the growth kinetics parameters $\log(D)$ and $\log(G)$ remain
similar between \textit{RD+Linearelastic} and \textit{RD+Hyperelastic}, the
stiffness field $\log(E)$ differs markedly
(Figure ~\ref{fig:params_hyperlinear_ratI}). 
Under the small-strain linear elastic assumption, $\log(E)$ progressively shifts
toward a broadly softened state across the domain. 
\blue{This pattern suggests that, as mass effect increases, the linear elastic model compensates for missing finite deformation kinematics by reducing the inferred stiffness to match the observed tumor expansion. In contrast, the hyperelastic formulation accommodates finite strain kinematics directly and yields a more spatially localized stiffness pattern near the tumor. Thus, although both models can produce similar tumor morphology, their inferred stiffness fields and mechanical responses are not equivalent.}
This helps explain why posterior plausibility can favor the hyperelastic formulation even when Dice and NTA differences are small, as the evidence reflects biomechanical consistency,
not merely agreement of tumor shape.

Overall, the dynamic model selection results show that, \blue{in the present longitudinal murine dataset, mechanically coupled formulations are favored over the uncoupled reaction-diffusion tumor growth model, and the relative support for linear elastic versus hyperelastic mechanics evolves across subjects and imaging times.} These findings motivate the use of posterior model plausibility, \blue{together with} morphology-based metrics, as a subject-specific criterion for evaluating biomechanical model adequacy in MRI-driven computational tumor growth prediction.

\section{Discussion and Conclusions}\label{sec:conclusions}

In this study, we developed an MRI-driven framework for sequential Bayesian
inference of biomechanical tumor growth models \blue{within a high-dimensional spatial parameterization}. Longitudinal imaging data of murine models of glioma were assimilated to update subject-specific
fields of diffusivity, proliferation rate, and elastic modulus within a
spatially varying parameter space. To investigate the role of mechanics in tumor
evolution, we compared three competing tumor growth models: a reaction–diffusion model without mechanical coupling, a reaction–diffusion model coupled with linear elasticity,
and a reaction–diffusion model coupled with hyperelastic mechanics. Posterior model plausibility, \blue{based on model evidence,} was evaluated sequentially as new MRI data were incorporated, enabling \blue{subject-specific dynamic selection of the most plausible candidate model for one-scan-ahead prediction at each assimilation time}.
The results show that, \blue{within the present longitudinal murine dataset}, mechanically coupled models were \blue{generally} more plausible than the uncoupled RD model. This finding is consistent with prior evidence that solid stresses generated by tumor expansion under tissue confinement can regulate proliferation, apoptosis, and invasion through mechanobiological pathways \cite{helmlinger1997solid,stylianopoulos2013coevolution,jain2014role,faghihi2020coupled}. 
The posterior plausibility trends indicate that diffusion--proliferation alone is less supported by the MRI time series than formulations that include mass effect and growth--stress feedback. Among the mechanically coupled models, the relative support for linear elastic and hyperelastic mechanics evolved across subjects and imaging times, with the hyperelastic formulation often favored at later scans when tumor burden and deformation were larger. This trend suggests that finite-deformation effects can become important during progression, while linear elasticity may remain competitive when deformation is modest. Importantly, Dice score and NTA often did not distinguish between the two mechanically coupled models, even when their displacement, stress, and inferred stiffness fields differed. Thus, posterior model plausibility provides a mechanics-informed complement to morphology-based metrics for evaluating biomechanical model adequacy.

This study has several limitations. 
First, although consistent trends were observed across the analyzed animals, the cohort remains small, and larger longitudinal studies are needed before drawing population level conclusions about identified mechanical regimes. 
\blue{Second, the present analysis was conducted in a two dimensional (2D) based on MRI slices to enable repeated high-dimensional Bayesian inference, gpCN comparison, and model evidence evaluation across competing PDE models. The slice-selection and adjacent slice analyses reported in the \textit{Supplementary Information} support the robustness of the main trends, but fully three dimensional (3D) Bayesian model selection remains an important extension.} To this end, reliable reduced-order or neural-operator surrogate models may help make such studies computationally feasible while preserving biomechanical fidelity \cite{singh2024framework,islam2025stochastic}. 
\blue{Third, the tumor model is intentionally parsimonious and does not explicitly include necrosis, apoptosis, or additional cellular compartments. Extending the framework to richer multiphase tumor-growth models \cite{faghihi2020coupled,lucci2022coupling} and biology-informed hyperelastic constitutive laws \cite{goriely2015mechanics} will require additional imaging biomarkers or histological data to avoid introducing poorly identifiable states.} 
\blue{Fourth, not all biophysical parameters can be reliably inferred as spatially varying from MRI-derived tumor volume fraction alone. Mechanical quantities are only indirectly observed through model coupling, and the fixed parameters $C$, $H$, and $\nu$ were therefore prescribed rather than inferred. Sensitivity analyses for $C$ and $H$ are provided in the \textit{Supplementary Information} showed that the main model-plausibility trends were preserved.} Future work should incorporate MRI-derived estimates of mass effect and deformation, as in \cite{tuncc2021modeling}, to more directly constrain displacement and stress fields and should explicitly propagate segmentation, registration, and ADC-processing uncertainty. 
\blue{Finally, the conclusions regarding mechanical coupling and constitutive plausibility are restricted to the observed 10--21 day progression window, and extrapolation beyond this period will require additional longitudinal imaging data.}

\blue{
The proposed sequential inference and model-selection framework provides a natural foundation for patient-specific biomechanical cancer digital twins, in the spirit of recent efforts \cite{chaudhuri2023predictive,pash2026predictive,kapteyn2026tumortwin}. 
In such a setting, clinical MRI scans would be assimilated over time to update not only model parameters, but also the plausibility of competing growth, treatment response, or post surgical progression models. This is particularly important under therapy, where adaptive treatment decisions can alter the observed tumor dynamics. One-scan-ahead prediction as demonstrated in this work would provide a clinically relevant short horizon forecast before the next imaging or treatment decision, while dynamic model selection allows the prediction model to adapt to the subject specific response under different treatment regimes. Coupled with risk-averse optimization methods developed in our prior work \cite{tan2024scalable,singh2025chance}, this framework could support patient specific ``what if'' simulations that balance tumor control against treatment induced toxicity. Realizing this goal will require prospective clinical validation and computational acceleration, for example through credible surrogate modeling strategies \cite{singh2024framework}, because repeated Bayesian inference and model evidence evaluation across multiple PDE models may be too costly for clinical decision timelines.
}

\blue{
In summary, this work introduces an MRI-informed framework for sequential, subject-specific evaluation of biomechanical tumor growth models. In the present murine glioma dataset, mechanically coupled formulations were favored over the uncoupled RD model, while the relative support for linear elastic versus hyperelastic mechanics changed across animals and imaging times as new scans were assimilated. These findings show the value of updating both spatial parameter fields and model plausibility from longitudinal imaging data, rather than prescribing a single biomechanical formulation a priori.
}

\section*{Acknowledgments}
\noindent
AAN was partially supported by the State University of New York (SUNY) System Administration under SUNY Research Seed Grant Award No.~1191358. 
PKS and DF acknowledge support from the U.S. National Science Foundation (NSF) through CAREER Award CMMI-2143662. 
DAH acknowledge support from the NSF through the award DMS-2436499.
The authors gratefully acknowledge computational support provided by the Center for Computational Research at the University at Buffalo.

\section*{Author contributions}
Danial Faghihi: Conceptualization, methodology, theoretical development, supervision, writing original draft, review and editing.
Abdullah Al Noman: Software development, formal analysis, implementation of computational framework, data analysis, visualization, writing results.
Pratyush Kumar Singh: Software support, code development, validation, and editing.
David A. Hormuth, II: Data curation (MRI data), resources, biological interpretation of results, review and editing.

All authors contributed to the discussion of the results and approved the final manuscript.

\section*{Ethics approval}
All procedures involving animals were approved by the appropriate Institutional Animal Care and Use Committee (IACUC) at University of Texas at Austin and were performed in accordance with institutional guidelines and applicable national regulations for the care and use of laboratory animals.

\section*{Consent to participate}
Not applicable.

\section*{Data availability}
The data supporting the findings of this study are available from the corresponding author upon reasonable request. The computational code used in this study is available at 
\url{https://github.com/PCELab/RecursiveGliomaInference}.

\section*{Competing interests}
The authors declare that they have no competing interests.



 \bibliographystyle{elsarticle-num} 
 \bibliography{references}

\end{document}